\documentclass[aps,nofootinbib, superscriptaddress, prx, twocolumn]{revtex4-2}
\usepackage{ORI_Group_style}
\usepackage{blindtext}
\usepackage[dvipsnames]{xcolor}
\usepackage{array}
\usepackage{subfigure}
\usepackage{float}
\usepackage{enumitem}
\usepackage{titlesec}
\usepackage[scr=boondoxo]{mathalfa}

\usepackage{array}
\newcolumntype{P}[1]{>{\centering\arraybackslash}p{#1}}
\newcolumntype{M}[1]{>{\centering\arraybackslash}m{#1}}
\newcommand*{\tensorbar}[1]{\overline{\overline{#1}}}
\newcommand*{\TOCentry}[1]{\hypersetup{linkcolor=black}{\nameref{#1}} 
\let\cleaders\leaders\dotfill
\hypersetup{linkcolor=red}{\hyperlink{#1}{\pageref{#1}}}}

\hypersetup{colorlinks,linkcolor=red,citecolor=blue,urlcolor=blue}

\begin{document}

\title{Towards a quantum interface between spin waves and paramagnetic spin baths}

\author{C. Gonzalez-Ballestero}
\email{c.gonzalez-ballestero@uibk.ac.at}
\affiliation{Institute for Quantum Optics and Quantum Information of the Austrian Academy of Sciences, 6020 Innsbruck, Austria.}
\affiliation{Institute for Theoretical Physics, University of Innsbruck, A-6020 Innsbruck, Austria.}

\author{Toeno van der Sar}
\affiliation{Department of Quantum Nanoscience, Kavli Institute of Nanoscience, Delft University of Technology, Lorentzweg 1, 2628 CJ, Delft, The Netherlands}

\author{O. Romero-Isart}
\affiliation{Institute for Quantum Optics and Quantum Information of the Austrian Academy of Sciences, 6020 Innsbruck, Austria.}
\affiliation{Institute for Theoretical Physics, University of Innsbruck, A-6020 Innsbruck, Austria.}

\begin{abstract}
Spin waves have risen as promising candidate information carriers for the next generation of information technologies. Recent experimental demonstrations of their detection using electron spins in diamond pave the way towards studying the back-action of a controllable paramagnetic spin bath on the spin waves. Here, we present a quantum theory describing the interaction between spin waves and paramagnetic spins. As a case study we consider an ensemble of nitrogen-vacancy spins in diamond in the vicinity of an Yttrium-Iron-Garnet thin film. We show how the back-action of the ensemble results in strong and tuneable modifications of the spin-wave spectrum and propagation properties. These modifications include the full suppression of spin-wave propagation and, in a different parameter regime, the enhancement of their propagation length by $\sim 50\%$. Furthermore, we show how the spin wave thermal fluctuations induce a measurable frequency shift of the paramagnetic spins in the bath. This shift results in a thermal dispersion force that can be measured optically and/or mechanically with a diamond mechanical resonator. In addition, we use our theory to compute the spin wave-mediated interaction between the spins in the bath. We show that all the above effects are measurable by state-of-the-art experiments. Our results provide the theoretical foundation for describing hybrid quantum systems of spin waves and spin baths, and establish the potential of quantum spins as active control, sensing, and interfacing tools for spintronics.
\end{abstract}


\maketitle

\section{Introduction}

In the last years spin waves have become the focus of intense research because of the following reasons. First, spin waves can be integrated into optical, microwave, and acoustic technological platforms~\cite{TabuchiPRL2014,ZhangPRL2014,HaighPRL2016,ZhangSciAdv2016,KusminskiyPRA2016,GonzalezBallesteroPRL2020,HueblPRL2013,LiNatComm2016,TabuchiScience2015,MarcosPRL2010,ZhuNature2011}. Second,  spin waves act as carriers for next-generation information processing in the field of magnon spintronics~\cite{LenkPhysRep2011,KruglyakJPhysD2010,ChumakNatPhys2015,StampsJPhysD2014}. This is due to their low loss as compared to electronic currents, especially in ferromagnetic insulators (e.g. YIG -- Yttrium-Iron-Garnet)~\cite{ChumakNatPhys2015,SergaJPhysD2010,ChumakNatComm2014}. Third, spin waves and their quanta (magnons) display a rich and tuneable phenomenology including exotic dispersion relations, non-reciprocity~\cite{ChumakNatPhys2015,Gurevich1996magnetization,StancilBook2009}, room temperature Bose-Einstein condensation~\cite{NakataPRB2014} and superfluidity~\cite{TakeiPRL2014}. These properties make spin waves very attractive in the context of hybrid quantum technologies~\cite{ChumakNatComm2014,LachanceQuirionAPE2019,StampsJPhysD2014}.

\begin{figure*}[tbh!]
	\centering
	\includegraphics[width=\linewidth]{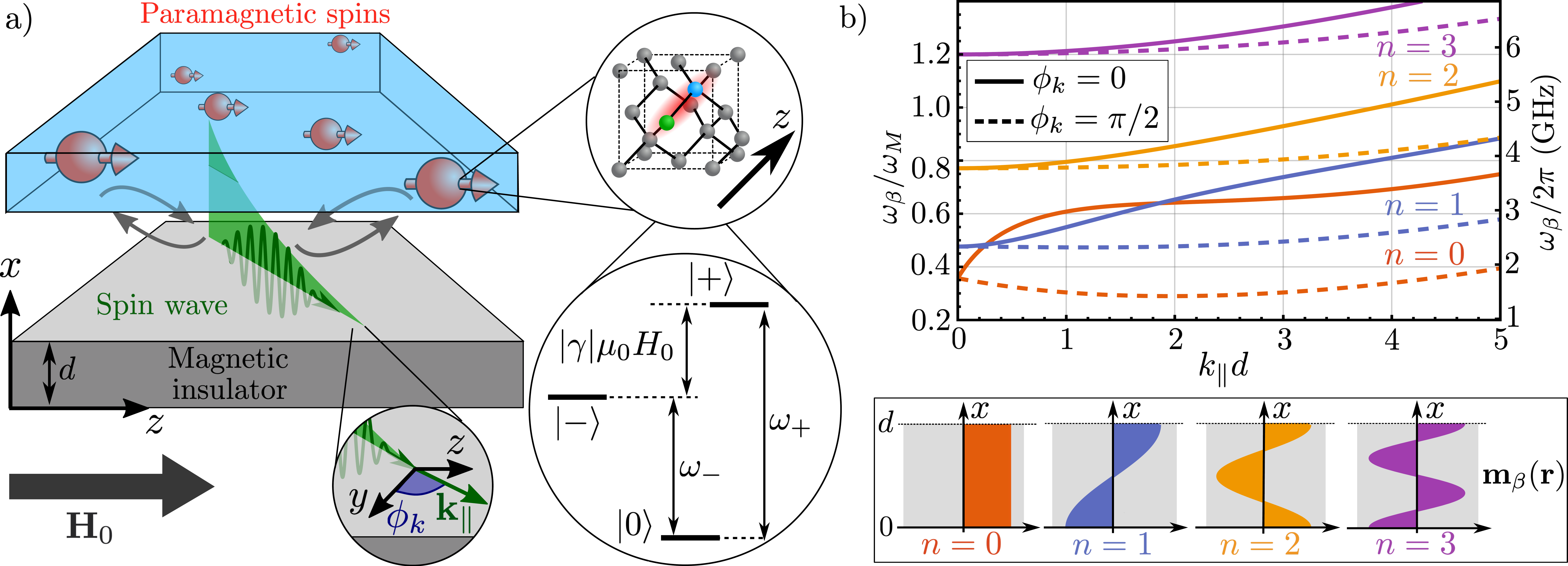}
	\caption{a) We consider paramagnetic spins with spin $S=1$ in the vicinity of a magnetic insulator supporting spin waves. As a case study we focus on NV centres (upper inset) near a YIG thin film. The wavevector of the spin waves in the $y-z$ plane is $\mathbf{k}_\parallel$. 
		b) Lowest four spin wave energy bands of a YIG film for the parameters in Table~\ref{tablePARAMS} and an applied field $\mu_0H_0 = 20$mT as a function of wavevector modulus, for $\phi_k=0$ (solid lines, Damon-Eshbach propagation) and $\phi_k=\pi/2$ (dashed lines, parallel propagation). The left and right axes show the spin wave frequency in units of $\omega_M\equiv\vert\gamma\vert\mu_0 M_S$ and in GHz, respectively. As indicated by the bottom panel, the band index $n$ corresponds to the number of nodes of the magnetization mode function across the film thickness $d$.}\label{FigureSystemBandStructure}
\end{figure*}

In addition, recent experiments~\cite{GieselerPRL2020,BertelliSciAdv2020,LeeWongNanoLetters2020,WolfePRB2014,AndrichNPJ2017,PageJAP2019,KikuchiAPExpress2017,ZhangPRB2020} have demonstrated the possibility of interfacing spin waves with solid-state paramagnetic spin baths. Inspired by this possibility, in this article we propose to explore the quantum phenomena stemming from the interaction between fields and quantum emitters and its potential applications in the field of spintronics. In particular, we will focus on the strong back-action exerted by ensembles of quantum emitters on fields and viceversa. Such backaction has been demonstrated in many platforms in the optical domain, including dye molecules in plasmonics~\cite{RodriguezPRL2013}, rare earth impurities in optical fibers~\cite{GehringScience2006} and cold and Rydberg atoms in free space~\cite{KangPRA2004,PayronelNature2012,MurrayInColl2016}. Harnessing a similar back-action for spin waves could, for instance, allow to dynamically mold the flow of spin currents without the need for material microstructuring in magnon spintronics. 
Furthermore and in analogy to plasmonics~\cite{CaoNanophotonics2018}, a fundamental understanding of spin wave-paramagnetic spin interfaces could bring a new degree of control to spin wave-based hybrid technologies and enable applications in sensing \cite{AndrichNPJ2017,GieselerPRL2020,CasolaNatReviews2018}, transduction \cite{AuAPL2012,ZhangPRB2020}, and
computing~\cite{KhitunJAP2011,ChumakNatPhys2015,ChumakNatComm2014,VogtNatComm2014}.

In this article we provide a quantum theory describing spin waves interacting with paramagnetic spin baths. We focus on paramagnetic spins with total spin $S=1$. As a case study (\figref{FigureSystemBandStructure}[a]), we will consider nitrogen-vacancy (NV) centres in diamond in the vicinity of a YIG thin film. The choice of NVs is motivated by current experiments and their controllability~\cite{DohertyPhysRep2013} and potential for hybrid quantum technologies~\cite{AharonovichNatPhot2011,AtatureNatRevMat2018,AlbrechtPRL2013,MazeNature2008,DoldeNatPhys2011,GieselerPRL2020,LaraouiNatComm2015,GraziosoAPL2013,ShaoPRApp2016}. We apply our theory to describe mutual back-action between the paramagnetic spin bath and the spin waves, and highlight two particular applications of relevance for magnon spintronics. First, the tailoring of the spin wave propagation properties via their controlled coupling to the paramagnetic spin bath. Second, a new method to probe spin waves based on the back-action of spin waves on the paramagnetic spins.  Our theory allows us to understand and predict exhaustively the interaction between spin waves and paramagnetic spins, both at the classical and at the quantum level. Our predictions can be tested with state-of-the-art experiments.

This article is organized as follows. First, we describe how we model the interaction of spin waves and a paramagnetic spin bath and summarize its coherent and dissipative dynamics in Sec.~\ref{SecSystem}. We then derive the effective spin wave dynamics induced by back-action of the paramagnetic spin bath in Sec.~\ref{SecSWmodification} and discuss the back-action-enabled possibility of modifying the spin-wave spectrum and propagation features. In Sec.~\ref{SecCasimirPolder} we derive the effective dynamics of a paramagnetic spin under back-action of the spin waves and its potential as a novel probing method for spin waves. Our conclusions and outlook are in Sec.~\ref{SecConclusion}. This article is complemented by six exhaustive appendices including derivations and additional results. The quantization of the spin wave eigenmodes of a thin film, a detailed analysis of their properties, and the computation of the magnetic field power spectral densities outside the film are contained in Appendix~\ref{appendixSpinWaves}. Appendix~\ref{AppendixNVcentres} contains the analysis of the quantum dynamics of NV centres both at thermal equilibrium and under optical pumping. The derivation of the interaction between paramagnetic spins and spin waves is contained in Appendix~\ref{AppendixInteraction}. In Appendix~\ref{Appendix_TracingOut} we summarize and give a practical formulation of the open quantum system approach to obtain effective equations of motion for a system coupled to a bath. We apply this procedure in the last two appendices: in Appendix~\ref{Appendix_EffectiveNVdynamics} we derive the spin wave-induced effective dynamics of an ensemble of paramagnetic spins, and analyze in detail the modification of their lifetimes, the induced frequency shifts and corresponding forces, and the induced coupling between different spins. In Appendix~\ref{Appendix_effectiveSWdynamics} we derive the effective spin wave dynamics induced by a paramagnetic spin bath and study the modification of the magnetic field power spectral densities outside the film.

\section{System and equations of motion}\label{SecSystem}

In this section we summarize the Hamiltonian and the dissipative dynamics governing the coupled system formed by paramagnetic spins and spin waves. 
The density matrix of the total system obeys the von Neumann equation
\begin{equation}\label{vonNeumanntotal}
\frac{d}{dt}\hat{\rho} = -\frac{i}{\hbar}\left[\hat{H},\hat{\rho}\right] + \mathcal{D}_{\rm sw}\left[\hat{\rho}\right]+\mathcal{D}_{\rm ps}\left[\hat{\rho}\right].
\end{equation}
Here, the total Hamiltonian is written as a sum of three contributions,
\begin{equation}\label{Htotal}
    \hat{H} = \hat{H}_{\rm sw} + \hat{H}_{\rm ps} + \hat{V},
\end{equation}
namely the free Hamiltonian of the spin waves, the free Hamiltonian of the paramagnetic spins, and their interaction. The last two terms in Eq.~\eqref{vonNeumanntotal} correspond to the independent dissipation of the spin waves and the paramagnetic spins. The detailed derivation of all the above terms is given in Appendices~\ref{appendixSpinWaves}, \ref{AppendixNVcentres}, and \ref{AppendixInteraction}. Let us summarize each of the contributions in Eq.~\eqref{vonNeumanntotal} separately. 

We first focus on the spin waves, namely magnetization waves supported by a saturated ferromagnetic insulator. In the presence of a static field  $\mathbf{H}_0 = H_0\mathbf{e}_z$, a ferromagnetic insulator of arbitrary geometry acquires a magnetization field given by $\mathbf{M}(\mathbf{r},t)= M_S\mathbf{e}_z+ \mathbf{m}(\mathbf{r},t)$, with $M_S$ the saturation magnetization. The dynamical component $\mathbf{m}(\mathbf{r},t)$
describes small oscillations $(\vert\mathbf{m}(\mathbf{r},t)\vert \ll M_S)$ above such fully magnetized state, namely spin waves~\cite{StancilBook2009}. The quantum Hamiltonian of the spin waves, namely the first contribution in Eq.~\eqref{Htotal}, reads
\begin{equation}\label{Hsw}
    \hat{H}_{\rm sw} = \hbar\sum_\beta \omega_\beta \hat{s}^\dagger_\beta\hat{s}_\beta.
\end{equation}
Here, $\beta$ is a multi-index labelling all the spin wave eigenmodes supported by the ferromagnetic structure, $\omega_\beta$ the corresponding mode frequency, and $\hat{s}_\beta^\dagger$ and $\hat{s}_\beta$ are bosonic ladder operators which describe creation and annihilation of spin wave quanta (magnons) in eigenmode $\beta$.

Hereafter we focus on the particular magnetic structure depicted in \figref{FigureSystemBandStructure}(a), i.e. a YIG thin film infinitely extended on the $y-z$ plane and occupying the region $0\le x\le d$. The static magnetic field $\mathbf{H}_0$ is applied along a direction parallel to the film, a configuration chosen in most experiments as it gives rise to rich spin wave dynamics~\cite{BertelliSciAdv2020,LeeWongNanoLetters2020,AndrichNPJ2017,WolfePRB2014,PageJAP2019,KikuchiAPExpress2017}. 
As shown in Appendix~\ref{appendixSpinWaves}, to derive $\hat{H}_{\rm sw}$ we first diagonalize the classical equations of motion for the spin waves, namely the linearized Landau-Lifshitz equations in the magnetostatic approximation~\cite{aharoni2000introduction,StancilBook2009}. In this way we obtain the corresponding  eigenfrequencies $\omega_\beta$ and dimensionless magnetization eigenmodes $\mathbf{m}_\beta(\mathbf{r})$ analytically. We follow the approach in Refs.~\cite{KalinikosSoviet1981,KalinikosJPHYSC1986}, where the exchange interaction is fully accounted for whereas the dipole-dipole interaction is included to first order in perturbation theory. 

For the YIG film geometry under study the spin wave eigenmodes and eigenfrequencies are fully characterized by five parameters, namely the gyromagnetic ratio $\gamma$, the film thickness $d$, the applied field $H_0$, the exchange stiffness $\alpha_x$, and the saturation magnetization $M_S$ or, equivalently, the natural frequency $\omega_M\equiv\vert\gamma\vert\mu_0M_S$~\cite{StancilBook2009}. 
The  eigenmodes are labelled by three mode indices $\beta \equiv \{\mathbf{k}_\parallel,n \}$, namely the wavevector on the film plane $\mathbf{k}_\parallel = k_\parallel\left[\mathbf{e}_y\cos(\phi_k) + \mathbf{e}_z\sin(\phi_k)\right]$ plus a discrete band index $n=0,1,2...$ indicating the number of nodes of the magnetization mode function across the film thickness, see \figref{FigureSystemBandStructure}(b). 
Among the many properties of the eigenmodes, three are of special interest regarding the interaction with paramagnetic spins: (i) The polarization of the magnetic field generated by a spin wave outside the film depends strongly on its propagation direction. As an example, a spin wave propagating in the Damon-Eshbach configuration~\cite{Kalinikos1994,PattonPHYSREP1984}, i.e. with wavevector $\mathbf{k}_\parallel = \pm k_\parallel\mathbf{e}_y$ or, equivalently, $\phi_k = 0$ or $\phi_k =\pi$, produces a circularly polarized field with polarization $\mathbf{e}_\mp \equiv(\mathbf{e}_x\mp i\mathbf{e}_y)/\sqrt{2}$ above the film and with the opposite polarization below it. 
(ii) The amplitude of the spin wave magnetic field can have different values above and below the film and, at certain values of $\mathbf{k}_\parallel$, can even completely vanish at one of the sides, a phenomenon known as \emph{modal-profile non-reciprocity}~\cite{KalinikosJPHYSC1986,KostylevJAP2013}. (iii) Outside the film, the magnetic field amplitude of a spin wave decays exponentially as $\exp(-k_\parallel l)$, with 
$l$ the absolute vertical separation from the surface of the film.

Once the classical eigenmodes have been obtained and characterized we quantize them~\cite{Mills2006}
to obtain both the Hamiltonian Eq.~\eqref{Hsw} and the spin wave magnetization operator,
\begin{equation}\label{moperator}
    \hat{\mathbf{m}}(\mathbf{r}) = \sum_\beta \mathcal{M}_{0\beta}\left[\mathbf{m}_\beta(\mathbf{r})\hat{s}_\beta + \text{H.c.}\right],
\end{equation}
where 
\begin{equation}\label{zeropointM}
    \mathcal{M}_{0\beta} \equiv \sqrt{\frac{\hbar\vert\gamma\vert M_S}{2L^2d }\frac{\omega_M}{\omega_\beta}}
\end{equation}
is the zero-point magnetization~\cite{Mills2006,GonzalezBallesteroPRB2020,GonzalezBallesteroPRL2020}, and $L\to\infty$ is a quantization length~\footnote{Note that, in analogy to the mode volume in quantum optics~\cite{CarmichaelBook}, no physical observable will depend on the quantization length $L$.}.

We now focus on the Hamiltonian of a single paramagnetic spin at a position $\mathbf{r}_0$ outside the magnetic structure and whose symmetry axis is oriented parallel to the $z-$axis, as depicted in \figref{FigureSystemBandStructure}(a). 
We describe the paramagnetic spin through the three states $\vert 0\rangle$, $\vert +\rangle$, and $\vert -\rangle$ corresponding to the eigenstates of the spin operator $\hat{S}_z$ with eigenvalue $m_S = 0,+\hbar$, and $-\hbar$, respectively. The Hamiltonian of the paramagnetic spin is~\cite{BarGillNatComm2013,AjisakaPRB2016,WangNJP2014,MarcosPRL2010,ZhuNature2011}
\begin{equation}\label{HNV}
    \hat{H}_{\rm ps} = \hbar^{-1}D_0\hat{S}_z^2+\omega_H\hat{S}_z
    \\=\hbar \sum_{\alpha=\pm}\omega_{\alpha} \hat{\sigma}_{\alpha\alpha},
\end{equation}
where we define the transition matrices $\hat{\sigma}_{\alpha\alpha'} \equiv \vert \alpha \rangle\langle \alpha'\vert$, and the frequencies $\omega_\pm \equiv D_0\pm\omega_H$ and $\omega_H \equiv \vert\gamma_s\vert\mu_0H_0$ with $\gamma_s$ the gyromagnetic ratio of the spin and $\mu_0$ the vacuum permeability.
The first term in Eq.~\eqref{HNV} describes the zero-field splitting between the $m_S=0$ and the $m_S=\pm 1$ states, quantified by a rate $D_0$. This splitting, absent in paramagnetic spins with $S=1/2$, is crucial for tuning spin waves in resonance with the transitions of the paramagnetic spin.
The second term in Eq.~\eqref{HNV}, proportional to $\omega_H$, corresponds to the Zeeman splitting induced by the applied field $H_0$ between the levels $\vert \pm \rangle$, which are degenerate at zero field. 
The Hamiltonian Eq.~\eqref{HNV} describes the dynamics of, among others, the ground-state manifold of negatively charged NV centres~\cite{BarGillNatComm2013,AjisakaPRB2016,WangNJP2014,MarcosPRL2010,ZhuNature2011}, with parameters given by Table~\ref{tablePARAMS}.

The third contribution in Eq.~\eqref{Htotal}, namely the interaction between the spin waves and the single paramagnetic spin at position $\mathbf{r}_0$, stems from the magnetic dipole interaction:
\begin{multline}\label{Vgeneral}
    \hat{V} = -\mu_0\hat{\boldsymbol{\mu}}_{\rm ps}\cdot\left[\hat{\mathbf{H}}(\mathbf{r}_0)-\mathbf{H}_0\right] 
    \\
    = \hbar\sum_\beta\left(g_\beta\hat{s}_\beta\hat{\sigma}_{-0} + \text{H.c.}\right) + \hat{S}_z\sum_{\beta \beta'}\tilde{g}_{\beta\beta'}\hat{s}_{\beta}^\dagger \hat{s}_{\beta'},
\end{multline}
with $\hat{\boldsymbol{\mu}}_{\rm ps} = -\vert\gamma_s\vert \hat{\mathbf{S}}$ the magnetic dipole moment of the paramagnetic spin and $\hat{\mathbf{S}}$ its total spin operator. 
The field generated by the spin waves, $\hat{\mathbf{H}}(\mathbf{r}_0)-\mathbf{H}_0$, is given by the total magnetic field operator minus the applied field $\mathbf{H}_0$, whose interaction with the paramagnetic spin has already been included in the spin Hamiltonian Eq.~\eqref{HNV}. 
As detailed in Appendix~\ref{AppendixInteraction}, the explicit expression in the second line of Eq.~\eqref{Vgeneral} is obtained by computing the spin wave magnetic field up to second order in magnon operators $\hat{s}_\beta$ and $\hat{s}_\beta^\dagger$ and undertaking a rotating wave approximation. The first contribution in Eq.~\eqref{Vgeneral} describes magnon-induced decay and absorption along the spin transition $\vert 0 \rangle \leftrightarrow \vert - \rangle$. It is characterized by a coupling rate
\begin{equation}\label{gfirstorder}
    g_{\beta} \equiv \mu_0 \vert \gamma_s\vert\mathcal{M}_{0\beta}\int d^3\mathbf{r}\left[\mathbf{e}^*_-\cdot \tensorbar{\mathcal{G}}(\mathbf{r}_0-\mathbf{r})\mathbf{m}_\beta(\mathbf{r})\right],
\end{equation}
where $\tensorbar{\mathcal{G}}(\mathbf{r})$ is the magnetostatic Green's tensor of the film, defined and analytically computed in Appendix~\ref{appendixSpinWaves}. Within the rotating wave approximation, the coupling between spin waves and the transition $\vert 0 \rangle\leftrightarrow\vert +\rangle$ can be neglected. The second contribution in Eq.~\eqref{Vgeneral} describes spin wave-induced dephasing of the paramagnetic spins. Despite being of second order, this contribution, characterized by a coupling rate
\begin{multline}\label{gsecondorder}
    \tilde{g}_{\beta\beta'} = -\mu_0\vert\gamma_s\vert\frac{\mathcal{M}_{0\beta}\mathcal{M}_{0\beta'}}{M_S}
    \\
    \times \int d^3\mathbf{r}
    \left[\mathbf{m}_{\beta}^*(\mathbf{r})\cdot\mathbf{m}_{\beta'}(\mathbf{r})\right]
    \text{Re}\left[\mathbf{e}_z\cdot\tensorbar{\mathcal{G}}(\mathbf{r}_0-\mathbf{r})\mathbf{e}_z\right],
\end{multline}
is resonant for an infinite set of magnon pairs $\{\beta,\beta'\}$. It can thus be as relevant as the first-order contribution which is resonant only for spin waves fulfilling $\omega_\beta=\omega_-$. This cumulative effect of second-order contributions has already been found to be relevant for e.g. acoustic phonon baths~\cite{HummerArxiv2020}.

\begin{table}[t]
	\centering
	\begin{tabular}{ p{4.7cm} | p{3.6cm} }
		\hline
		\hline
		Parameter & Value \\
		\hline
		YIG film thickness
		 & $d = 200$ nm \\
		YIG gyromagnetic ratio &  $\gamma=-1.76\times 10^{11}$T$^{-1}$ s$^{-1}$ \\
		YIG magnetization & $M_S=1.39\times 10^5$A m$^{-1}$ \\
		YIG exchange stiffness & $\alpha_x = 2.14\times 10^{-4}(\mu\text{m})^{-2}$ \\
		YIG Gilbert damping parameter & $\alpha_G = 10^{-4}$ \\
		\hline
		NV zero-field splitting &  $D_0 = 2\pi\times 2.877$ GHz\\
		NV gyromagnetic ratio &  $\gamma_s=-1.76\times 10^{11}$T$^{-1}$ s$^{-1}$ \\
		NV occupation lifetime & $T_1=3$ms \\
		NV coherence lifetime & $T_2^*=1\mu$s \\
		\hline
		\hline
	\end{tabular}
	\caption{Chosen values for the relevant parameters of YIG and of the NV centres across the main text. 
	Note that other relevant parameters whose value is not fixed, such as the field $H_0$ or the distance $l$ between the NV centres and the film, are not included in this table.} \label{tablePARAMS}
\end{table}

We now focus on the dissipative contributions to the von Neumann equation Eq.~\eqref{vonNeumanntotal}. We first consider the spin waves, which undergo dissipation through Gilbert damping~\cite{StancilBook2009,Gurevich1996magnetization,KalarickalJAP2006}. Gilbert damping is modelled through the dissipator~\cite{breuer2002theory}
\begin{equation}\label{dissipatorSW}
    \mathcal{D}_{\rm sw}[\hat{\rho}]=\sum_\beta \gamma_\beta\left(\bar{n}_\beta\mathcal{L}_{\hat{s}_\beta\hat{s}_\beta^\dagger}[\hat{\rho}]+(1+\bar{n}_\beta)\mathcal{L}_{\hat{s}_\beta^\dagger\hat{s}_\beta}[\hat{\rho}]\right),
\end{equation}
which describes absorption and decay into a thermal bath at temperature $T$, in terms of Lindblad superoperators defined as
\begin{equation}\label{Lindbladiandefinition}
    \mathcal{L}_{\hat{a}\hat{b}}[\hat{\rho}]\equiv\hat{a}\hat{\rho}\hat{b}-\frac{1}{2}\{\hat{b}\hat{a},\hat{\rho}\},
\end{equation}
with $\gamma_\beta$ the magnon decay rate and $\bar{n}_\beta = [\exp(\hbar\omega_\beta/k_B T)-1]^{-1}$ the Bose-Einstein distribution at the spin wave frequency.  The magnon decay rate $\gamma_\beta$, typically in the $\sim$MHz range, is computed analytically using the expression from phenomenological loss theory~\cite{StancilBook2009,ChumakInBook,StancilJAP1986}, 
\begin{equation}\label{gammabeta}
    \gamma_\beta = \frac{2\alpha_G\omega_\beta}{\vert\gamma\vert\mu_0}\frac{\partial \omega_\beta}{\partial H_0},
\end{equation}
which is known to be a good description of propagation loss in thin films and stripes \cite{ChumakInBook}. Here, we have introduced the additional Gilbert damping parameter $\alpha_G$ characterizing the spin wave losses (see Table~\ref{tablePARAMS}).

The dissipation of the paramagnetic spin is modelled through the following dissipator~\cite{AjisakaPRB2016,BarGillNatComm2013,WangNJP2014,TetiennePRB2013}:
\begin{multline}\label{DissipatorNVcentres}
    \mathcal{D}_{\rm ps}[\hat{\rho}]=\frac{\kappa_2}{\hbar^2}\mathcal{L}_{\hat{S}_{z}\hat{S}_{z}}[\hat{\rho}]\\+\kappa_1\sum_{\alpha=\pm}\left(\bar{n}_\alpha\mathcal{L}_{\hat{\sigma}_{\alpha0}\hat{\sigma}_{0\alpha}}[\hat{\rho}]+(\bar{n}_\alpha+1)\mathcal{L}_{\hat{\sigma}_{0\alpha}\hat{\sigma}_{\alpha0}}[\hat{\rho}]\right),
\end{multline}
The first term above describes dephasing at a rate $\kappa_2$, whereas the second line describes decay and absorption at a rate $\kappa_1$ along the two spin transitions, namely $\vert 0 \rangle \leftrightarrow \vert \pm \rangle$, induced by a bosonic thermal reservoir at temperature $T$, with $\bar{n}_\alpha = [\exp(\hbar\omega_\alpha/k_B T)-1]^{-1}$.
The rates $\kappa_1$ and $\kappa_2$ are related to the experimentally measured values for the two decoherence timescales $T_1$ and $T_2^*$ of the paramagnetic spin. These timescales are defined through the decay of the occupations $\langle \hat{\sigma}_{\alpha\alpha}\rangle\sim\exp(-t/T_1)$ ($\alpha=0,\pm$) and the coherences $\langle \hat{\sigma}_{0 \pm}\rangle \sim\exp(-t/T_2^*)$ in the zero-field limit $H_0\to 0$~\cite{MyersPRL2017,TetiennePRB2013}, and are given by
\begin{equation}\label{t1t2def}
    T_1^{-1} \equiv \kappa_1(1+3\bar{n}_0) \hspace{0.4cm} ; \hspace{0.4cm} \left(T_2^*\right)^{-1} \equiv   \frac{T_1^{-1} +\kappa_2}{2}  
\end{equation}
with $\bar{n}_0 = [\exp(\hbar D_0/k_B T)-1]^{-1}$. For the specific case of NV centres these timescales typically lie on the range  $T_1\sim$ms and $T_2^*\sim\mu$s~\cite{MyersPRL2017,HerbschlebNatComm2019,BarGillNatComm2013,AndrichNPJ2017,HansonScience2008,deLangeScience2010}~\footnote{Although, as indicated by Table~\ref{tablePARAMS}, throughout this work we take $T_2^*=1\mu$s, this value can be significantly increased using isotopically pure samples~\cite{BarGillNatComm2013,HerbschlebNatComm2019} or dynamical decoupling techniques~\cite{HansonScience2008,DohertyPhysRep2013,BarGillNatComm2013,deLangeScience2010}.}.

The above expressions can be generalized to an ensemble of paramagnetic spins situated at positions $\mathbf{r}_j$, $j=1,...N$ (see Appendix~\ref{AppendixInteraction}). We assume the density of paramagnetic spins is low enough such that the spins are independent from each other, that is, any interaction between them can be neglected. In the specific case of NV centres, this approximation holds for all the densities considered in this work (at most $10^5$($\mu$m)$^{-3}$)~\cite{KleinsasserAPL2016,BauchPRX2018}. Under this assumption both the Hamiltonian and the dissipator of the paramagnetic spins are simply written as a sum of the independent Hamiltonian and dissipator, Eqs.~\eqref{HNV} and \eqref{DissipatorNVcentres}, over all the spins in the ensemble. Furthermore, the above expressions can be extended to include external pumping of the paramagnetic spins, i.e. any mechanism whereby the spins are initialized near their ground state $\vert 0 \rangle$. This external pumping results in an enhancement of the mutual back-action between spin waves and paramagnetic spins as we will see below. A particular case, namely the optical pumping of NV centres~\cite{RobledoNJP2011,WangNJP2015,MeirzadaPRB2018,DohertyPhysRep2013,MildrenBook,RobertsPRB2019,TetienneNJP2012}, is modelled and analyzed in detail in Appendix~\ref{AppNV_OpticalPumping}. 
Further generalizations of our model, for instance to different magnetic materials and geometries and different spin pumping mechanisms, could be carried out following the same procedure.

In the following two sections we study the back-action of the paramagnetic spins on the spin waves (Sec.~\ref{SecSWmodification}) and, conversely, the back-action of spin waves onto the paramagnetic spins (Sec.~\ref{SecCasimirPolder}). In both cases we follow the approach in open quantum systems~\cite{breuer2002theory}, namely we trace out the degrees of freedom of one component (the bath) in order to derive an effective equation of motion for the second component (the system).
By doing so, we obtain a reduced master equation describing only the dynamics of the system degrees of freedom, and including the back-action of the bath in the form of effective dynamical terms such as additional dissipation or frequency shifts. 
This master equation allows to characterize the full dynamics of the system under the back-action of the bath.
A brief general summary of the involved techniques and approximations is given in Appendix~\ref{Appendix_TracingOut}.

\section{Modification of the spin wave properties}\label{SecSWmodification}

The strong back-action experienced by optical frequency electromagnetic fields due to their interaction with high-density quantum emitter ensembles has been demonstrated in many different platforms~\cite{RodriguezPRL2013,GehringScience2006,KangPRA2004,PayronelNature2012,MurrayInColl2016}.
For spin waves, however, the back-action of paramagnetic spins has remained practically unexplored and is not yet well understood, in spite of the significant advantages that an engineered back-action could provide for spintronics, from molding the flow of spin currents to patterning the spin waves without the need for material microstructuring.  In this section we derive and study the effective spin wave dynamics induced by a bath of paramagnetic spins and discuss its potential applications. The detailed derivation of the results in this section is provided in Appendix~\ref{Appendix_effectiveSWdynamics}.

\begin{figure}[t!]
	\centering
	\includegraphics[width=\linewidth]{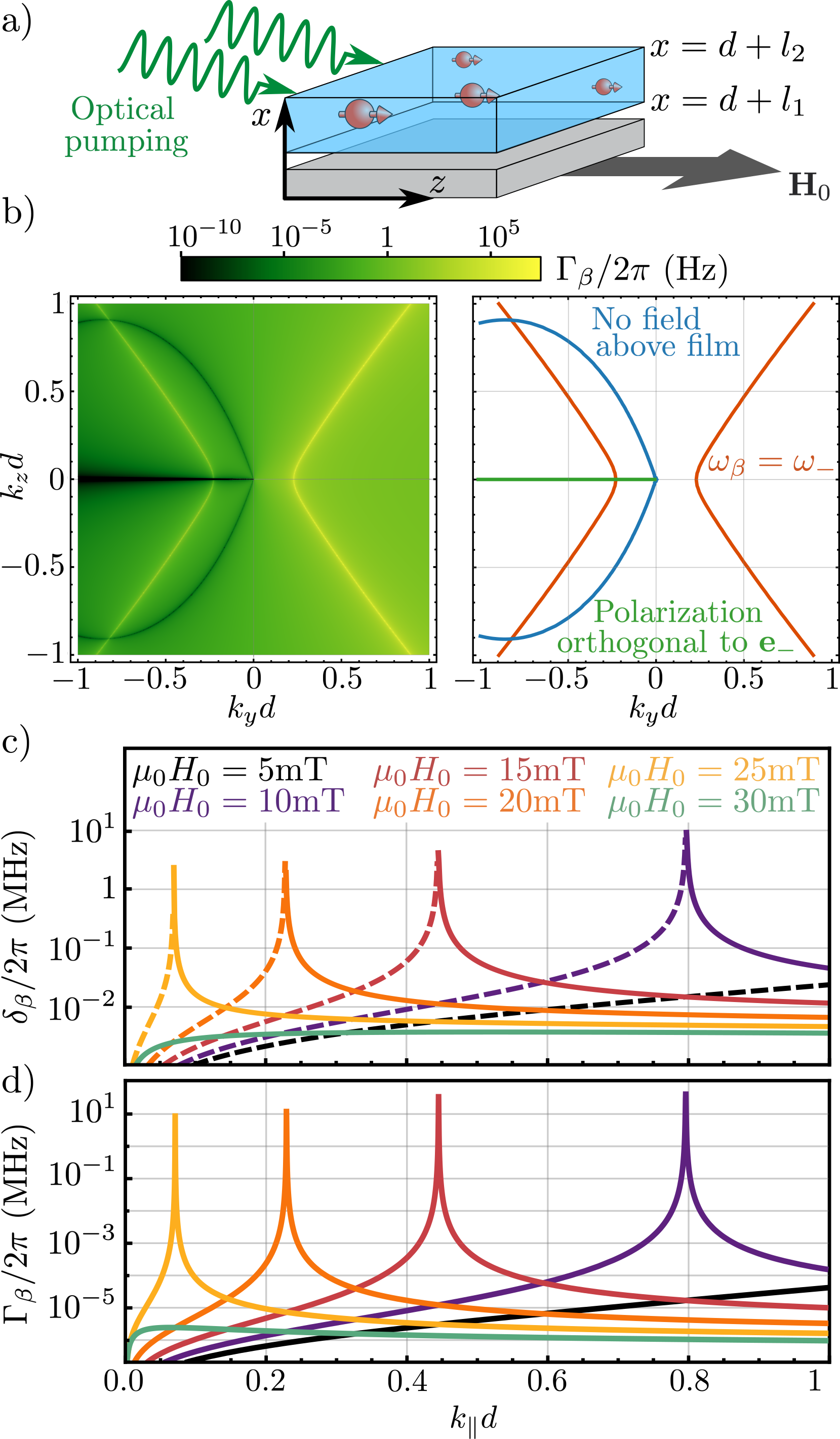}
	\caption{a) We study the effective dynamics of spin waves in the presence of a diamond slab containing optically pumped NV centres. b) Left panel: increase of the linewidth of $n=0$ spin waves at $\mu_0H_0=20$mT as a function of their parallel wavevector. Right panel: iso-lines corresponding to spin waves with $\omega_\beta=\omega_-$ (red), spin waves with polarization orthogonal to $\mathbf{e}_-$ (green), and spin waves generating zero magnetic field above the film (blue). c-d) Frequency and linewidth modification for $n=0$ spin wave modes propagating along the $+y$ direction as a function of wavenumber, for different values of the applied field $H_0$. Dashed and solid lines indicate negative and positive sign, respectively.
	In all the panels we take $l_1=0$, $l_2=\infty$, $\varrho_{\rm nv} = 10^4\text{($\mu$m)}^{-3}$, $T=300$K, optimum optical pumping parameters, and the parameters in Table~\ref{tablePARAMS}. }\label{Figure_band_modifications_1}
\end{figure}

We focus on the particular configuration depicted in \figref{Figure_band_modifications_1}(a), which is used in most experiments~\cite{BertelliSciAdv2020,LeeWongNanoLetters2020,WolfePRB2014,PageJAP2019,KikuchiAPExpress2017}. Specifically, we consider an ensemble of $N\gg1$ identical paramagnetic spins~\footnote{
Our results can be directly extended to  include spin-dependent frequencies and lifetimes, as well as to
an arbitrary number and spatial distribution of paramagnetic spins. The latter extension is carried out in Appendix~\ref{Appendix_effectiveSWdynamics} where we show that, for a slab below the YIG film, the shifts $\delta_\beta$ and $\Gamma_\beta$ are given by Eq.~\eqref{deltabetaGammabetaexpressions} under the substitution $\phi_k\to\phi_k+\pi$.} hosted, at randomly distributed positions $\mathbf{r}_j$, inside an infinite diamond slab parallel to the YIG film. The diamond slab has a width $l_2-l_1$ and is placed at a height $l_1$ above the film. 
Under these assumptions, and after tracing out the bath of paramagnetic spins,
we obtain the following master equation for the reduced density matrix of the spin waves, $\hat{\rho}_{\rm sw}$:
\begin{equation}\label{MasterEquationMagnons}
    \frac{d}{dt}\hat{\rho}_{\rm sw} = -\frac{i}{\hbar}\left[\hat{H}_{\rm sw}' ,\hat{\rho}_{\rm sw}\right] 
    \\
    + \mathcal{D}_{\rm sw}[\hat{\rho}_{\rm sw}]+\mathcal{D}_{\rm e}[\hat{\rho}_{\rm sw}],
\end{equation}
The first term in Eq.~\eqref{MasterEquationMagnons} describes the modified spin wave Hamiltonian and includes a paramagnetic spin-induced frequency shift,
\begin{equation}
    \hat{H}_{\rm sw}' =\hbar\sum_\beta (\omega_\beta+\delta_\beta)\hat{s}^\dagger_\beta\hat{s}_\beta \equiv \hbar\sum_\beta \omega_\beta'\hat{s}^\dagger_\beta\hat{s}_\beta.
\end{equation}
The dissipative part is composed by the original dissipator $\mathcal{D}_{\rm sw}[\hat{\rho}]$, given by Eq.~\eqref{dissipatorSW}, and an additional absorption and decay for each spin wave mode,
\begin{equation}\label{dissipatorextramagnons}
    \mathcal{D}_{\rm e}[\hat{\rho}]\equiv\sum_\beta \Gamma_{d\beta}\mathcal{L}_{\hat{s}_{\beta}\hat{s}_\beta^\dagger}[\hat{\rho}] + \Gamma_{a\beta}\mathcal{L}_{\hat{s}_{\beta}^\dagger\hat{s}_\beta}[\hat{\rho}].
\end{equation}
General expressions for the rates $\delta_\beta$, $\Gamma_{d\beta}$, and $\Gamma_{a\beta}$ are given in Appendix~\ref{Appendix_effectiveSWdynamics} (see also Eq.~[\ref{deltabetaGammabetaexpressions}]).
The above dissipator Eq.~\eqref{dissipatorextramagnons} results in an increase of the linewidth of the spin waves, namely
\begin{equation}
    \gamma_\beta \longrightarrow \gamma_\beta + \Gamma_{d\beta}-\Gamma_{a\beta} \equiv \gamma_\beta + \Gamma_\beta.
\end{equation}
We remark that the full master equation includes an additional interaction between different spin wave modes. The corresponding coupling rates, as opposed to the shift $\delta_\beta$ and the linewidth increase $\Gamma_\beta$, depend on the parallel coordinates $(y_j,z_j)$ of each paramagnetic spin and can be shown to average out to zero in the limit of $N \gg 1$ randomly distributed paramagnetic spins.

The rates $\delta_\beta$ and $\Gamma_\beta$ are given by
\begin{multline}\label{deltabetaGammabetaexpressions}
    \left[
    \begin{array}{c}
         \delta_\beta  \\
         \Gamma_\beta 
    \end{array}
    \right] = 
    \langle\hat{\sigma}_{00}-\hat{\sigma}_{--}\rangle_{\rm ss}
    \frac{\mu_0^2\vert\gamma\vert\gamma_s^2\hbar M_S\varrho_{\rm ps}}{\Delta_\beta^2 + (\kappa_T/2)^2}
    \left[
    \begin{array}{c}
         \Delta_\beta  \\
         \kappa_T 
    \end{array}
    \right]
    \\
    \times \frac{\cos^4(\phi_k/2)}{\omega_\beta/\omega_M}h_{\beta+0}^2\int_{l_1/d}^{l_2/d}dx e^{-2(k_\parallel d)x}
\end{multline}
where the sub-index ``ss'' indicates the steady state of the paramagnetic spins and $h_{\beta+0}$ is a real and dimensionless mode amplitude of order unity, defined in Appendix~\ref{appendixSpinWaves}. In the above expression we have defined the detuning $\Delta_\beta \equiv \omega_\beta-\omega_-$, the total decay rate of the paramagnetic spin coherence $\langle\hat{\sigma}_{0-}\rangle$, namely $\kappa_T/2$, and the volumetric density of paramagnetic spins,
$\varrho_{\rm ps} \equiv N L^{-2} (l_2-l_1)^{-1}$. Note that although we take the limit $N,L\to\infty$, the density $\varrho_{\rm ps}$ remains constant. 
For a given spin wave mode $\beta$, the rates in Eq.~\eqref{deltabetaGammabetaexpressions} are maximized in the thick slab limit $k_\parallel l_2 \gg 1$ which, for the parameters in this work, is effectively achieved at $l_2\gtrsim 5\mu$m. In this limit the rates decay exponentially with the slab-film distance $l_1$, and are thus maximized for a slab lying directly on top of the YIG film ($l_1=0$). Furthermore, the rates $\delta_\beta$ and $\Gamma_\beta$ depend crucially on the spin wave propagation direction through $\phi_k$, its amplitude and polarization through the mode amplitude $h_{\beta + 0}$, its frequency through the sharply peaked factor $[\Delta_\beta^2+(\kappa_T/2)^2]^{-1}$, and temperature and spin pumping through the occupation factor $\langle\hat{\sigma}_{00}-\hat{\sigma}_{--}\rangle_{\rm ss}$.

\begin{figure*}[t!]
	\centering
	\includegraphics[width=\linewidth]{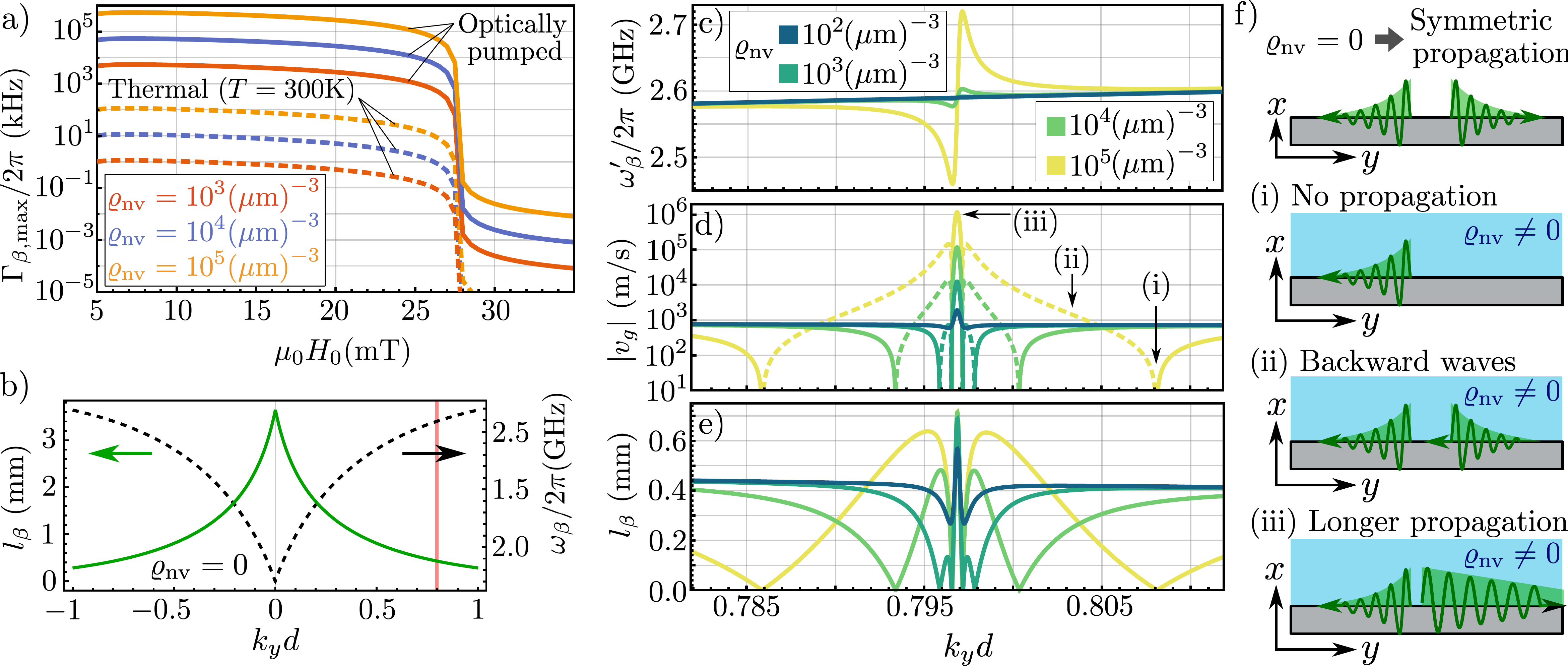}
	\caption{a) Maximum linewidth increase induced by optically pumped NV centres (assuming optimal pumping conditions, solid lines) and NV centres at room-temperature thermal equilibrium (dashed lines), as a function of applied field $H_0$ and for different densities of NV centres.  b) Propagation length (green) and dispersion relation (dashed black) of $n=0$ Damon-Eshbach spin waves in the absence of NV centres. The red shaded area indicates the spin waves which are significantly affected by the NV centres. c-e) Spin wave dispersion relation, absolute value of group velocity, and propagation length respectively, for optically pumped NV centres (optimal pumping conditions). The horizontal axis spans the red area indicated in panel (b), and different colors indicate different NV centre densities. In panel (d) solid/dashed curves indicate positive/negative values of $v_g$. f) Illustration of the modification of the spin wave properties induced by the NV centres, at the points (i), (ii), and (iii) marked in panel (d).
	In all the panels we take $\mu_0 H_0=10$mT, $l_1=0$, $l_2=\infty$, and the parameters in Table~\ref{tablePARAMS}.}\label{Figure_propagation_length}
\end{figure*}

Let us analyze the shifts $\delta_\beta$ and $\Gamma_\beta$ in detail
for the fundamental spin wave band $n=0$, where they are most relevant. We focus on the specific case of an ensemble of optically pumped NV centres, where $\kappa_T/2= \kappa_1(1+2\bar{n}_-+\bar{n}_+)/2+\kappa_2/2+\Omega \approx [T_2^*]^{-1}$ with $\Omega$ the optical pumping rate, proportional to the pumping intensity (see Appendix~\ref{AppendixNVcentres}). These ensembles can reach NV centre densities $\varrho_{\rm nv}$ as high as $\varrho_{\rm nv}\sim 10^5 (\mu\text{m})^{-3}$~\cite{KleinsasserAPL2016,HatanoPhysStatusSolidi2018,BauchPRX2018,OzawaJapJAPhys2019} and even larger~\cite{AcostaPRB2009,KucskoPRL2018}, although in such cases the extreme concentrations prove harmful for the coherence and decay times of NV centres.
The shift $\Gamma_\beta$ is shown in the left panel of \figref{Figure_band_modifications_1}(b) as a function of the wave vectors $k_y$ and $k_z$, for an NV centre density $\varrho_{\rm nv} = 10^4\text{($\mu$m)}^{-3}$. We assume optimal optical pumping conditions, i.e., maximally polarized NV centres,  $\langle\hat{\sigma}_{00}-\hat{\sigma}_{--}\rangle_{\rm ss} \approx 0.7$. The dark areas in \figref{Figure_band_modifications_1}(b), which indicate $\Gamma_\beta=0$, correspond to spin waves that do not couple to the paramagnetic spin bath. This is certified by the green and blue curves in \figref{Figure_band_modifications_1}(b, right panel) which indicate, respectively, the spin wave modes whose polarization is orthogonal to that of the spin transition $\vert 0 \rangle \leftrightarrow \vert - \rangle$ and the spin wave modes whose field amplitude vanishes above the YIG film. Conversely, the sharp maxima displayed by $\Gamma_\beta$ and indicated by yellow lines in the left panel of \figref{Figure_band_modifications_1}(b) stem from the factor $\kappa_T/[\Delta_\beta^2+(\kappa_T/2)^2]^{-1}$ in Eq.~\eqref{deltabetaGammabetaexpressions}, and occur for spin waves in resonance with the paramagnetic spin transition ($\Delta_\beta=0$). This is evidenced by the red curves in the right panel of \figref{Figure_band_modifications_1}(b), which are isolines indicating the resonant spin wave condition $\Delta_\beta=0$.
The behavior of the frequency shift $\delta_\beta$, not shown in the figure, is very similar to \figref{Figure_band_modifications_1}(b). In this case, however, the different dependence with the detuning,  $\delta_\beta\propto\Delta_\beta/[\Delta_\beta^2+(\kappa_T/2)^2]$, results in two sharp extremal points, namely a maximum and a minimum at $\omega_\beta = \omega_- + \kappa_{T}/2$ and $\omega_\beta = \omega_- - \kappa_{T}/2$, respectively. For spin waves exactly on resonance, $\omega_\beta = \omega_-$, the frequency shift is exactly zero.
Finally, note that the spin waves that experience the largest shifts $\delta_\beta$ and $\Gamma_\beta$ are the ($\mathbf{e}_-$-polarized) Damon-Eshbach modes propagating along the $+y$ direction, for which the coupling to the paramagnetic spins is maximized. In the following we will focus on these modes.

Since the most relevant feature of
the shifts $\delta_\beta$ and $\Gamma_\beta$ is their sharp modifications for near-resonant spin waves, $\omega_\beta=\omega_-(H_0)$, these shifts depend critically on the applied magnetic field $H_0$. This is illustrated in \figref{Figure_band_modifications_1}(c) and (d) where we display $\delta_\beta$ and $\Gamma_\beta$ respectively for spin wave modes propagating along the $+y$ direction and for different values of the applied field. 
Both shifts are suppressed for large fields ($\mu_0 H_0 \gtrsim 30$mT) as the resonance condition is never met ($\omega_\beta > \omega_-$ $\forall \beta$, see e.g. \figref{FigRWAappendix}[a]). They are also suppressed at small fields ($\mu_0 H_0 \lesssim 5$mT) as the resonance condition is met only for spin waves with large wavenumber $k_\parallel$, which are very weakly coupled to the paramagnetic spin transition due to their exponentially reduced magnetic field amplitude above the film (see the discussion above Eq.~[\ref{ldefinition}]). In order to analyze the maximum possible value for the shifts $\delta_\beta$ and $\Gamma_\beta$, we display in \figref{Figure_propagation_length}(a) the maximum linewidth modification
 $\Gamma_{\beta,\rm max} =\max_{\mathbf{k}_\parallel}\Gamma_\beta\vert_{n=0}$ as a function of applied field $H_0$ for three values of the NV centre density, both for
 optically pumped NV centres (solid lines) and for NV centres in the absence of optical pumping, i.e., at room temperature thermal equilibrium (dashed lines). The maximum frequency shift $\delta_\beta$  fulfills $\delta_{\beta,\rm max} \approx \Gamma_{\beta,\rm max}/4$~\footnote{This is to be expected from the ratio between the corresponding peaked functions appearing on $\Gamma_\beta$ and $\delta_\beta$,
$\max[2a(x^2+a^2)^{-1}]/\max[x (x^2+a^2)^{-1}] = 4$.}. According to \figref{Figure_propagation_length}(a), the highest possible values for
 $\Gamma_\beta$ are achieved for $\mu_0 H_0 \approx 10-15$mT where the resonant spin waves have moderate wavenumbers $k_\parallel d \sim 0.4-0.8$ (compare with \figref{Figure_band_modifications_1}[c-d]). 
 Indeed, for higher or lower wavenumbers $k_\parallel$ the spin wave field is either too tightly or too loosely confined to the YIG film, resulting in a weaker coupling to the paramagnetic spins.
 In the absence of pumping (dashed curves in \figref{Figure_propagation_length}[a]) the rate $\Gamma_{\beta, \rm max}$ is largely suppressed as the NV transition is practically saturated, i.e., $\langle\hat{\sigma}_{00}-\hat{\sigma}_{--}\rangle_{\rm ss} \approx 10^{-4}$. In this case, the maximum attainable value at large NV densities $\varrho_{\rm nv} \gtrsim 10^{4}\text{($\mu$m)}^{-3}$, $\Gamma_{\beta, \rm max} \approx 2\pi\times 10-100$kHz, represents at best a correction of roughly $10\%$ to the bare spin wave linewidth $\gamma_\beta \approx 2\pi\times 600$kHz. On the other hand, at optimal optical pumping conditions (solid curves in \figref{Figure_propagation_length}[a]) the NV centres are near their ground state and the change in the linewidth can be much larger, e.g. $\Gamma_{\beta, \rm max} \approx 2\pi\times 20\text{MHz} \approx 30 \gamma_\beta$ at $\varrho_{\rm nv} = 10^{4}\text{($\mu$m)}^{-3}$. These results suggest that, in this regime, a significant modification of the spin wave properties should be observed. 
 
To analyze the modification of the spin wave properties, we compute three characteristics particularly relevant for applications in spin-wave based information processing~\cite{ChumakInBook,BertelliSciAdv2020}, namely the modified spin wave frequency $\omega_\beta'\equiv \omega_\beta+\delta_\beta$, the modified group velocity defined as $v_g \equiv \partial\omega_\beta'/\partial k_y$, and the propagation length defined as $l_\beta \equiv v_g\tau_\beta$, where $\tau_\beta=2/(\gamma_\beta+\Gamma_\beta)$ is the spin wave lifetime. For the sake of illustration we display in \figref{Figure_propagation_length}(b) both the frequency $\omega_\beta$ and the propagation length $l_\beta$ in the absence of NV centres ($\varrho_{\rm nv}=0$). In the presence of the paramagnetic spins, the spin wave properties experience strong modifications within a narrow wavenumber range (indicated by the red shaded area in \figref{Figure_propagation_length}[b]), corresponding to the frequency range $\omega_--\kappa_T/2 \lesssim \omega_\beta \lesssim \omega_-+\kappa_T/2$. The spin wave frequencies, group velocities (in absolute value), and propagation lengths within this narrow range are shown in \figref{Figure_propagation_length}(c-e) respectively, for different densities of NV centres. As evidenced by the figure, although the modification of the frequency $\omega_\beta'$ is relatively small ($\sim4\%$), it displays very abrupt slope changes resulting in variations of several orders of magnitude in the group velocity. The propagation length does not display such a strong increase as the effective  spin wave loss rate $\gamma_\beta + \Gamma_\beta$ becomes also very large. 

Three especially interesting possibilities, schematically illustrated in \figref{Figure_propagation_length}(f), can be identified in \figref{Figure_propagation_length}(c-e): (i) Full suppression of spin wave propagation at wavenumbers for which $v_g = l_\beta=0$; (ii) backward-wave propagation~\cite{MojahediIEEE2003} for a wide range of wavenumbers where the group velocity becomes negative, $v_g<0$; (iii) enhancement of the propagation length for near-resonant spin waves. This enhancement, a priori counterintuitive due to the very large effective decay rate $\gamma_\beta+\Gamma_\beta$, is enabled by the corresponding increase in the group velocity. Note that all of these modifications occur for Damon-Eshbach modes propagating with $k_y>0$, whereas the modes with $k_y<0$ remain unchanged as their polarization is orthogonal to the paramagnetic spin transition (see \figref{Figure_band_modifications_1}[b]). The presence of the paramagnetic spins as a bath thus induces, within a narrow but tuneable range of wave-vectors, a further symmetry breaking for the Damon-Eshbach spin waves.
The exotic phenomenology shown in \figref{Figure_propagation_length} should be observable in current NV centre-spin wave hybrid platforms, for which changes as small as $\sim 10\%$ in e.g. the spin wave propagation length can be resolved~\cite{BertelliSciAdv2020}. Similar or even stronger modifications of the propagation properties could be achieved with other paramagnetic spins (e.g. allowing for higher spin densities than NV centre ensembles) and/or more efficient pumping schemes able to reach maximum polarization, $\langle\hat{\sigma}_{00}-\hat{\sigma}_{--}\rangle_{\rm ss} \approx 1$.

\section{Back-action-based spin wave sensing}\label{SecCasimirPolder}

\begin{figure}[tbh!]
	\centering
	\includegraphics[width=\linewidth]{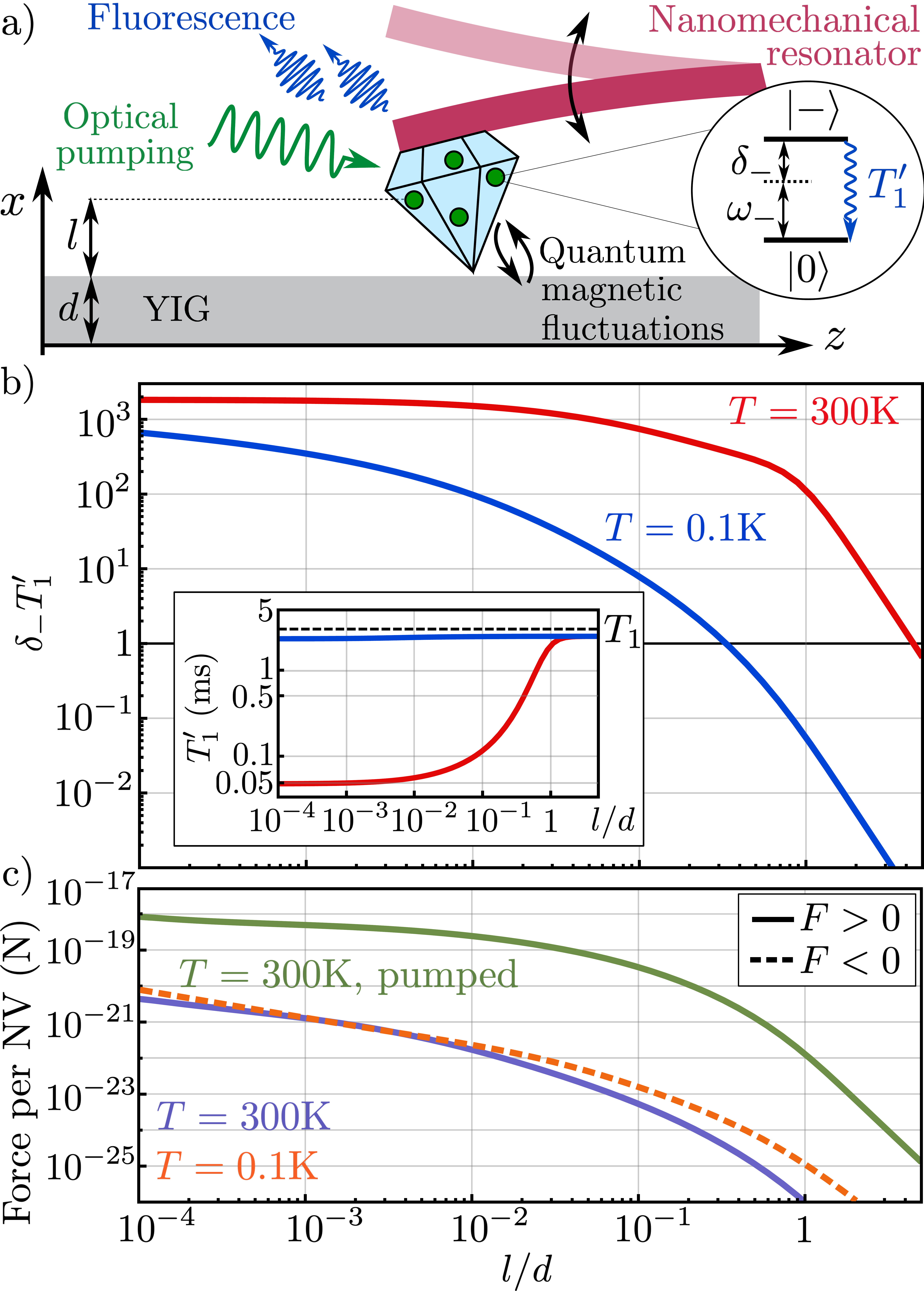}
	\caption{a) The spin waves modify the frequency and lifetime of the $\vert 0 \rangle \to \vert - \rangle$ transition of the (potentially optically pumped) NV centres. The frequency shift and corresponding force can be detected through fluorescence or nanomechanical force sensing.
	 b) Product of the frequency shift $\delta_-$ and the modified transition lifetime $T_1'$ of the $\vert 0 \rangle\leftrightarrow\vert - \rangle$ transition of a single NV centre above the YIG film, at room (red) and cryogenic (blue) temperatures. The inset shows the corresponding modified lifetimes $T_1'$, the dashed line indicating the bare lifetime in the absence of spin waves, $T_1$. c) Spin-wave induced force per NV centre, at room temperature (purple), cryogenic temperature (dashed orange), and room temperature under optimal optical pumping conditions (green). Solid (dashed) lines indicate positive (negative) forces. In all panels we choose $\mu_0H_0=35$mT and the parameters in Table~\ref{tablePARAMS}.  }\label{FigureNV_LambShift}
\end{figure}

The possibility of modifying spin wave properties through the passive back-action effected by the paramagnetic spins is far from the full potential of these spins for spintronics. Indeed, a natural step forward consists on using the paramagnetic spins as \emph{active} components to probe and control spin waves~\cite{BertelliSciAdv2020,LeeWongNanoLetters2020,PageJAP2019,KikuchiAPExpress2017,WolfePRB2014,ZhangPRB2020}. An essential capability toward this goal is that of measuring
the back-action exerted by the spin waves on the paramagnetic spins.
In this section we characterize such back-action and show the possibility of detecting it both optically and mechanically, should the paramagnetic spins be embedded in a micromechanical oscillator (e.g. a diamond cantilever), see \figref{FigureNV_LambShift}(a). 
Aside from spin wave probing, this optical and/or mechanical detection of spin wave back-action could allow for a new generation of flexible spin wave hybrid platforms integrating optical and mechanical degrees of freedom.

The spin wave back-action induces the following effective dynamics on an ensemble of paramagnetic spins: (i) A shift of the transition frequencies $\omega_\pm$ and a corresponding mechanical force, see details below; (ii) a modification of the lifetimes $T_1$ and $T_2^*$; (iii) an effective interaction between  different paramagnetic spins, which inherits the direction-dependent character of the spin waves. All these effects are studied in detail in Appendix~\ref{Appendix_EffectiveNVdynamics}. Hereafter we consider a single paramagnetic spin at an arbitrary position outside the YIG film. The resulting master equation for its reduced density matrix, $\hat{\rho}_{\rm ps}$, is given by 
\begin{multline}\label{MEQNVcentres}
    \frac{d}{dt}\hat{\rho}_{\rm ps} = -\frac{i}{\hbar}\left[\hat{H}_{\rm ps}',\hat{\rho}_{\rm ps}\right]+
    \mathcal{D}_{\rm ps}[\hat{\rho}_{\rm ps}] + \mathcal{D}_{\rm d}[\hat{\rho}_{\rm ps}].
\end{multline}
The coherent contribution, given by the Hamiltonian
\begin{equation}
    \hat{H}_{\rm ps}' = \hbar\sum_{\alpha=\pm}(\omega_\alpha + \delta_\alpha)\hat{\sigma}_{\alpha\alpha},
\end{equation}
describes the modification of the two transition frequencies by a shift $\delta_\alpha$, in analogy to the AC Stark effect experienced by electric dipoles in an electromagnetic bath~\cite{CarmichaelBook}. Although within our approximations the coupling between the transition $\vert0 \rangle \to \vert + \rangle$ and spin waves is neglected (see Eq.~[\ref{Vgeneral}]), the frequency of this transition is also modified due to the energy shift of the state $\vert 0 \rangle$. 
Moreover, the spin waves introduce additional dissipative dynamics, given in Eq.~\eqref{MEQNVcentres} by the dissipator
\begin{equation}\label{NVadditionaldissipator}
    \mathcal{D}_{\rm d}[\hat{\rho}] = \kappa_{a}\mathcal{L}_{\hat{\sigma}_{-0}\hat{\sigma}_{0-}}[\hat{\rho}]+\kappa_{d}\mathcal{L}_{\hat{\sigma}_{0-}\hat{\sigma}_{-0}}[\hat{\rho}]+\frac{\kappa_2'}{\hbar^2}\mathcal{L}_{\hat{S}_{z}\hat{S}_{z}}[\hat{\rho}].
\end{equation}
Here, the first two terms represent additional absorption and decay along the $\vert0 \rangle\to\vert -\rangle$ transition, with rates $\kappa_a$ and $\kappa_d$ respectively, whereas the last term represents an additional dephasing of both excited states $\vert + \rangle$ and $\vert - \rangle$ at a rate $\kappa_2'$. Analytical expressions for the rates $\delta_\alpha$, $\kappa_a$, $\kappa_d$, and $\kappa_2'$ are given in Appendix~\ref{Appendix_EffectiveNVdynamics}.

Let us discuss the frequency shifts $\delta_\pm$. These shifts fulfill $\vert\delta_-\vert\ge 2\vert\delta_+\vert$
~\footnote{Specifically at room temperature $\delta_-\approx 2\delta_+$ whereas at cryogenic temperatures $\delta_+\approx 0$.}, allowing us to focus on the main shift, namely that of the coupled transition $\vert 0\rangle\leftrightarrow\vert - \rangle$ given by
\begin{equation}\label{deltaM}
         \delta_{-} = -\sum_\beta\frac{\Delta_\beta}{\Delta_\beta^2+(\gamma_\beta/2)^2}\vert g_{\beta}\vert^2\left(
         1+2\bar{n}_\beta 
    \right).
\end{equation}
The above frequency shift is measurable, for instance by fluorescence in the case of an NV centre, provided that it is larger than the linewidth of the transition $\vert 0\rangle\leftrightarrow\vert - \rangle$, i.e., provided that $\vert\delta_- T_1'\vert >1$. Here, $T_1' < T_1$ is the lifetime of the transition $\vert 0\rangle\leftrightarrow\vert - \rangle$, which is also modified by the back-action of the spin waves. The definition of $T_1'$ is not straightforward since, in the presence of spin waves, the occupations $\langle \hat{\sigma}_{--}\rangle(t)$ do not obey a simple- but a multi-exponential decay characterized by more than one rate~\footnote{Adding to the difficulty, the rates associated to the decays of the two occupations $\langle \hat{\sigma}_{--}\rangle(t)$ and $\langle \hat{\sigma}_{++}\rangle(t)$ are generally different.}. We define $T_1'$ as the shortest timescale of this evolution or, conversely, as the inverse of the largest decay rate.
Both $\delta_-$ and $T_1'$ are computed numerically by expressing them in integral form. In such form it can be shown using symmetry arguments that both quantities are independent on the parallel ($y,z$) coordinates of the paramagnetic spin or on the side of the film on which the paramagnetic spin is placed, and are a function exclusively of the ratio $l/d$.

Let us consider the specific case of NV centres.
The product $\vert\delta_- T_1'\vert$ for these paramagnetic spins is displayed in \figref{FigureNV_LambShift}(b) as a function of vertical separation $l/d$, for $\mu_0H_0=35$mT and thermal equilibrium at room (red) and cryogenic (blue) temperatures. The corresponding inset shows the modified lifetime $T_1'$ for the same parameters. The spin wave-induced effects are stronger both at room temperature and at short separations $l$, as the amplitude of the thermal spin waves is larger. At room temperature the large values of $\delta_-$ (up to $\sim2\pi\times8$MHz) are more significant than the reduction of the lifetime $T_1'$, resulting in a product $\vert\delta_- T_1'\vert > 1$ for NV-YIG film separations as large as $l=4d=800$nm. This should allow for the experimental observation of the spin wave-induced frequency shift of the transition $\vert 0\rangle\leftrightarrow\vert - \rangle$ in current setups, e.g. via fluorescence of the NV centre~\cite{GieselerPRL2020,BertelliSciAdv2020,LeeWongNanoLetters2020,WolfePRB2014,AndrichNPJ2017,PageJAP2019,KikuchiAPExpress2017,ZhangPRB2020,MyersPRL2017}. Note that the results in \figref{FigureNV_LambShift}(b) do not require optical pumping of the NV centres.
 More details on the behavior of the frequency shift $\delta_-$ and of the lifetime $T_1'$, including their dependence with the applied field $H_0$, are provided in Appendix~\ref{Appendix_EffectiveNVdynamics}.

A consequence of a position-dependent frequency shift, that is, a position-dependent transition frequency, is the presence of a net force acting on the paramagnetic spin, namely the magnetic thermal Casimir-Polder force~\cite{CasimirPhysRev1948} which has been extensively studied in quantum electrodynamics~\cite{CasimirPhysRev1948,BuhmannPQE2007,BuhmannPRL2008}.
The net force experienced by a paramagnetic spin lying above the YIG film~\footnote{For a spin below the film the force has opposite sign, see Appendix~\ref{Appendix_EffectiveNVdynamics}.} in its steady state is given by $\mathbf{F} = -\langle \nabla \hat{H}_{\rm ps}'\rangle_{\rm ss}$~\cite{BuhmannPQE2007,BuhmannPRL2008}. Here we focus on the high- and low-temperature limits, where the force is given by
\begin{multline}\label{CasimirPolderForce}
    \mathbf{F} = -\mathbf{e}_x\frac{\hbar}{2} \frac{d\delta_-(l)}{dl}
    \\
    \times
    \left\lbrace
    \begin{array}{cc}
        \langle\hat{\sigma}_{--}- \hat{\sigma}_{00}\rangle_{\rm ss} & \text{ for } k_B T \gg \hbar\omega_M \\
        2\langle\hat{\sigma}_{--}\rangle_{\rm ss} & \text{ for } k_B T \ll \hbar\omega_M.
    \end{array}
    \right.
\end{multline}
Note the qualitative difference between room temperature, where both spin transitions contribute to the force ($\delta_+\approx\delta_-/2$), and cryogenic temperatures where only the $\vert 0 \rangle \leftrightarrow\vert-\rangle$ transition does as $\delta_+\to 0$.
As opposed to the frequency shifts $\delta_\pm$, the force depends on the state of the paramagnetic spin.

The force Eq.~\eqref{CasimirPolderForce} is displayed in \figref{FigureNV_LambShift}(c) for an NV centre at $T=300$K under optimal optical pumping (green line), at $T=300K$ without optical pumping (purple line), and at $T=100$mK without optical pumping (orange line). At thermal equilibrium at room temperature the force Eq.~\eqref{CasimirPolderForce} is smaller, as the steady-state occupations are very close to their zero-field values, i.e. $\langle \hat{\sigma}_{\pm\pm}\rangle_{\rm ss}\approx\langle \hat{\sigma}_{00}\rangle_{\rm ss}\approx 1/3$. This force can, however, be increased by several orders of magnitude via optical pumping, as evidenced by the green curve in \figref{FigureNV_LambShift}(c). At cryogenic temperatures (orange curve in \figref{FigureNV_LambShift}[c]), despite the paramagnetic spin being close to its ground state, the force remains of the same order as at room temperature as the field amplitude of the thermal spin waves, and hence their back-action on the paramagnetic spins, is largely reduced. In this case, moreover, the force cannot be significantly increased via conventional optical pumping. The spin wave-induced forces are mostly repulsive at room temperature, a well-known feature of fluctuational forces on magnetic dipole transitions~\cite{SkagerstamPRA2009,HaakhPRA2009}, and attractive at cryogenic temperatures.
A detailed analysis and characterization of all these forces, including their dependence with the applied field $H_0$, is given in Appendix~\ref{Appendix_EffectiveNVdynamics}.

The possibility of pumping the paramagnetic spin paves the way toward the detection of the spin wave-induced force at room temperature. Measurements of such force (and hence indirect measurements of the frequency shift) could be performed using nanomechanical sensing devices. For instance, a nanodiamond containing either a single NV centre or an ensemble of NV centres could be attached to a high-Q cantilever as schematically depicted in \figref{FigureNV_LambShift}(a). In this setup, the spin wave-induced force could be tracked through the shifts in the mechanical frequency of the cantilever~\cite{VinanteNatComm2011}. For a single, optically pumped NV centre at a distance $l\lesssim d/5 \approx 40$nm, the forces in \figref{FigureNV_LambShift}(c) lie within the sensitivity range ($10^{-21}-10^{-18}$N Hz$^{1/2}$) of current ultra-sensitive force detectors, based on membranes~\cite{BunchScience2007,WeberNatComm2016,FischerNJP2019} or nanowire resonators and cantilevers~\cite{LiNatNano2007,SazonovaNature2004,MoserNatNano2013,NicholPRB2012,GloppeNatNano2014,RossiNatNano2017,BraakmanNanotechnology2019} of diverse materials, including single-crystal diamond~\cite{HeritierNanoLett2018}.
Moreover, the additive force experienced by nano-diamonds containing ensembles of as few as $N\sim 10^3$ NV centres could reach current experimental sensitivities even for NV centres in thermal equilibrium, or for optically pumped NV centres at distances as large as $l\approx d =200$nm. Last but not least, larger mechanical forces could be achieved for different paramagnetic spins and/or more efficient pumping schemes.

\section{Conclusion}\label{SecConclusion}

We have developed a comprehensive quantum theory of hybrid interfaces based on spin waves magnetically coupled to paramagnetic spins. In the first part of our work, we have applied this theory to characterize the effective spin wave dynamics induced by the back-action of the paramagnetic spins. This back-action results in a tuneable modification of the spin wave propagation properties. Specifically, it can induce full cancellation or enhancement of spin wave propagation length, as well as backward wave propagation for Damon-Eshbach modes. All these modifications are strong and measurable in state of the art setups. We have also quantified the impact of the paramagnetic spins back-action on the magnetic field fluctuations, specifically on the magnetic field power spectral density, outside the magnetic structure. Our results show the potential of electron spins as passive tools to engineer spin wave properties.

In the second part of our work, we have studied the opposite effect, namely the back-action exerted by spin waves on nearby paramagnetic spins. This back-action results in a frequency shift of the paramagnetic spins transitions, a modification of the paramagnetic spin decay and decoherence lifetimes, and a spin wave-mediated interaction between different paramagnetic spins within an ensemble. All the above effects have been characterized in detail. The frequency shift of single paramagnetic spins can be measured in current experiments, with usual fluorescence techniques. In addition, this shift is accompanied by a force that can be detected with state of the art mechanical sensing devices. These results evidence the further role of paramagnetic spins as active tools for probing and controlling spin waves.

The results presented in this work have many applications in spin wave-based technologies. The suppression of the spin wave propagation lengths could be used to devise spin wave mirrors, polarization filters, optically-gated spin wave transistors, or magnonic crystals, among others. Similarly, the enhancement of the spin wave propagation could help reduce losses in spin wave information processing devices. All these capabilities can in principle be spatially tailored with state of the art techniques, such as by distributing the paramagnetic spins in a convenient spatial arrangement~\cite{AndrichNPJ2017,WolfePRB2014,PageJAP2019} or by using selective optical pumping with nanometric resolution~\cite{JaskulaOptExpress2017}. Our results could thus pave the way toward back-action based reconfigurable spin wave circuits. On the other hand, the reverse back-action of spin waves on paramagnetic spins evidences their potential for spin wave detection and for accurate measurement of distances. Moreover, it paves the way toward using paramagnetic spins as mediators between spin waves and electromagnetic or mechanical fields. A particularly interesting prospect is to enhance the spin wave-induced force (for instance through coherent driving of spin waves resonant with the paramagnetic spin transition) hence increasing the magneto-mechanical coupling between spin waves and a mechanical resonator.
This magneto-mechanical coupling, whose promising quantum applications are only starting to be explored~\cite{GieselerPRL2020,GonzalezBallesteroPRL2020,GonzalezBallesteroPRB2020,ZhangSciAdv2016,ZoepflPRL2020,RablPRB2009,ViaPRL2015}, might allow to integrate magnonic degrees of freedom in micro- and opto-mechanical systems, bringing new degrees of flexibility to technology-oriented spin wave platforms.

\newpage
\clearpage
\pagebreak

\appendix

\textbf{\large Appendix: table of contents}

\vspace{0.2cm}
\begin{enumerate}[leftmargin=*]
    \item[] \textbf{Appendix A:} \TOCentry{appendixSpinWaves}
    \begin{enumerate}
        \item[] A1. \TOCentry{AppSW_EOM}%
        \vspace{-0.1cm}
        \item[] A2. \TOCentry{AppSW_EigenmodeEquation}\vspace{-0.1cm}
        \item[] A3. \TOCentry{AppSW_Perturbationtheory}\vspace{-0.1cm}
        \item[] A4. \TOCentry{AppSW_Modeproperties}\vspace{-0.1cm}
        \item[] A5. \TOCentry{AppSW_quantizationEOMgamma}\vspace{-0.1cm}
        \item[] A6. \TOCentry{AppSW_PSD}%
        \vspace{-0.1cm}
        \item[] A7. \TOCentry{AppSW_2ndorderfield}%
        \vspace{-0.1cm}
    \end{enumerate}
    \item[] \textbf{Appendix B:} \TOCentry{AppendixNVcentres}
    \begin{enumerate}
        \item[] B1. \TOCentry{AppNV_baredynamics}\vspace{-0.1cm}
        \item[] B2. \TOCentry{AppNV_OpticalPumping}\vspace{-0.1cm}
    \end{enumerate}
    \item[] \textbf{Appendix C:} \TOCentry{AppendixInteraction}
    \item[] \textbf{Appendix D:} \TOCentry{Appendix_TracingOut}
    \begin{enumerate}
        \item[] D1. \TOCentry{AppTraceOut_BornMarkov}\vspace{-0.1cm}
        \item[] D2.
        \TOCentry{AppTracingOut_FrozenBath}\vspace{-0.1cm}
    \end{enumerate}
    \item[] \textbf{Appendix E:} \TOCentry{Appendix_EffectiveNVdynamics}\vspace{-0.1cm}
    \begin{enumerate}
        \item[] E1.  \TOCentry{AppNVdynamics_generalMEq}\vspace{-0.1cm}
        \item[] E2. \TOCentry{AppNVdynamics_T1T2modified}\vspace{-0.1cm}
        \item[] E3. \TOCentry{AppNVdynamics_LambShift}\vspace{-0.1cm}
        \item[] E4. \TOCentry{AppNVdynamics_Couplings}\vspace{-0.1cm}
    \end{enumerate}
    \item[] \textbf{Appendix F:} \TOCentry{Appendix_effectiveSWdynamics}
    \begin{enumerate}
        \item[] F1. \TOCentry{AppSWdynamics_generalMEq}\vspace{-0.1cm}
        \item[] F2. \TOCentry{AppSWdynamics_largeNrandom}\vspace{-0.1cm}
        \item[] F3. \TOCentry{AppSWdynamics_PSDs}\vspace{-0.1cm}
    \end{enumerate} 
\end{enumerate}

\section{Spin waves in a slab}\label{appendixSpinWaves}

In this Appendix we give further details on the spin waves supported by a thin film. First, in Sec.~\ref{AppSW_EOM}, we
describe the Landau-Lifshitz equation governing the magnetization dynamics and its linearization. We then derive, in Sec.~\ref{AppSW_EigenmodeEquation}, the spin wave eigenmode equation and cast it in a convenient form for a perturbative treatment, which we summarize in Sec.~\ref{AppSW_Perturbationtheory}. After, we discuss the spin wave eigenmode properties in detail in Sec.~\ref{AppSW_Modeproperties}. We complete the introduction of the spin wave formalism in Sec.~\ref{AppSW_quantizationEOMgamma} by summarizing the spin wave quantization, the introduction of loss rates, and the computation of spin wave observables. In the last two sections of this appendix, we extend our study by computing and analyzing the magnetic field power spectral densities outside the film (Sec.~\ref{AppSW_PSD}), and by computing the magnetization and magnetic field operators to second order in magnon operators (Sec.~\ref{AppSW_2ndorderfield}), which are required in Sec.~\ref{AppendixInteraction}.

\subsection{Equations of motion and linearization}\label{AppSW_EOM}

We consider first a lossless insulating ferromagnet in the presence of a static magnetic field $\mathbf{H}_0$. The fundamental equation governing the dynamics of the magnetization field $\mathbf{M}(\mathbf{r},t)$ is the lossless Landau-Lifshitz equation~\cite{StancilBook2009},
\begin{equation}\label{EqLLEappendix}
    \frac{d}{dt}\mathbf{M}(\mathbf{r},t) = -\vert \gamma\vert\mu_0\mathbf{M}(\mathbf{r},t)\times\left[\mathbf{H}_0 + \mathbf{H}'(\mathbf{M},\mathbf{r},t) \right],
\end{equation}
with $\gamma$ the gyromagnetic ratio and $\mu_0$ the vacuum permittivity. This equation is identical to the equation of motion for the orientation of a single fixed magnetic dipole with moment $\mathbf{M}$, except for the additional nonlinear contribution  $\mathbf{H}'(\mathbf{M},\mathbf{r},t)$, known as the effective field, which describes the many complex interactions arising inside the magnetic material. The effective field can be written as \cite{Gurevich1996magnetization,StancilBook2009}
\begin{multline}\label{Hprimedecomposition}
    \mathbf{H}'(\mathbf{M},\mathbf{r},t) = \mathbf{H}_d(\mathbf{M},\mathbf{r},t)\\+\mathbf{H}_x(\mathbf{M},\mathbf{r},t)+\mathbf{H}_a(\mathbf{M},\mathbf{r},t).
\end{multline}
The first term above, namely the demagnetizing field, accounts for the dipole-dipole interaction. Under the magnetostatic approximation $\nabla\times \mathbf{H}_d(\mathbf{r},t)\approx 0$, valid for spin wave wavelengths much shorter or much longer than the vacuum wavelength~\cite{StancilBook2009}, the demagnetizing field is given by~\cite{Jackson1975classical}
\begin{equation}\label{Hdemag}
    \mathbf{H}_d(\mathbf{M},\mathbf{r},t) = \frac{1}{4\pi} \nabla\int d^3\mathbf{r}' \frac{\nabla'\cdot \mathbf{M}(\mathbf{r}',t)}{\vert \mathbf{r}-\mathbf{r}'\vert}.
\end{equation}
The remaining two contributions in Eq.~\eqref{Hprimedecomposition} describe the effect of exchange and magnetocrystalline anisotropy on the magnetization dynamics. Since they are effective terms, they do not correspond to genuine Maxwell fields, i.e. the total field appearing in Maxwell equations is given by $\mathbf{H}(\mathbf{r},t) = \mathbf{H}_0 + \mathbf{H}_d(\mathbf{M},\mathbf{r},t)$. On the one hand, the exchange field $\mathbf{H}_x$ accounting for the exchange interaction is given, for a material with a cubic lattice such as YIG, by the expression~\cite{StancilBook2009}
\begin{equation}\label{Hx}
    \mathbf{H}_x(\mathbf{M},\mathbf{r},t)=\alpha_{x}\nabla^2\mathbf{M}(\mathbf{r},t)
\end{equation}
with $\alpha_x$ the exchange stiffness. On the other hand, $\mathbf{H}_a$ accounts for the magnetocrystalline anisotropy interaction, and its form is not only sample- and geometry-dependent but widely tuneable e.g. through rare earth dopants \cite{wohlfarth1986handbook,KrysztofikJPhysD2017}. For undoped YIG thin films, anisotropy is generally small~\cite{KalinikosJPHYSC1986,KostylevJAP2013,BertelliSciAdv2020,LeeWongNanoLetters2020} and can be neglected for describing spin waves. Hereafter we will thus take $\mathbf{H}_a\approx 0$, although the effect of different kinds of anisotropies, such as e.g. uniaxial or cubic, could be included~\cite{KalinikosJPhysCondMat1990}.

From the definitions above it is evident that a homogeneous magnetization profile $\mathbf{M}(\mathbf{r},t)=M_S\mathbf{e}_z$ is a solution of the Landau-Lifshitz Equation Eq.~\eqref{EqLLEappendix}. We seek spin wave solutions, i.e., solutions describing small fluctuations above such fully magnetized state,
\begin{equation}\label{spinwaveapprox}
    \mathbf{M}(\mathbf{r},t) = M_S\mathbf{e}_z +\mathbf{m}(\mathbf{r},t),
\end{equation}
with $\vert\mathbf{m}(\mathbf{r},t)\vert \ll M_S$. By introducing the above expression into Eq.~\eqref{EqLLEappendix} and neglecting the small quadratic terms $\mathcal{O}(\mathbf{m}^2)$ we obtain the linearized Landau-Lifhsitz equations, namely
\begin{multline}\label{linearizedLLEappendix}
    \left[\frac{d}{dt}-\mathbf{e}_z\times \left(\omega_H-\omega_M\alpha_x\nabla^2\right)\right]\mathbf{m}(\mathbf{r},t) =
    \\
    =-\omega_M\mathbf{e}_z\times\mathbf{h}(\mathbf{r},t).
\end{multline}
where, by definition, the magnetic field associated to the magnetization field $\mathbf{m}(\mathbf{r},t)$, i.e., the magnetic field of the spin wave, is given by
\begin{multline}\label{hgreenfunctionmappendix}
    \mathbf{h}(\mathbf{r},t) = \frac{1}{4\pi} \nabla\int d^3\mathbf{r}' \frac{\nabla'\cdot \mathbf{m}(\mathbf{r}',t)}{\vert \mathbf{r}-\mathbf{r}'\vert}
    \\
    \equiv\int d^3\mathbf{r}'\tensorbar{\mathcal{G}}(\mathbf{r}-\mathbf{r}')\mathbf{m}(\mathbf{r}',t).
\end{multline}
The Green's tensor above is defined as
\begin{equation}\label{Ggeneral}
    \tensorbar{\mathcal{G}}_{ij}(\mathbf{r}-\mathbf{r}')=\frac{-1}{4\pi \vert\mathbf{r}-\mathbf{r}'\vert^3}(\mathbf{r}-\mathbf{r}')_i\frac{\partial}{\partial r_j'},
\end{equation}
where the sub-indices $i,j=x,y,z$ indicate Cartesian components. The above expression is usually symmetrized by (i) making use of the vector identity $\nabla\cdot(f\mathbf{A}) = f\nabla\cdot\mathbf{A}+(\nabla f)\cdot\mathbf{A}$ and integrating by parts; (ii) applying the divergence theorem, and  (iii) taking an integration volume enclosed by an area infinitely far away from the magnetic material. The resulting symmetrized expression reads
\begin{equation}\label{Gsymetrized}
    \tensorbar{\mathcal{G}}_{ij}(\mathbf{r}-\mathbf{r}')=\frac{-1}{4\pi}\frac{\partial}{\partial r_i}\frac{\partial}{\partial r_j'} \frac{1}{\vert\mathbf{r}-\mathbf{r}'\vert}.
\end{equation}
It is convenient, given our planar geometry, to express the Green's tensor in a plane-wave representation in the parallel coordinates. Using the identity
\begin{equation}
    \frac{1}{\vert \mathbf{r}-\mathbf{r}'\vert}=\frac{1}{2\pi}\int d^2\mathbf{q}_\parallel\frac{e^{i\mathbf{q}_\parallel(\mathbf{r}_\parallel-\mathbf{r}_\parallel')}}{q_\parallel}e^{-q_\parallel\vert x-x'\vert},
\end{equation}
with $\mathbf{q}_\parallel \equiv q_y\mathbf{e}_y + q_z\mathbf{e}_z$ and $\mathbf{r}_\parallel \equiv y\mathbf{e}_y + z\mathbf{e}_z$,
we can write the Green's tensor as
\begin{multline}\label{Greenstensorplanewaves}
    \tensorbar{\mathcal{G}}(\mathbf{r}-\mathbf{r}') = -\frac{1}{8\pi^2}\int d^2\mathbf{q}_\parallel e^{i\mathbf{q}_\parallel(\mathbf{r}_\parallel-\mathbf{r}_\parallel')}e^{-q_\parallel\vert x-x'\vert}
    \\
    \left[
    \begin{array}{ccc}
        2\delta(x-x')-q_\parallel & -iq_y\xi_{xx'} & -iq_z\xi_{xx'} \\
         -iq_y\xi_{xx'} & q_y^2/q_\parallel & q_yq_z/q_\parallel\\
          -iq_z\xi_{xx'} & q_yq_z/q_\parallel & q_z^2/q_\parallel
    \end{array}
    \right]
\end{multline}
where $\xi_{xx'}\equiv\text{sign}[x-x']$. Note that the above expression is still general, but expressed in a convenient form for a film.

The linearized Landau-Lifshitz equations Eq.~\eqref{linearizedLLEappendix} together with the magnetostatic relation Eq.~\eqref{hgreenfunctionmappendix} forms an integro-differential equation for the magnetization field $\mathbf{m}(\mathbf{r},t)$. It is complemented by two sets of boundary conditions. On the one hand, the usual Maxwell boundary conditions, namely the continuity of $\mathbf{e}_z\times \mathbf{h}(\mathbf{r},t)$ and of $\mathbf{e}_z\cdot\mathbf{b}(\mathbf{r},t) \equiv \mu_0\mathbf{e}_z\cdot\left[\mathbf{h}(\mathbf{r},t)+\mathbf{m}(\mathbf{r},t)\right]$ across the two film boundaries (in our thin film geometry, at $x=0$ and $x=d$). On the other hand, the boundary conditions imposed by the exchange interaction, which are geometry- and material-dependent. We choose free-pinning boundary conditions~\cite{RadoJPCS1959,Mills2006},
\begin{equation}\label{freepinningbc}
    \frac{\partial}{\partial x}\mathbf{m}(\mathbf{r},t)\big\vert_{x=0,d}=0,
\end{equation}
which are best suited to thin films with $d\lesssim 500$nm~\cite{WangPRL2019}. Our results can be generalized to other pinning conditions \cite{KalinikosJPHYSC1986,Kalinikos1994,PattonPHYSREP1984} with little change in the final mode structure, especially for thin films \cite{deWamesAPS1969,DeWamesJAP1970,deWamesPRL1971,WolframPRL1970,YuPRB2019,KostylevJAP2013}. In some limiting cases the above integro-differential equation can be solved easily, for instance when exchange is negligible (dipolar spin waves, corresponding to $\alpha_x\ll \lambda_{\rm sw}^2$, with $\lambda_{\rm sw}^2$ the wavelength of the spin wave) or when it largely dominates over the dipole-dipole interaction (exchange waves, $\alpha_x\gg \lambda_{\rm sw}^2$). However, in many structures, including the thin films under study, both exchange and dipole-dipole interaction are relevant and their competition is responsible for the complex phenomenology of the so-called magnetostatic dipole-exchange spin waves \cite{PattonPHYSREP1984,Kalinikos1994}.

\subsection{Eigenmode equation}\label{AppSW_EigenmodeEquation}

The dipole-exchange eigenmodes of the linearized Landau-Lifshitz equation Eq.~\eqref{linearizedLLEappendix} can be computed in different forms \cite{PattonPHYSREP1984,Kalinikos1994}, even exactly~\cite{DeWamesJAP1970,deWamesPRL1971}, although the exact solution is numerically demanding. Here we follow the perturbative approach by Kalinikos in Ref.~\cite{KalinikosJPHYSC1986} (see also Refs.~\cite{YuPRB2019,KalinikosSoviet1981,KostylevJAP2013}). 
The first part of the derivation consists on manipulating the equations into a suitable form for the perturbative expansion.
We start by decomposing the magnetization field in terms of eigenmodes,
\begin{equation}
    \mathbf{m}(\mathbf{r},t) = \sum_\beta \mathbf{m}_\beta(\mathbf{r}) e^{-i\omega t} + \text{C.c.}
\end{equation}
and making use of the translational invariance to write the magnetization mode functions as
\begin{equation}\label{mdecompose}
    \mathbf{m}_\beta(\mathbf{r}) = \mathbf{m}_\beta(x)e^{i\mathbf{k}_\parallel \mathbf{r}_\parallel}.
\end{equation}
By introducing the above expressions into Eq.~\eqref{hgreenfunctionmappendix} and using the Green's tensor Eq.~\eqref{Greenstensorplanewaves} we find the same dependence for the associated field of the spin waves, namely
\begin{equation}
    \mathbf{h}(\mathbf{r},t) = \sum_\beta \mathbf{h}_\beta(\mathbf{r}) e^{-i\omega t} + \text{C.c.}
\end{equation}
with
\begin{equation}\label{hdecompose}
    \mathbf{h}_\beta(\mathbf{r}) = \mathbf{h}_\beta(x)e^{i\mathbf{k}_\parallel \mathbf{r}_\parallel},
\end{equation}
and the following one-dimensional relation between the transverse profiles,
\begin{equation}
    \mathbf{h}_\beta(x)=\int_0^d dx'\tensorbar{\mathcal{G}}_0(x-x')\mathbf{m}_\beta(x').
\end{equation}
The one-dimensional Green's function of the slab can be calculated by direct integration and reads
\begin{multline}\label{G0onedim}
    \tensorbar{\mathcal{G}}_0(x-x')=\frac{e^{-k_\parallel\vert x-x'\vert}}{2}
    \\
    \left[
    \begin{array}{ccc}
        k_\parallel-2\delta(x-x') & -ik_y\xi_{xx'} & -ik_z\xi_{xx'} \\
        -ik_y\xi_{xx'} & -k_y^2/k_\parallel & -k_yk_z/k_\parallel\\
        -ik_z\xi_{xx'}&-k_yk_z/k_\parallel& -k_z^2/k_\parallel
    \end{array}
    \right].
\end{multline}
Using the above relations, the linearized Landau-Lifshitz equation Eq.~\eqref{linearizedLLEappendix} can be cast as the following matrix equation for the two-dimensional vector $\mathbf{m}_\beta(x)=m_{\beta x}(x)\mathbf{e}_x+m_{\beta y}(x)\mathbf{e}_y$:
\begin{equation}\label{eqform}
    \hat{N}\mathbf{m}_\beta(x)=i\frac{\omega}{\omega_M}\tensorbar{T}\mathbf{m}_\beta(x)+\int_0^d dx'\tensorbar{\mathcal{G}}_0(x-x')\mathbf{m}_\beta(x')
\end{equation}
where we define the linear operator
\begin{equation}
    \hat{N} \equiv \left[\frac{\omega_H}{\omega_M}+\alpha_xk_\parallel^2-\alpha_x\frac{\partial^2}{\partial_x^2}\right]\mathbb{1}_{3\times3},
\end{equation}
and the matrix 
\begin{equation}
    \tensorbar{T} \equiv \left[\begin{array}{ccc}
        0 & -1 &0 \\
        1 &  0 &0 \\
        0 & 0 & 0
    \end{array}\right].
\end{equation}
Note that the equations do not couple modes with different values of $\mathbf{k}_\parallel$, thus confirming that the parallel wavevector is a good mode index.
The above system is combined with the free-spin boundary conditions
\begin{equation}\label{boundarycondition}
    \partial_x\mathbf{m}_\beta(x)\vert_{x=0,d}=0.
\end{equation}
In the above form the problem has been reduced from a four-variable ($\mathbf{r},t$) to a one-variable ($x$) integro-differential equation.

The second step is to transform the above simplified integro-differential equation Eq.~\eqref{eqform} into a system of algebraic equations, by expressing it in a suitable orthogonal basis. Since in thin films exchange interaction is the most relevant, we choose as a basis the vector eigenmodes of the operator $\hat{N}$ that satisfy the free-spin boundary conditions, i.e., $\mathbf{S}_n^p(x)=S_n(x)\mathbf{e}_p$ ($p=x,y$) with
\begin{equation}\label{basisfunctionsexchange}
    S_n(x)\equiv\sqrt{\frac{2}{(1+\delta_{n0})}}\cos\left(n\pi\frac{x}{d}\right).
\end{equation}
It can be easily checked that the above functions are eigenmodes of $\hat{N}$ with eigenvalue
\begin{equation}\label{Nnvalue}
    N_n(k_\parallel) \equiv \frac{\omega_H}{\omega_M}+\alpha_x\left[k_\parallel^2 + \left(\frac{n\pi}{d}\right)^2\right],
\end{equation}
that they fulfill the boundary conditions Eq.~\eqref{boundarycondition}, and that they form an orthogonal basis and are normalized to the film thickness $d$,
\begin{equation}
    \int_0^d dx \mathbf{S}_n^p(x)\cdot\mathbf{S}_{n'}^{p'}(x)=d\delta_{nn'}\delta_{pp'}.
\end{equation}
We now expand the magnetization vector in this basis as
\begin{equation}\label{mxexpansioneigenmodesN}
    \mathbf{m}_\beta(x) = \sum_{n}m_n^x\mathbf{S}_n^x(x)+m_n^y\mathbf{S}_n^y(x),
\end{equation}
and seek an equation for the unknown two mode amplitudes $\boldsymbol{\mathscr{m}}_n \equiv (m_n^x,m_n^y)^T$. Specifically, by using the orthogonality relation between the basis elements, we cast the integro-differential equation Eq.~\eqref{eqform} into an infinite algebraic system of coupled equations for the  amplitude vectors:
\begin{equation}\label{matrixEq1}
    \tensorbar{D}^{(nn)}\boldsymbol{\mathscr{m}}_n+\sum_{n'\ne n} \tensorbar{R}^{(nn')}\boldsymbol{\mathscr{m}}_{n'}=0,
\end{equation}
where the $2\times2$ matrices above, which depend on the parallel wavevector $\mathbf{k}_\parallel$, are given by
\begin{equation}\label{Dnndefinition}
    \tensorbar{D}^{(nn)} =\left[
    \begin{array}{cc}
        N_n(k_\parallel)-\Gamma_{nn}^{xx} &  -\Gamma_{nn}^{xy}+i(\omega/\omega_M)\\
        -\Gamma_{nn}^{yx}-i(\omega/\omega_M) & N_n(k_\parallel)-\Gamma_{nn}^{yy}
    \end{array}
    \right]
\end{equation}
\begin{equation}
    \tensorbar{R}^{(nn')} =\left[
    \begin{array}{cc}
         -\Gamma_{nn'}^{xx} &  -\Gamma_{nn'}^{xy}\\
        -\Gamma_{nn'}^{yx} & -\Gamma_{nn'}^{yy}
    \end{array}
    \right]
\end{equation}
Note that the sub-indices $n$ and $n'$ in the matrices $\tensorbar{D}^{(nn)}$ and $\tensorbar{R}^{(nn')}$ are mode labels and do not represent the matrix indices. 
In the above expressions, 
\begin{equation}\label{Gammannpp}
    \Gamma_{nn'}^{pp'}(\mathbf{k}_\parallel)\equiv\frac{1}{d}\int_0^ddx\int_0^ddx'S_n(x)S_{n'}(x')\mathcal{G}_0^{pp'}(x-x')
\end{equation}
represents the dipole-dipole interaction term expanded in the basis $\mathbf{S}^p_n(x)$. The dipole-dipole interaction appears in both matrices $\tensorbar{D}^{(nn)}$ and $ \tensorbar{R}^{(nn')}$. On $\tensorbar{D}^{(nn)}$, it accounts for the self-dipole-dipole interaction of each eigenmode of $\hat{N}$. This contribution is important and will be taken into account fully, i.e., non-perturbatively. Conversely, the dipole-dipole interaction between different eigenmodes of $\hat{N}$ is fully captured by the term $ \tensorbar{R}^{(nn')}$, and will be treated perturbatively.

\subsection{Perturbative eigenmode calculation}\label{AppSW_Perturbationtheory}

Let us briefly sketch the perturbative treatment of the above system of equations. By combining all the two-dimensional vectors $\boldsymbol{\mathscr{m}}_n$ into an infinite-dimensional vector $\boldsymbol{\mathscr{M}} \equiv (\boldsymbol{\mathscr{m}}_0,\boldsymbol{\mathscr{m}}_1,...)^T$, we can cast the system of equations
Eq.~\eqref{matrixEq1} as 
\begin{equation}
    \mathfrak{L}\boldsymbol{\mathscr{M}}=0,
\end{equation}
where $\mathfrak{L}$ is an infinite-dimensional matrix. Note that throughout this section we reserve the double-bar tensor notation $\tensorbar{(*)}$ for $2\times 2$ matrices for the sake of clarity. The $2\times2$ block entries of the matrix $\mathfrak{L}$ are given by
 \begin{equation}\label{Lblockentries}
     \tensorbar{\mathfrak{L}}_{\{nn'\}} = \delta_{nn'}\tensorbar{D}^{(nn)} + (1-\delta_{nn'})\tensorbar{R}^{(nn')}.
\end{equation}
Here, the indices given inside curly brackets denote $2\times2$ block entries, i.e. $\tensorbar{\mathfrak{L}}_{\{nn'\}}$ is the $2\times 2$ matrix occupying the $\{n,n'\}$ block of $\mathfrak{L}$.
The spin wave eigenmodes are in general a linear combination of all the modes $\boldsymbol{\mathscr{m}}_n$. Their dispersion relations are given by the implicit equation
\begin{equation}\label{detLequal0}
    \det[\mathfrak{L}]=0.
\end{equation}
We aim at keeping the diagonal $2\times 2$ block entries of Eq.~\eqref{Lblockentries} as our unperturbed matrix, and apply perturbation theory to include the blocks $\tensorbar{R}^{(nn')}$. We will do this through block-diagonalization.

Let us assume it is possible to block-diagonalize the matrix $\mathfrak{L}$, i.e. that we can solve the following generalized block eigenvalue equation:
\begin{equation}\label{generalblockdiagequation}
    \mathfrak{L}\vert \mathbf{V}_{l\alpha}\rangle = \Lambda_l\vert \mathbf{V}_{l\alpha}\rangle
\end{equation}
Here, $l$ is an index labelling the block eigenvalues and the block eigenvectors. We define the infinite block-diagonal matrix $\Lambda_l$, whose $2\times2$ block entries are given by
\begin{equation}
    \tensorbar{\left[\Lambda_l\right]}_{\{ij\}}=\delta_{ij}\tensorbar{\lambda}_l,
\end{equation}
i.e., the diagonal blocks are all identical and equal to the $2\times2$ eigenblock $\tensorbar{\lambda}_l$. 
The block eigenvectors $\vert \mathbf{V}_{l\alpha}\rangle$, expressed in Dirac notation for convenience, are matrices formed by $2$ infinite column vectors. Because in a block diagonalization there might be more than one linearly independent vector corresponding to the same block eigenvalue, we have introduced an auxiliary index $\alpha$ labelling all the eigenvectors within each manifold $l$. 
If the block eigenvalue problem Eq.~\eqref{generalblockdiagequation} is solved, one can transform the coefficient matrix of our system of equations into a block-diagonal form,
\begin{equation}
    \mathfrak{L} \longrightarrow \mathfrak{L}'.
\end{equation}
The block $2\times 2$ entries of the block-diagonal matrix $\mathfrak{L}'$ are given by
\begin{equation}
     \tensorbar{\mathfrak{L}'}_{\{ll'\}} = \delta_{ll'}\tensorbar{\lambda}_l.
\end{equation}
Due to this convenient structure the dispersion relations for the eigenmodes, Eq.~\eqref{detLequal0}, are easily obtained as the determinant factorizes into a product of determinants of $2\times2$ matrices:
 \begin{equation}
     \text{det}[\mathfrak{L}]=\text{det}[\mathfrak{L}'] = \prod_{l=0}^\infty\det\left[\tensorbar{\lambda}_l\right]=0.
 \end{equation}
 Moreover, we can identify $l$ as an additional good mode index labelling the eigenmodes. The magnetization mode functions corresponding to mode $l$
can analogously be computed from the corresponding eigenvectors, $\vert \mathbf{V}_{l\alpha}\rangle$. The problem of finding the spin wave eigenmodes and eigenfrequencies is thus reduced to block-diagonalizing the original matrix $\mathfrak{L}$.

The block-diagonalization of the matrix $\mathfrak{L}$ is carried out by dividing it into an unperturbed diagonal part $\mathfrak{L}_d$ and an off-diagonal perturbation $\mathfrak{L}_o$, given in block form by
\begin{equation}
    \tensorbar{\left[\mathfrak{L}_d\right]}_{\{nn'\}} = \delta_{nn}\tensorbar{D}^{(nn)},
\end{equation}
 \begin{equation}
    \tensorbar{\left[\mathfrak{L}_o\right]}_{\{nn'\}} =  (1-\delta_{nn'})\tensorbar{R}^{(nn')}.
\end{equation}
The diagonal part is already in a block-diagonal form, and hence the zero-th order solution in the perturbation $\tensorbar{R}^{(nn')}$, namely the block-eigenvalue matrix $\Lambda_l^{(0)}$, is simply given by
\begin{equation}
    \tensorbar{\left[\Lambda_l^{(0)}\right]}_{\{ij\}}=\delta_{ij}\tensorbar{\lambda}_l^{(0)}=\delta_{ij}\tensorbar{D}^{(ll)}.
\end{equation}
The zero-th order block eigenvectors are similarly given by $\vert \mathbf{V}_{l\alpha}^{(0)}\rangle = (\tensorbar{0},\tensorbar{0},\hdots,\tensorbar{V}_{l\alpha}^{(0)},\hdots)^T$, i.e. they are zero except for a $2\times 2$ block matrix at position $l$. Since this matrix is arbitrary we can identify four linearly independent block eigenvectors labelled by the index $\alpha=1,...4$. Hereafter and for simplicity, we choose these eigenvectors such that they form an orthonormal basis according to the inner product
\begin{multline}\label{orthogonalityVecs0}
    \langle  \mathbf{V}_{l\alpha}^{(0)} \vert \mathbf{V}_{l'\beta}^{(0)}\rangle\equiv 
    \\
    \equiv \text{Tr}\left(\tensorbar{0},\tensorbar{0},\hdots,\tensorbar{V}_{l\alpha}^{(0)\dagger},\hdots\right)
    \left(\begin{array}{c}
         \tensorbar{0}  \\
         \vdots \\
         \tensorbar{V}_{l'\beta}^{(0)}\\
         \vdots
    \end{array}\right)
   = \delta_{ll'}\delta_{\alpha\beta}. 
\end{multline}
Note that the above condition amounts to choosing the matrices $\tensorbar{V}_{l\alpha}^{(0)}$ to form an orthonormal basis in the space of complex $2\times2$ matrices. 

Once the above orthogonality relations and the zero-th order solution are defined, we
can proceed in an analogous way as in usual perturbation theory. First, we expand the block-diagonalization equation Eq.~\eqref{generalblockdiagequation} in orders of the perturbation as
\begin{multline}\label{expansionorders}
    \left(\mathfrak{L}_d+\mathfrak{L}_o\right)\left(\vert\mathbf{V}_{l\alpha}^{(0)}\rangle+\vert\mathbf{V}_{l\alpha}^{(1)}\rangle+...\right) = \\
    =\left(\Lambda_l^{(0)}+\Lambda_l^{(1)}+\Lambda_l^{(2)}...\right)\left(\vert\mathbf{V}_{l\alpha}^{(0)}\rangle+\vert\mathbf{V}_{l\alpha}^{(1)}\rangle...\right),
\end{multline}
where the terms with super-index $n$ are of $n-$th order. To obtain the first-order correction to the equation we discard all the terms of order $2$ or larger, obtaining
\begin{equation}
    \mathfrak{L}_d\vert\mathbf{V}_{l\alpha}^{(1)}\rangle + \mathfrak{L}_o\vert\mathbf{V}_{l\alpha}^{(0)}\rangle = \Lambda_l^{(0)}\vert\mathbf{V}_{l\alpha}^{(1)}\rangle + \Lambda_l^{(1)}\vert\mathbf{V}_{l\alpha}^{(0)}\rangle.
\end{equation}
We can now overlap on the left with $\langle\mathbf{V}_{l\beta}^{(0)}\vert$ and use the hermiticity properties of the zero-th order solution and the orthogonality relation Eq.~\eqref{orthogonalityVecs0} to obtain
\begin{equation}
    \langle\mathbf{V}_{l\beta}^{(0)}\vert\Lambda_l^{(1)}\vert\mathbf{V}_{l\alpha}^{(0)}\rangle = \langle\mathbf{V}_{l\beta}^{(0)}\vert\mathfrak{L}_o\vert\mathbf{V}_{l\alpha}^{(0)}\rangle = 0,
\end{equation}
where the last equality stems from the block-off-diagonal character of $\mathfrak{L}_o$. Since the above equation holds for any $\alpha$ and $\beta$ and the matrices $\tensorbar{V}_{l\alpha}^{(0)}$ form an orthonormal basis, we conclude that the first-order correction to the block eigenvalues is zero,
\begin{equation}
    \Lambda_l^{(1)} = 0 \hspace{0.3cm}\rightarrow \hspace{0.3cm} \tensorbar{\lambda}_{l}^{(1)} = \tensorbar{0}.
\end{equation}
By following the usual route in perturbation theory one can easily find the first correction to the eigenvectors as
\begin{equation}\label{finalV1decomposition}
    \vert\mathbf{V}_{l\alpha}^{(1)}\rangle = \sum_{l'\ne l,\beta}\frac{\langle \mathbf{V}_{l'\beta}^{(0)}\vert\mathfrak{L}_o\vert\mathbf{V}_{l\alpha}^{(0)}\rangle}{\det\Big[\tensorbar{D}^{(ll)}\Big]-\det\Big[\tensorbar{D}^{(l'l')}\Big]}\vert\mathbf{V}_{l'\beta}^{(0)}\rangle.
\end{equation}
In a similar way, it is possible to obtain eigenvalues and eigenvectors up to any order in perturbation theory \cite{KalinikosJPHYSC1986}.

\subsection{Mode properties}\label{AppSW_Modeproperties}

To describe the properties of the spin waves in thin films, where exchange interaction dominates over dipole-dipole coupling, it is usually sufficient to include the dipole-dipole coupling up to first order in perturbation theory~\cite{BertelliSciAdv2020,KalinikosJPHYSC1986,Kalinikos1994,LeeWongNanoLetters2020}. 
Here, since the first-order correction to the block eigenvalues is zero, the eigenfrequency equation reads
\begin{equation}
     \text{det}[\mathfrak{L}'] = \prod_{n=0}^\infty\det\Big[\tensorbar{D}^{(nn)}\Big] = 0 + \mathcal{O}\Big(\Big[\tensorbar{R}^{(nn')}\Big]^2\Big),
 \end{equation}
 and hence we can still identify the index $n$ introduced in Eq.~\eqref{basisfunctionsexchange} as a good mode index labelling different spin wave bands. 
 We can now group the three mode indices of a spin wave, namely $n$ and the parallel wavevector $\mathbf{k}_\parallel$, into a single compound index
 \begin{equation}
     \beta \equiv \{n,\mathbf{k}_\parallel\}.
 \end{equation}
 The dispersion relation of band $n$ is given by
 \begin{equation}\label{Dnnredef}
     \det\Big[\tensorbar{D}^{(nn)}\Big] = \det\left[
     \begin{array}{cc}
          \nu_{\beta x} & i\omega/\omega_M \\
        -i\omega/\omega_M & \nu_{\beta y}
     \end{array}
     \right] = 0.
 \end{equation}
 Here, we have used the expression of the matrix $\tensorbar{D}^{(nn)}$ Eq.~\eqref{Dnndefinition} and calculated the elements $\Gamma_{nn}^{pp'}$ (Eq.~\eqref{Gammannpp}) explicitly. Specifically, $\Gamma_{nn}^{xy}=0$, and we define
 \begin{equation}\label{nux}
     \nu_{\beta x} \equiv N_n(k_\parallel)-\Gamma_{nn}^{xx}(k_\parallel) = N_n(k_\parallel) +1-P_{nn}(k_\parallel),
 \end{equation}
  \begin{equation}\label{nuy}
     \nu_{\beta y} \equiv N_n(k_\parallel)-\Gamma_{nn}^{yy}(\mathbf{k}_\parallel) = N_n(k_\parallel) +\frac{k_y^2}{k_\parallel^2}P_{nn}(k_\parallel) ,
 \end{equation}
 with
\begin{multline}\label{Pnn}
    P_{nn}(k_\parallel)\equiv\frac{k_\parallel}{d}\int_0^ddx\int_0^ddx'\frac{e^{-k_\parallel\vert x-x'\vert}}{2}S_n(x)S_n(x')
    \\
    = \frac{\delta_{n0}}{2} + \left[\frac{k_\parallel d}{(k_\parallel d)^2 + (n\pi)^2}\right]^2
    \\
        \times \frac{\left[(k_\parallel d)^2 + (n\pi)^2-2k_\parallel d(1-(-1)^ne^{-k_\parallel d})\right]}{1+\delta_{n0}}.
\end{multline}
The dispersion relation of band $n$ is thus compactly written as
\begin{equation}\label{omegaSW}
    \omega_\beta = \omega_n(\mathbf{k}_\parallel) = \omega_M\sqrt{\nu_{\beta x}\nu_{\beta y}}.
\end{equation}
These frequencies correspond to the curves in \figref{FigureSystemBandStructure}(b).

\begin{figure}[t!]
	\centering
	\includegraphics[width=\linewidth]{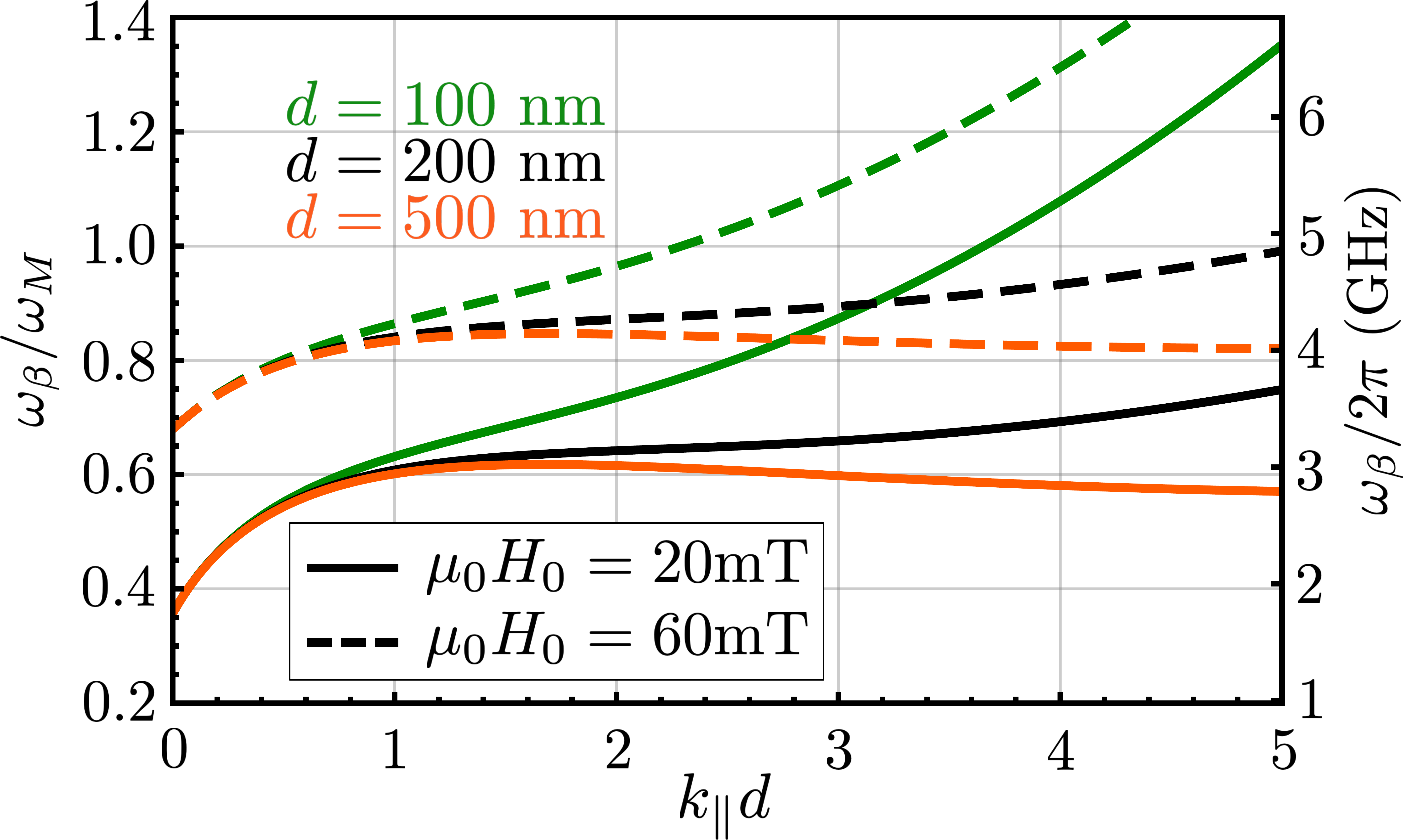}
	\caption{$n=0$ spin wave band of a YIG film for $\mu_0H_0 = 20$mT (solid lines) and $\mu_0H_0=60$mT (dashed lines), and for three values of the film thickness, namely $d=100$nm (green), $d=200$nm (black), $d=500$nm (orange). }\label{FigbanddependenceB0d}
\end{figure}

Let us briefly explore the properties of the spin wave frequencies. First, we consider the insightful long- and short-wavelength limits $k_\parallel d \ll 1$ and $k_\parallel d \gg 1$, which correspond to $P_{nn}\approx 0$ and $P_{nn} \approx 1$, respectively. In the long wavelength limit, the dispersion relation in the case of negligible exchange (i.e., moderate $n$ and $\alpha_x/d^2 \lesssim 1$) tends to the expression of the uniformly precessing mode in the absence of exchange~\cite{StancilBook2009},
\begin{equation}\label{omegalongwavelengthlimit}
    \omega_n(\mathbf{k}_\parallel) \approx \sqrt{\omega_H(\omega_H+\omega_M)}.
\end{equation}
On the other hand, if exchange is relevant (i.e., large $n$ and $\alpha_x/d^2 \gtrsim 1$), then $N_n (k_\parallel) \gg 1$ and the long-wavelength limit becomes a quadratic function of the wavenumber~\cite{YuPRB2019} (see also Eq.~[\ref{Nnvalue}])
\begin{equation}\label{omegalongwavelengthlimit2}
    \omega_n(\mathbf{k}_\parallel) \approx \omega_M\left[N_n(k_\parallel)+\frac{1}{2}\right].
\end{equation}
For short wavelength modes, the exchange interaction always dominates and one obtains, for $n \ll k_\parallel d/\pi$, the dispersion of exchange spin waves in a bulk material~\cite{PattonPHYSREP1984},
\begin{multline}
    \omega_n(\mathbf{k}_\parallel) \\\approx \left(\omega_H + \omega_M\alpha_x k_\parallel^2\right)\sqrt{1+\frac{\omega_M\sin^2\phi_k}{\omega_H + \omega_M\alpha_x k_\parallel^2}}.
\end{multline}
Second, note that, aside from the three dimensionless mode indices $n,k_\parallel d, \phi_k$, the spin wave eigenfrequencies depend explicitly on $H_0$ and the film thickness $d$. As shown by \figref{FigbanddependenceB0d}, increasing $H_0$ leads to the bands shifting to higher frequencies (see Eq.~\eqref{omegalongwavelengthlimit}), while decreasing $d$ leads to the band becoming increasingly parabolic as exchange interaction becomes more relevant.

\begin{figure}[tbh!]
	\centering
	\includegraphics[width=\linewidth]{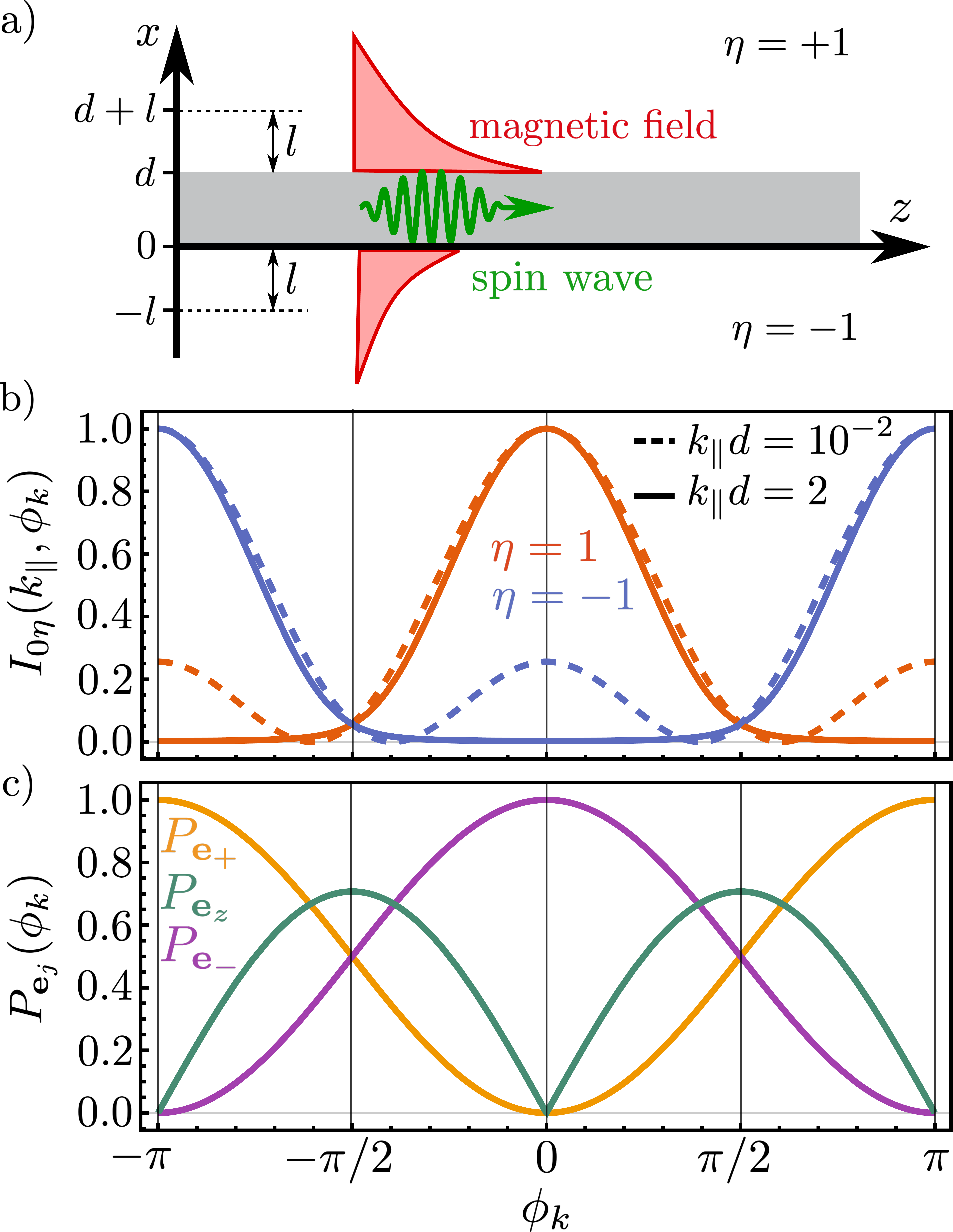}
	\caption{a) The magnetic field of a spin wave decays exponentially with the vertical distance from the film, $l$, and is different above ($\eta=+1$) and below ($\eta=-1$) the film. b) Normalized field intensity (see text for details) of the spin wave above/below a YIG film (red/blue lines, respectively) as a function of propagation direction $\phi_k$, for $n=0$, $\mu_0H_0 = 20$mT, and $k_\parallel d=2$ (solid lines) and $k_\parallel d=10^{-2}$ (dashed lines). c) Degree of polarization of the spin wave field above the film along the cylindrical basis vectors Eq.~\eqref{cylindricalvectors}. }\label{FigureModalFields}
\end{figure}

Let us now focus on the magnetization eigenmodes. Since for thin films the different bands are usually very spaced in energies (see e.g.  \figref{FigureSystemBandStructure}[b]), the first-order corrections Eq.~\eqref{finalV1decomposition}, which are inversely proportional to $\omega_n(\mathbf{k}_\parallel) - \omega_{n'}(\mathbf{k}_\parallel')$, are typically small and become only relevant at the band crossings where they result in a lifting of the degeneracy via mode splitting~\cite{KalinikosJPHYSC1986}. Here we will work far from these particular points and thus describe the eigenmodes through their zero-th order expression~\cite{YuPRB2019,BertelliSciAdv2020}, given by the diagonal part of Eq.~\eqref{matrixEq1}
\begin{equation}
    \tensorbar{D}^{(nn)}\mathbf{m}_n = 0.
\end{equation}
Combining the above equation with Eqs.~\eqref{Dnnredef} and \eqref{omegaSW} we obtain the following relation for the spin wave amplitudes,
\begin{equation}
    \mathbf{m}_n \propto \left[\begin{array}{c}
         -i\sqrt{\nu_{\beta y}}  \\
         \sqrt{\nu_{\beta x}} 
    \end{array}\right].
\end{equation}
The magnetization mode function of an eigenmode is thus obtained directly from its definition Eq.~\eqref{mdecompose}, together with Eqs.~\eqref{basisfunctionsexchange} and \eqref{mxexpansioneigenmodesN}:
\begin{equation}\label{mmodefunctions}
    \mathbf{m}_\beta(\mathbf{r}) = \sqrt{\frac{2}{(1+\delta_{n0})}}\cos\left(n\pi\frac{x}{d}\right)e^{i\mathbf{k}_\parallel\mathbf{r}_\parallel}\left[\begin{array}{c}
         \sqrt{\nu_{\beta y}}  \\
         i\sqrt{\nu_{\beta x}} \\
         0
    \end{array}\right]
\end{equation}
in Cartesian coordinates, up to an arbitrary normalization constant.
Finally, the magnetic field mode functions are obtained using the identities Eqs.~\eqref{hdecompose}-\eqref{G0onedim}. Here we focus on the fields outside the film, and introduce an additional index $\eta \equiv \text{sign}[x]$ to differentiate between the field above and below the film ($\eta=+1$ and $\eta=-1$ respectively, see \figref{FigureModalFields}[a]). The mode functions outside the film can be written in a compact form as
\begin{multline}\label{hmodefunctions}
    \mathbf{h}_{\beta\eta}(\mathbf{r}) = \mathbf{h}_{n\eta}(\mathbf{k}_\parallel,\mathbf{r})=
    \\
    =h_{\beta \eta 0} e^{i\mathbf{k}_\parallel\mathbf{r}_\parallel} e^{-k_\parallel  l}\left[\begin{array}{c}
         1  \\
         - i \eta k_y/k_\parallel \\
         - i \eta k_z/k_\parallel
    \end{array}\right]
\end{multline} 
where we have defined the absolute distance from the surface of the film,
\begin{equation}\label{ldefinition}
    l \equiv \left\lbrace
    \begin{array}{cc}
    x-d & \text{ for } x>d \\
    -x & \text{ for } x<0.
    \end{array}
    \right.
\end{equation}
The amplitudes in Eq.~\eqref{hmodefunctions} are given by
 \begin{multline}\label{hfieldamplitude}
     h_{\beta \eta 0} = \sqrt{\frac{2}{(1+\delta_{n0})}}
    \left(\sqrt{\nu_{\beta y}}+\eta \frac{k_y}{k_\parallel} \sqrt{\nu_{\beta x}}\right)
    \frac{\eta^n}{2}
    \\
    \times\frac{(k_\parallel d)^2}{(k_\parallel d)^2 + (n\pi)^2}
    (-e^{- k_\parallel d} + (-1)^n).
 \end{multline}
 Note that $h_{
 \beta\eta 0}$ is real and fulfills the symmetry properties
 \begin{equation}\label{h0symkz}
    h_{n + 0}(k_y,k_z)= h_{n + 0}(k_y,-k_z)
 \end{equation}
 and
 \begin{equation}\label{h0symky}
    h_{n + 0}(k_y,k_z)= (-1)^nh_{n - 0}(-k_y,k_z)
 \end{equation}
 which will be useful in the following sections.
 
It is insightful, especially regarding the coupling of spin waves to paramagnetic spins, to explore the properties of the magnetic field mode functions. 
According to Eq.~\eqref{hmodefunctions} the modal field has a plane-wave dependence on the $y$ and $z$ coordinates and decays exponentially with the distance to the film $l$ (see \figref{FigureModalFields}(a) or Eq.~\eqref{ldefinition} in the main text). Both its polarization and its amplitude depend on wavenumber $k_\parallel$, propagation direction $\phi_k$, and film side $\eta$, giving rise to rich phenomenology \cite{Kalinikos1994,PattonPHYSREP1984}. To gain insight into it, we define two position-independent quantities describing the field intensity and direction. First, the normalized field intensity $I_{n\eta}(k_\parallel,\phi_k) \equiv \vert h_{\beta\eta 0}\vert^2/\text{max}_{k_\parallel,\phi_k}\vert h_{\beta\eta 0}\vert^2$, displayed in \figref{FigureModalFields}(b) for the fundamental band $n=0$ (similar results are obtained for higher $n$). As shown by this figure, the field intensity is different on both sides of the film except in the parallel propagation case, $\phi_k=\pm\pi/2$. Additionally, as evidenced by the differences between solid and dashed lines, the ratio between the field intensities on each side of the film depends on $k_\parallel d$. 
This so-called \emph{modal-profile non-reciprocity} is well known in the literature \cite{KalinikosJPHYSC1986,KostylevJAP2013} and, for spin waves fulfilling $\sqrt{\nu_{\beta y}} = -\eta\cos(\phi_k)\sqrt{\nu_{\beta x}}$, reaches its most pronounced manifestation as the field completely vanishes on one side of the film. In the relevant limits $k_\parallel d \gtrsim 1$ (solid lines in \figref{FigureModalFields}(b)) or $d^2/\alpha_x \lesssim k_\parallel d \lesssim 1$ \footnote{Note that the regime $d^2/\alpha_x \lesssim k_\parallel d \lesssim 1$ is not realizable with the parameters in Table~\ref{tablePARAMS} as $d^2/\alpha_x \approx 130$.} this maximum asymmetry is reached for Damon-Eshbach propagation, $\phi_k=0,\pi$ (see e.g. \cite{YuPRB2019,KostylevJAP2013}). To quantify the direction of the modal field, we define its degree of polarization along a unit vector $\mathbf{e}_j$ above the film $(\eta=+1)$ as $P_{\mathbf{e}_j}(\phi_k) \equiv \vert \mathbf{e}_j^*\cdot \mathbf{h}_{\beta +}(\mathbf{r})\vert/[\mathbf{h}^*_{\beta +}(\mathbf{r})\cdot \mathbf{h}_{\beta +}(\mathbf{r})]^{1/2}$, which by definition depends exclusively on the angle $\phi_k$. Note that the above quantity completely determines the polarization also below the film, since from Eq.~\eqref{hmodefunctions} we can deduce that the degree of polarization along $\mathbf{e}_j$ below the film is given by $P_{\mathbf{e}_j^*}(-\phi_k)$.
A further useful property derived from such equation is $\mathbf{h}_{\beta+}^*(\mathbf{r})\cdot \mathbf{h}_{\beta-}(\mathbf{r}) = 0$, i.e., the field polarizations above and below the film are always orthogonal. In \figref{FigureModalFields}(c) we display the degree of polarization as a function of $\phi_k$ for the three cylindrical basis vectors
\begin{equation}\label{cylindricalvectors}
    \{\mathbf{e}_+,\mathbf{e}_-,\mathbf{e}_z\} \equiv \left\{\frac{\mathbf{e}_x+i\mathbf{e}_y}{\sqrt{2}},\frac{\mathbf{e}_x-i\mathbf{e}_y}{\sqrt{2}},\mathbf{e}_z \right\}.
\end{equation}
Whereas for parallel propagation ($\phi_k=\pm\pi/2$) the modal field has components along all the basis vectors~\footnote{As a matter of fact, for parallel propagation the spin wave is circularly polarized in the $x-z$ plane, i.e., it has a longitudinal component of the same magnitude as its transverse component.}, in the Damon-Eshbach configuration the field is exactly polarized along one of the cylindrical vectors $\mathbf{e}_\pm$, with exactly the opposite helicity on the other side of the film. This polarization structure has a strong impact on the spin-wave dynamics in the presence of spin qubits as we discuss in the main text.

\subsection{Quantization, magnon dynamics, and magnon decay rates}\label{AppSW_quantizationEOMgamma}

We carry out the quantization of the spin waves following the procedure established by Mills for finite samples~\cite{Mills2006}. We do not show the details of this quantization as it has been discussed in detail in the literature~\cite{GonzalezBallesteroPRB2020}. For an infinitely extended film, we can extend this formalism following the quantization-in-a-box procedure common in quantum optics. First, we assume the film has an extension $L$ along $y$ and along $z$, and obeys periodic boundary conditions at the film edges $y=\pm L/2$ and $z=\pm L/2$. We then proceed with the quantization and take the continuum limit $L\to\infty$ when computing observables. In this limit, no observable will depend on the quantization length $L$. Following this procedure, we write the magnetization of the spin wave in the Schr\"odinger picture as
\begin{equation}\label{moperatorAppendix}
    \hat{\mathbf{m}}(\mathbf{r}) = \sum_\beta \mathcal{M}_{0\beta}\left[\mathbf{m}_\beta(\mathbf{r})\hat{s}_\beta + \text{H.c.}\right],
\end{equation}
where the ladder operators, which describe the creation and annihilation of spin wave quanta (magnons), obey bosonic commutation relations,
\begin{equation}\label{commutationmagnons}
    \left[\hat{s}_\beta,\hat{s}_{\beta'}^\dagger\right]=\delta_{\beta\beta'} \hspace{0.3cm} ; \hspace{0.3cm}  \left[\hat{s}_\beta,\hat{s}_{\beta'}\right]=0,
\end{equation}
and the zero-point magnetization is given by
\begin{equation}\label{zeropointMAppendix}
    \mathcal{M}_{0\beta} \equiv \sqrt{\frac{\hbar\vert\gamma\vert M_S}{2L^2d }\frac{\omega_M}{\omega_\beta}}.
\end{equation}
The magnetic field operator has a similar expression, namely
\begin{equation}\label{hoperator}
    \hat{\mathbf{h}}(\mathbf{r}) = \sum_\beta \mathcal{M}_{0\beta}\left[\mathbf{h}_\beta(\mathbf{r})\hat{s}_\beta + \text{H.c.}\right].
\end{equation}
Finally, the Hamiltonian operator for the spin waves reads
\begin{equation}\label{Hswappendix}
    \hat{H}_{\rm sw} = \hbar\sum_\beta\omega_\beta \hat{s}^\dagger_\beta\hat{s}_\beta.
\end{equation}

\begin{figure}[t!]
	\centering
	\includegraphics[width=\linewidth]{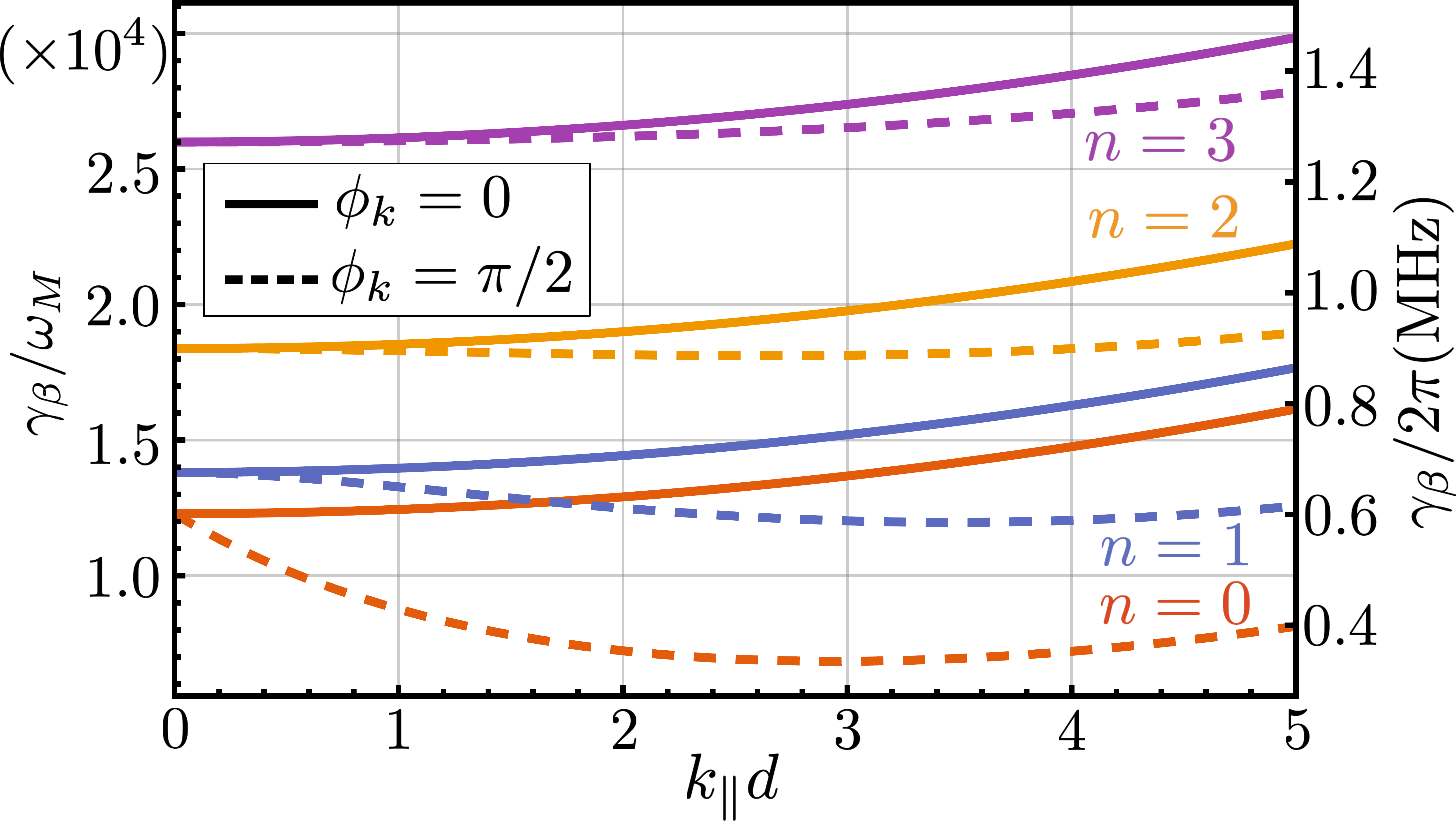}
	\caption{ b) Magnon decay rate Eq.~\eqref{gammabetaexpression} for the first four bands of YIG and the parameters of Table~\ref{tablePARAMS}. Solid and dashed lines correspond to Damon-Eshbach and parallel propagation, respectively. }\label{FigDecayRates}
\end{figure}

To properly account for the dynamics of spin waves, it is necessary to include their dissipation.
In a quantum framework, losses are included after quantization of the lossless classical system. In particular, we add a dissipative term $\mathcal{D}[\hat{\rho}_{\rm sw}]$ to the von Neumann equation for the spin wave density matrix, $\rho_{\rm sw}$ \cite{breuer2002theory},
\begin{equation}\label{MasterEquationOnlyMagnonsAppendix}
    \frac{d}{dt}\hat{\rho}_{\rm sw} = -\frac{i}{\hbar}\left[\hat{H}_{\rm sw},\hat{\rho}_{\rm sw}\right] + \mathcal{D}_{\rm sw}[\hat{\rho}_{\rm sw}].
\end{equation}
We choose the most common form for the dissipator, namely
\begin{equation}\label{dissipatorSWAppendix}
    \mathcal{D}_{\rm sw}[\hat{\rho}]=\sum_\beta \gamma_\beta\left(\bar{n}_\beta\mathcal{L}_{\hat{s}_\beta\hat{s}_\beta^\dagger}[\hat{\rho}]+(1+\bar{n}_\beta)\mathcal{L}_{\hat{s}_\beta^\dagger\hat{s}_\beta}[\hat{\rho}]\right),
\end{equation}
which describes absorption and decay into a thermal bath at temperature $T$, in terms of Lindblad superoperators defined as
\begin{equation}\label{LindbladiandefinitionAppendix}
    \mathcal{L}_{\hat{a}\hat{b}}[\hat{\rho}]\equiv\hat{a}\hat{\rho}\hat{b}-\frac{1}{2}\{\hat{b}\hat{a},\hat{\rho}\},
\end{equation}
with $\gamma_\beta$ the magnon decay rate and $\bar{n}_\beta = [\exp(\hbar\omega_\beta/k_B T)-1]^{-1}$ the Bose-Einstein distribution at the magnon frequency.
This dissipator can be rigorously derived by tracing out a reservoir coupled to spin waves via particle-conserving interactions, e.g. a phononic reservoir or a bath of two-level impurities~\cite{breuer2002theory,GonzalezBallesteroPRB2020,KusturaToAppear2020}.

 All the spin wave phenomenology is captured by the Master Equation Eq.~\eqref{dissipatorSWAppendix}. First, it can be shown that the steady-state of the spin waves is thermal, $\hat{\rho}_{\rm sw,ss} \propto \exp(-\hat{H}_{\rm sw}/k_B T)$.  Second, regarding the spin wave dynamics, we can compute the equation of motion for the expected value of any magnon operator $\hat{O}$ as
\begin{equation}
    \frac{d}{dt}\langle\hat{O}\rangle(t) = \frac{d}{dt}\text{Tr}[\hat{O}\hat{\rho}_{\rm sw}(t)].
\end{equation}
Furthermore, 
if the master equation is quadratic, as is the case for Eq.~\eqref{MasterEquationOnlyMagnonsAppendix}, the above equations of motion form a closed system of differential equations and can thus be solved exactly~\cite{breuer2002theory,CarmichaelBook}.
Here we compute the equations of motion for single magnon operators and for two-operator products, which are given by
\begin{equation}\label{EOMsfree}
    \frac{d}{dt}\langle\hat{s}_\beta\rangle(t) = \left(-i\omega_\beta-\frac{\gamma_\beta}{2}\right)\langle\hat{s}_\beta\rangle(t),
\end{equation}
\begin{multline}\label{EOMssfree}
    \frac{d}{dt}\langle\hat{s}_\beta\hat{s}_{\beta'}\rangle(t) =
    \\\left[-i(\omega_\beta+\omega_{\beta'})-\frac{\gamma_\beta+\gamma_{\beta'}}{2}\right]\langle\hat{s}_\beta\hat{s}_{\beta'}\rangle(t),
\end{multline}
\begin{multline}\label{EOMsdagsfree}
    \frac{d}{dt}\langle\hat{s}_\beta^\dagger\hat{s}_{\beta'}\rangle(t) = \\\left[i(\omega_\beta-\omega_{\beta'})-\frac{\gamma_\beta+\gamma_{\beta'}}{2}\right]\langle\hat{s}_\beta^\dagger\hat{s}_{\beta'}\rangle(t) + \gamma_\beta\bar{n}_\beta\delta_{\beta\beta'}.
\end{multline}
Finally, we can use the above results, together with the quantum regression formula~\cite{CarmichaelBook}, to calculate the steady-state two-time correlators of the magnon operators, which determine the effective dynamics of NV centres as shown in Appendix~\ref{Appendix_EffectiveNVdynamics}.  Specifically, for a thermal state, correlators between two creation or two annihilation operators vanish, i.e.,
\begin{equation}\label{ssquaredcorrelator}
    \langle\hat{s}_\beta(t)\hat{s}_{\beta'}(t-s)\rangle_{\rm ss} = 0 \hspace{0.3cm} \forall s.
\end{equation}
On the other hand, correlators between a creation and an annihilation operator are given, for $s\ge 0$, by
\begin{equation}\label{sdagscorrelator}
    \langle\hat{s}^\dagger_\beta(t)\hat{s}_{\beta'}(t-s)\rangle_{\rm ss} = \delta_{\beta\beta'}\bar{n}_\beta e^{(i\omega_\beta-\gamma_\beta/2)s} ,
\end{equation}
\begin{equation}\label{ssdagcorrelator}
    \langle\hat{s}_\beta(t)\hat{s}^\dagger_{\beta'}(t-s)\rangle_{\rm ss} = \delta_{\beta\beta'}(1+\bar{n}_\beta) e^{(-i\omega_\beta-\gamma_\beta/2)s} .
\end{equation}
The corresponding correlators for negative $s$ are obtained from the above expressions using the time-translational definitory property of the steady-state, i.e.,
\begin{equation}\label{steadystateinvariance}
    \langle \hat{O}_1(t)\hat{O}_2(t-s)\rangle_{\rm ss} = \langle \hat{O}_1(t+s)\hat{O}_2(t)\rangle_{\rm ss}
\end{equation}
for any two operators $\hat{O}_1$ and $\hat{O}_2$.

Once we have shown how to extract all the dynamical information of the spin waves from the master equation, the only step left is to relate the quantum decay rate, $\gamma_\beta$, to the measured spin wave lifetimes which are usually defined and measured in the classical limit. Classically, the losses of spin waves are modelled by adding the Gilbert damping term $(\alpha_G/M_S)\mathbf{M}(\mathbf{r},t)\times (d/dt){\mathbf{M}}(\mathbf{r},t)$ to the right-hand side of the Landau Lifshitz equation Eq.~\eqref{EqLLEappendix}, with $\alpha_G$ the adimensional Gilbert damping parameter~\cite{StancilBook2009,Gurevich1996magnetization}. Together with the mode frequency $\omega_\beta$, the damping $\alpha_G$ determines the lifetime of a given spin wave mode $\beta$~\cite{StancilBook2009,KalarickalJAP2006,Gurevich1996magnetization}. Classically, the lifetime or relaxation time $\tau_\beta$ of a spin wave eigenmode $\beta$ is defined as the time required for the amplitude of its magnetization to decay by a factor $1/e$~\cite{StancilBook2009}. Here we use the expression provided by phenomenological loss theory~\cite{StancilBook2009,ChumakInBook,StancilJAP1986},
\begin{equation}\label{taubeta}
    \frac{1}{\tau_\beta} = \alpha_G\omega_\beta\frac{\partial \omega_\beta}{\partial \omega_H},
\end{equation}
which applies for small $\alpha_G$ (for YIG, $\alpha_G=10^{-4}$~\cite{ChumakInBook}) and is known to be a good description of propagation loss in thin films and stripes \cite{ChumakInBook}. The relaxation time is usually inferred
by coherently driving a selected spin wave mode through diverse methods and observing the decay of their amplitude~\cite{KalarickalJAP2006,LiuPRL2007,BertelliSciAdv2020,LeeWongNanoLetters2020,SebastianFrontiers2015,AcremannScience2000}. We thus compute, using the equations of motion Eqs.~\eqref{EOMsfree}-\eqref{EOMsdagsfree}, the expected value of the magnetization for a coherently populated single mode $\beta_0$, i.e. for an initial magnon state fulfilling $\langle \hat{s}_\beta(t=0)\rangle = \alpha_0 \delta_{\beta\beta_0}$:
\begin{equation}
    \langle\hat{\mathbf{m}}(\mathbf{r},t)\rangle_{\rm coh} \propto e^{-\gamma_{\beta_0}t/2}\text{Re}\left[\alpha_0\mathbf{m}_{\beta_0}(\mathbf{r})e^{-i\omega_{\beta_0}t}\right].
\end{equation}
Combining this equation, the definition of $\tau_\beta$, and its expression Eq.~\eqref{taubeta}, we establish a correspondence between the spin wave relaxation time $\tau_\beta$ and its decay rate through\footnote{The factor $2$ in this definition stems from our use of the usual convention in open quantum systems, according to which the ``natural'' or most fundamental decay rate is the decay rate of the energy or, equivalently, of the occupation $\langle\hat{s}_\beta^\dagger\hat{s}_\beta\rangle \propto \exp(-\gamma_\beta t)$.}
\begin{equation}\label{gammabetaAppendix}
    \gamma_\beta = \frac{2}{\tau_\beta}.
\end{equation}
The following explicit expression can be obtained from Eqs.~\eqref{taubeta} and \eqref{omegaSW},
\begin{equation}\label{gammabetaexpression}
    \gamma_\beta = \alpha\omega_M\left(\nu_{\beta x}+\nu_{\beta y}\right).
\end{equation}
This expression is displayed in  \figref{FigDecayRates} for Damon-Eshbach ($\phi_k=0$) and parallel ($\phi_k=\pi/2$) propagation modes. As evidenced by the figure, $\gamma_\beta$ lies in the MHz range, corresponding to a spin wave lifetime of around $\tau_\beta \sim 300$ns which agrees with experiments. Finally, note that $\gamma_\beta$ displays the relevant symmetries  $\gamma_n(k_y,k_z)=\gamma_n(-k_y,k_z)=\gamma_n(k_y,-k_z)$, which will be used in the next sections.

\subsection{Magnon power spectral densities.}\label{AppSW_PSD}

The potential of spin waves is conditioned to the degree of precision with which they can be measured. Usually, measurements are carried out on the spin wave magnetic field, and thus the figure of merit for sensing is the magnetic field power spectral density. We define the power spectral density of the magnetic field as\footnote{Although for magnetic field sensing we are interested on the power spectral density for the magnetic field evaluated at same positions, one can also define the nonlocal power spectral densities $S_{\mathbf{e}_j\mathbf{e}_k}(\mathbf{r}_1,\mathbf{r}_2,\omega)$ and compute them in an analogous way.}
\begin{multline}\label{twosidedPSDdef}
    S_{\mathbf{e}_j\mathbf{e}_k}(\mathbf{r},\omega)
    \equiv\frac{1}{2\pi}\int_{-\infty}^\infty ds e^{i\omega s}
    \\
    \times\left\langle\left(\mathbf{e}_j^\dagger\cdot\hat{\mathbf{h}}(\mathbf{r},t)\right)\left(\hat{\mathbf{h}}(\mathbf{r},t-s)\cdot\mathbf{e}_k\right)\right\rangle_{\rm ss},
\end{multline}
i.e., as the Fourier transform of the steady-state two-time correlator of two arbitrary vector components of the spin wave magnetic field. Hereafter we will focus on the steady state of Eq.~\eqref{MasterEquationOnlyMagnonsAppendix}, namely a thermal state, but we remark that the power spectral density can be computed for arbitrary steady states such as e.g. a spin wave coherent state. Note that, as can be easily shown, the power spectral density of the total magnetic field $\hat{\mathbf{H}}(\mathbf{r}) \equiv \mathbf{H}_0+\hat{\mathbf{h}}(\mathbf{r})$ is also given by Eq.~\eqref{twosidedPSDdef} for all $\omega\ne 0$, as the only effect of the homogeneous component $\mathbf{H}_0$ appears at $\omega=0$~\footnote{This can be easily proven for time-independent Liouvillians using the time-translational invariance of the steady state. }.

To compute the power spectral densities, we first calculate the two-time correlators of two arbitrary components of the spin wave magnetic field using Eqs.~\eqref{hoperator}, \eqref{hmodefunctions}, and \eqref{hfieldamplitude},
\begin{widetext}
\begin{multline}\label{PSDintermsofcorrelators}
    \left\langle\left(\mathbf{e}_j^\dagger\cdot\hat{\mathbf{h}}(\mathbf{r},t)\right)\left(\hat{\mathbf{h}}(\mathbf{r},t-s)\cdot\mathbf{e}_k\right)\right\rangle_{\rm ss}= 
    \\
    = \sum_\beta \mathcal{M}_{0\beta}^2 h_{\beta\eta0}^2 e^{-2k_\parallel l} \Big[\bar{\Lambda}_{jk}(\eta,\mathbf{k}_\parallel)\langle\hat{s}_\beta(t)\hat{s}^\dagger_{\beta}(t-s)\rangle_{\rm ss}
    +\bar{\Lambda}_{jk}(-\eta,\mathbf{k}_\parallel)\langle\hat{s}^\dagger_\beta(t)\hat{s}_{\beta}(t-s)\rangle_{\rm ss}\Big].
\end{multline}
Here, we have defined the matrix elements $\bar{\Lambda}_{jk}(\eta,\mathbf{k}_\parallel) \equiv [\mathbf{e}_j^*\cdot \mathbf{v}_{\eta}(\mathbf{k}_\parallel)][ \mathbf{v}_{\eta}^*(\mathbf{k}_\parallel)\cdot\mathbf{e}_k]$, with
$\mathbf{v}_{\eta}(\mathbf{k}_\parallel) \equiv [1,-i\eta k_y/k_\parallel,-i\eta k_z/k_\parallel]^T$.  By explicitly introducing the magnon correlators Eqs.~\eqref{sdagscorrelator} and \eqref{ssdagcorrelator}, integrating over the delay $s$, and using the symmetry $\bar{\Lambda}_{jk}(\eta,\mathbf{k}_\parallel)=\bar{\Lambda}^*_{kj}(\eta,\mathbf{k}_\parallel)$, we compute the power spectral density as
\begin{equation}\label{PSDnoNVs}
    S_{\mathbf{e}_j\mathbf{e}_k}(l,\omega) = \frac{1}{2\pi}\sum_\beta  \mathcal{M}_{0\beta}^2 h_{\beta\eta0}^2 e^{-2k_\parallel l} \gamma_\beta
    \left[\bar{\Lambda}_{jk}(\eta,\mathbf{k}_\parallel)\frac{\bar{n}_\beta+1}{(\omega-\omega_\beta)^2+(\gamma_\beta/2)^2}+\bar{\Lambda}_{jk}(-\eta,\mathbf{k}_\parallel)\frac{\bar{n}_\beta}{(\omega+\omega_\beta)^2+(\gamma_\beta/2)^2}\right].
\end{equation}
The above expression is valid for any two vectors $\mathbf{e}_j$ and $\mathbf{e}_k$. Here, we are interested on the (relatively simpler) power spectral density expressed in the cylindrical basis $ \{\mathbf{e}_+,\mathbf{e}_-,\mathbf{e}_z\}$ defined by Eq.~\eqref{cylindricalvectors}. In this basis the above matrix takes the form
\begin{equation}\label{BarLambdaMatrix}
    \bar{\Lambda}(\eta,\mathbf{k}_\parallel) = \frac{1}{2 k_\parallel^2}\left[
    \begin{array}{ccc}
        (k_\parallel-\eta k_y)^2 & k_z^2 & i\eta\sqrt{2}k_z (k_\parallel-\eta k_y)\\
        k_z^2 & (k_\parallel+\eta k_y)^2 & i\eta\sqrt{2}k_z (k_\parallel+\eta k_y)\\
        -i\eta\sqrt{2}k_z (k_\parallel-\eta k_y) & -i\eta\sqrt{2}k_z (k_\parallel+\eta k_y) & 2k_z^2
    \end{array}
    \right].
\end{equation}
\end{widetext}
Using this expression, and expressing the sum in mode indices $\beta$ explicitly in integral form as
\begin{equation}\label{continuummagnons}
    \sum_\beta \to \left(\frac{L}{2\pi}\right)^2\sum_n\int d^2\mathbf{k}_\parallel,
\end{equation}
we can numerically compute the power spectral densities. 

\begin{figure*}[tbh!]
	\centering
	\includegraphics[width=\linewidth]{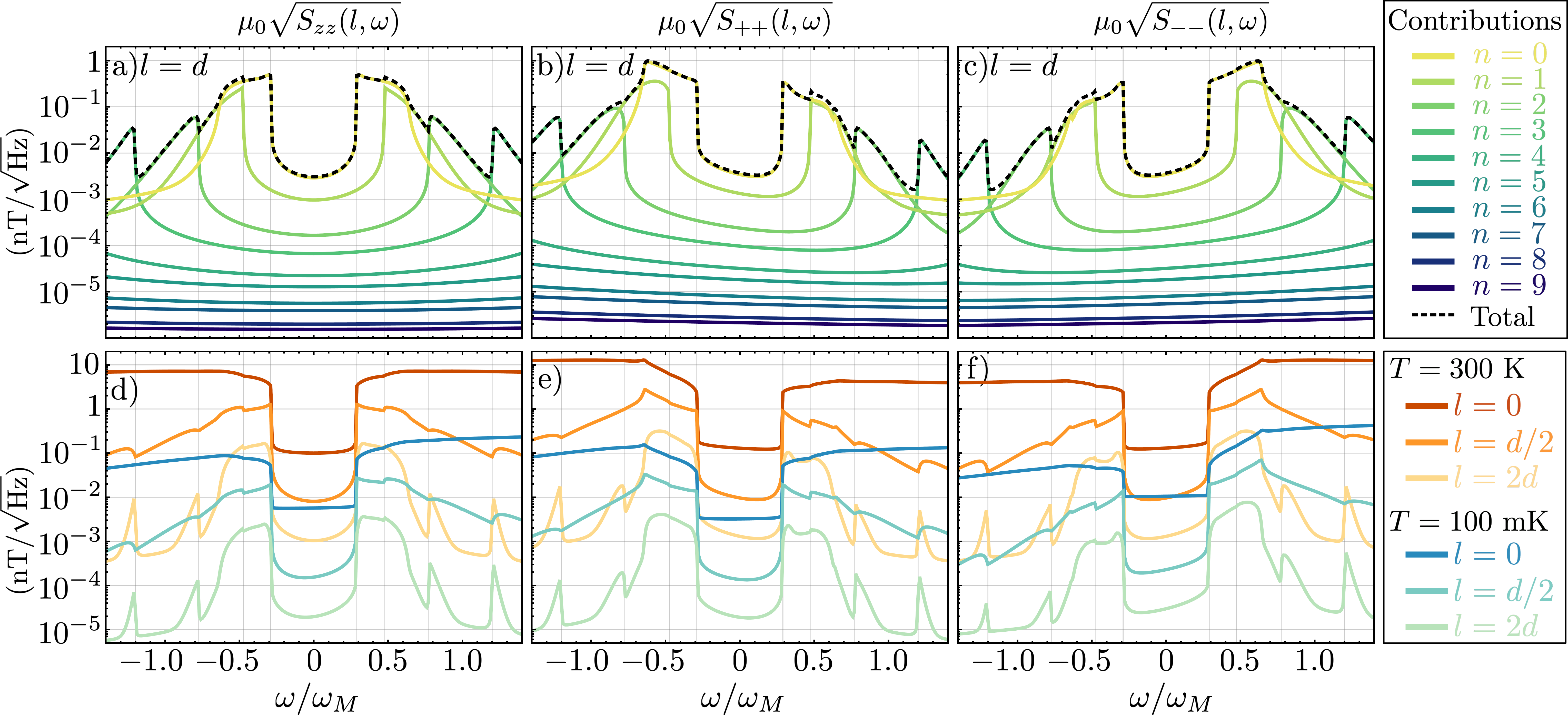}
	\caption{(a-c) Power spectral densities of the spin wave magnetic field components along $\mathbf{e}_z$ (a), $\mathbf{e}_+$ (b), and $\mathbf{e}_-$ (c), at a distance $l=d=200$nm above the YIG film, $T=300$K and $\mu_0H_0=20$mT. The dashed line shows the total power spectral density, whereas the colored lines depict the contribution of each band $n$, i.e. the power spectral density computed using only the spin wave modes from a single band $n$. (d-f) Total power spectral densities at room ($T=300$K, warm-coloured lines) and cryogenic temperatures ($T=100$mK, cool-coloured lines), at three diferent distances $l$ above the film.}\label{FigureMagnonPSDs}
\end{figure*}

Due to the symmetries Eq.~\eqref{h0symkz}-\eqref{h0symky} and the invariance of $\gamma_\beta$ and $\omega_\beta$ under change of sign of $k_y$ or $k_z$, the following properties for the power spectral densities can be proven:
\begin{enumerate}
    \item The power spectral densities depend only on the vertical distance to the film, $l$ (see \figref{FigureModalFields}), and are independent on the $y$ and $z$ coordinates.
    \item On each side of the film, the power spectral densities fulfill
    \begin{equation}\label{PSDprops1}
         S_{\mathbf{e}_\pm\mathbf{e}_z}(l,\omega)=S_{\mathbf{e}_z\mathbf{e}_\pm}(l,\omega)=0,
    \end{equation}
    \begin{equation}\label{PSDprops2}
        S_{\mathbf{e}_+\mathbf{e}_-}(l,\omega)=S_{\mathbf{e}_-\mathbf{e}_+}(l,\omega)=\frac{1}{2}S_{\mathbf{e}_z\mathbf{e}_z}(l,\omega),
    \end{equation}
    \item In the high temperature limit ($\bar{n}_\beta\gg 1$), the power spectral density along $\mathbf{e}_z$ is an even function,
    \begin{equation}\label{PSDprops3a}
        S_{\mathbf{e}_z\mathbf{e}_z}(l,\omega)=S_{\mathbf{e}_z\mathbf{e}_z}(l,-\omega) \hspace{0.3cm}\text{ for }\bar{n}_\beta \gg 1.
    \end{equation}
    Conversely, the functions $S_{\mathbf{e}_+\mathbf{e}_+}$ and $S_{\mathbf{e}_-\mathbf{e}_-}$ are not even due to the non-reciprocal modal properties discussed above. However, they are related through
    \begin{equation}\label{PSDprops3b}
        S_{\mathbf{e}_+\mathbf{e}_+}(\mathbf{r},-\omega) = S_{\mathbf{e}_-\mathbf{e}_-}(\mathbf{r},\omega) \hspace{0.3cm}\text{ for }\bar{n}_\beta \gg 1.
    \end{equation}
    \item The power spectral densities on different sides of the film are related through
    \begin{equation}\label{PSDpropslast}
        \left[
        \begin{array}{c}
            S_{\mathbf{e}_+\mathbf{e}_+}(l,\omega)  \\
            S_{\mathbf{e}_-\mathbf{e}_-}(l,\omega)  \\
            S_{\mathbf{e}_z\mathbf{e}_z}(l,\omega)
        \end{array}
        \right]_{\eta=-1}
        =
        \left[
        \begin{array}{c}
            S_{\mathbf{e}_-\mathbf{e}_-}(l,\omega)  \\
            S_{\mathbf{e}_+\mathbf{e}_+}(l,\omega)  \\
            S_{\mathbf{e}_z\mathbf{e}_z}(l,\omega)
        \end{array}
        \right]_{\eta=1}
    \end{equation}
\end{enumerate}

Because of the above relations, we can focus on the three independent power spectral densities $S_{\mathbf{e}_+\mathbf{e}_+}(\mathbf{r},\omega)$, $S_{\mathbf{e}_-\mathbf{e}_-}(\mathbf{r},\omega)$, and $S_{\mathbf{e}_z\mathbf{e}_z}(\mathbf{r},\omega)$ above the film ($\eta=1$).
In \figref{FigureMagnonPSDs}(a-c) we display the square root of these three quantities at room temperature ($T=300$K), for $\mu_0H_0=20$mT and $l=d=200$nm. The involved form of the total power spectral densities (dashed lines) is a consequence of adding the contributions from each spin wave band $n$, displayed by the different colored lines. 
At frequencies below the cutoffs of each spin wave band, the power spectral densities are suppressed due to the absence of resonant eigenmodes. At the cutoff frequency of each band, indicated by the vertical grid lines in the figure, the power spectral density shows a sharp jump due to the sudden increase in the density of states $\propto(d\omega_\beta/dk_\parallel)^{-1}$, followed by a smooth decrease as such density of states is reduced. Finally, note that as discussed above the power spectral density $S_{zz}(l,\omega)$ is an even function of frequency, whereas the functions $S_{\pm \pm}(l,\omega)$ are strongly asymmetric and, above cutoff, their values at positive and negative frequencies differ by roughly an order of magnitude. This asymmetry stems directly from the non-reciprocal spin wave mode properties. The above basic features of the power spectral densities remain unchanged at low temperature or at different positions $l$ above the film, as evidenced by \figref{FigureMagnonPSDs}(d-f). 
As shown by this figure, the power spectral densities decrease both at lower temperature and at larger distances from the film $l$, following the reduction of the thermal amplitude of the spin wave magnetic field. 
As a final remark, note that the power spectral densities shown in \figref{FigureMagnonPSDs} are experimentally measurable as they lie well above the sensitivity limit ($\sim 10^{-4}-10^{-6}$nT/$\sqrt{\text{Hz}}$) of ultra-sensitive room-temperature magnetometry techniques based e.g. on ensembles of NV centres~\cite{WolfPRX2015,WrachtrupJMR2016} or on atomic vapor cells~\cite{WasilewskiPRL2010,LedbetterPNAS2008,WrachtrupJMR2016}.

\subsection{Second-order correction to the spin wave fields}\label{AppSW_2ndorderfield}

In order to evaluate the spin-wave induced dephasing of the NV centres in the following sections, the expressions of the spin-wave magnetization and magnetic fields have to be computed to second order in magnon operators. 
This is equivalent to considering the lowest-order nonlinear terms in the spin wave approximation Eq.~\eqref{spinwaveapprox}. We proceed in the standard way~\cite{ZhangSciAdv2016,GonzalezBallesteroPRB2020} by writing the total magnetization as
\begin{multline}\label{Mquadratic}
    \mathbf{M}(\mathbf{r},t) = m_x(\mathbf{r},t)\mathbf{e}_x + m_y(\mathbf{r},t)\mathbf{e}_y+
    \\
    +\sqrt{M_S^2-m_x^2(\mathbf{r},t)-m_y^2(\mathbf{r},t)}\mathbf{e}_z,
 \end{multline}
and expanding to second order in the spin wave amplitudes $m_x$ and $m_y$,
\begin{multline}
    \mathbf{M}(\mathbf{r},t) = M_S\mathbf{e}_z + \mathbf{m}(\mathbf{r},t) \\-\frac{\mathbf{m}(\mathbf{r},t)\cdot \mathbf{m}(\mathbf{r},t)}{2M_S}\mathbf{e}_z +\mathcal{O}(\mathbf{m}^3).
\end{multline}
Here, $\mathbf{m}(\mathbf{r},t)=m_x(\mathbf{r},t)\mathbf{e}_x + m_y(\mathbf{r},t)\mathbf{e}_y$ is the first-order spin wave amplitude, i.e. the solution of the linearized Landau-Lifshitz equation Eq.~\eqref{linearizedLLEappendix}. By substituting it by the corresponding quantum operator $\hat{\mathbf{m}}(\mathbf{r})$, Eq.~\eqref{moperator}, we obtain the expression for the total magnetization operator,
\begin{equation}
    \hat{\mathbf{M}}(\mathbf{r}) = M_S\mathbf{e}_z + 
    \hat{\mathbf{m}}^{(2)}(\mathbf{r})+\mathcal{O}(\hat{s}_\beta^3)
  ,
\end{equation}
where $\hat{\mathbf{m}}^{(2)}(\mathbf{r})$ is the spin-wave magnetization up to second order in magnon operators, namely
\begin{equation}\label{m2ndorder}
    \hat{\mathbf{m}}^{(2)}(\mathbf{r}) = \hat{\mathbf{m}}(\mathbf{r}) + \delta \hat{\mathbf{m}}(\mathbf{r}),
\end{equation}
with a second-order correction given by
\begin{multline}
    \delta \hat{\mathbf{m}}(\mathbf{r}) =- \mathbf{e}_z\sum_{\beta\beta'}\frac{\mathcal{M}_{0\beta}\mathcal{M}_{0\beta'}}{2M_S}
    \Big[\mathbf{m}_\beta(\mathbf{r})\cdot \mathbf{m}_{\beta'}(\mathbf{r})\hat{s}_\beta\hat{s}_{\beta'}  \\+
    \mathbf{m}_\beta(\mathbf{r})\cdot \mathbf{m}_{\beta'}^*(\mathbf{r})\hat{s}_\beta\hat{s}^\dagger_{\beta'} + \text{H.c.}
    \Big].
\end{multline}
From Eq.~\eqref{m2ndorder} we can compute the corresponding spin wave magnetic field as, within the magnetostatic approximation, it is always related to the magnetization through the Green's tensor by Eq.~\eqref{hgreenfunctionmappendix}. Thus, to second order in magnon operators, the spin wave field operator reads
\begin{equation}\label{h2definition}
    \hat{\mathbf{h}}^{(2)}(\mathbf{r}) = \hat{\mathbf{h}}(\mathbf{r}) + \delta \hat{\mathbf{h}}(\mathbf{r}),
\end{equation}
with $\hat{\mathbf{h}}(\mathbf{r})$ being the first-order magnetic field given by Eq.~\eqref{hoperator}, and a second order correction given by
\begin{equation}\label{hsecondorder}
    \delta \hat{\mathbf{h}}(\mathbf{r}) = \sum_{\beta\beta'}
    \Big[\mathbf{X}^{+}_{\beta\beta'}(\mathbf{r})\hat{s}_\beta\hat{s}_{\beta'}  +
    \mathbf{X}^{-}_{\beta\beta'}(\mathbf{r})\hat{s}_\beta\hat{s}^\dagger_{\beta'} + \text{H.c.}
    \Big],
\end{equation}
where we define the vectors
\begin{multline}\label{Xvectorsdefinition}
    \left[
    \begin{array}{c}
         \mathbf{X}^{+}_{\beta\beta'}(\mathbf{r})  \\
         \mathbf{X}^{-}_{\beta\beta'}(\mathbf{r}) 
    \end{array}
    \right]
    \equiv -\frac{\mathcal{M}_{0\beta}\mathcal{M}_{0\beta'}}{2M_S}
    \\
    \times \int d^3\mathbf{r}'\tensorbar{\mathcal{G}}(\mathbf{r}-\mathbf{r}')\mathbf{e}_z
    \left[
    \begin{array}{c}
     \mathbf{m}_{\beta}(\mathbf{r}')\cdot \mathbf{m}_{\beta'}(\mathbf{r}')  \\
         \mathbf{m}_{\beta}(\mathbf{r}')\cdot \mathbf{m}_{\beta'}^*(\mathbf{r}')
    \end{array}
    \right].
\end{multline}

In order to compute the coefficients $\mathbf{X}^{\pm}_{\beta\beta'}(\mathbf{r})$, we introduce the expression of the magnetization mode functions, Eq.~\eqref{mmodefunctions}, into Eq.~\eqref{Xvectorsdefinition} and rearrange the terms to write in a compact form
\begin{equation}\label{Xvectorsdecomposition}
    \mathbf{X}_{\beta\beta'}^\pm = \frac{\mathcal{M}_{0\beta}\mathcal{M}_{0\beta'}}{M_S} X^\pm_{0\beta\beta'}\mathbf{x}_{nn'}(\mathbf{r},\mathbf{k}_\parallel \pm \mathbf{k}_\parallel'),
\end{equation}
where
\begin{multline}\label{Xvectorsamplitude}
    X^\pm_{0\beta\beta'}=\frac{-1}{\sqrt{(1+\delta_{n0})(1+\delta_{n'0})}}
    \\
    \times\left(\sqrt{\nu_{\beta y}\nu_{\beta'y}}\mp \sqrt{\nu_{\beta x}\nu_{\beta'x}}\right)
\end{multline}
with $\nu_{\beta x}$ and $\nu_{\beta y}$ given by Eqs.~\eqref{nux} and \eqref{nuy}, and
\begin{multline}
    \mathbf{x}_{nn'}(\mathbf{r},\mathbf{q}) \equiv \int_{\rm slab} d^3\mathbf{r}'\tensorbar{\mathcal{G}}(\mathbf{r}-\mathbf{r}')\mathbf{e}_z
    \\
    \times\cos\left[n\pi\frac{x'}{d}\right]\cos\left[n'\pi\frac{x'}{d}\right]e^{i\mathbf{q}\mathbf{r}_\parallel'}.
\end{multline}
We now focus on a position $\mathbf{r}$ outside the slab, and introduce the Green's tensor Eq.~\eqref{Greenstensorplanewaves} to find the following expression for the above integral:
\begin{multline}\label{xnnprimeanalytical}
    \mathbf{x}_{nn'}(\mathbf{r},\mathbf{q}) =e^{i\mathbf{q}_\parallel\mathbf{r}_\parallel}e^{-q_\parallel l}(-1)^{n+n'}\frac{q_zd}{4}\left[
    \begin{array}{c}
         i\eta q_\parallel d  \\
         -q_yd \\
         -q_z d
    \end{array}
    \right]
    \\
    \times \left(1-e^{-q_\parallel d}\right)\sum_{\lambda=\pm}\frac{1}{\pi^2(n'+\lambda n)^2+(q_\parallel d)^2}
    .
\end{multline}
where we have introduced the absolute distance to the slab, $l$, as in Eq.~\eqref{ldefinition}, and the variable $\eta$ to differentiate between positions $\mathbf{r}$ above and below the slab.
Note that, importantly, the above function decays exponentially with the modulus of $\mathbf{q}_\parallel$, and
\begin{equation}
     \mathbf{x}_{nn'}(\mathbf{r},0) \to 0.
\end{equation}
This, combined with Eqs.~\eqref{Xvectorsdecomposition} and \eqref{Xvectorsamplitude}, leads to the result
\begin{equation}\label{Xbetabetaequal0}
    \mathbf{X}_{\beta\beta}^- = 0,
\end{equation}
which will be important in the following sections.


\section{Nitrogen-Vacancy centres}\label{AppendixNVcentres}

This Appendix is devoted to the analysis of the NV centres and their dynamics. First, in Sec.~\ref{AppNV_baredynamics}, we theoretically describe the Hamiltonian and dissipative mechanisms governing the dynamics of a single NV centre in thermal equilibrium. We derive the equations of motion for all the observables and rigorously define the dissipation rates in terms of the lifetimes $T_1$ and $T_2^*$, and we compute the two-time correlators for NV centre observables. Then, in Sec.~\ref{AppNV_OpticalPumping}, we extend our analysis to optically pumped NV centres. We compute the dynamics of the NV observables and determine the optimal pumping rate. Finally, we compute both the two-time correlators  and the NV correlation time in the presence of optical pumping.

\subsection{Dynamics of a single isolated NV centre in thermal equilibrium}\label{AppNV_baredynamics}

In this section we study the dynamics of a single NV centre in thermal equilibrium. As in the main text, we assume the symmetry axis of the NV centre is oriented parallel to the $z-$axis, and a static magnetic field $\mathbf{H}_0=H_0\mathbf{e}_z$ is applied along this axis. The ground state manifold is a spin triplet with total spin $S=1$, thus containing three states $\vert 0\rangle$, $\vert +\rangle$, and $\vert -\rangle$ corresponding to the eigenstates of the spin operator $\hat{S}_z$ with eigenvalue $m_S = 0,+\hbar$, and $-\hbar$, respectively. In the absence of optical pumping, excited states are uncoupled and we can describe the NV centre through the Hamiltonian of the ground state manifold~\cite{BarGillNatComm2013,AjisakaPRB2016,WangNJP2014,MarcosPRL2010,ZhuNature2011}:
\begin{equation}\label{HNVappendix}
    \hat{H}_{\rm ps} = \hbar^{-1}D_0\hat{S}_z^2+\omega_H\hat{S}_z
    \\=\hbar \sum_{\alpha=\pm}\omega_{\alpha} \hat{\sigma}_{\alpha\alpha},
\end{equation}
where we define the transition matrices $\hat{\sigma}_{\alpha\alpha'} \equiv \vert \alpha \rangle\langle \alpha'\vert$ and the frequencies $\omega_\pm \equiv D_0\pm\omega_H$ and $\omega_H \equiv \vert\gamma_s\vert\mu_0H_0$, and the $z-$component of the spin operator is given by $\hat{S}_z/\hbar = \hat{\sigma}_{++}-\hat{\sigma}_{--}$. For an NV centre the gyromagnetic factor is $\gamma_s=\gamma$ (see Table~\ref{tablePARAMS}). 
The density matrix of the NV centre obeys the von Neumann equation
\begin{equation}\label{MasterEqNVcentresAppendix}
    \frac{d}{dt}\hat{\rho}_{\rm ps} = -\frac{i}{\hbar}\left[\hat{H}_{\rm ps},\hat{\rho}_{\rm ps}\right]
    +\mathcal{D}_{\rm ps}[\hat{\rho}_{\rm ps}],
\end{equation}
where the second term accounts for dissipation, namely decay and dephasing. In principle, there are different approaches to describe these processes~\cite{BarGillNatComm2013,AjisakaPRB2016,WangNJP2014,TetiennePRB2013}. Here we choose the description given by the dissipator
\begin{multline}\label{DissipatorEqNVcentresAppendix}
    \mathcal{D}_{\rm ps}[\hat{\rho}]=\frac{\kappa_2}{\hbar^2}\mathcal{L}_{\hat{S}_{z}\hat{S}_{z}}[\hat{\rho}]\\+\kappa_1\sum_{\alpha=\pm}\left(\bar{n}_\alpha\mathcal{L}_{\hat{\sigma}_{\alpha0}\hat{\sigma}_{0\alpha}}[\hat{\rho}]+(\bar{n}_\alpha+1)\mathcal{L}_{\hat{\sigma}_{0\alpha}\hat{\sigma}_{\alpha0}}[\hat{\rho}]\right).
\end{multline}
The first term above describes dephasing at a rate $\kappa_2$, whereas the second line describes decay and absorption at a rate $\kappa_1$ along the two NV transitions, namely $\vert 0 \rangle \leftrightarrow \vert \pm \rangle$, induced by a bosonic thermal reservoir at temperature $T$, with $\bar{n}_\alpha = [\exp(\hbar\omega_\alpha/k_B T)-1]^{-1}$.

The equations of motion for the spin observables can be obtained from the above equation.  On the one hand, the two independent quantities describing the level occupations $\langle\hat{\sigma}_{\alpha\alpha}\rangle$ obey the coupled system of equations\cite{MyersPRL2017,AjisakaPRB2016,TetiennePRB2013}
\begin{equation}\label{EOMbareNVdiagonal0}
    \frac{d }{dt}\left[
    \begin{array}{c}
         \langle \hat{\sigma}_{00}\rangle  \\
         \langle\hat{S}_z/\hbar\rangle 
    \end{array}
    \right]=
    \left[
    \begin{array}{c}
         \kappa_1+\gamma_+  \\
         \gamma_- 
    \end{array}
    \right]
    -\tensorbar{M}_0\left[
    \begin{array}{c}
         \langle \hat{\sigma}_{00}\rangle  \\
         \langle\hat{S}_z/\hbar\rangle 
    \end{array}
    \right],
\end{equation}
with
\begin{equation}
    \tensorbar{M}_0=\left[
    \begin{array}{cc}
        \kappa_1+3\gamma_+ & \gamma_- \\
        3\gamma_- & \kappa_1+\gamma_+
    \end{array}
    \right]
\end{equation}
and
\begin{equation}\label{gammaPMdef}
    \gamma_\pm \equiv\frac{\kappa_1}{2}(\bar{n}_-\pm\bar{n}_+).
\end{equation}
 The general solution of the above equation is given by
\begin{multline}\label{EOMbareNVdiagonal}
    \left[
    \begin{array}{c}
         \langle \hat{\sigma}_{00}(t)\rangle  \\
         \langle\hat{S}_z(t)/\hbar\rangle 
    \end{array}
    \right] = \left[
    \begin{array}{c}
         \langle \hat{\sigma}_{00}\rangle_{\rm ss}  \\
         \langle\hat{S}_z/\hbar\rangle_{\rm ss} 
    \end{array}
    \right] 
    \\
    + \frac{e^{-(\kappa_1+2\gamma_+)t}}{3\gamma_-}\tensorbar{M}(t)
    \left[
    \begin{array}{c}
         \langle \hat{\sigma}_{00}\rangle(0)-\langle \hat{\sigma}_{00}\rangle_{\rm ss}  \\
         \langle\hat{S}_z/\hbar\rangle(0)-\langle\hat{S}_z/\hbar\rangle_{\rm ss}
    \end{array}
    \right] ,
\end{multline}
with
\begin{equation}\label{barMdefinition}
    \tensorbar{M}(t) \equiv \left[
    \begin{array}{cc}
        (\gamma_+-s_d)e^{s_d t} & (\gamma_++s_d)e^{-s_d t} \\
        3\gamma_-e^{s_d t} & 3\gamma_- e^{-s_d t}
    \end{array}
    \right]
\end{equation}
and $s_d\equiv\sqrt{\gamma_+^2+3\gamma_-^2}$. The steady-state solution reads
\begin{multline}
    \left[
    \begin{array}{c}
         \langle \hat{\sigma}_{00}\rangle_{\rm ss}  \\
         \langle\hat{S}_z/\hbar\rangle_{\rm ss} 
    \end{array}
    \right] =
    \\
    \frac{1}{(\kappa_1+\gamma_+)(\kappa_1+3\gamma_+)-3\gamma_-^2}\left[
    \begin{array}{c}
         (\kappa_1+\gamma_+)^2-\gamma_-^2  \\
         -2\gamma_- \kappa_1
    \end{array}
    \right].
\end{multline}
On the other hand, the system coherences remain uncoupled and obey the equations
\begin{equation}\label{EOMbareNVsigma0pm}
    \frac{d\langle\hat{\sigma}_{0 \pm}\rangle}{dt}=\left[-i\omega_\pm-\frac{\kappa_1+3\gamma_+\mp\gamma_-}{2}-\frac{\kappa_2}{2}\right]\langle\hat{\sigma}_{0\pm}\rangle,
\end{equation}
\begin{equation}\label{EOMbareNVsigmapm}
    \frac{d\langle\hat{\sigma}_{+-}\rangle}{dt}=\left[i(\omega_+-\omega_-)-(\gamma_++\kappa_1)-2\kappa_2\right]\langle\hat{\sigma}_{+-}\rangle.
\end{equation}
From the above equations it is evident that all coherences are zero in the steady-state, and that they decay and dephase at different rates.

\subsubsection{Definition of $T_1$ and $T_2^*$}

Since the rates $\gamma_\pm$ depend on the externally applied field through the frequencies $\omega_\pm$, the decay time of the expected values $\langle\hat{\sigma}_{\alpha\alpha'}\rangle$ also does. For this reason, the intrinsic $T_1$ and $T_2^*$ are defined in the zero-field limit $H_0\to 0$~\cite{MyersPRL2017,TetiennePRB2013}. In this limit $\gamma_-\to 0$ and
\begin{equation}
    \gamma_+ \to \kappa_1\bar{n}_0 
\end{equation}
with $\bar{n}_0 = [\exp(\hbar D_0/k_B T)-1]^{-1}$ the Bose-Einstein distribution evaluated at the zero-field splitting frequency $D_0$.
The decay time $T_1$ is defined as the inverse of the equilibration rate of the NV occupations after initialization in the ground state $\vert 0 \rangle$~\cite{MyersPRL2017,TetiennePRB2013}. For such an initial state $\langle \hat{S}_z/\hbar\rangle (t)=0$ and the occupations of all levels obey a single exponential behavior,
\begin{equation}
    \langle \hat{\sigma}_{00}(t)\rangle = \left[1-\langle \hat{\sigma}_{00}\rangle_{\rm ss}\right]e^{-(\kappa_1+3\gamma_+)t}+\langle \hat{\sigma}_{00}\rangle_{\rm ss},
\end{equation}
\begin{equation}
    \langle \hat{\sigma}_{++}(t)\rangle=\langle \hat{\sigma}_{--}(t)\rangle=\frac{1-\langle \hat{\sigma}_{00}(t)\rangle}{2}.
\end{equation}
From the above expressions the relaxation time $T_1$ is defined as
\begin{equation}\label{defT1appendix}
    T_1^{-1} \equiv \kappa_1+3\gamma_+ = \kappa_1(1+3\bar{n}_0).
\end{equation}
The above definition allows to fix a value for $\kappa_1$ from experimentally measured values of $T_1$.

The second relevant timescale, namely the decoherence time $T_2^*$, is defined through the decay of the coherences $\langle \hat{\sigma}_{0,\pm}\rangle$, which in the zero-field limit obey the same exponential decay,
\begin{equation}
    \langle \hat{\sigma}_{0,\pm}\rangle(t) = \langle \hat{\sigma}_{0,\pm}(0)\rangle e^{[-i\omega_{\pm}-(\kappa_2+\kappa_1+3\gamma_+)/2]t}.
\end{equation}
From the above expression we define
\begin{equation}
    [T_2^*]^{-1} = \frac{\kappa_2}{2} + \frac{\kappa_1+3\gamma_+}{2} = \frac{T_{1}^{-1}}{2} + \frac{\kappa_2}{2}.
\end{equation}
Note that $T_2^*$ has contributions from decay and dephasing rates $\kappa_1$ and $\kappa_2$, which gives rise to the fundamental limit $T_2^* \le 2 T_1$~\cite{MyersPRL2017,HerbschlebNatComm2019,BarGillNatComm2013}.

\subsubsection{Two-time correlators of NV operators}

In a similar fashion as for the magnon operators, we can compute the two-time correlators of products of NV operators, which will determine the effective spin wave dynamics as we will see below. We compute such correlators from the equations of motion Eqs.~\eqref{EOMbareNVdiagonal0},\eqref{EOMbareNVsigma0pm}, and \eqref{EOMbareNVsigmapm} and the quantum regression theorem~\cite{CarmichaelBook}.
The simplest correlators are those involving the coherences, which for a general operator $\hat{O}$ and for $s\ge 0$ read
\begin{multline}\label{CorrelatorSigma0PM}
    \langle\hat{\sigma}_{0\pm}(t)\hat{O}(t-s)\rangle_{\rm ss} =
    \\
    =e^{-s[i\omega_\pm +(\kappa_1+3\gamma_+\mp \gamma_- +\kappa_2)/2]} \langle\hat{\sigma}_{0\pm}\hat{O}\rangle_{\rm ss},
\end{multline}
\begin{multline}\label{CorrelatorSigmaPM0}
    \langle\hat{\sigma}_{\pm 0}(t)\hat{O}(t-s)\rangle_{\rm ss} = \\
    =e^{-s[-i\omega_\pm +(\kappa_1+3\gamma_+\mp \gamma_- +\kappa_2)/2]} \langle\hat{\sigma}_{\pm 0}\hat{O}\rangle_{\rm ss}.
\end{multline}
\begin{multline}
    \langle\hat{\sigma}_{+-}(t)\hat{O}(t-s)\rangle_{\rm ss} =
    \\
    =e^{s[i(\omega_+-\omega_-) -\kappa_1-\gamma_+-2\kappa_2]} \langle\hat{\sigma}_{+-}\hat{O}\rangle_{\rm ss},
\end{multline}
\begin{multline}\label{correlationNVPMthermal2}
    \langle\hat{\sigma}_{-+}(t)\hat{O}(t-s)\rangle_{\rm ss} =
    \\
    =e^{s[-i(\omega_+-\omega_-) -\kappa_1-\gamma_+-2\kappa_2]} \langle\hat{\sigma}_{+-}\hat{O}\rangle_{\rm ss}.
\end{multline}

Regarding the occupations, the quantities of interest are usually not the operators $\langle\hat{\sigma}_{00}\rangle$ or $\langle\hat{S}_{z}\rangle$ themselves, but the fluctuations over their expected value, defined as
\begin{equation}\label{tildedoperatorsdef}
    \left[
    \begin{array}{c}
         \langle \tilde{\sigma}_{00}\rangle  \\
         \langle\tilde{S}_z/\hbar\rangle 
    \end{array}
    \right]\equiv \left[
    \begin{array}{c}
         \langle \hat{\sigma}_{00}\rangle  \\
         \langle\hat{S}_z/\hbar\rangle 
    \end{array}
    \right]-\left[
    \begin{array}{c}
         \langle \hat{\sigma}_{00}\rangle_{\rm ss}  \\
         \langle\hat{S}_z/\hbar\rangle_{\rm ss} 
    \end{array}
    \right].
\end{equation}
For any operator $\hat{O}$ and any $s\ge 0$ the correlators involving these operators are coupled,
\begin{multline}\label{CorrelatorSigmaDIAG}
    \left[
    \begin{array}{c}
         \langle \tilde{\sigma}_{00}(t)\hat{O}(t-s)\rangle  \\
         \langle(\tilde{S}_z/\hbar)(t)\hat{O}(t-s)\rangle 
    \end{array}
    \right] = 
    \\
    =
    \frac{e^{-(\kappa_1+2\gamma_+)s}}{3\gamma_-}\tensorbar{M}(s)\left[
    \begin{array}{c}
         \langle \tilde{\sigma}_{00}\hat{O}\rangle_{\rm ss}  \\
         \langle(\tilde{S}_z/\hbar)\hat{O}\rangle_{\rm ss} 
    \end{array}
    \right]
\end{multline}
where the matrix $\tensorbar{M}(s)$ is defined in Eq.~\eqref{barMdefinition}.

\subsection{Dynamics of a single optically pumped NV centre}\label{AppNV_OpticalPumping}

\begin{figure}[h!]
	\centering
	\includegraphics[width=\linewidth]{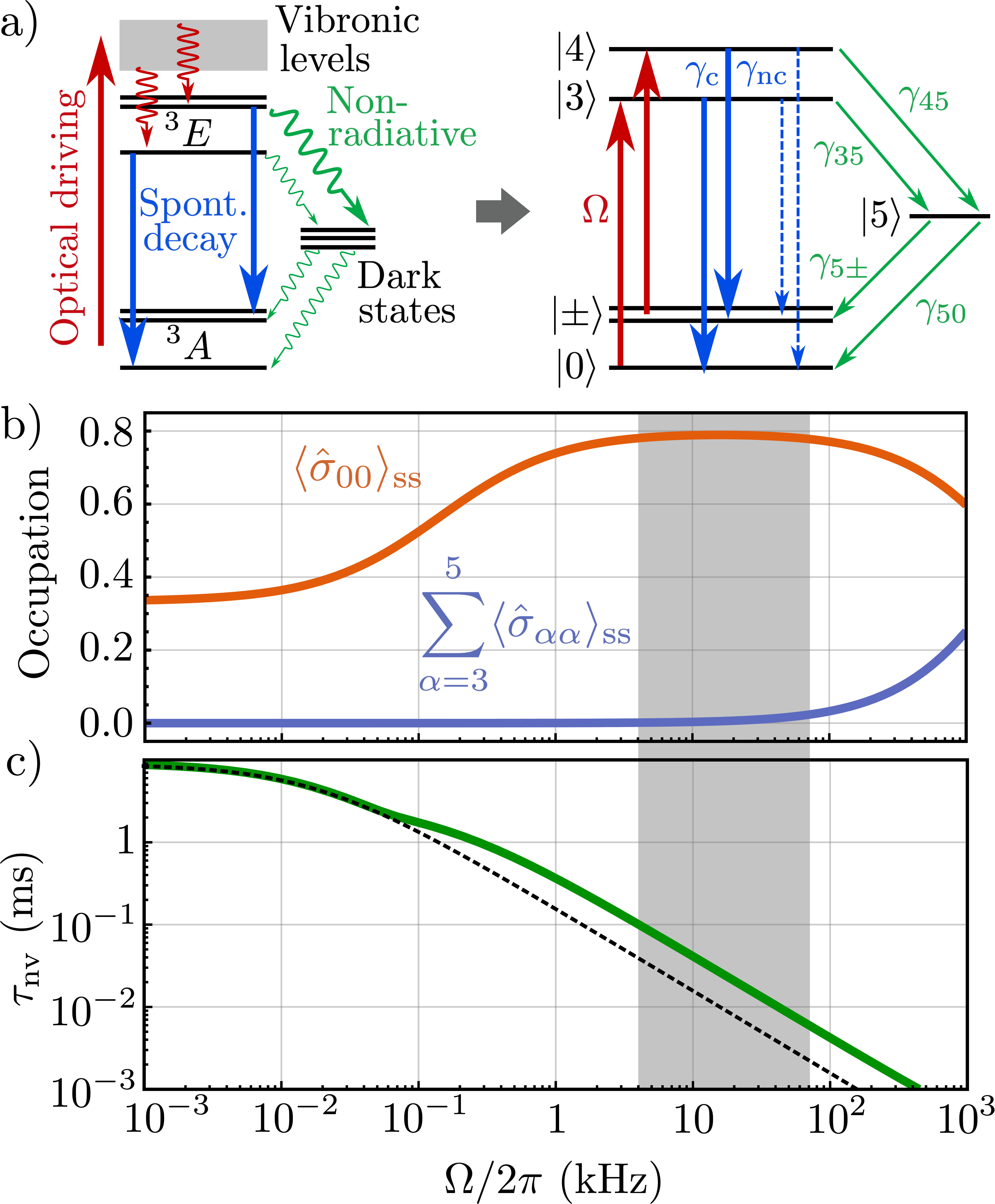}
	\caption{a) Left: illustration of the main processes involved in optical pumping. Straight and wiggly lines indicate radiative and non-radiative processes respectively. Right: modelling of the optical pumping and the relevant states. b) Ground-state occupation (red) and total occupation outside the ground-state manifold (blue) as a function of pumping rate. c) Correlation time of the NV centre occupations, Eq.~\eqref{tauNVdefinition} as a function of pumping rate. The dashed curve displays the timescale $(\kappa_1+\gamma_+ + \Omega)^{-1}$. The gray area indicates the optimal region for optical pumping. For panels b-c we take $\mu_0H_0=20$mT, and the results do not change appreciably within the range $0\le \mu_0H_0\le 100$mT.}\label{FigOpticalPumping}
\end{figure}

One of the multiple advantages of NV centres is the possibility to optically initialize them at room temperature, by a procedure known as optical pumping~\cite{RobledoNJP2011,WangNJP2015,MeirzadaPRB2018,DohertyPhysRep2013,MildrenBook,RobertsPRB2019,TetienneNJP2012}. In order to describe and understand this procedure one must consider the extended energy level structure of NV centres depicted in \figref{FigOpticalPumping}(a). The states $\vert 0\rangle$ and $\vert \pm\rangle$ introduced in the previous sections correspond to the ground-state manifold ${}^3A$. The excited state manifold ${}^3E$ is also a spin triplet with zero-field splitting $D_1\ne D_0$ (see Table~\ref{tableRATESopticalpumping}), here described as two states $\vert 3\rangle$ and $\vert 4\rangle$ for the sake of simplicity~\cite{RobledoNJP2011,MeirzadaPRB2018,RobertsPRB2019}. These two states spontaneously decay into the ground-state manifold, with decay rates $\gamma_{\rm c}$ and $\gamma_{\rm nc}\ll\gamma_{\rm c}$ for spin-conserving and spin non-conserving transitions, respectively. For optical pumping a coherent, linearly polarized (i.e. $m_S$-conserving) driving is applied to the transition ${}^3A\to {}^3E$. Since this driving mostly excites vibronic transitions which rapidly and stochastically decay to the zero-phonon line, the effective pumping of occupation into the excited-state manifold is incoherent~\cite{MeirzadaPRB2018,RobertsPRB2019}. 
Finally, in addition to their spontaneous decay, the states $\vert 3\rangle$ and $\vert 4\rangle$ can decay non-radiatively (incoherently) into a manifold of intermediate ``dark'' states, here modelled as a single state $\vert 5 \rangle$~\cite{TetienneNJP2012,MeirzadaPRB2018,RobertsPRB2019}, which in turn decay non-radiatively to the ground state manifold. This decay is largely imbalanced: specifically, the decay rate from state $\vert 4\rangle$ into state $\vert 5\rangle$ is much faster than any other decay rates into or out of the dark state $\vert 5\rangle$, i.e., $\gamma_{45} \gg \gamma_{35}, \gamma_{5\pm}, \gamma_{50} $~(see Table~\ref{tableRATESopticalpumping}). As a result of this imbalance, part of the initial occupation of states with non-zero spin, $\vert \pm \rangle$, is \emph{pumped} into the zero-spin state $\vert 0\rangle$, thus driving the NV centre toward its ground state.

We describe optical pumping of NV centres through the following master equation for the six-level system illustrated in \figref{FigOpticalPumping}(a):
\begin{equation}\label{MasterEqNVOpticalPumping}
    \frac{d\hat{\rho}}{dt} = -i\left[\hat{H}_{\rm ps} + \hat{H}_{\rm ps}^{(2)},\hat{\rho}\right]+
    \mathcal{D}_{\rm ps}[\hat{\rho}]  +
    \mathcal{D}_{\rm op}[\hat{\rho}]  
    .
\end{equation}
Here, $\hat{H}_{\rm ps}$ and $\mathcal{D}_{\rm ps}$ represent the Hamiltonian and the dissipator of the ground-state manifold, and are given by Eqs.~\eqref{HNVappendix} and \eqref{DissipatorEqNVcentresAppendix}. The second coherent contribution corresponds to the Hamiltonian of the additional levels, i.e.,
\begin{equation}
    \hat{H}_{\rm ps}^{(2)} = \hbar\sum_{\alpha=3}^5\omega_\alpha\hat{\sigma}_{\alpha\alpha}.
\end{equation}
The values of the frequencies $\omega_{3,4,5}$ are irrelevant for the dynamics of the ground state manifold as we will see below.
Regarding the additional dissipator, it can be split into three contributions,
\begin{equation}\label{DissipatorOP}
     \mathcal{D}_{\rm op}[\hat{\rho}] = \mathcal{D}_{\rm p}[\hat{\rho}] + \mathcal{D}_{\rm s}[\hat{\rho}] + \mathcal{D}_{\phi}[\hat{\rho}].
\end{equation}
The first term above represents the spin-conserving incoherent pumping from all the states in the ground state manifold into all states in the excited state manifold,
\begin{equation}
    \mathcal{D}_{\rm p}[\hat{\rho}] = \hbar\Omega\Big(\mathcal{L}_{\hat{\sigma}_{30}\hat{\sigma}_{03}}[\hat{\rho}]+\sum_{\alpha=\pm}\mathcal{L}_{\hat{\sigma}_{4\alpha}\hat{\sigma}_{\alpha4}}[\hat{\rho}]\Big).
\end{equation}
The incoherent driving rate $\Omega$ is related to the applied optical intensity $I_d$ through $\Omega= \alpha_d I_d$, where $\alpha_d=0.2-0.7$mHz W$^{-1}$m$^2$ is the optical pumping parameter (see Table~\ref{tableRATESopticalpumping}). The second contribution in the dissipator Eq.~\eqref{DissipatorOP} describes all the spontaneous decay processes illustrated in \figref{FigOpticalPumping}(a),
\begin{multline}
    \mathcal{D}_{\rm s}[\hat{\rho}] = \gamma_{\rm c}\Big(\mathcal{L}_{\hat{\sigma}_{03}\hat{\sigma}_{30}}[\hat{\rho}]
    +
    \frac{1}{2}\sum_{\alpha=\pm} \mathcal{L}_{\hat{\sigma}_{\alpha4}\hat{\sigma}_{4\alpha}}[\hat{\rho}]\Big)
    \\
    + \gamma_{\rm nc}\Big(\mathcal{L}_{\hat{\sigma}_{04}\hat{\sigma}_{40}}[\hat{\rho}]
    +
    \frac{1}{2}\sum_{\alpha=\pm} \mathcal{L}_{\hat{\sigma}_{\alpha3}\hat{\sigma}_{3\alpha}}[\hat{\rho}]\Big)
    \\
    + \gamma_{50}\mathcal{L}_{\hat{\sigma}_{05}\hat{\sigma}_{50}}[\hat{\rho}]
    + \gamma_{5\pm}\sum_{\alpha=\pm}\mathcal{L}_{\hat{\sigma}_{\alpha5}\hat{\sigma}_{5\alpha}}[\hat{\rho}]
    \\
    + \gamma_{45}\mathcal{L}_{\hat{\sigma}_{54}\hat{\sigma}_{45}}[\hat{\rho}]
   + \gamma_{35}\mathcal{L}_{\hat{\sigma}_{53}\hat{\sigma}_{35}}[\hat{\rho}].
\end{multline}
The factors $1/2$ in the first two lines ensure that the total radiative decay rate of both states $\vert 3\rangle$ and $\vert 4\rangle$ is the same, namely $\gamma_{\rm c}+\gamma_{\rm nc}$. Finally, the last term in the dissipator Eq.~\eqref{DissipatorOP} describes the additional fast dephasing of the excited-state manifold~\cite{WangNJP2015},
\begin{equation}
    \mathcal{D}_{\phi}[\hat{\rho}] = \frac{\kappa_\phi}{\hbar^2}\mathcal{L}_{\hat{\sigma}_{44}\hat{\sigma}_{44}}[\hat{\rho}].
\end{equation}
Room temperature values for all the relevant additional rates appearing in the above master equation are given in Table~\ref{tableRATESopticalpumping}. Note that we neglect spontaneous (thermal) decay and absorption between the excited states $\vert 3\rangle$ and $\vert 4 \rangle$, as these processes are negligible in comparison with the much faster spontaneous emission described by $\gamma_{\rm c}$ and $\gamma_{\rm nc}$.

\begin{table}[t]
	\centering
	\begin{tabular}{ p{2.0cm} | p{5.6cm} }
		\hline
		\hline
		Rate & Value \\
		\hline
		$D_1$
		 & $2\pi\times 1.423$ GHz \\
		$\gamma_{\rm c}$ & $2\pi\times 9.7$ MHz \\
		$\gamma_{\rm nc}$ & $\approx 10^{-2}\gamma_{\rm c}$ \\
		$\gamma_{45}$ & $2\pi\times 12$ MHz
		 \\
		 $\gamma_{35}$ & $2\pi\times 1.74$ MHz
		 \\
		 $\gamma_{5\pm}$ & $2\pi\times 0.21$ MHz
		 \\
		 $\gamma_{50}$ & $2\pi\times 0.49$ MHz
		 \\
		 $\kappa_{\phi}$ & $2\pi\times 10^6$ MHz
		 \\
		  $\alpha_d$ & $0.45$ mHz W$^{-1}$m$^2$
		 \\
		\hline
		\hline
	\end{tabular}
	\caption{Room temperature values for the rates involved in the optical pumping cycle of an NV centre~\cite{RobledoNJP2011,WangNJP2015,MeirzadaPRB2018,RobertsPRB2019}.} \label{tableRATESopticalpumping}
\end{table}

From the master equation Eq.~\eqref{MasterEqNVOpticalPumping} we can obtain the equations of motion for the expected values of transition matrices, $\langle\hat{\sigma}_{\alpha\alpha'}\rangle = \rho_{\alpha\alpha'}$. Since all the processes involved in the dynamics are dissipative, the diagonal elements $\langle\hat{\sigma}_{\alpha\alpha}\rangle$ decouple from the coherences
and form a closed $5\times5$ system which we can write as
\begin{equation}\label{EOMoccupationsPumping}
     \frac{d}{dt}
    \left[
    \begin{array}{c}
         \langle\hat{\sigma}_{00}\rangle  \\
         \langle\hat{S}_z/\hbar\rangle \\
         \langle\hat{\sigma}_{33}\rangle \\
         \langle\hat{\sigma}_{44}\rangle\\
         \langle\hat{\sigma}_{55}\rangle
    \end{array}
    \right] = \tensorbar{M}_{\rm op} \left[
    \begin{array}{c}
         \langle\hat{\sigma}_{00}\rangle  \\
         \langle\hat{S}_z/\hbar\rangle \\
         \langle\hat{\sigma}_{33}\rangle \\
         \langle\hat{\sigma}_{44}\rangle\\
         \langle\hat{\sigma}_{55}\rangle
    \end{array}
    \right] +
    \left[
    \begin{array}{c}
         \kappa_1 + \gamma_+ \\
         \gamma_- \\
         0 \\
         \Omega\\
         0
    \end{array}
    \right]
\end{equation}
with a coefficient matrix
\begin{widetext}
\begin{equation}\label{MdynOpticalPumping}
    \tensorbar{M}_{\rm op}
    =
    \left[
    \begin{array}{ccccc}
         -\kappa_1-3\gamma_+-\Omega &-\gamma_- & \gamma_c -\kappa_1-\gamma_+ & \gamma_{nc}-\kappa_1-\gamma_+ & \gamma_{50}-\kappa_1-\gamma_+  \\
         -3\gamma_- & -\kappa_1-\gamma_+-\Omega & -\gamma_- &-\gamma_- &-\gamma_- \\
         \Omega & 0 &-\gamma_{\rm c}-\gamma_{\rm nc}-\gamma_{35} &0 &0 \\
         -\Omega & 0 & -\Omega &-\gamma_{\rm c}-\gamma_{\rm nc}-\gamma_{45}-\Omega & -\Omega \\
         0 & 0 & \gamma_{35} & \gamma_{45} & -\gamma_{50} - 2\gamma_{5\pm}
    \end{array}
    \right].
\end{equation}
\end{widetext}
The occupation of the levels $\vert+\rangle$ and $\vert - \rangle$ is obtained from the solution of the above equations through the norm conservation identity
\begin{equation}
    \langle\hat{\sigma}_{\pm\pm}\rangle = \frac{1}{2}\bigg(1-\langle\hat{\sigma}_{00}\rangle-\sum_{\alpha=3}^5\langle\hat{\sigma}_{\alpha\alpha}\rangle \pm\langle\hat{S}_z/\hbar\rangle\bigg).
\end{equation}
Note that the occupations depend neither on the frequencies $\omega_\alpha$ of the states $\vert 3\rangle$, $\vert 4\rangle$, and $\vert 5\rangle$, nor on the dephasing rate $\kappa_\phi$. Since the spontaneous decay rates are fixed material parameters (see Table~\ref{tableRATESopticalpumping}), the only additional parameter determining the occupation dynamics is the driving strength $\Omega$. As evidenced by the red curve in \figref{FigOpticalPumping}(b), the steady-state occupation of the ground state $\vert 0 \rangle$ increases as a function of $\Omega$ up to a maximum value $\langle\hat{\sigma}_{00}\rangle_{\rm ss} \approx 0.8$, consistent with experimental observations~\cite{MildrenBook,DohertyPhysRep2013}. For too large driving strengths the efficiency of the optical pumping decreases since the higher states $\vert 3\rangle$, $\vert 4\rangle$, and $\vert 5\rangle$ become occupied in the steady-state, as evidenced by the blue curve in \figref{FigOpticalPumping}(b). The optimal driving strength for optical pumping is $\Omega \approx 2\pi\times 10$kHz, corresponding to an optical intensity of $I_d\approx 0.13$GW m$^{-2}$~\footnote{Although not the case for the results in the main text, note that in the presence of spin waves the lifetimes of the NV centres, and hence the rates in the master equation Eq.~\eqref{MasterEqNVOpticalPumping}, might be significantly altered (see next sections). As a consequence the optimum driving intensity might be very different or, in extreme situations, optical pumping might even become unfeasible. }, or, equivalently, to an optical
power of $P_d \approx 3\mu$W at $532$nm focused through a lens of numerical aperture $\sim 1.4$.

Equations of motion for the expected values of the coherences can analogously be obtained. Here we focus on the coherences of interest regarding the interaction with spin waves, namely the coherences of the ground-state manifold whose evolution is given by
\begin{multline}\label{EOMpumpedNVsigma0pm}
    \frac{d\langle\hat{\sigma}_{0 \pm}\rangle}{dt}=
    \\\left[-i\omega_\pm-\frac{\kappa_1+3\gamma_+\mp\gamma_-}{2}-\frac{\kappa_2}{2}-\Omega\right]\langle\hat{\sigma}_{0\pm}\rangle,
\end{multline}
\begin{multline}\label{EOMpumpedNVsigmapm}
    \frac{d\langle\hat{\sigma}_{+-}\rangle}{dt}=
    \\\left[i(\omega_+-\omega_-)-(\gamma_++\kappa_1)-2\kappa_2-\Omega\right]\langle\hat{\sigma}_{+-}\rangle.
\end{multline}
These equations are, except for the factor proportional to $\Omega$, identical to the corresponding equations in the absence of optical pumping, namely Eqs.~\eqref{EOMbareNVsigma0pm}-\eqref{EOMbareNVsigmapm}. Indeed, since the driving is incoherent, it contributes only to the decay of the coherences and does not result in any steady-state coherence, i.e.,
\begin{equation}
    \langle\hat{\sigma}_{0\pm}\rangle_{\rm ss} = \langle\hat{\sigma}_{+-}\rangle_{\rm ss}=0,
\end{equation}
in analogy to an NV centre in thermal equilibrium discussed in the previous section. Note that as a consequence, and regarding only the coherences above, an optically pumped NV centre can be interpreted as a thermal NV centre at an effective, lower, $\Omega-$dependent temperature. 

\subsubsection{Two-time correlators of NV operators}

Let us finally derive the two-time correlation functions of an optically pumped NV centre. We consider first the correlation functions involving the coherences which, due to the convenient structure of Eqs.~\eqref{EOMpumpedNVsigma0pm}-\eqref{EOMpumpedNVsigmapm}, can be easily derived using the quantum regression theorem:
\begin{multline}\label{CorrelatorSigma0PMpumping}
    \langle\hat{\sigma}_{0\pm}(t)\hat{O}(t-s)\rangle_{\rm ss} =
    \\
    =e^{-s[i\omega_\pm +(\kappa_1+3\gamma_+\mp \gamma_- +\kappa_2)/2 +\Omega]} \langle\hat{\sigma}_{0\pm}\hat{O}\rangle_{\rm ss},
\end{multline}
\begin{multline}\label{CorrelatorSigmaPM0pumping}
    \langle\hat{\sigma}_{\pm 0}(t)\hat{O}(t-s)\rangle_{\rm ss} = \\
    =e^{-s[-i\omega_\pm +(\kappa_1+3\gamma_+\mp \gamma_- +\kappa_2)/2+\Omega]} \langle\hat{\sigma}_{\pm 0}\hat{O}\rangle_{\rm ss}.
\end{multline}
\begin{multline}
    \langle\hat{\sigma}_{+-}(t)\hat{O}(t-s)\rangle_{\rm ss} =
    \\
    =e^{s[i(\omega_+-\omega_-) -\kappa_1-\gamma_+-2\kappa_2-\Omega]} \langle\hat{\sigma}_{+-}\hat{O}\rangle_{\rm ss},
\end{multline}
\begin{multline}
    \langle\hat{\sigma}_{-+}(t)\hat{O}(t-s)\rangle_{\rm ss} =
    \\
    =e^{s[-i(\omega_+-\omega_-) -\kappa_1-\gamma_+-2\kappa_2-\Omega]} \langle\hat{\sigma}_{+-}\hat{O}\rangle_{\rm ss}.
\end{multline}
Regarding the correlators involving the occupations, they obey the following equation of motion, which can be derived from Eq.~\eqref{EOMoccupationsPumping} using the quantum regression theorem:
\begin{multline}
    \frac{d}{ds} \langle \hat{O}_1(t)\hat{v}_j(t+s)\hat{O}_2(t)\rangle_{\rm ss} 
    \\
    = \tensorbar{M}_{\rm op} \langle \hat{O}_1(t)\hat{v}_j(t+s)\hat{O}_2(t)\rangle
\end{multline}
In this equation, valid for $s\ge 0$, $\hat{O}_1$ and $\hat{O}_2$ are two arbitrary NV centre operators, and $\hat{v}_j$ $(j=1,...5)$ is the $j-$component of the vector
\begin{equation}
    \hat{v} \equiv \left[
    \begin{array}{c}
         \hat{\sigma}_{00}-\langle\hat{\sigma}_{00}\rangle_{\rm ss}  \\
         \hat{S}_z/\hbar-\langle\hat{S}_z/\hbar\rangle_{\rm ss} \\
         \hat{\sigma}_{33}-\langle\hat{\sigma}_{33}\rangle_{\rm ss} \\
         \hat{\sigma}_{44}-\langle\hat{\sigma}_{44}\rangle_{\rm ss}\\
         \hat{\sigma}_{55}-\langle\hat{\sigma}_{55}\rangle_{\rm ss}
    \end{array}
    \right].
\end{equation}
The above differential equation has the following general solution,
\begin{multline}
    \langle \hat{O}_1(t)\hat{v}_j(t+s)\hat{O}_2(t)\rangle_{\rm ss}  
    \\=\sum_{i=1}^5 e^{\lambda_i s}\Lambda_{ji}\left[\tensorbar{\Lambda}^{-1}\langle\hat{O}_1\hat{v}_j\hat{O}_2\rangle_{\rm ss}\right]_j.
\end{multline}
Here, $\lambda_i$ are the eigenvalues of $\tensorbar{M}_{\rm op}$ and $\tensorbar{\Lambda}$ a matrix whose $i-$th column contains the corresponding eigenvector. The inverses of the real part of the eigenvalues, $ 1/\text{Re}[\lambda_i]$, determine the timescales at which the correlators decay to zero. We define the correlation time of the NV centres $\tau_{\rm nv}$ as the largest of these timescales,
\begin{equation}\label{tauNVdefinition}
    \tau_{\rm nv} \equiv \text{max}_i\left\vert\frac{1}{\text{Re}[\lambda_i]}\right\vert \sim (\kappa_1+\gamma_+ + \Omega)^{-1}
\end{equation}
where the absolute value is taken to ensure positiveness, as $\text{Re}[\lambda_i]<0$. The right-hand side is an estimation based on direct inspection of the decay rates of $\langle\hat{\sigma}_{00}\rangle$ and $\langle\hat{S}_{z}\rangle$
in Eqs.~\eqref{EOMoccupationsPumping}-\eqref{MdynOpticalPumping}.
The correlation time defined by Eq.~\eqref{tauNVdefinition} represents an upper bound to the decay time of the two-time correlators involving the occupations $\hat{\sigma}_{\alpha\alpha}$, and its value is relevant for deriving the reduced dynamics of the spin waves in Appendix~\ref{Appendix_effectiveSWdynamics}. In \figref{FigOpticalPumping}(c) we display the correlation time (green curve) as a function of the driving strength $\Omega$ at $\mu_0H_0=20$mT and for the parameters in Tables~\ref{tablePARAMS}-\ref{tableRATESopticalpumping}. The results do not change appreciably within the range of magnetic fields considered in this work. As evidenced by \figref{FigOpticalPumping}(c), optical pumping of the NV centres results in a reduction of the correlation time for the occupations. At optimal optical pumping conditions, $\Omega \approx 2\pi\times 10$kHz the correlation times take values $\tau_{\rm nv} \approx 50\mu$s, a decrease of three orders of magnitude with respect to a NV centre in thermal equilibrium. Finally, the approximation on the right hand side of Eq.~\eqref{tauNVdefinition}, namely $\tau_{\rm nv} \sim (\kappa_1+\gamma_++\Omega)^{-1}$, is a reasonable order-of-magnitude estimation
as shown by the dashed line in \figref{FigOpticalPumping}(c).

\section{Derivation of the resonant spin wave-paramagnetic spin interaction and extension to an ensemble of paramagnetic spins}\label{AppendixInteraction}

In this section we derive the resonant interaction Hamiltonian between a paramagnetic spin at a position $\mathbf{r}_0$ and the spin waves, Eq.~\eqref{Vgeneral}, from the general form of the magnetic dipole interaction. 
We start by writing the magnetic dipole interaction between the paramagnetic spin and the spin wave magnetic field in the general form
\begin{equation}\label{VddgeneralAppendix}
    \hat{V} = -\mu_0\hat{\boldsymbol{\mu}}_{\rm ps}\cdot\left[\hat{\mathbf{H}}(\mathbf{r}_0)-\mathbf{H}_0\right].
\end{equation}
Here, $\hat{\mathbf{H}}(\mathbf{r}_0)$ represents the total magnetic field operator. We substract the applied field $\mathbf{H}_0$ whose effect, namely the Zeeman splitting of the NV transitions, has already been included in $\hat{H}_{\rm ps}$, Eq.~\eqref{HNVappendix}. The resulting field $\hat{\mathbf{H}}(\mathbf{r}_0)-\mathbf{H}_0$ corresponds to the field generated by the spin waves at position $\mathbf{r}_0$, in principle up to arbitrary order in magnon operators. As we will see below, for studying the spin wave-induced modification of \emph{both} paramagnetic spin timescales, namely $T_1$ and $T_2^*$, the field must be included up to second order in magnon operators, i.e.,
\begin{equation}
    \hat{\mathbf{H}}(\mathbf{r})-\mathbf{H}_0\approx \hat{\mathbf{h}}^{(2)}(\mathbf{r})= \hat{\mathbf{h}}(\mathbf{r}) + \delta \hat{\mathbf{h}}(\mathbf{r}),
\end{equation}
where we have used the definition of the spin wave field up to second order, $\hat{\mathbf{h}}^{(2)}(\mathbf{r})$, as given by Eq.~\eqref{h2definition}. The first and second field contributions above are linear and quadratic in magnon operators, respectively, and have been calculated in Appendix~\ref{appendixSpinWaves}. We can consider each contribution separately by splitting the interaction Hamiltonian into linear and quadratic parts as
\begin{equation}\label{VdecomposeV1V2}
    \hat{V} = \hat{V}_{1} +  \hat{V}_{2},
\end{equation}
with
\begin{equation}\label{Vddappendix}
    \hat{V}_{1} \equiv -\mu_0\hat{\boldsymbol{\mu}}_{\rm ps}\cdot\hat{\mathbf{h}}(\mathbf{r}_0)
\end{equation}
and
\begin{equation}\label{deltaVddappendix}
    \hat{V}_{2} \equiv -\mu_0\hat{\boldsymbol{\mu}}_{\rm ps}\cdot\delta \hat{\mathbf{h}}(\mathbf{r}_0).
\end{equation}

Let us focus on the linear interaction term $\hat{V}_{1}$. By using the identity $\hat{\boldsymbol{\mu}}_{\rm ps} = -\vert\gamma\vert \hat{\mathbf{S}}$, we cast this interaction as
\begin{equation}
    \hat{V}_{1} = \sum_\beta \hat{s}_\beta\left[g_{\beta}^{+}\frac{\hat{S}_+}{\sqrt{2}}+g_{\beta}^{-}\frac{\hat{S}_-}{\sqrt{2}}+ g_{\beta}^{z}\hat{S}_z\right]+\text{H.c.}
\end{equation}
Here, we have introduced the expression for the spin wave field $\hat{\mathbf{h}}(\mathbf{r})$, Eq.~\eqref{hoperator}, and defined the raising and lowering spin operators
\begin{equation}
    \hat{S}_{\pm}\equiv\hat{S}_x\pm i\hat{S}_y =\hbar\sqrt{2}\left(\hat{\sigma}_{\pm0}+\hat{\sigma}_{0\mp}\right)
\end{equation}
as well as the coupling rates
\begin{equation}
    g_{\beta}^{k} \equiv \mu_0\vert\gamma_s\vert \mathcal{M}_{0\beta}\mathbf{e}_k^\dagger\cdot\mathbf{h}_\beta(\mathbf{r}_0).
\end{equation}
In the interaction picture with respect to the free Hamiltonian $\hat{H}_{\rm ps}+\hat{H}_{\rm sw}$ (see Eq.~\eqref{Htotal}), the above interaction takes the form
\begin{multline}
    \frac{\hat{V}_{1}(t)}{\hbar} = \sum_\beta \hat{s}_\beta e^{-i\omega_\beta t}\bigg[g_\beta^+\left(\hat{\sigma}_{+0}e^{i\omega_+t}+\hat{\sigma}_{0-}e^{-i\omega_-t}\right)
    \\
    +g_\beta^-\left(\hat{\sigma}_{0+}e^{-i\omega_+t}+\hat{\sigma}_{-0}e^{i\omega_-t}\right) +g_\beta^z\hat{S}_{z}\bigg]+\text{H.c.}
\end{multline}
We now undertake a rotating wave approximation~\cite{CarmichaelBook,breuer2002theory}, i.e., we neglect all the rapidly oscillating terms to obtain
\begin{multline}\label{V1RWA}
    \frac{\hat{V}_{1}(t)}{\hbar} \approx \sum_\beta \hat{s}_\beta e^{-i\omega_\beta t}
    \\
    \left[g_\beta^+\hat{\sigma}_{+0}e^{i\omega_+t}
    +g_\beta^-\hat{\sigma}_{-0}e^{i\omega_-t}+g_\beta^z\hat{S}_{z}\right]+\text{H.c.}
\end{multline}
This approximation is valid provided that
\begin{equation}\label{RWAjustification}
     g_{\beta}^{\pm},  \vert \omega_\beta-\omega_\pm\vert \ll \vert\omega_\beta + \omega_\pm\vert.
\end{equation}
These conditions are well satisfied in our system due to the lower cutoff for the spin wave bands (see \figref{FigureSystemBandStructure}[b]), as opposed to other reservoirs where it might be compromised by the presence of low-frequency modes~\cite{HummerArxiv2020}.

\begin{figure}[tbh!]
	\centering
	\includegraphics[width=\linewidth]{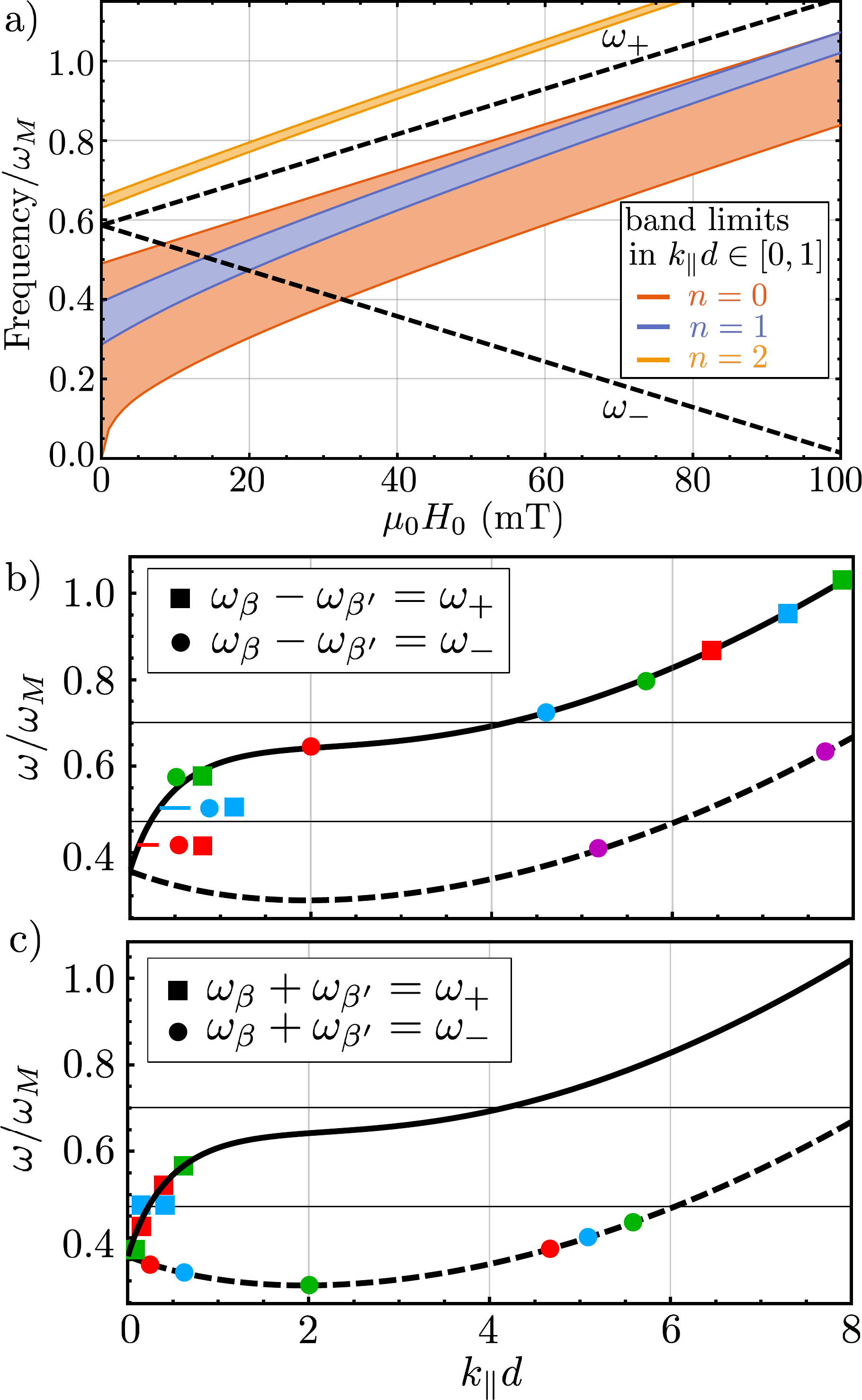}
	\caption{a) Frequencies of the two transitions of the NV centre (dashed lines) as a function of external field $H_0$. The three colored stripes represent the range of spin wave eigenfrequencies spanned by the bands $n=0,1$, and $2$ corresponding to significant coupling to the NV centre, i.e. in the low-wavenumber region $0\le k_\parallel d\le 1$. Specifically, the stripes correspond to the regions $\min_{\phi_k,k_\parallel}\omega_n(k_\parallel,\phi_k)\vert_{k_\parallel   d\in[0,1]}\le \omega \le \max_{\phi_k,k_\parallel}\omega_n(k_\parallel,\phi_k)\vert_{k_\parallel   d\in[0,1]}$, for $d=200$nm. b-c) Illustration of the three-operator resonance conditions appearing in the nonlinear interaction term $\hat{V}_2$. In each panel, symbols of the same color mark the energies of magnon pairs $\beta$ and $\beta'$ whose frequency difference (panel a) or sum (panel b) is equal to $\omega_+$ (squares) or $\omega_-$ (circles). The solid and dashed lines represent the $n=0$ spin wave bands in Damon-Eshbach and parallel propagation, respectively, whereas the horizontal gridlines mark the frequencies $\omega_+$ and $\omega_-$. For this figure we choose $d=200$nm and $\mu_0H_0= 20$mT.}\label{FigRWAappendix}
\end{figure}

The Hamiltonian can be further simplified by noting that only the spin waves 
with a low wavenumber $k_\parallel$ will be significantly coupled to the transitions of the paramagnetic spin, as the coupling strengths $g_\beta^k \propto \mathbf{e}_k^\dagger\cdot\mathbf{h}_\beta(\mathbf{r}_0)$ decay exponentially with the spin wave wavenumber (see e.g. Eq.~\eqref{hmodefunctions}). 
For $\gamma_s \sim\gamma$, and in the range of slab thicknesses
and applied fields we focus on in this work, namely $d\gtrsim 50\text{nm}$ and $\mu_0H_0\sim 10-30$mT, the spin wave modes with low wavenumber, say for definiteness $k_\parallel d <1$, are far detuned with respect to the $\vert 0 \rangle \leftrightarrow \vert + \rangle$ transition frequency, by a detuning of at least $0.1\omega_M \sim 2\pi\times 0.5$GHz.
This is illustrated in \figref{FigRWAappendix}(a), where we display the transition frequencies $\omega_\pm$ for the specific case of NV centres as a function of applied field, and compare them with the frequency range spanned by the first three spin wave bands in the low-wavenumber range $k_\parallel d\in[0,1]$.
As evidenced by this figure, the spin wave modes near resonance with $\omega_+$ are high-wavenumber modes, thus resulting in very weak coupling rates.  Conversely, the lower NV transition $\vert 0 \rangle \leftrightarrow \vert - \rangle$ is resonant with low-wavenumber spin waves for a wide range of applied fields $H_0$. We thus conclude that 
\begin{equation}
    \vert\omega_\beta-\omega_-\vert_{(k_\parallel d \lesssim 1)} \ll \vert\omega_\beta-\omega_+\vert_{(k_\parallel d \lesssim 1)}, \omega_\beta\vert_{(k_\parallel d \lesssim 1)}.
\end{equation}
Noting that the condition Eq.~\eqref{RWAjustification} is valid also for $g_{\beta}^z$, i.e., $g_{\beta}^z \ll\omega_\beta$, this allows us to discard the negligible terms $\propto\hat{S}_z, \hat{\sigma}_{+0},\hat{\sigma}_{0+}$ in the interaction Hamiltonian Eq.~\eqref{V1RWA}.
Back in the Schr\"odinger picture, we can finally write the first-order interaction as
\begin{equation}\label{V1RWAappendix}
    \hat{V}_1 \approx\hbar\sum_\beta\left(g_\beta^-\hat{s}_\beta\hat{\sigma}_{-0} + \text{H.c.}\right).
\end{equation}

We now follow a similar procedure for the nonlinear interaction term $\hat{V}_2$. By introducing the expression for the second-order field correction Eq.~\eqref{hsecondorder} we cast this contibution as
\begin{multline}\label{V2expanded}
    \hat{V}_2=\sum_{\beta\beta'}\hat{s}_\beta\hat{s}_{\beta'}\left[\tilde{g}_{\beta\beta'}^{'+}\frac{\hat{S}_+}{\sqrt{2}}+\tilde{g}_{\beta\beta'}^{'-}\frac{\hat{S}_-}{\sqrt{2}}+ \tilde{g}_{\beta\beta'}^{'z}\hat{S}_z\right]
    \\
    + \sum_{\beta\beta'}\hat{s}_\beta\hat{s}_{\beta'}^\dagger\left[\tilde{g}_{\beta\beta'}^{+}\frac{\hat{S}_+}{\sqrt{2}}+\tilde{g}_{\beta\beta'}^{-}\frac{\hat{S}_-}{\sqrt{2}}+ \tilde{g}_{\beta\beta'}^{z}\hat{S}_z\right]
    \\+\text{H.c.}
\end{multline}
where
\begin{equation}
    \tilde{g}_{\beta\beta'}^{'k}\equiv \mu_0\vert\gamma_s\vert\mathbf{e}_k^\dagger\cdot\mathbf{X}^+_{\beta\beta'}(\mathbf{r}_0),
\end{equation}
and
\begin{equation}
    \tilde{g}_{\beta\beta'}^{k}\equiv \mu_0\vert\gamma_s\vert\mathbf{e}_k^\dagger\cdot\mathbf{X}^-_{\beta\beta'}(\mathbf{r}_0),
\end{equation}
with the vectors $\mathbf{X}^\pm_{\beta\beta'}(\mathbf{r})$ defined in Eq.~\eqref{Xvectorsdefinition}.
We now transform $\hat{V}_2$ to the interaction picture with respect to $\hat{H}_{\rm sw}+\hat{H}_{\rm ps}$, and undertake the rotating wave approximation:
\begin{enumerate}
    \item First, we neglect the rapidly oscillating terms at frequencies $\omega_\beta + \omega_{\beta'}$ and $\omega_\beta + \omega_{\beta'}+\omega_\pm$. This is justified on the basis of the first rotating wave approximation Eq.~\eqref{RWAjustification} and on the observation that all the coupling rates in the above Hamiltonian are second order and thus smaller than the first-order rates $g_\beta^k$.
    \item Second, we neglect the terms oscillating at frequencies $\omega_\beta - \omega_{\beta'} +\omega_\pm$, since they are resonant only for magnon pairs $\beta$ and $\beta'$ whose frequency difference is equal to the transition frequencies $\omega_\pm$ of the paramagnetic spin. It can be checked that, in the range of parameters we focus on in this work, magnon pairs fulfilling this resonance condition have very different wavenumbers, i.e., $(k_\parallel - k_\parallel')d \gg 1$, see \figref{FigRWAappendix}(b) for an example in the case of NV centres. These terms will thus be very weakly coupled since the coupling rates decay exponentially with this difference, i.e., since $\tilde{g}_{\beta\beta'}^k\propto \exp(-l\vert \mathbf{k}_\parallel - \mathbf{k}_\parallel'\vert)$ (see Eqs.~\eqref{Xvectorsdecomposition} and \eqref{xnnprimeanalytical}).
    \item Third, we neglect the terms oscillating at frequencies $\omega_\beta + \omega_{\beta'} -\omega_-$, since they are resonant only for magnon pairs $\beta$ and $\beta'$ whose frequencies add up to the paramagnetic spin transition frequency $\omega_-$. As evidenced by the circles in \figref{FigRWAappendix}(c), which show the specific case of NV centres, the magnon pairs fulfilling this resonance condition also fulfill $(k_\parallel+k_\parallel')d \gg1$, their coupling rate   $\tilde{g}_{\beta\beta'}^{'k}\propto \exp(-l\vert \mathbf{k}_\parallel + \mathbf{k}_\parallel'\vert)$ thus becoming negligible (see Eqs.~\eqref{Xvectorsdecomposition} and \eqref{xnnprimeanalytical}).
    \item Two kind of terms remain at this point: on the one hand, terms $\propto \hat{s}_\beta\hat{s}_{\beta'}\hat{\sigma}_{+0}$, oscillating at frequencies $\omega_\beta + \omega_{\beta'} -\omega_+$, whose resonance condition is met for several low-wavenumber magnon pairs (square symbols in \figref{FigRWAappendix}[c]). On the other hand, terms $\propto \hat{s}_\beta\hat{s}_{\beta'}^\dagger\hat{S}_{z}$, oscillating at  $\omega_\beta - \omega_{\beta'}$, whose resonance condition is met for near-resonant magnon pairs of arbitrary wavenumber. The terms oscillating at $\omega_\beta + \omega_{\beta'} -\omega_+$ are neglected for the following reasons: a) their coupling rate
     $\tilde{g}_{\beta\beta'}^{'k}\propto \exp(-l\vert \mathbf{k}_\parallel + \mathbf{k}_\parallel'\vert)$ decays much faster than the coupling rate $\tilde{g}_{\beta\beta'}^{k}\propto \exp(-l\vert \mathbf{k}_\parallel - \mathbf{k}_\parallel'\vert)$ of the terms $\propto \hat{s}_\beta\hat{s}_{\beta'}^\dagger\hat{S}_{z}$, thus making the interaction relevant over a much narrower wavenumber range. b) While the terms proportional to $\hat{S}_z$ describe a new spin wave-induced decoherence mechanism (dephasing), the terms containing $\hat{\sigma}_{+0}$ represent a small correction to the decay lifetime $T_1$, whose main effect is already accounted for at first order (i.e. by $\hat{V}_1$). 
\end{enumerate}

Under the above simplifications, the second-order correction to the spin wave-paramagnetic spin interaction Eq.~\eqref{V2expanded} reduces to
\begin{equation}\label{V2RWAappendix}
    \hat{V}_2=\hat{S}_z\sum_{\beta\beta'}\left[\hat{s}_\beta\hat{s}_{\beta'}^\dagger \tilde{g}_{\beta\beta'}^{z}
    +\text{H.c.}\right]
\end{equation}
Finally, by using the commutation relations Eq.~\eqref{commutationmagnons} and the property Eq.~\eqref{Xbetabetaequal0} we can write the above expression in the form
\begin{equation}\label{V2RWAappendixprocessed}
    \hat{V}_2=\hat{S}_z\sum_{\beta \beta'}\hat{s}_{\beta}^\dagger \hat{s}_{\beta'}\tilde{g}_{\beta\beta'}
\end{equation}
where we define
\begin{equation}\label{g2norderfinal}
    \tilde{g}_{\beta\beta'} = \tilde{g}_{\beta'\beta}^* \equiv \tilde{g}_{\beta'\beta}^{z}
    + \tilde{g}_{\beta\beta'}^{z*}.
\end{equation}
By introducing Eqs.~\eqref{V1RWAappendix} and \eqref{V2RWAappendixprocessed} into Eq.~\eqref{VdecomposeV1V2} we obtain the interaction Hamiltonian given by Eq.~\eqref{Vgeneral} in the main text.

The equations of motion derived for a single paramagnetic spin in Appendices~\ref{AppendixNVcentres}-\ref{AppendixInteraction} can be generalized to an ensemble of paramagnetic spins situated at positions $\mathbf{r}_j$, $j=1,...N$. We assume the paramagnetic spins to be independent, that is, we neglect any interaction between them, either direct (e.g. dipole-dipole) or mediated by their corresponding thermal reservoirs. This approximation is valid for the densities of NV centres considered in this work ($\le 10^5\text{($\mu$m)}^{-3}$), as experiments show no significant dipole-dipole-induced modification of the coherence times $T_2^*$ for ensembles of NV centres within this range of densities~\cite{KleinsasserAPL2016,BauchPRX2018}.
Under this assumption the Hamiltonian of the paramagnetic spins and the interaction with the spin waves are written as
\begin{equation}
    \hat{H}_{\rm ps} = \sum_{j=1}^N\hat{H}_{\rm ps}^{(j)}
    \hspace{0.3cm};\hspace{0.3cm}
    \hat{V} = \sum_{j=1}^N \hat{V}^{(j)}.
\end{equation}
The expressions of $\hat{H}_{\rm ps}^{(j)}$ and $\hat{V}^{(j)}$ are identical to Eqs.~\eqref{HNV} and \eqref{Vgeneral} respectively, i.e.,
\begin{equation}
    \hat{H}_{\rm ps}^{(j)}=\hbar \sum_{\alpha=\pm}\omega_{\alpha}^{(j)} \hat{\sigma}_{\alpha\alpha}^{(j)}
\end{equation}
and
\begin{multline}
    \hat{V}^{(j)}= \hbar\sum_\beta\left(g_{j\beta}\hat{s}_\beta\hat{\sigma}^{(j)}_{-0} + \text{H.c.}\right) +
    \\+ \hat{S}_z^{(j)}\sum_{\beta \beta'}\tilde{g}_{j\beta\beta'}\hat{s}_{\beta}^\dagger \hat{s}_{\beta'},
\end{multline}
this time with paramagnetic spin-dependent
transition matrices $\hat{\sigma}^{(j)}_{\alpha\alpha'} \equiv \vert \alpha\rangle_j{}_j\langle \alpha'\vert$, transition frequencies $\omega_{\alpha}^{(j)}$, and coupling rates $g_{j\beta}$ and $\tilde{g}_{j\beta\beta'}$. These coupling rates are given respectively by identical expressions as Eqs.~\eqref{gfirstorder} and \eqref{gsecondorder} under the substitution $\mathbf{r}_0\to\mathbf{r}_j$, i.e.,
\begin{equation}\label{gfirstorderAppendix}
    g_{j\beta} \equiv \mu_0 \vert \gamma_s\vert\mathcal{M}_{0\beta}\int d^3\mathbf{r}\left[\mathbf{e}^*_-\cdot \tensorbar{\mathcal{G}}(\mathbf{r}_j-\mathbf{r})\cdot\mathbf{m}_\beta(\mathbf{r})\right],
\end{equation}
\begin{multline}\label{gsecondorderAppendix}
    \tilde{g}_{j\beta\beta'} = -\mu_0\vert\gamma_s\vert\frac{\mathcal{M}_{0\beta}\mathcal{M}_{0\beta'}}{M_S}
    \\
    \times \int d^3\mathbf{r}
    \left[\mathbf{m}_{\beta}^*(\mathbf{r})\cdot\mathbf{m}_{\beta'}(\mathbf{r})\right]
    \text{Re}\left[\mathbf{e}_z\cdot\tensorbar{\mathcal{G}}(\mathbf{r}_j-\mathbf{r})\cdot\mathbf{e}_z\right].
\end{multline}
The dissipation of the NV centres is generalized in a similar way,
\begin{multline}\label{DissipatorNVcentresENSEMBLE}
    \mathcal{D}_{\rm ps}[\hat{\rho}]=\sum_j\frac{\kappa_{2j}}{\hbar^2}\mathcal{L}_{\hat{S}_{z}^{(j)}\hat{S}_{z}^{(j)}}[\hat{\rho}] +\sum_j \sum_{\alpha=\pm}\\
    \kappa_{1j}\left(\bar{n}_\alpha^{(j)}\mathcal{L}_{\hat{\sigma}_{\alpha0}^{(j)}\hat{\sigma}_{0\alpha}^{(j)}}[\hat{\rho}]+(\bar{n}_\alpha^{(j)}+1)\mathcal{L}_{\hat{\sigma}_{0\alpha}^{(j)}\hat{\sigma}_{\alpha0}^{(j)}}[\hat{\rho}]\right),
\end{multline}
with $\bar{n}_\alpha^{(j)} = [\exp(\hbar\omega_\alpha^{(j)}/k_B T)-1]^{-1}$. Note that, although in the above expressions we have allowed the characteristic rates $\omega_{\pm}$, $\kappa_1$, and $\kappa_2$ to be different for each paramagnetic spin, in this article we consider identical paramagnetic spins for the sake of simplicity.

\section{Tracing out procedure}\label{Appendix_TracingOut}

In this appendix we summarize briefly the open quantum systems techniques employed to obtain the effective dynamics in the successive appendices. We begin by considering a system and a bath with free Hamiltonians $\hat{H}_S$ and $\hat{H}_B$ respectively, which interact through a general interaction potential $\hat{V}$. The total Hamiltonian of the compound system+bath is
\begin{equation}
    \hat{H} = \hat{H}_S + \hat{H}_B + \hat{V}.
\end{equation}
Here we assume for simplicity that $\hat{H}$ is time-independent. We also assume that system and bath form a closed system, so that the dynamics of the total density matrix $\hat{\rho}_T$ is given by the Von Neumann equation
\begin{equation}\label{Meqtotappendix}
    \dot{\hat{\rho}}_T = -\frac{i}{\hbar}\left[\hat{H},\hat{\rho}_T\right].
\end{equation}
The procedure below can be generalized to the case where system and/or bath are open, i.e. they undergo additional dissipative dynamics, in the case where the baths generating this dynamics are independent~\footnote{This is achieved by means of projection operator techniques~\cite{breuer2002theory}. For a derivation using a similar system as ours, see e.g. Ref~\cite{KusturaToAppear2020}.}.  In the interaction picture with respect to $\hat{H}_S + \hat{H}_B$ we rewrite the von Neumann equation in integro-differential form as
\begin{multline}
    \dot{\hat{\rho}}_T = -\frac{i}{\hbar}\left[\hat{V}(t),\hat{\rho}_T(0)\right]\\-\frac{1}{\hbar^2}\int_0^t ds\left[\hat{V}(t),\left[\hat{V}(t-s),\hat{\rho}_T(t-s)\right]\right].
\end{multline}
The above equation is still exact, but cast in a suitable form for approximations.

\subsection{Born-Markov master equation}\label{AppTraceOut_BornMarkov}

The most usual tracing out procedure~\cite{breuer2002theory} consists on the following steps: (i)
undertaking the weak-coupling or Born approximation $\hat{\rho}_T(t) \approx \hat{\rho}(t)\otimes\hat{\rho}_{\rm ss}$ with $\hat{\rho}(t)$ the reduced density matrix of the system and $\hat{\rho}_{\rm ss}$ the steady-state density matrix of the bath. (ii) undertaking the Markov approximation by taking $\hat{\rho}(t-s) \approx\hat{\rho}(t)$ inside the integral.  The Markov approximation is justified if the two-time correlators of bath operators, i.e. the correlators appearing in Eq.~\eqref{onesidedPSDsdefinition}, decay with the delay $s$ much faster than the timescale associated to the evolution of the density matrix in the interaction picture~\cite{breuer2002theory}. In the following sections we will check the validity of the Markov approximation in our system. (iii) taking the partial trace over the bath modes and assuming $\text{Tr}_B[\hat{V}(t),\hat{\rho}_T(0)]=0$~\footnote{This condition can always be fulfilled if one writes $\hat{V} = \hat{V}-\text{Tr}_B[\hat{V}]+\text{Tr}_B[\hat{V}] \equiv \hat{V}'+\text{Tr}_B[\hat{V}]$, and re-absorbs the second term into the system Hamiltonian. One can then work with the interaction $\hat{V}'$ which by definition fulfills $\text{Tr}_B[\hat{V}']=0$.}. As a result the following equation of motion for the system density matrix is obtained,
\begin{equation}\label{BornMarkovgeneral}
    \dot{\hat{\rho}} \approx -\frac{1}{\hbar^2}\text{Tr}_B\int_0^\infty ds\left[\hat{V}(t),\left[\hat{V}(t-s),\hat{\rho}(t)\otimes\hat{\rho}_{\rm ss}\right]\right].
\end{equation}

The above equation can then be manipulated by writing the interaction potential in the form
\begin{equation}\label{interactiongeneral}
	\frac{1}{\hbar}\hat{V}(t) = \sum_{\alpha}\hat{O}_{s,\alpha}(t)\otimes\hat{B}_\alpha(t) + \text{H.c.},
\end{equation}
with $\hat{O}_{s,\alpha}(t)$ and $\hat{B}_\alpha(t)$ being system and bath operators, respectively. We choose the above representation such that the system operators $\hat{O}_{s,\alpha}(t)$ in the interaction picture always evolve with a trivial phase, i.e.,
	\begin{equation}
	\hat{O}_{s,\alpha}(t) = \hat{O}_{s,\alpha} e^{-i\Omega_\alpha t}
	\end{equation}
with $ \hat{O}_{s,\alpha}$ the corresponding Schr\"odinger Picture operator and $\Omega_\alpha \in \mathbb{R}$. By introducing Eq.~\eqref{interactiongeneral} into Eq.~\eqref{BornMarkovgeneral}  and assuming $[\hat{H}_B,\hat{\rho}_{\rm ss}] = 0$, we can cast the master equation in the convenient Lindblad form~\cite{breuer2002theory}. In the Schr\"odinger picture, this Lindblad master equation reads
\begin{multline}\label{MEqsupergeneral}
    \dot{\hat{\rho}} = -\frac{i}{\hbar}\left[\hat{H}_S+\hat{H}_{\rm eff},\hat{\rho}\right] 
    + \sum_{\alpha\alpha'}\Big\lbrace \Gamma_{\alpha\alpha'}^{(n)}\mathcal{L}_{\hat{O}_{s,\alpha'}\hat{O}_{s,\alpha}^\dagger} [\hat{\rho}]
    \\+ 
    \Gamma_{\alpha\alpha'}^{(a)}\mathcal{L}_{\hat{O}_{s,\alpha'}^\dagger\hat{O}_{s,\alpha}}[\hat{\rho}]
    +\left(\Gamma_{\alpha\alpha'}^{(s)}\mathcal{L}_{\hat{O}_{s,\alpha'}\hat{O}_{s,\alpha}}[\hat{\rho}]+\text{H.c.}\right)
    \Big\rbrace.
\end{multline}
The bath-induced coherent dynamics is captured by the effective Hamiltonian
\begin{multline}
    \hat{H}_{\rm eff} = \hbar\sum_{\alpha\alpha'}G_{\alpha\alpha'}^a\hat{O}_{s,\alpha}\hat{O}_{s,\alpha'}^\dagger + G_{\alpha\alpha'}^n\hat{O}_{s,\alpha}^\dagger\hat{O}_{s,\alpha'}
    \\+\left(
    G_{\alpha\alpha'}^s\hat{O}_{s,\alpha}\hat{O}_{s,\alpha'} + \text{H.c.},
    \right),
\end{multline}
whereas the dissipative dynamics is represented by the Lindblad superoperators, defined in Eq.~\eqref{Lindbladiandefinition}.
The rates involved in the master equation can be written in compact form as
\begin{equation}
    \left[
    \begin{aligned}
        \Gamma_{\alpha\alpha'}^{(n)}  \\
        \Gamma_{\alpha\alpha'}^{(a)}   \\
        \Gamma_{\alpha\alpha'}^{(s)}
    \end{aligned}
    \right]
    \equiv
    2\pi\left[
    \begin{aligned}
        \bar{S}_{\alpha\alpha'}^{(n)}(\Omega_{\alpha'})+\bar{S}_{\alpha'\alpha}^{(n)*}(\Omega_{\alpha})  \\
        \bar{S}_{\alpha\alpha'}^{(a)}(-\Omega_{\alpha'})+\bar{S}_{\alpha'\alpha}^{(a)*}(-\Omega_{\alpha})   \\
        \bar{S}_{\alpha\alpha'}^{(-)}(\Omega_{\alpha'})+\bar{S}_{\alpha\alpha'}^{(+)*}(-\Omega_{\alpha})
    \end{aligned}
    \right]
\end{equation}
and
\begin{equation}
    \left[
    \begin{aligned}
        G_{\alpha\alpha'}^{n}  \\
        G_{\alpha\alpha'}^{a}   \\
        G_{\alpha\alpha'}^{s}
    \end{aligned}
    \right]
    \equiv
    -i\pi\left[
    \begin{aligned}
        \bar{S}_{\alpha\alpha'}^{(n)}(\Omega_{\alpha'})-\bar{S}_{\alpha'\alpha}^{(n)*}(\Omega_{\alpha})  \\
        \bar{S}_{\alpha\alpha'}^{(a)}(-\Omega_{\alpha'})-\bar{S}_{\alpha'\alpha}^{(a)*}(-\Omega_{\alpha})   \\
       \bar{S}_{\alpha\alpha'}^{(-)}(\Omega_{\alpha'})-\bar{S}_{\alpha\alpha'}^{(+)*}(-\Omega_{\alpha})
    \end{aligned}
    \right].
\end{equation}
Here, we define the one-sided power spectral densities of the bath, that capture the whole effect of the bath on the system dynamics, as
\begin{equation}\label{onesidedPSDsdefinition}
    \left[
    \begin{aligned}
         \bar{S}^{(n)}_{\alpha\alpha'}(\omega)  \\
         \bar{S}^{(a)}_{\alpha\alpha'}(\omega) \\
         \bar{S}^{(+)}_{\alpha\alpha'}(\omega) \\
         \bar{S}^{(-)}_{\alpha\alpha'}(\omega) \\
    \end{aligned}
    \right]
    \equiv
    \int_0^\infty \frac{ds}{2\pi} e^{i\omega s}
    \left[
    \begin{aligned}
         \left\langle\hat{B}_\alpha^\dagger(t)\hat{B}_{\alpha'}(t-s)\right\rangle_{\rm ss}  \\
         \left\langle\hat{B}_\alpha(t)\hat{B}_{\alpha'}^\dagger(t-s)\right\rangle_{\rm ss} \\
         \left\langle\hat{B}_\alpha^\dagger(t)\hat{B}_{\alpha'}^\dagger(t-s)\right\rangle_{\rm ss} \\
         \left\langle\hat{B}_\alpha(t)\hat{B}_{\alpha'}(t-s)\right\rangle_{\rm ss} \\
    \end{aligned}
    \right].
\end{equation} 
The above expressions are very general and, in the following subsections, we will apply them to the two cases detailed in the main text.

\subsection{Frozen bath model}\label{AppTracingOut_FrozenBath}

The Born-Markov master equation derived above is valid when the bath correlation times are much shorter than the typical timescales associated with the system-bath interaction. It is also possible to obtain effective dynamics in the opposite limit, the so-called ``frozen bath'' regime~\cite{MontoyaCastilloJchemPhys2015}. In this regime, the evolution of the bath is much slower than the timescale associated to the system evolution in the interaction picture. 
In this limit, the effective dynamics can be obtained directly from Eq.~\eqref{Meqtotappendix}, by undertaking the Born approximation and by substituting~\cite{MontoyaCastilloJchemPhys2015,CampisiPRA2012}
\begin{equation}
    \hat{V}(t) \approx \text{Tr}_{B}[\hat{V}(t)].
\end{equation}
Within this approximation the bath is perceived, from the point of view of the system, as a stationary driving term given by the time-independent (``frozen'') expected value of the interaction potential. The frozen bath regime is not common in theory of open quantum systems as, in this limit, the bath does not induce any dissipation in the system as we will see below. However, we will make use of this regime to include the effect of the slow occupation dynamics of the NV centre on the spin wave master equation, in Sec.~\ref{SecSWmodification} and Appendix~\ref{Appendix_effectiveSWdynamics}.

\section{Effective dynamics of paramagnetic spins}\label{Appendix_EffectiveNVdynamics}

In this section we apply the general formalism derived in Appendix~\ref{Appendix_TracingOut} to derive the effective dynamics of a system of paramagnetic spins interacting with a spin wave bath. Since the derivation, specifically the justification of the Markov approximation, relies on the particular values of $T_1$ and $T_2^*$, we take the specific values for NV centres and will for simplicity refer to the paramagnetic spins as ``NV centres'' in this section. 
The results derived below hold for any species of paramagnetic spin with similar values of $T_1$ and $T_2^*$. Extending our derivation  to other paramagnetic spins is also possible.  This Appendix is organized as follows: first, we provide the master equation for the NV centres and give the expressions for all the involved rates in Sec.~\ref{AppNVdynamics_generalMEq}. In the following sections we analyze the different effective dynamics separately: in Sec.~\ref{AppNVdynamics_T1T2modified} we study the modification of the occupation ($T_1$) and coherence ($T_2^*$) lifetimes for a single NV centre. Then, in Sec.~\ref{AppNVdynamics_LambShift} we derive the expressions for the frequency shift and spin wave-induced force analyzed in the main text. Finally, in Sec.~\ref{AppNVdynamics_Couplings} we analyze the spin wave-mediated couplings between different NV centres in an ensemble.

\subsection{Derivation of the general master equation and equations of motion}\label{AppNVdynamics_generalMEq}

We consider for generality an ensemble of NV centres at arbitrary positions $\mathbf{r}_j$ outside the YIG film.
Our starting point is the interaction Hamiltonian in the Interaction Picture, given by
\begin{multline}\label{VtotalAPP}
    \hat{V} = \hat{V}_1 + \hat{V}_2 = \sum_{j\beta} g_{j\beta}\hat{s}_\beta(t)\hat{\sigma}^{(j)}_{-0}e^{i\omega_-t} + \text{H.c.}
    \\
    +
    \sum_{j}\frac{\hat{S}_{z}^{(j)}}{\hbar}\frac{1}{2}\sum_{\beta\beta'}\tilde{g}_{j\beta\beta'}\hat{s}^\dagger_\beta(t)\hat{s}_{\beta'}(t) + \text{H.c.}.
\end{multline}
with
$\hat{s}_\beta(t) = \hat{s}_\beta\exp(-i\omega_\beta t)$. As shown in Appendix~\ref{AppSW_quantizationEOMgamma}, the spin waves are governed by the master equation Eq.~\eqref{MasterEquationOnlyMagnonsAppendix}, whose steady state is thermal. By applying the quantum regression formula~\cite{CarmichaelBook} one can prove that the two-time correlators between three magnon operators cancel out in a thermal state:
\begin{equation}
    \left\langle \hat{s}_\beta(t) \hat{s}_{\beta'}^\dagger(t-s)\hat{s}_{\beta''}(t-s)\right\rangle_{\rm ss}=0 \hspace{0.3cm} \forall s,
\end{equation}
\begin{equation}
    \left\langle \hat{s}_\beta^\dagger(t) \hat{s}_{\beta'}^\dagger(t-s)\hat{s}_{\beta''}(t-s)\right\rangle_{\rm ss}=0 \hspace{0.3cm} \forall s.
\end{equation}
As a consequence, the two terms in the interaction, $\hat{V}_1$ and $\hat{V}_2$, are independent from each other and can be treated separately. The two-time correlators of the magnon operators appearing on each of these terms decay at similar rates $\sim\gamma_\beta$, whereas the system evolution in the interaction picture is dominated by the timescale $\sim T_2^*$.
We treat both interaction terms within the Born-Markov approximation, valid for $\gamma_\beta T_2^* \gg 1$. This condition is well fulfilled for all spin wave bands with $n>0$. The lowest band $n=0$ lies at the boundary of this validity regime $\gamma_\beta T_2^*\approx 1$, and hence small non-Markovian corrections to the effective dynamics of the NV centres could be expected, especially for NV transition frequencies near the lower cutoff of a spin wave band~\cite{LiuNatPhys2017}. Although outside the scope of our work, exploring these dynamics beyond the Markov approximation is an interesting outlook to our work.
\footnote{Note that these corrections are expectedly small, because the band $n=0$ represents only one among many contributions to the effective dynamics of the NV centres (specifically, to the rates appearing in the master Eq.~[\ref{fullMEqNV}]). The contributions from higher bands ($n=1,2,...$), which are significant especially at fields $H_0 \lesssim 15$mT, are well within the validity regime of the Markov approximation.}

Under the above approximations, the resulting master equation for the ensemble of NV centres in the Schr\"odinger picture reads
\begin{multline}\label{fullMEqNV}
    \frac{d}{dt}\hat{\rho}_{\rm ps} = -\frac{i}{\hbar}\left[\hat{H}_{\rm ps} + \hat{H}_{\rm eff,1}+ \hat{H}_{\rm eff,2},\hat{\rho}_{\rm ps}\right]
    \\+\mathcal{D}_{\rm ps}[\hat{\rho}_{\rm ps}]+
    \mathcal{D}_{\rm op}[\hat{\rho}_{\rm ps}]  +\mathcal{D}_{1}[\hat{\rho}_{\rm ps}]+\mathcal{D}_{2}[\hat{\rho}_{\rm ps}].
\end{multline}
where the sub-indices $1$ and $2$ indicate the contributions stemming from $\hat{V}_1$ and $\hat{V}_2$, respectively.
Let us first focus on the contribution from $\hat{V}_1$. By applying the 
general formalism of Appendix~\ref{Appendix_TracingOut} we obtain an effective Hamiltonian
\begin{multline}\label{Hprime1Appendix}
    \frac{1}{\hbar}\hat{H}_{\rm eff,1} = \sum_{j\ne j'}G_{jj'}^a\hat{\sigma}_{-0}^{(j)}\hat{\sigma}_{0-}^{(j')} + G_{jj'}^n\hat{\sigma}_{0-}^{(j)}\hat{\sigma}_{-0}^{(j')}
    \\
    +\sum_j \sum_{\alpha=\pm}\delta_{1\alpha}^{(j)} \hat{\sigma}_{\alpha\alpha}^{(j)},
\end{multline}
with
\begin{equation}\label{Gjjanexpression}
    \left[
    \begin{array}{c}
          G_{jj'}^a  \\
          G_{jj'}^n 
    \end{array}
    \right]
    =\sum_\beta\frac{-\Delta_\beta}{\Delta_\beta^2+(\gamma_\beta/2)^2}
    \left[
    \begin{array}{c}
         (1+\bar{n}_\beta) g_{j\beta}g_{j'\beta}^*   \\
          -\bar{n}_\beta g_{j\beta}^*g_{j'\beta} 
    \end{array}
    \right],
\end{equation}
where $\Delta_\beta \equiv\omega_\beta-\omega_-$,
and with first-order frequency shifts given by
\begin{equation}\label{delta1minusfromG}
    \delta_{1-}^{(j)} \equiv G_{jj}^a-G_{jj}^n \hspace{0.3cm} ;  \hspace{0.3cm} \delta_{1+}^{(j)} \equiv -G_{jj}^n.
\end{equation}
Note that in writing down the Hamiltonian Eq.~\eqref{Hprime1Appendix} we have neglected a constant term $-\sum_j\delta_{1+}^{(j)}(\mathbf{r}_j)$ that we must include when computing the induced force.
The corresponding dissipator reads
\begin{multline}
    \mathcal{D}_1[\hat{\rho}] = \sum_{j}\kappa_{a}^{(j)}\mathcal{L}_{\hat{\sigma}_{-0}^{(j)}\hat{\sigma}_{0-}^{(j)}}[\hat{\rho}]+\kappa_{d}^{(j)}\mathcal{L}_{\hat{\sigma}_{0-}^{(j)}\hat{\sigma}_{-0}^{(j)}}[\hat{\rho}]
    \\+ \sum_{j\ne j'}\Gamma_{jj'}^{(n)}\mathcal{L}_{\hat{\sigma}_{-0}^{(j')}\hat{\sigma}_{0-}^{(j)}}[\hat{\rho}]+\Gamma_{jj'}^{(a)}\mathcal{L}_{\hat{\sigma}_{0-}^{(j')}\hat{\sigma}_{-0}^{(j)}}[\hat{\rho}]
\end{multline}
with
\begin{equation}
    \left[
    \begin{array}{c}
          \Gamma_{jj'}^{(a)}  \\
          \Gamma_{jj'}^{(n)} 
    \end{array}
    \right]
    =\sum_\beta\frac{\gamma_\beta}{\Delta_\beta^2+(\gamma_\beta/2)^2}
    \left[
    \begin{array}{c}
         (1+\bar{n}_\beta) g_{j\beta}g_{j'\beta}^*   \\
          \bar{n}_\beta g_{j\beta}^*g_{j'\beta} 
    \end{array}
    \right],
\end{equation}
and decay and absorption rates given by
\begin{equation}
    \kappa_a^{(j)} \equiv \Gamma_{jj}^{(n)} \hspace{0.3cm} ;  \hspace{0.3cm}  \kappa_d^{(j)} \equiv \Gamma_{jj}^{(a)}.
\end{equation}

Regarding the contribution $\hat{V}_2$, one can largely simplify it by noting that, in the representation Eq.~\eqref{interactiongeneral}, the bath operator
\begin{equation}
    \hat{B}_\alpha(t)\to\hat{B}_j(t) \equiv\frac{1}{2}\sum_{\beta\beta'}\tilde{g}_{j\beta\beta'}\hat{s}^\dagger_\beta(t)\hat{s}_{\beta'}(t)
\end{equation}
is Hermitian, and hence all the one-sided power spectral densities in Eq.~\eqref{onesidedPSDsdefinition} are equal. Their value at zero frequency reads
\begin{equation}\label{PSDV2zerofreq}
    \bar{S}_{jj'}(0) = \frac{-1}{8\pi}\sum_{\beta \beta'}\frac{\bar{n}_\beta(1+\bar{n}_{\beta'})\tilde{g}_{j\beta\beta'}\tilde{g}_{j'\beta'\beta}}{i(\omega_\beta-\omega_\beta')-(\gamma_\beta+\gamma_{\beta'})/2}.
\end{equation}
Furthermore, the system operator $\hat{O}_{s,\alpha}$ is also Hermitian, allowing us to simplify the final master equation. Specifically, the coherent contribution reads
\begin{equation}
    \frac{\hat{H}_{\rm eff,2}}{\hbar} =  \sum_{j}\sum_{\alpha=\pm}\delta_{2\alpha}^{(j)}\hat{\sigma}_{\alpha\alpha}^{(j)}+\sum_{j\ne j'}\frac{G_{2,jj'}}{ \hbar^2}\hat{S}_{z}^{(j)}\hat{S}_{z}^{(j')},
\end{equation}
with
\begin{equation}\label{G2jjprime}
    G_{2,jj'} \equiv -2\pi i\left[\bar{S}_{jj'}(0) - \bar{S}_{j'j}^*(0)\right]
\end{equation}
and second-order frequency shifts given by
\begin{equation}\label{delta2fromG}
    \delta_{2\pm}^{(j)} \equiv G_{2,jj}=4\pi\text{Im}[\bar{S}_{jj}(0)].
\end{equation}
On the other hand, the dissipative contribution reads
\begin{equation}
    \mathcal{D}_2[\hat{\rho}]= \sum_{j}\frac{\kappa_{2}'^{(j)}}{\hbar^2}\mathcal{L}_{\hat{S}_{z}^{(j)}\hat{S}_{z}^{(j)}}[\hat{\rho}]+\sum_{j\ne j'}\frac{\Gamma_{2,jj'}}{\hbar^2}\mathcal{L}_{\hat{S}_{z}^{(j')}\hat{S}_{z}^{(j)}}[\hat{\rho}],
\end{equation}
with the rates
\begin{multline}
    \Gamma_{2,jj'} \equiv 2\pi\left[3\bar{S}_{jj'}(0)+3\bar{S}_{j'j}^*(0)+\bar{S}_{j'j}(0) + \bar{S}_{jj'}^*(0)\right],
\end{multline}
and the additional dephasing rate
\begin{equation}\label{kappa2prime}
    \kappa_{2}'^{(j)} \equiv \Gamma_{2,jj} = 16\pi\text{Re}[\bar{S}_{jj}(0)].
\end{equation}

It is insightful to compute the equations of motion for the expected value of an arbitrary transition operator of a given NV centre, $\langle \hat{\sigma}_{\alpha\alpha'}^{(j)}\rangle$, as given by the complete master equation Eq.~\eqref{fullMEqNV}. For simplicity we will assume NV centres in thermal equilibrium, i.e. $\Omega = 0$. The extension to optically pumped NV centres can be done following the derivation in Appendix~\ref{AppNV_OpticalPumping}. The equations of motion for the ground state manifold can be written in compact form in the following way,
\begin{widetext}
\begin{multline}\label{EOMNVsA}
    \frac{d}{dt}\left\langle \hat{\sigma}_{00}^{(j)} \right\rangle = - \left(2\gamma_++\kappa_a^{(j)}\right)\left\langle \hat{\sigma}_{00}^{(j)} \right\rangle
    +
    \kappa_1(1+\bar{n}_+)\left\langle \hat{\sigma}_{++}^{(j)} \right\rangle
    +
    \left[\kappa_1(1+\bar{n}_-)+\kappa_d^{(j)}\right]\left\langle \hat{\sigma}_{--}^{(j)} \right\rangle
    -
    \sum_{k\ne j}\left[\Gamma_{d,jk}\left\langle \hat{\sigma}_{0-}^{(k)}\hat{\sigma}_{-0}^{(j)} \right\rangle +\text{c.c.}\right]  
\end{multline}
\begin{equation}
    \frac{d}{dt}\left\langle \hat{\sigma}_{++}^{(j)} \right\rangle = 
    \kappa_1\bar{n}_+\left\langle \hat{\sigma}_{00}^{(j)} \right\rangle - \kappa_1(1+\bar{n}_+)\left\langle \hat{\sigma}_{++}^{(j)} \right\rangle
\end{equation}
\begin{equation}
    \frac{d}{dt}\left\langle \hat{\sigma}_{--}^{(j)} \right\rangle = 
    \left[\kappa_1\bar{n}_-+\kappa_a^{(j)}\right]\left\langle \hat{\sigma}_{00}^{(j)} \right\rangle - \left[\kappa_1(1+\bar{n}_-)+\kappa_d^{(j)}\right]\left\langle \hat{\sigma}_{--}^{(j)} \right\rangle+\sum_{k\ne j}\left[\Gamma_{d,jk}\left\langle \hat{\sigma}_{0-}^{(k)}\hat{\sigma}_{-0}^{(j)} \right\rangle +\text{c.c.}\right]  
\end{equation}
\begin{multline}\label{EOMsigma0plustotal}
    \frac{d}{dt}\left\langle \hat{\sigma}_{0+}^{(j)} \right\rangle =\left[-i\left(\omega_+ + \delta_{+}^{(j)}\right)-\frac{\kappa_2+\kappa_2'^{(j)}}{2}-\frac{\kappa_1+3\gamma_+-\gamma_-+\kappa_a^{(j)}}{2}\right]\left\langle \hat{\sigma}_{0+}^{(j)} \right\rangle
    \\
    -\sum_{k\ne j}\frac{\Gamma_{z,jk}}{\hbar}\left\langle \hat{\sigma}_{0+}^{(j)}\hat{S}_z^{(k)} \right\rangle
     -\sum_{k\ne j}\Gamma_{d,jk}\left\langle \hat{\sigma}_{-+}^{(j)}\hat{\sigma}_{0-}^{(k)} \right\rangle
\end{multline}
\begin{multline}\label{EOMsigma0minustotal}
    \frac{d}{dt}\left\langle \hat{\sigma}_{0-}^{(j)} \right\rangle =\left[-i\left(\omega_- + \delta_{-}^{(j)}\right)-\frac{\kappa_2+\kappa_2'^{(j)}}{2}-\frac{\kappa_1+3\gamma_++\gamma_-+\kappa_d^{(j)}+\kappa_a^{(j)}}{2}\right]\left\langle \hat{\sigma}_{0-}^{(j)} \right\rangle
    \\+\sum_{k\ne j}\frac{\Gamma_{z,jk}}{\hbar}\left\langle \hat{\sigma}_{0-}^{(j)}\hat{S}_z^{(k)} \right\rangle
     -\sum_{k\ne j}\Gamma_{d,jk}\left\langle \left(\hat{\sigma}_{--}^{(j)}-\hat{\sigma}_{00}^{(j)}\right)\hat{\sigma}_{0-}^{(k)} \right\rangle
\end{multline}
\begin{multline}\label{EOMNVsZ}
    \frac{d}{dt}\left\langle \hat{\sigma}_{+-}^{(j)} \right\rangle =
    \left[i\left(\omega_++\delta_{+}^{(j)}-\omega_- - \delta_{-}^{(j)}\right)-2(\kappa_2+\kappa_2'^{(j)})-\frac{2\kappa_1+2\gamma_++\kappa_d^{(j)}}{2}\right]\left\langle \hat{\sigma}_{+-}^{(j)} \right\rangle
     \\+2\sum_{k\ne j}\frac{\Gamma_{z,jk}}{\hbar}\left\langle \hat{\sigma}_{+-}^{(j)}\hat{S}_z^{(k)} \right\rangle
     +\sum_{k\ne j}\Gamma_{d,jk}\left\langle \hat{\sigma}_{+0}^{(j)}\hat{\sigma}_{0-}^{(k)} \right\rangle
\end{multline}
\end{widetext}
where $\gamma_+$ and $\gamma_-$ are defined in Eq.~\eqref{gammaPMdef}. We define the total frequency shifts
\begin{equation}\label{Lambshiftgrouping}
    \delta^{(j)}_{\pm} \equiv  \delta^{(j)}_{1\pm}+ \delta^{(j)}_{2\pm}
\end{equation}
and the total coupling rates between NV centres,
\begin{equation}\label{GammaPMdefinition}
    \Gamma_{d,jk}\equiv\frac{\Gamma_{jk}^{(n)}-\Gamma_{kj}^{(a)}}{2}- i\left(G_{jk}^n+G_{kj}^a\right),
\end{equation}
\begin{equation}\label{GammaZdefinition}
    \Gamma_{z,jk}\equiv\frac{\Gamma_{2,jk}-\Gamma_{2,kj}}{2}+iG_{2,jk}+iG_{2,kj}.
\end{equation}
In the above equations of motion it is evident that all the interaction terms between different NV centres, both dissipative and coherent, combine into the single interaction rates $\Gamma_{d,jk}$ and $\Gamma_{z,jk}$, which are the relevant quantities describing the spin wave-induced interaction.
Let us analyse each of the spin wave-induced dynamics separately.

\subsection{Spin-wave induced modification of $T_1$ and $T_2^*$ for a single NV centre in thermal equilibrium}\label{AppNVdynamics_T1T2modified}

Here we focus on the modification of the decay and dephasing rates, which are defined in thermal equilibrium (i.e. no optical pumping). We thus consider a single NV centre by making $\Gamma_{d,jk}=\Gamma_{z,jk}=0$ in the equations of motion Eqs.~\eqref{EOMNVsA}-\eqref{EOMNVsZ}, and take $\Omega=0$. The resulting equations of motion have a similar form as for the NV centre in the absence of spin waves, Eqs.~\eqref{EOMbareNVdiagonal0} and \eqref{EOMbareNVsigma0pm}-\eqref{EOMbareNVsigmapm}, allowing us to define the new decay and dephasing times. 
Let us first discuss the dynamics of the occupations of the NV levels, namely $\langle\hat{\sigma}_{\alpha\alpha}\rangle (t)$, which are independent on the dephasing rates $\kappa_2$ and $\kappa_2'$. 
The dynamics of these occupations are given by
\begin{multline}\label{EOMcupledNVdiagonal0}
    \frac{d }{dt}\left[
    \begin{array}{c}
         \langle \hat{\sigma}_{00}\rangle  \\
         \langle\hat{S}_z/\hbar\rangle 
    \end{array}
    \right]=
    \left[
    \begin{array}{c}
         \kappa_1+\gamma_++(\Gamma_++\Gamma_-)/2  \\
         \gamma_- +(\Gamma_++\Gamma_-)/2
    \end{array}
    \right]
    \\
    -\left(\tensorbar{M}_0+\tensorbar{M'}\right)\left[
    \begin{array}{c}
         \langle \hat{\sigma}_{00}\rangle  \\
         \langle\hat{S}_z/\hbar\rangle 
    \end{array}
    \right],
\end{multline}
with
\begin{equation}
    \tensorbar{M'}=\frac{1}{2}\left[
    \begin{array}{cc}
       3\Gamma_+-\Gamma_- & \Gamma_++\Gamma_- \\
        3\Gamma_+-\Gamma_- & \Gamma_++\Gamma_-
    \end{array}
    \right].
\end{equation}
Here, we have defined the sum and difference of the spin-wave induced decay rates as
\begin{multline}\label{GammaPMdef}
    \Gamma_\pm\equiv\frac{\kappa_d\pm\kappa_a}{2} = v^2\sum_n\int d^2\mathbf{k}_\parallel \frac{\omega_M}{\omega_\beta}\left[1+\eta\frac{k_y}{k_\parallel}\right]^2
    \\
    \times\frac{\gamma_\beta \vert h_{\beta\eta0}\vert^2e^{-2k_\parallel l}}{(\omega_--\omega_\beta)^2+(\gamma_\beta/2)^2}\left[  1+\bar{n}_\beta\pm\bar{n}_\beta\right],
\end{multline}
where the last expression has been obtained by substituting the coupling rate from Eq.~\eqref{gfirstorder}, the zero-point magnetization from Eq.~\eqref{zeropointM}, the modal fields from Eq.~\eqref{hmodefunctions}, and taking the continuum limit Eq.~\eqref{continuummagnons}. We have also defined the coefficient
\begin{equation}\label{vconstantdefinition}
    v^2\equiv \frac{\mu_0^2\vert\gamma\vert\gamma_s^2\hbar M_S}{16\pi^2 d},
\end{equation}
which has dimensions of velocity squared.
Note that at not too low temperatures ($T\gtrsim 10$K) the above decay rates fulfill $\vert\Gamma_+ \vert\gg \vert\Gamma_-\vert$ due to the difference in the thermal factors appearing on each quantity. 

\begin{figure}[tbh!]
	\centering
	\includegraphics[width=\linewidth]{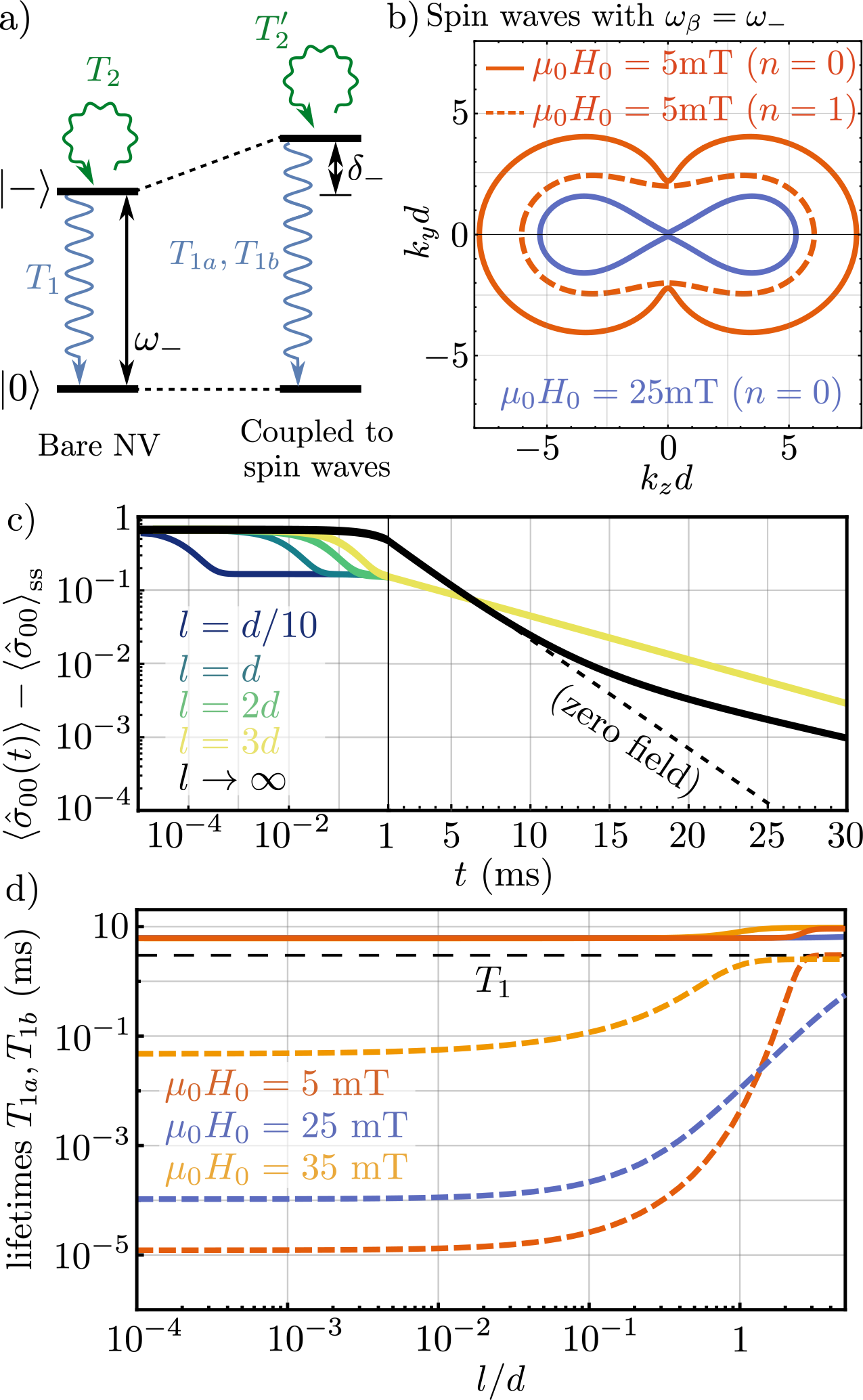}
	\caption{a) Spin waves cause a frequency shift $\delta_-$ of the NV transition frequency and decrease of the lifetimes $T_1$ and $T_2^*$.  b) Spin waves resonant with the $\vert 0\rangle\leftrightarrow\vert-\rangle$ transition of the NV: at $\mu_0H_0=5$mT spin waves in both $n=0$ (solid red line) and $n=1$ bands (dashed red line) are resonant, whereas for $\mu_0H_0=25$mT (blue lines) only the $n=0$ spin waves are, and for $\mu_0H_0\gtrsim 30$mT no spin wave is resonant.
	c) Evolution of the ground state occupation of the NV after initialization in the $\vert 0 \rangle$ state, for different NV-film separations and $\mu_0H_0=25$mT. The dashed line shows the single exponential decay for an isolated NV at $H_0=0$, which defines the timescale $T_1$. d) The two timescales describing the evolution of the NV occupations in the presence of spin waves (solid and dashed lines, respectively) as a function of NV-film distance $l$, for the three values of static fields $H_0$ considered in panel (b). The horizontal dashed line indicates the original $T_1$. In all three panels we take $T=300$K and the parameters in Table~\ref{tablePARAMS}.}\label{FigT1}
\end{figure}

As detailed in Appendix~\ref{AppNV_baredynamics}, in the absence of spin waves, at zero applied field $H_0$ and for an initially polarized NV centre ($\langle \hat{S}_z\rangle(t=0)=0$), the system of equations Eq.~\eqref{EOMcupledNVdiagonal0} decouples and hence all the occupations decay to their steady state value exponentially with a lifetime $T_1$ given by Eq.~\eqref{defT1appendix}. This behavior is altered in the presence of spin waves or at $H_0\ne 0$, where the evolution of the occupations follows a double exponential decay
\begin{equation}\label{doubleexponential}
    \langle\hat{\sigma}_{\alpha\alpha}\rangle (t) \sim ae^{-t/T_{1a}} + be^{-t/T_{1b}},
\end{equation}
for any initial state (see \figref{FigT1}[a]). The inverse rates $T_{1a}^{-1}$ and $T_{1b}^{-1}$ are given by the eigenvalues of the $2\times2$ matrix $\tensorbar{M}_0+\tensorbar{M'}$ appearing in Eq.~\eqref{EOMcupledNVdiagonal0}, namely
\begin{multline}\label{lambdaPMSW}
    \lambda_\pm = 2\gamma_++\kappa_1+\Gamma_+ \pm\Big[\gamma_+^2+3\gamma_-^2 
    \\+ \Gamma_+^2 + \Gamma_+(\gamma_++\gamma_-)+\Gamma_-(\gamma_+-\gamma_-)\Big]^{1/2}.
\end{multline}
Note that from Eq.~\eqref{GammaPMdef}, Eq.~\eqref{h0symky}, and the even parity of $\omega_\beta$ and $\gamma_\beta$ under the reflection $k_y\to -k_y$, one can easily show that the eigenvalues $\lambda_\pm$, and thus the modification of the occupation dynamics, are the same on both sides of the slab and depend only on the vertical separation from its surface, $l$.
 In \figref{FigT1}(c) we show the evolution of the ground-state occupation of the NV centre at $\mu_0H_0=25$mT, for an NV at different separations from the YIG film. In the absence of spin waves, i.e. at $l\to \infty$ (solid black line), the evolution of $\langle\hat{\sigma}_{00}\rangle (t)$ is very well approximated by a single exponential decay with slope $T_1$, as it only deviates from the zero-field case (dashed line) at very long times where the steady state is practically reached. This justifies the use of the timescale $T_1$, technically defined only at $H_0=0$, to describe also NV centres in the presence of weak applied fields. However, when the NV centre is placed close to the YIG film, the deviation from an exponential decay becomes very relevant. The evolution of $\langle\hat{\sigma}_{00}\rangle$ is in this case characterized by a fast decay to a metastable value $\langle\hat{\sigma}_{00}\rangle \approx \langle\hat{\sigma}_{00}\rangle_{\rm ss} + 0.2$, followed by a slow decay toward the steady-state.
 Since both these processes are relevant, one must consider the two relaxation times $T_{1a}$ and $T_{1b}$ separately.

The two relaxation times $T_{1a}$ and $T_{1b}$ are displayed in~\figref{FigT1}(d), in solid and dashed lines, for three different values of the applied field $H_0$ and for the parameters in Table~\ref{tablePARAMS}. The most striking effect is given by the lowest lifetime (dashed lines), which is orders of magnitude smaller than the original lifetime $T_1$.
This strong modification, known as Purcell enhancement in nanophotonics~\cite{novotny2006principles}, originates from the film-induced modification of the electromagnetic density of states. Indeed, the dashed lines in~\figref{FigT1}(d) display the typical behavior observed for Purcell-enhanced lifetimes associated to exponentially localized surface modes such as thin-film~\cite{MarocicoPRA2011} or graphene~\cite{HuidobroPRB2012} surface plasmons.
The lowest lifetime (dashed lines) tends to the original one, $T_1$, as $l\to\infty$, i.e. in the absence of spin waves. This motivates us to define the spin-wave-reduced lifetime of the transition $\vert 0 \rangle \leftrightarrow \vert - \rangle$ as
\begin{equation}
    T_1' \equiv \min(T_{1a}, T_{1b}) = \max\left(\frac{1}{\lambda_+},\frac{1}{\lambda_-}\right).
\end{equation}
This quantity, more details of which are given below, is the modified lifetime used in the main text.

The dependence of the dashed lines in \figref{FigT1}(d) on the applied field $H_0$ can be understood
from \figref{FigT1}(b), which indicates the resonant spin waves (i.e. the spin waves fulfilling the resonance condition $\omega_\beta=\omega_-$) for each value of $H_0$ considered in \figref{FigT1}(d): on the one hand, at high fields, $\mu_0 H_0 \gtrsim 30$mT, no spin waves are resonant with the NV transition and the effect of the spin waves is strongly suppressed, resulting in a lifetime closer to $T_1$. On the other hand, at low fields $\mu_0 H_0 \gtrsim 5$mT, the lifetime modification is maximized as many spin wave modes fulfill the resonance condition $\omega_\beta=\omega_-$. Finally, the intermediate field case $\mu_0 H_0 \gtrsim 25$mT is characterized by an intermediate lifetime enhancement in between the above two extreme cases. In this situation, moreover, the lifetime remains significantly different from   the value $T_1$ even at large distances $l\sim 5d$, as the resonance condition $\omega_\beta=\omega_-$ is fulfilled also for spin waves of very low wavenumbers (see blue line in \figref{FigT1}[b]). 
Note finally that the lifetimes on \figref{FigT1}(d) remain significantly different from $T_1$ even at easily achievable NV-film distances $l\approx d$, suggesting that the modified dynamics exemplified in \figref{FigT1}(c) could be measured through the fluorescence of the NV centre~\cite{WolfePRB2014,MyersPRL2017,AndrichNPJ2017,PageJAP2019,LeeWongNanoLetters2020,BertelliSciAdv2020}. Indeed, the spin wave-induced modification of the lifetime $T_1$ has already been proposed as a potential explanation for the observed modifications in the optically detected magnetic resonance spectrum of NV centres close to magnetic structures ~\cite{WolfePRB2014,AndrichNPJ2017,PageJAP2019}. Conversely, the strong dependence of the lifetimes on the NV separation from the film, $l$, could be exploited to optically measure such separation.

Let us finally focus on the lifetime of the coherences, particularly of the coherences $\langle\hat{\sigma}_{0\pm}\rangle$, whose lifetime at $H_0 = 0$ defines the coherence lifetime $T_2^*$. In the presence of spin waves, the definition of a single coherence time is not possible, since the lifetime of each of these coherences becomes different, see Eqs.~\eqref{EOMsigma0plustotal} and \eqref{EOMsigma0minustotal}. Specifically, from these equations we can extract the lifetimes of these two coherences as
\begin{multline}\label{coherencelifetimesMOD}
    \left[\begin{array}{c}
         \tau_{0+}^{-1}  \\
         \tau_{0-}^{-1} 
    \end{array}\right] = 
    \\
    =\frac{\kappa_2 +\kappa_2'+\kappa_1+3\gamma_+}{2} -\frac{1}{2}\left[\begin{array}{c}
         \gamma_- - \Gamma_++\Gamma_-  \\
         -\gamma_- - 2\Gamma_+
    \end{array}\right].
\end{multline}
The calculation of these rates requires determining the rate $\kappa_2'$ (Eq.~\eqref{kappa2prime}) which, by following similar steps as above, can be cast in integral form as
\begin{multline}\label{kappa2primeexplicit}
    \kappa_2' = v^4\sum_{nn'}\int d^2\mathbf{k}_\parallel \int d^2\mathbf{k}_\parallel' \left[1-e^{-d\vert\mathbf{k}_\parallel-\mathbf{k}_\parallel'\vert}\right]^2
    \\
    \!\!\times   \! \frac{\bar{n}_\beta(\bar{n}_{\beta'}+1)}{\omega_\beta\omega_{\beta'}}\frac{\gamma_\beta+\gamma_{\beta'}}{(\omega_\beta-\omega_{\beta'})^2+[(\gamma_\beta+\gamma_{\beta'})/2]^2}
    \\
  \times \!\! \left(X_{0\beta\beta'}^-\right)^2
    \sum_{\xi=\pm}\frac{(k_z-k_z')^4d^4 e^{-2l\vert\mathbf{k}_\parallel-\mathbf{k}_\parallel'\vert}}{\pi^2(n'+\xi n)^2+(\vert\mathbf{k}_\parallel-\mathbf{k}_\parallel'\vert d)^2},
\end{multline}
From the above expression it is evident that $\kappa_2'$, and therefore the spin wave-induced dephasing, is the same on both sides of the slab as it does not depend on $\eta$. Furthermore, $\kappa_2'$ only depends on the vertical separation from its surface, $l$.

\begin{figure}[t]
	\centering
	\includegraphics[width=\linewidth]{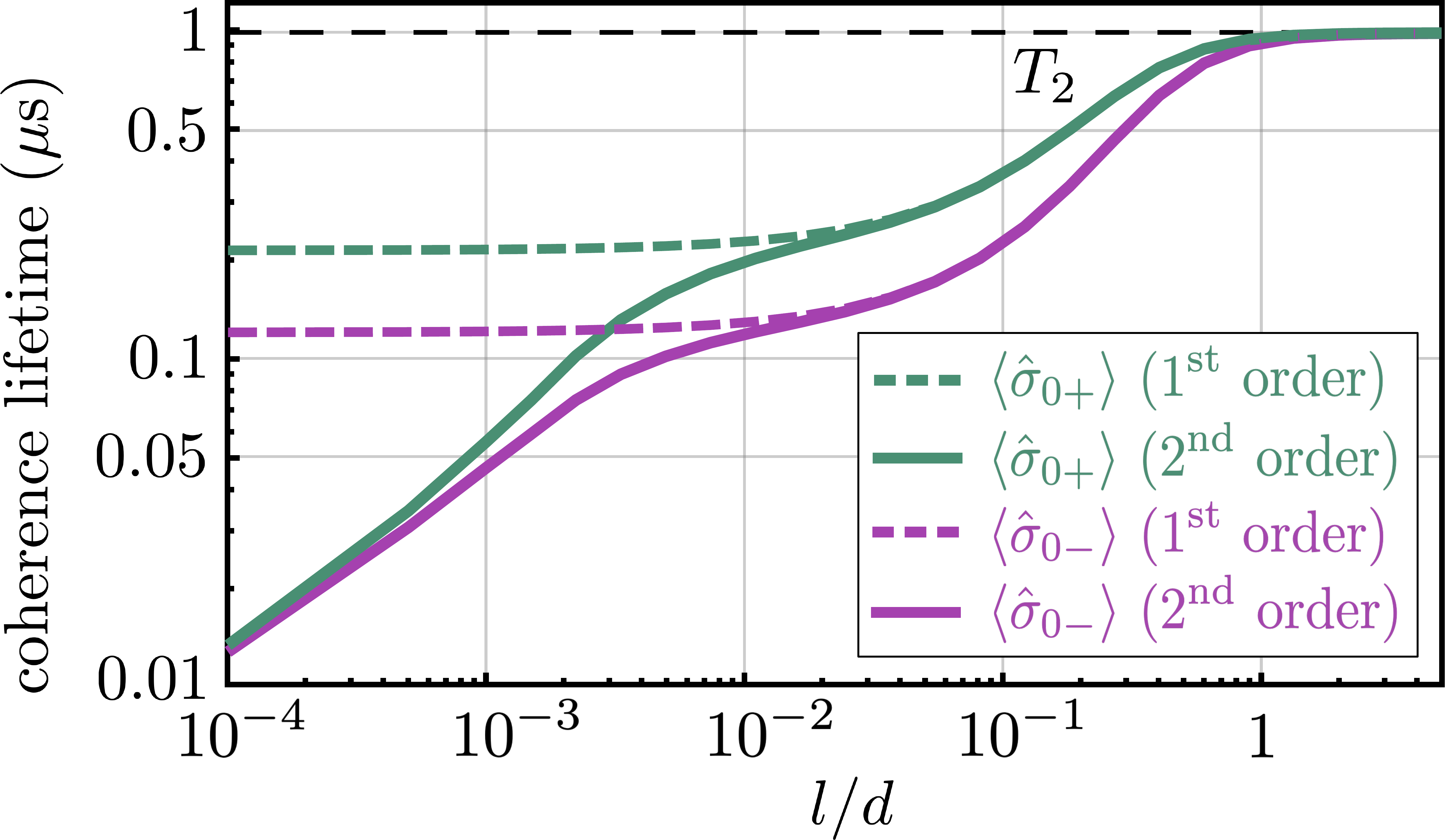}
	\caption{ Lifetimes of the two NV coherences $\langle\hat{\sigma}_{0\pm}\rangle$,  Eq.~\eqref{coherencelifetimesMOD}, at $\mu_0H_0=20$mT as a function of NV-film distance $l$. Dashed lines indicate the result including only the first-order contribution to the potential $\hat{V}_1$ in Eq.~\eqref{VtotalAPP}, whereas the solid lines show the full result including the second-order term $\hat{V}_2$. In all three panels we take $T=300$K and the parameters in Table~\ref{tablePARAMS}.}\label{FigT2}
\end{figure}

The lifetimes of the coherences $\tau_{0\pm}$, given by Eq.~\eqref{coherencelifetimesMOD}, are displayed in \figref{FigT2} (green and purple lines, respectively) as a function of the distance from the YIG surface. 
 In order to isolate the effect of the additional decay rates $\kappa_a$ and $\kappa_d$ from that of the additional dephasing rate $\kappa_2'$, in \figref{FigT2} we plot these lifetimes calculated only from the first order contribution $\hat{V}_1$, i.e. for $\kappa_2'=0$ (dashed lines) and calculated up to second order, i.e. including $\kappa_2'$ (solid lines). As evidenced by the figure, the effect of $\kappa_2'$ is negligible for $l\gtrsim 10^{-2} d$, where the deviation of the coherence lifetimes from their original value $T_2^*$ is dominated by the decay rates $\kappa_a$ and $\kappa_d$, i.e., by $T_1'$. In this regime the coherence lifetimes are Purcell-reduced in a similar way as the occupation lifetimes in~\figref{FigT1}(d). 
The lifetime of the coherence $\langle\hat{\sigma}_{0+}\rangle$ is not modified as strongly as the lifetime of $\langle\hat{\sigma}_{0-}\rangle$ as the corresponding NV transition  $\vert 0 \rangle\leftrightarrow \vert +\rangle$ is uncoupled from the spin wave modes. As a result there is a strong difference between these two lifetimes, that can differ by up to a factor $2$ at low separations $l\sim 10^{-2}d$.
At even lower separations, $l/d\lesssim 10^{-2}$, the effect of the additional dephasing $\kappa_2'$ becomes relevant and eventually dominant over the first-order contribution. In the low separation limit $l\ll d$ the two lifetimes become equal and much smaller than the original $T_2^*$ of the NV centre. This large increase on the dephasing rate suggests that placing the NV centres as close as possible to the YIG film is not always advisable, as the corresponding rise of the NV-spin wave coupling rate could be overcome by the unavoidable increase in decoherence. Finally, note that both the stark difference between the two lifetimes at $l\approx 10^{-2} d$ and their further decrease at $l\lesssim 10^{-2}d$ are strong effects as compared to the original decoherence time $T_2^*$, and could in principle be resolved by spin echo experiments~\cite{MyersPRL2017,AndrichNPJ2017,HansonScience2008,deLangeScience2010}. Such experiments, combined with independent measurements of the occupation dynamics displayed in \figref{FigT1}(c), could provide valuable information of the spin wave structure of the film.

\subsection{Frequency shift and spin wave-induced force for a single NV centre}\label{AppNVdynamics_LambShift}

Let us now focus on the frequency shift experienced by the two transitions of a single NV centre. According to Eq.~\eqref{Lambshiftgrouping}, the total frequency shift experienced by the transitions of a single NV reads
\begin{equation}
    \delta_\pm = \delta_{1,\pm} + \delta_{2,\pm}.
\end{equation}
The two contributions stem from the first and second order interaction potentials $\hat{V}_1$ and $\hat{V}_2$, respectively. The expression for the former, given by Eqs.~\eqref{Gjjanexpression}, reads
\begin{equation}\label{delta1pm}
    \left[\begin{array}{c}
         \delta_{1,-}  \\
         \delta_{1,+} 
    \end{array}\right] = \sum_\beta\frac{-\Delta_\beta}{\Delta_\beta^2+(\gamma_\beta/2)^2}\vert g_{j\beta}\vert^2\left[
    \begin{array}{c}
         1+2\bar{n}_\beta  \\
         \bar{n}_\beta 
    \end{array}
    \right].
\end{equation}
Regarding the second-order contribution, we obtain it from combining Eqs.~\eqref{PSDV2zerofreq} and Eqs.~\eqref{G2jjprime}-\eqref{delta2fromG},
\begin{multline}\label{delta2pm}
    \delta_{2,\pm} = \frac{1}{2}\sum_{\beta \beta'}\bar{n}_\beta(1+\bar{n}_{\beta'})\vert \tilde{g}_{j\beta\beta'}\vert^2
    \\
    \times\frac{(\omega_\beta-\omega_{\beta'})}{(\omega_\beta-\omega_{\beta'})^2+[(\gamma_\beta+\gamma_{\beta'})/2]^2}.
\end{multline}
By writing the above second-order contribution in integral form as done in Eq.~\eqref{kappa2primeexplicit} and making use of parity arguments it can be shown that the term proportional to $\bar{n}_\beta\bar{n}_{\beta'}$ is exactly zero. Furthermore, the same parity properties allow us to cast the second-order contribution as
\begin{multline}\label{delta2pmprocessed}
    \delta_{2,\pm} = \frac{1}{4}\sum_{\beta \beta'}(\bar{n}_\beta-\bar{n}_{\beta'})\vert \tilde{g}_{j\beta\beta'}\vert^2
    \\
    \times\frac{(\omega_\beta-\omega_{\beta'})}{(\omega_\beta-\omega_{\beta'})^2+[(\gamma_\beta+\gamma_{\beta'})/2]^2}.
\end{multline}
The above expression is, in general, much smaller than the first-order contributions Eq.~\eqref{delta1pm} for two reasons: first, the function in the second line is non-negligible only in a narrow region of frequency space, namely in the region $\omega_\beta -\gamma_\beta\lesssim \omega_{\beta'} \lesssim \omega_\beta +\gamma_\beta$. In these regions, however, the thermal factors fulfill $\bar{n}_\beta\approx \bar{n}_{\beta'}$ as $\gamma_\beta \sim $ MHz (see \figref{FigDecayRates}). Second, even in these regions the coupling rates $\vert \tilde{g}_{j \beta\beta'}\vert$ are usually much smaller than the first-order rates $\vert g_{j\beta}\vert$ since they stem from a second-order correction to the interaction potential. These arguments allow us to safely neglect this shift, i.e.,
\begin{equation}
    \delta_{2,\pm} \ll\delta_{1,\pm} \hspace{0.3cm}\rightarrow\hspace{0.3cm} \delta_{\pm} \approx  \delta_{1,\pm}.
\end{equation}
The frequency shifts of the two NV transitions can thus be written in integral form as
\begin{multline}\label{deltaminusintegralform}
    \left[\begin{array}{c}
         \delta_-  \\
         \delta_+ 
    \end{array}\right] \approx 
     -v^2\sum_n\int d^2\mathbf{k}_\parallel
   \frac{\omega_M}{\omega_\beta}
    \frac{\Delta_\beta}{\Delta_\beta^2+(\gamma_\beta/2)^2} 
    \\
    \times\vert h_{\beta\eta 0}\vert^2 e^{-2k_\parallel l}\left(1+\eta\frac{k_y}{k_\parallel}\right)^2
    \left[\begin{array}{c}
         2\bar{n}_\beta+1  \\
         \bar{n}_\beta 
    \end{array}\right]
    .
\end{multline}
From the above expression it is evident that $\vert\delta_-\vert\ge2\vert\delta_+\vert$. Specifically, in the low- and high-temperature limits, valid respectively at cryogenic and room temperature, we find
\begin{equation}\label{lowThighTdelta}
    \delta_+ \to\left\lbrace
    \begin{array}{cc}
        0 & \text{ for } k_B T \ll \hbar \omega_M \\
        \delta_-/2 & \text{ for } k_B T \gg \hbar \omega_M.
    \end{array}
    \right.
\end{equation}
The above relations allow us to focus on the more relevant frequency shift, namely that of the coupled transition $\vert 0 \rangle \leftrightarrow \vert - \rangle$. 

\begin{figure}[htp]
	\centering
	\includegraphics[width=\linewidth]{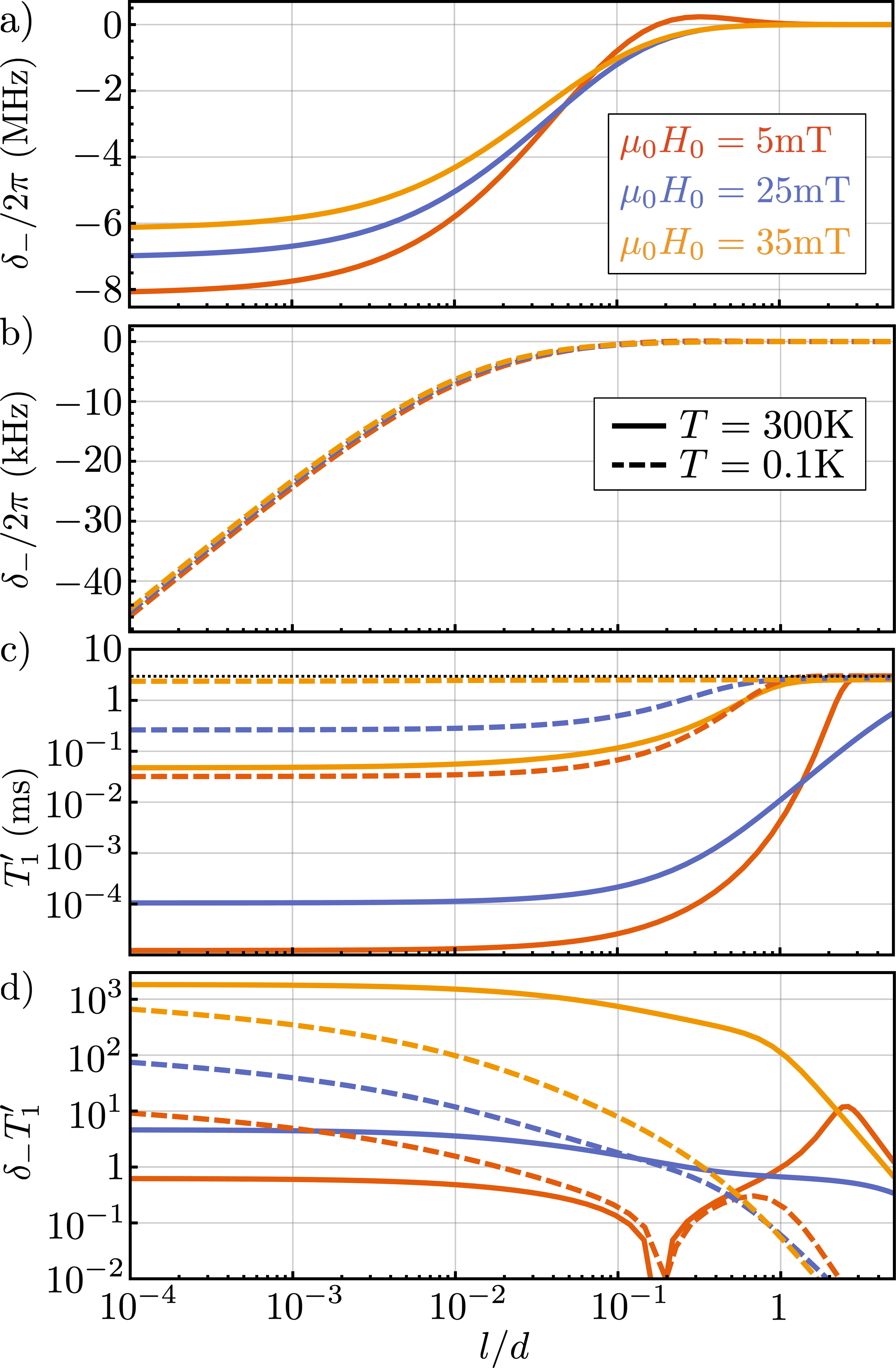}
	\caption{a-b) Frequency shift $\delta_-$ at room (a) and cryogenic (b) temperatures as a function of the distance between the NV centre and the YIG film, for three values of the applied field $H_0$. c-d) Lifetime of the $\vert 0 \rangle\leftrightarrow\vert - \rangle$ transition, $T_1'$, and product $\delta_-T_1'$, respectively, for the same magnetic fields as in panels (a-b), and for room (solid) and cryogenic (dashed) temperatures. The black dotted line in panel (c) indicates the original NV lifetime in the absence of spin waves, $T_1$. The yellow solid and dashed lines in panel (d) correspond to the red and blue curves in \figref{FigureNV_LambShift}(b), respectively.
	 In all panels we choose the parameters as in Table~\ref{tablePARAMS}. }\label{FigureNV_LambShift_Appendix}
\end{figure}

The frequency shift $\delta_-$ is displayed in \figref{FigureNV_LambShift_Appendix}(a-b) as a function of vertical separation $l/d$, for three representative values of the applied magnetic field and at room (panel a) and cryogenic (panel b) temperatures. Regardless of the applied field $H_0$, the frequency shift becomes negligible at sufficiently large distances $l\sim d$, as the amplitude of the exponentially localized ($\exp(-k_\parallel l)$) spin waves vanishes. 
The different behavior of the frequency shift at different values of the field $H_0$
can be understood from Eq.~\eqref{deltaminusintegralform} and the resonance iso-lines in panel \figref{FigT1}(b). Two limiting cases can be identified: first, the high-field case $\mu_0H_0 \gtrsim 30$mT (yellow lines in \figref{FigureNV_LambShift_Appendix}[a-b]) where no spin wave is resonant with the NV transition (hence no curve corresponding to these fields is displayed in \figref{FigT1}(b)). Specifically, as evidenced also by \figref{FigRWAappendix}(a), in this case all magnons fulfill $\omega_\beta>\omega_-$, resulting in a negative frequency shift for all values of the separation $l$. 
Second, the weak field case exemplified by $\mu_0H_0 = 5$mT (red lines in \figref{FigureNV_LambShift_Appendix}[a-b]), where spin waves in both $n=0$ and $n=1$ bands  are resonant with the transition $\vert 0 \rangle\leftrightarrow\vert-\rangle$. The phenomenology in this case depends critically on the distance to the film, $l$: specifically, for large $l$ only the spin waves with a very low wavenumber contribute significantly due to the exponential decay of the coupling rate $g_\beta \propto \exp(-k_\parallel l)$. Since all these spin wavemodes fulfill $\omega_\beta<\omega_-$ the frequency shift becomes positive. On the other hand, for sufficiently small $l$ the spin waves with higher wavenumber, namely those outside the curves in \figref{FigT1}(b), start to contribute with a negative shift as they fulfill $\omega_\beta>\omega_-$. For sufficiently short distances $l$ the negative contribution overcomes the positive one and the frequency shift turns negative. Finally, for intermediate values of the field such as $\mu_0H_0=25$mT
(blue lines in \figref{FigureNV_LambShift_Appendix}[a-b]) most of the spin wave modes fulfill $\omega_\beta>\omega_-$, even in the low-wavenumber limit. As a consequence the frequency shift remains negative for all separations $l$ despite the resonance condition being fulfilled by some spin waves in the $n=0$ band. All these behaviors are similar for room and cryogenic temperatures, with the room-temperature case displaying much higher shifts due to the increased field amplitude of the thermal spin waves. Finally, note that the frequency shift is measurable for all three values of the field chosen in \figref{FigureNV_LambShift_Appendix}(a-b), since the corresponding spin wave-reduced lifetimes $T_1'$, shown in \figref{FigureNV_LambShift_Appendix}(c), remain high enough for the condition $\vert \delta_- T_1'\vert >1$ to be fulfilled (see \figref{FigureNV_LambShift_Appendix}[d]). This measurement becomes less challenging at room temperature where the product $\vert \delta_- T_1'\vert$ is enhanced.

We now focus on the force exerted by the spin waves on a single NV centre. This force is calculated as the steady-state expected value of the gradient of the total Hamiltonian,
\begin{equation}\label{FCasimirGeneralAppendix}
    \mathbf{F} = -\left\langle \nabla\left[\hat{H}_{\rm ps} + \hat{H}_{\rm eff,1}+ \hat{H}_{\rm eff,2} \right]\right\rangle_{\rm ss}.
\end{equation}
As detailed above, we can neglect the second-order frequency shifts $\delta_{2,\pm}$ and write
\begin{multline}
    \frac{\hat{H}_{\rm eff,1}+ \hat{H}_{\rm eff,2}}{\hbar} \approx
    -\delta_+\hat{\sigma}_{00}+(\delta_--\delta_+)\hat{\sigma}_{--}
    ,
\end{multline}
where we have added the constant (but position-dependent) correction that we neglected in Eq.~\eqref{Hprime1Appendix}.
Using this expression in combination with Eq.~\eqref{FCasimirGeneralAppendix} one can calculate the force in a general situation. In this work, however, we focus either on the high- or on the low-temperature limit, where the expression for the force can be simplified, using the identities Eq.~\eqref{lowThighTdelta}, to
\begin{multline}
    \mathbf{F} = -\eta\frac{\hbar}{2}\mathbf{e}_x \frac{d\delta_-(l)}{dl}\times
    \\\times
    \left\lbrace
    \begin{array}{cc}
        2\langle\hat{\sigma}_{--}\rangle_{\rm ss} & \text{ for } k_B T \ll \hbar \omega_M \\
        \langle\hat{\sigma}_{--}- \hat{\sigma}_{00}\rangle_{\rm ss} & \text{ for } k_B T \gg \hbar \omega_M.
    \end{array}
    \right.
\end{multline}
The force given by Eq.~\eqref{CasimirPolderForce} in the main text results from particularizing the above expression to an NV centre above the YIG film ($\eta=1$). 

\begin{figure}[t!]
	\centering
	\includegraphics[width=\linewidth]{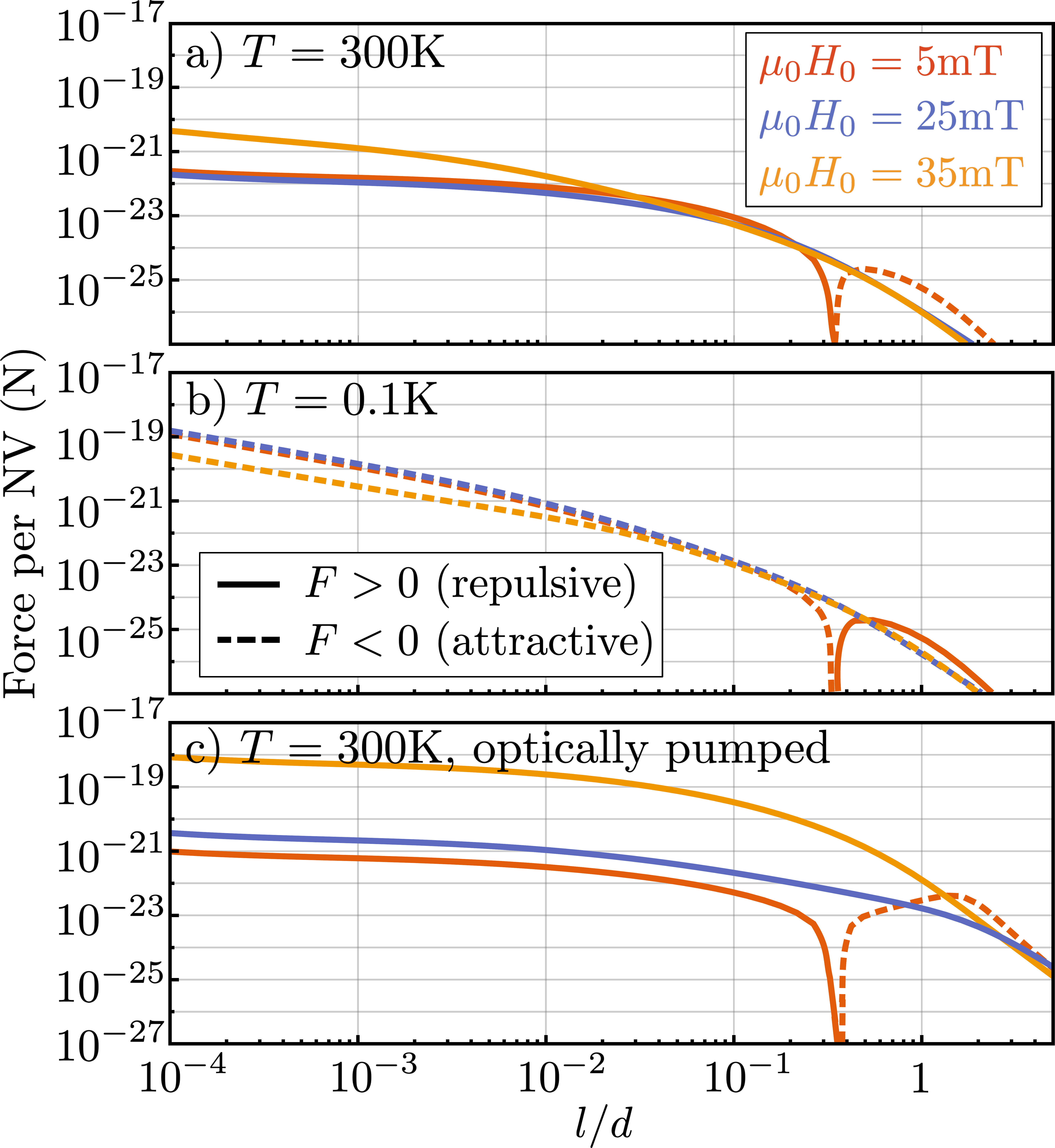}
	\caption{Spin-wave induced force per NV centre as a function of its distance to the YIG film, for three values of the applied field $H_0$. a) System at room temperature. b) System at cryogenic temperatures. c) System at room temperature with the NV centre being optically pumped (optimal pumping conditions).
	The yellow curves in each panel correspond to the curves in \figref{FigureNV_LambShift}(c).
	 In all panels we choose the parameters as in Table~\ref{tablePARAMS}. }\label{FigureNV_Force_Appendix}
\end{figure}

The force Eq.~\eqref{FCasimirGeneralAppendix} is shown in \figref{FigureNV_Force_Appendix} for a single NV centre at room temperature (panel a), cryogenic temperature (panel b) and room temperature under optical pumping at optimal pumping conditions (panel c). At cryogenic temperatures optical pumping does not significantly change the force as the NV centre is already near its ground state. At room temperature the forces are very small since the steady-state occupations are generally very close to their zero-field values, i.e. $\langle\hat{\sigma}_{--}\rangle_{\rm ss}\approx \langle \hat{\sigma}_{00}\rangle_{\rm ss}\approx 1/3$. The resulting force is thus significantly weaker than the usual Casimir-Polder force for atoms with optical transitions (for which $\langle \hat{\sigma}_{00}\rangle_{\rm ss}\approx 1$ even at room temperature), and lies in the $10^{-23}-10^{-21}$ range. This reduced Casimir-Polder force is common for magnetic dipole transitions where the transition frequencies are comparable to thermal energy scales~\cite{HaakhPRA2009}. Moreover, room temperature forces are mostly repulsive, except at very low fields $\mu_0H_0 = 5$mT (red curve), where the force becomes attractive for $l\gtrsim 0.3d\approx 60$nm following the corresponding change of slope in the frequency shift (compare with \figref{FigureNV_LambShift_Appendix}[a]). Repulsive Casimir-Polder forces are a common feature of magnetic dipole transitions~\cite{SkagerstamPRA2009,HaakhPRA2009}.  Conversely, forces become attractive at cryogenic temperatures.
Regarding optical pumping, it is clearly more efficient at high fields as shown by \figref{FigureNV_Force_Appendix}(c). The reason behind this is the relatively large lifetime $T_1'$ of the NV transition at these fields (see \figref{FigureNV_LambShift_Appendix}[c]), which facilitates the pumping of occupation into the ground state. Moreover, in this case the pumping intensity used for optimal pumping in the absence of spin waves, $I_d=0.13$GW m$^{-2}$, still results in very efficient initialization of the NV centre, since $T_1$ and $T_1'$ are relatively similar. Conversely, at fields for which the resonance condition $\omega_\beta=\omega_-$ is fulfilled (i.e. red and blue curves \figref{FigureNV_Force_Appendix}), the very short lifetimes $T_1'$ make optical pumping inefficient. We emphasize that in these cases the force can still be enhanced by employing higher optical pumping intensities $I_d$.

\subsection{Coupling between different NV centres}\label{AppNVdynamics_Couplings}

\begin{figure*}[tbh!]
	\centering
	\includegraphics[width=\linewidth]{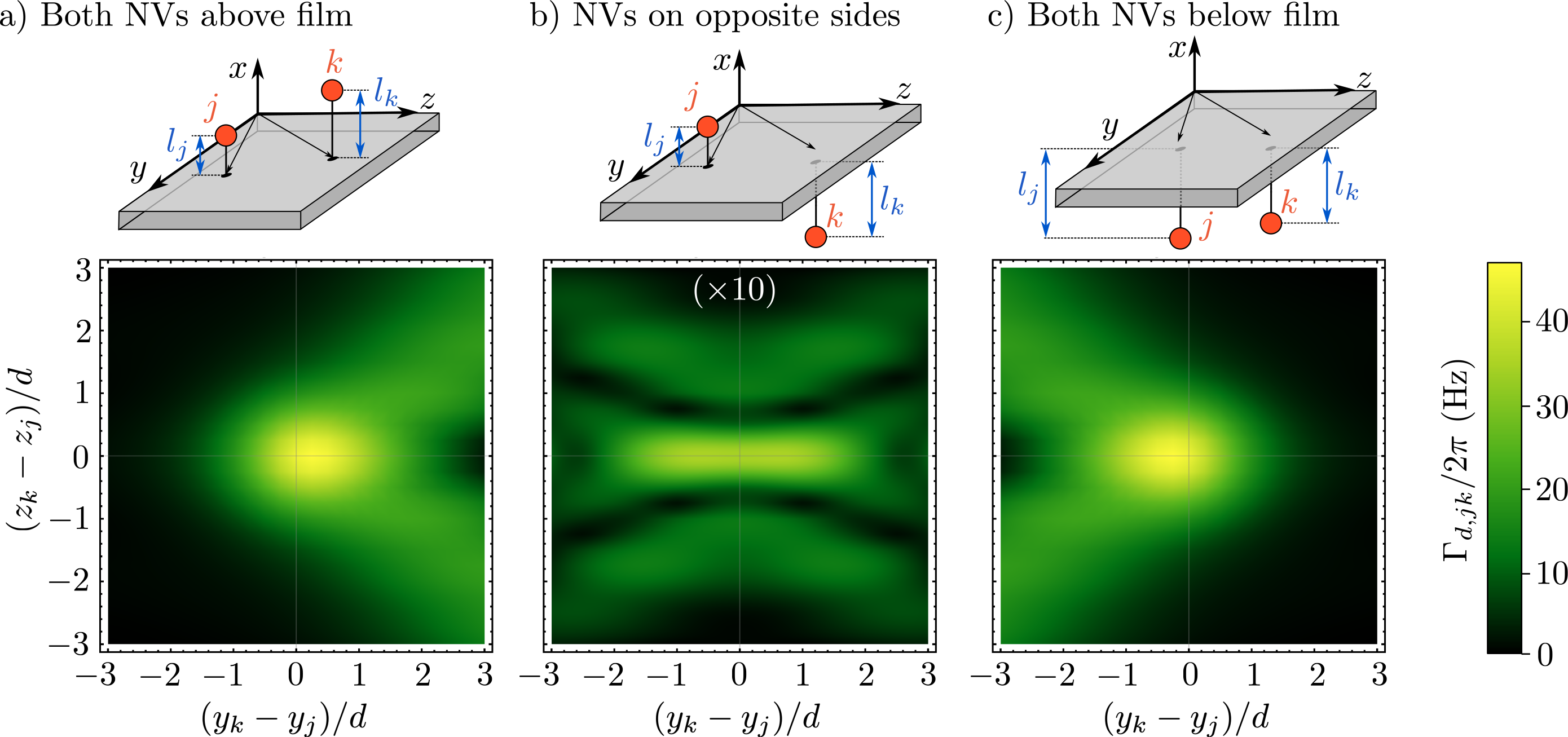}
	\caption{Modulus of the coupling rate between NV centre $j$ and NV centre $k$ as appearing in the equation of motion for NV centre $j$, i.e. $\vert\Gamma_{d,jk}\vert$, as a function of the relative separation in the parallel plane. The total vertical separation is $L_{kj}= l_j+l_k = d = 200$nm, the applied field $\mu_0 H_0 = 20$mT and the NV centres are in thermal equilibrium at room temperature. Panels (a), (b), and (c) show, respectively, the cases of the two NV centres lying above the film, one NV centre above and a second below the film, and the two NV centres below the film.}\label{FigureNVNVCouplings}
\end{figure*}

Let us finally consider the effective spin-wave dynamics induced on an ensemble of NV centres. On the one hand, each NV centre in the ensemble will experience the same frequency shift and the same lifetime modifications as in the single-NV case analyzed in previous sections. On the other hand, however, different NV centres will be coupled through the compound coupling rates appearing in the equations of motion, namely 
\begin{equation}
    \Gamma_{d,jk} = \sum_\beta \frac{g_{j\beta}^*g_{k\beta}}{ i(\omega_--\omega_\beta)-\gamma_\beta/2},
\end{equation}
and
\begin{multline}
    \Gamma_{z,jk} = i\sum_{\beta\beta'}\bar{n}_\beta (\bar{n}_{\beta'}+1)\tilde{g}_{j\beta\beta'}\tilde{g}_{k\beta'\beta} \\\times\frac{\omega_\beta-\omega_{\beta'}}{(\omega_\beta-\omega_{\beta'})^2+[(\gamma_\beta-\gamma_{\beta'})/2]^2}.
\end{multline}
The latter contribution, $\Gamma_{z,jk}$, can be manipulated analogously to the second-order frequency shifts $\delta_{2,\pm}$, see Eqs.~\eqref{delta2pm} and \eqref{delta2pmprocessed}, and following the same argumentation it can also be neglected. Hereafter we thus focus on the dominant coupling rate which in integral form reads
\begin{multline}
    \Gamma_{d,jk} = 
    v^2
    \sum_n\int d^2\mathbf{k}_\parallel\frac{\omega_M}{\omega_\beta}e^{-k_\parallel(l_j+l_k)}e^{i\mathbf{k}_\parallel(\mathbf{r}_{\parallel k}-\mathbf{r}_{\parallel j})}
    \\\times \left[1+\eta_j\frac{k_y}{k_\parallel}\right]\left[1+\eta_k\frac{k_y}{k_\parallel}\right] \frac{h_{\beta \eta_j 0}h_{\beta \eta_k 0}}{ i(\omega_--\omega_\beta)-\gamma_\beta/2}.
\end{multline}
From the above expression we deduce that the coupling rates only depend on three variables, namely the difference between the parallel coordinates of the two involved spins, $\mathbf{r}_{\parallel k}-\mathbf{r}_{\parallel j} \equiv y_{kj}\mathbf{e}_y + z_{kj}\mathbf{e}_z$, and the sum of their distances to the slab, $l_k+l_j \equiv L_{kj}$:
\begin{equation}
    \Gamma_{d,jk}\equiv\Gamma_d(L_{kj},y_{kj},z_{kj}).
\end{equation}
Moreover, by using the symmetries of the frequencies and loss rates under sign change of $k_y$ and $k_z$, together with the symmetries Eq.~\eqref{h0symkz} and \eqref{h0symky}, one can show that the couplings are symmetric under permutation of the $z-$coordinate of the two NV centres,
\begin{equation}
    \Gamma_d(L_{kj},y_{kj},z_{kj})=\Gamma_d(L_{kj},y_{kj},z_{jk}).
\end{equation}
Using the same symmetry arguments we can also derive relations between the coupling rates corresponding to different positions of the NV centres with respect to the slab, quantified by the respective indices $\eta_j$ and $\eta_k$. Specifically,
\begin{enumerate}
    \item For NVs on different sides of the slab ($\eta_j=-\eta_k$) the rate $\Gamma_d$ becomes independent on $\eta_j$ and $\eta_k$, i.e. it does not depend on which side each Nv centre is on. Furthermore, in this case the coupling rate is symmetric under permutation of the two Nv centres:
    \begin{equation}
        \Gamma_d(L_{kj},y_{kj},z_{kj})=\Gamma_d(L_{kj},y_{jk},z_{kj}). 
    \end{equation}
    \item For NVs on the same side of the slab ($\eta_j=\eta_k=\eta$), the coupling remains $\eta-$dependent but obeys the symmetry
    \begin{equation}
        \Gamma_{d}(L_{kj},y_{kj},z_{kj})\big\vert_{\eta=+} = \Gamma_{d}(L_{kj},y_{jk},z_{kj})\big\vert_{\eta=-},
    \end{equation}
\end{enumerate}

The coupling rates between two NV centres $j$ and $k$, as appearing in the equations of motion for NV centre $j$, are shown in \figref{FigureNVNVCouplings} for three cases, namely the two NV centres above the slab (panel a), one NV centre above and a second below the slab (panel b), and the two NV centres below the slab (panel c). In general the most relevant coupling rates correspond to the two NV centres lying on the same side of the slab (panels a and c). These couplings also display a strong directionality which originates from the combination of the non-reciprocity of the bath modes, i.e. the spin waves, and the polarization-selective transition of the NV centre, $\vert 0\rangle \leftrightarrow \vert - \rangle$. To understand this asymmetry, let us focus on the case of the two NV centres above the slab (panel a). 
The NV centre $j$, assumed to lie at the origin of coordinates for simplicity, interacts only with spin waves polarized along $\mathbf{e}_-$, which propagate on the positive $y$ direction (see \figref{FigureModalFields}[c]). This enhances the interaction with NV centres $k$ placed on the right half of the $y-z$ plane ($y_k>y_j$) while suppressing the interaction with those placed on the left half ($y_k<y_j$). The emitted spin waves are resonant with the frequency of the NV transition, $\omega_-$. Thus, emitted spin waves propagating near the $y-$axis (i.e. $\phi_k\ll 1$) have much lower wavenumbers than emitted spin waves propagating along a larger angle $\phi_k$. Since the field amplitude, and thus the coupling rate, becomes very small also at low wavenumbers, the coupling is maximized for NV centres $k$ placed along a certain angle along the $y-z$ plane. Note that this directional coupling mechanism is analogous to the phenomenon of chirality in quantum nanophotonics, which has attracted significant attention lately~\cite{LodahlNature2017} due to its potential applications in quantum computing and processing, among others. Most of these applications, however, are not directly extensible to the NV-spin wave interfaces as the typical coupling strengths (several Hz in  \figref{FigureNVNVCouplings}, potentially several kHz for NV centres near the surface of the film) are much weaker than the dissipation rates of the NV centres. 
 An interesting outlook of our work, and a direct application of our theoretical model, consists on exploring the enhancement of the interaction rates, for instance through coherent driving of particular spin wave modes, as well as the potential of these highly directional interactions e.g. for studying unconventional many-body physics~\cite{LodahlNature2017}.

\section{Effective spin wave dynamics}\label{Appendix_effectiveSWdynamics}

In this section we apply the general formalism derived in Appendix~\ref{Appendix_TracingOut} to derive the effective dynamics of a system of spin waves induced by a bath of paramagnetic spins.
As in the previous section we focus on NV centres, but our results hold for any paramagnetic spin with similar values of $T_1$ and $T_2^*$.
First, we summarize the derivation of the master equation for the spin waves and provide analytical expressions for all the involved rates in Sec.~\ref{AppSWdynamics_generalMEq}. We then particularize these rates to a bath formed by a large number $N\gg1$ of NV centres at random positions within the diamond slab in Sec.~\ref{AppSWdynamics_largeNrandom}. Finally, in Sec.~\ref{AppSWdynamics_PSDs}, we compute the magnetic field power spectral density in the presence of the NV centres and discuss its experimental detection.

\subsection{Derivation of the effective dynamics}\label{AppSWdynamics_generalMEq}

Deriving the effective spin wave dynamics requires some special attention due to the drastic difference between the
slow timescale associated to the second-order contribution $\hat{V}_2$, namely $T_1$ (or $\tau_{\rm nv}$ for optically pumped NV centres, see Appendix~\ref{AppNV_OpticalPumping}), and the fast timescales associated to both the first-order contribution $\hat{V}_1$ and the bare dissipation of the spin waves, namely $\sim T_2^*$ and  $\gamma_\beta^{-1}$. 
Note that from the equations of motion Eq.~\eqref{EOMbareNVdiagonal0} and \eqref{EOMbareNVsigma0pm}, and from the correlators of the NV centre operators, Eqs.~\eqref{CorrelatorSigma0PM}, \eqref{CorrelatorSigmaPM0}, and \eqref{CorrelatorSigmaDIAG} one can show that
\begin{equation}\label{crosscorrelatorsarezeroA}
    \langle \hat{\sigma}_{-0}^{(j)}(t)\hat{S}_z^{(k)}(t-s)\rangle_{\rm ss}= \langle \hat{S}_z^{(k)}(t)\hat{\sigma}_{-0}^{(j)}(t-s)\rangle_{\rm ss}=0
\end{equation}
and
\begin{equation}\label{crosscorrelatorsarezeroB}
    \langle \hat{\sigma}_{0-}^{(j)}(t)\hat{S}_z^{(k)}(t-s)\rangle_{\rm ss}= \langle \hat{S}_z^{(k)}(t)\hat{\sigma}_{0-}^{(j)}(t-s)\rangle_{\rm ss}=0.
\end{equation}
The same identities hold in the presence of optical pumping, i.e., for NV centres governed by Eqs.~\eqref{EOMpumpedNVsigma0pm}, \eqref{CorrelatorSigma0PMpumping}, and \eqref{CorrelatorSigmaPM0pumping}.
As a consequence of the above expressions, the master equation does not contain crossed terms involving both $\hat{V}_1$ and $\hat{V}_2$, which can thus be treated separately as independent reservoirs. The master equation for the magnons can thus be cast in the following form,
\begin{multline}\label{MasterEquationOnlyMagnonsAppendix2}
    \frac{d}{dt}\hat{\rho}_{\rm sw} = -\frac{i}{\hbar}\left[\hat{H}_{\rm sw} +\hat{H}_{\rm sw, eff,1}+\hat{H}_{\rm sw, eff,2},\hat{\rho}_{\rm sw}\right] 
    \\+ \mathcal{D}_{\rm sw}[\hat{\rho}_{\rm sw}]+ \mathcal{D}_{\rm sw, 1}[\hat{\rho}_{\rm sw}]+\mathcal{D}_{\rm sw,2}[\hat{\rho}_{\rm sw}],
\end{multline}
namely a modification of the original magnon master equation Eq.~\eqref{MasterEquationOnlyMagnonsAppendix} which includes the independent contributions of $\hat{V}_1$ and $\hat{V}_2$, i.e., the contributions originating from the effect of the NV coherences and from the NV occupations, respectively.

Let us first analyze the contribution of the coherences, namely of the first-order term $\hat{V}_1$. We include this term within the Born-Markov formalism introduced in Appendix~\ref{Appendix_TracingOut}. Since the two-time correlators $\langle\hat{\sigma}_{0\pm}(t)\hat{\sigma}_{\mp}(t-s)\rangle_{\rm ss}$ decay on a timescale $2/(\kappa_1+\kappa_2+3\gamma_++\gamma_- +2\Omega) \approx T_2^*$ (see Eqs.~\eqref{CorrelatorSigma0PM}-\eqref{CorrelatorSigmaPM0} or Eqs.~\eqref{CorrelatorSigma0PMpumping}-\eqref{CorrelatorSigmaPM0pumping} and note that for efficient optical pumping $\Omega\ll\kappa_2$), the Markov approximation is valid for $T_2^*\gamma_\beta \ll 1$. Regarding the spin wave dynamics, all the results in this article focus on the $n=0$ spin wave band, for which the Markov approximation is justified as $\gamma_\beta T_2^* \lesssim 1$ (especially at low fields $H_0\lesssim 15$mT). For higher order ($n\ge 1$) spin wave bands, non-Markovian corrections could be expected. Exploring the small non-Markovian corrections expectable for higher order ($n\ge 1$) bands could be an interesting outlook of our work.

Within the Markov approximation and using the correlators Eqs.~\eqref{CorrelatorSigma0PM} and \eqref{CorrelatorSigmaPM0} or, for optically pumped NV centres, the correlators \eqref{CorrelatorSigma0PMpumping} and \eqref{CorrelatorSigmaPM0pumping}, we can compute the contributions stemming from $\hat{V}_1$. On the one hand, the coherent contribution 
\begin{equation}
    \hat{H}_{\rm sw, eff,1} = \hbar\sum_\beta \delta_\beta\hat{s}^\dagger_\beta\hat{s}_\beta + \hbar\sum_{\beta,\beta'\ne \beta}W_{\beta\beta'}\hat{s}^\dagger_\beta\hat{s}_{\beta'}
\end{equation}
which describes both a spin wave frequency shift and an effective interaction between spin waves mediated by the NV centres. Under the assumption of identical NV centres, these rates are given by
\begin{multline}
    W_{\beta\beta'} \equiv -\frac{G_{\beta\beta'}^2}{2}\langle \hat{\sigma}_{--}-\hat{\sigma}_{00}\rangle_{\rm ss}\\\times\frac{\Delta_\beta + \Delta_{\beta'}}{(i\Delta_\beta+\kappa_T/2)(-i\Delta_{\beta'}+\kappa_T/2)}
\end{multline}
and
\begin{equation}\label{deltabetadefAppendix}
    \delta_\beta \equiv W_{\beta\beta} = -G_{\beta\beta}^2\langle \hat{\sigma}_{--}-\hat{\sigma}_{00}\rangle_{\rm ss}\frac{\Delta_\beta}{\Delta_\beta^2+(\kappa_T/2)^2},
\end{equation}
where we define the total coupling rates as
\begin{equation}
    G_{\beta\beta'}^2\equiv\sum_jg_{j\beta}^*g_{j\beta'}
\end{equation}
and the total decay rate of the coherence $\langle\hat{\sigma}_{0-}\rangle$ as (see Eq.~\eqref{EOMpumpedNVsigma0pm})
\begin{equation}
    \kappa_T \equiv \kappa_1+\kappa_2+3\gamma_++\gamma_-+2\Omega.
\end{equation}
On the other hand, the dissipator
\begin{multline}
    \mathcal{D}_{\rm sw, 1}[\hat{\rho}] =\sum_\beta \Gamma_{d\beta}\mathcal{L}_{\hat{s}_{\beta}\hat{s}_\beta^\dagger}[\hat{\rho}] + \Gamma_{a\beta}\mathcal{L}_{\hat{s}_{\beta}^\dagger\hat{s}_\beta}[\hat{\rho}]
    \\
    +\sum_{\beta,\beta'\ne\beta} \Gamma_{\beta\beta'}^{(n)}\mathcal{L}_{\hat{s}_{\beta'}\hat{s}_\beta^\dagger}[\hat{\rho}] + \Gamma_{\beta\beta'}^{(a)}\mathcal{L}_{\hat{s}_{\beta'}^\dagger\hat{s}_\beta}[\hat{\rho}]
\end{multline}
which represents incoherent interaction between different spin wave modes at rates
\begin{multline}
    \Gamma_{\beta\beta'}^{(n)} = \Gamma_{\beta'\beta}^{(a)}\frac{\langle\hat{\sigma}_{00}\rangle_{\rm ss}}{\langle\hat{\sigma}_{--}\rangle_{\rm ss}}\equiv \langle\hat{\sigma}_{00}\rangle_{\rm ss} G_{\beta\beta'}^2\\\times\frac{i(\Delta_\beta-\Delta_{\beta'})+\kappa_T}{(i\Delta_\beta+\kappa_T/2)(-i\Delta_{\beta'}+\kappa_T/2)},
\end{multline}
as well as additional decay and absorption of each spin wave mode at rates
\begin{equation}
    \Gamma_{d\beta}\equiv \Gamma_{\beta\beta}^{(n)}\hspace{0.3cm};\hspace{0.3cm} \Gamma_{a\beta}\equiv \Gamma_{\beta\beta}^{(a)}.
\end{equation}
It is convenient to define the difference between the above ratios as
\begin{multline}\label{GammabetadefAppendix}
    \Gamma_\beta \equiv \Gamma_{d\beta}-\Gamma_{a\beta} =  G_{\beta\beta}^2\langle\hat{\sigma}_{00}-\hat{\sigma}_{--}\rangle_{\rm ss}
    \\
    \times\frac{\kappa_T}{\Delta_\beta^2+(\kappa_T/2)^2}
\end{multline}
which, as we will see below, represents the increase in the decay rate of magnon mode $\beta$.

Let us now focus on the contribution from the occupations of the NV centres, namely of the term $\hat{V}_2$. For this term the Markov approximation is not applicable, as the correlators $\langle\tilde{S}_z(t)\tilde{S}_z(t-s)\rangle$ decay at a very slow rate $\gtrsim \tau_{\rm nv}^{-1}$ (see Eq.~\eqref{tauNVdefinition} and \figref{FigOpticalPumping}[c]). However, since the spin wave observables decay at a much faster rate $\gamma_\beta \gg \tau_{\rm nv}^{-1}$, we can compute the contribution $\hat{V}_2$ within the frozen bath model described in Appendix~\ref{AppTracingOut_FrozenBath}. The resulting contributions to the master equation are thus
\begin{equation}
    \mathcal{D}_{\rm sw, 2}[\hat{\rho}] = 0,
\end{equation}
\begin{equation}
    \hat{H}_{\rm sw, eff,2} =\sum_{\beta\beta'}\Omega_{\beta\beta'}\hat{s}_\beta^\dagger\hat{s}_{\beta'},
\end{equation}
with a rate given, assuming identical NV centres, by
\begin{equation}
    \Omega_{\beta\beta'}\equiv \langle\hat{S}_z/\hbar\rangle_{\rm ss}\sum_j\tilde{g}_{j\beta\beta'}.
\end{equation}
As expected from a bath with ``frozen'' fluctuations, the contribution $\hat{V}_2$ does not generate dissipation on the spin waves. Note that since $\tilde{g}_{j\beta\beta}=0$ (see Eqs.~\eqref{Xbetabetaequal0} and \eqref{g2norderfinal}) the above contribution does not generate a frequency shift on the spin wave modes.

Having obtained the master equation for the spin waves, we can compute the equations of motion for the expected values of the spin observables in a similar way as for Eqs.~\eqref{EOMsfree}, \eqref{EOMssfree}, and \eqref{EOMsdagsfree}. These equations of motion read
\begin{multline}\label{EOMsEFF}
    \frac{d}{dt}\langle\hat{s}_\beta\rangle = \left[-i(\omega_\beta+\delta_\beta)-\frac{\gamma_\beta+\Gamma_\beta}{2}\right]\langle\hat{s}_\beta\rangle
    \\-\sum_{\beta'\ne\beta} K_{\beta'\beta}^*\langle\hat{s}_{\beta'}\rangle,
\end{multline}
\begin{multline}\label{EOMssEFF}
    \frac{d}{dt}\langle\hat{s}_\beta\hat{s}_{\beta'}\rangle = \bigg[-i(\omega_\beta+\omega_{\beta'}+\delta_\beta+\delta_{\beta'})
    \\-\frac{\gamma_\beta+\gamma_{\beta'}+\Gamma_\beta+\Gamma_{\beta'}}{2}\bigg]\langle\hat{s}_\beta\hat{s}_{\beta'}\rangle
    \\
    -\sum_{\beta''\ne\beta'} K_{\beta'\beta''}\langle\hat{s}_\beta\hat{s}_{\beta''}\rangle  -\sum_{\beta''\ne\beta} K_{\beta\beta''}\langle\hat{s}_{\beta'}\hat{s}_{\beta''}\rangle ,
\end{multline}
\begin{multline}\label{EOMsdagsEFF}
    \frac{d}{dt}\langle\hat{s}_\beta^\dagger\hat{s}_{\beta'}\rangle = \bigg[i(\omega_\beta+\delta_\beta-\omega_{\beta'}-\delta_{\beta'}) \\-\frac{\gamma_\beta+\gamma_{\beta'}+\Gamma_\beta+\Gamma_{\beta'}}{2}\bigg]\langle\hat{s}_\beta^\dagger\hat{s}_{\beta'}\rangle(t) 
    \\+ \Gamma_{\beta\beta'}^{(a)}+\delta_{\beta\beta'}\left(\gamma_\beta\bar{n}_\beta+\Gamma_{a\beta}\right)
     \\
    -\sum_{\beta''\ne\beta'} K_{\beta'\beta''}\langle\hat{s}_\beta^\dagger\hat{s}_{\beta''}\rangle  -\sum_{\beta''\ne\beta} K_{\beta\beta''}^*\langle\hat{s}_{\beta''}^\dagger\hat{s}_{\beta'}\rangle ,
\end{multline}
where we have defined a compound interaction rate
\begin{multline}
    K_{\beta\beta'} \equiv i\Omega_{\beta\beta'} + iW_{\beta\beta'} + \frac{\Gamma_{\beta\beta'}^{(n)}-\Gamma_{\beta'\beta}^{(a)}}{2}
    \\
    =
    i\Omega_{\beta\beta'}+\frac{G_{\beta\beta'}^2}{i\Delta_\beta + \kappa_T/2}\langle\hat{\sigma}_{--}-\hat{\sigma}_{00}\rangle_{\rm ss}
\end{multline}
Note that the above system of equations is closed as the master equation is quadratic.

\subsection{Master equation rates for large $N$ and randomly distributed spins}\label{AppSWdynamics_largeNrandom}

All the rates in the master equation or in the equations of motion are given in terms of the frequencies $G_{\beta\beta'}$ and $\Omega_{\beta\beta'}$. 
Let us derive an expression for the former in the limit of interest, namely a large number $N\gg1$ of NV centres at random positions. A similar argument follows for the rate $\Omega_{\beta\beta'}$.
By using the definitions of $G_{\beta\beta'}$ and the expressions for the modal fields and the couplings in Appendices~\ref{appendixSpinWaves} and \ref{AppendixNVcentres}, we can write this rate explicitly as
\begin{multline}\label{Gbetabetaprimeexpanded}
    G_{\beta\beta'}^2 = \left(\frac{2\pi}{L}\right)^2\frac{v^2\omega_M}{\sqrt{\omega_\beta\omega_{\beta'}}} \left(1+\eta\frac{k_y}{k_\parallel}\right)\left(1+\eta\frac{k_y'}{k_\parallel'}\right)
    \\\times h_{\beta\eta0}h_{\beta'\eta0}\sum_{j=1}^N e^{-i(\mathbf{k}_\parallel-\mathbf{k}_\parallel')\mathbf{r}_{\parallel j}} e^{-(k_\parallel+k_\parallel')l_j},
\end{multline}
where the coefficient $v^2$ is defined in Eq.~\eqref{vconstantdefinition}, and $h_{\beta\eta0}$ are the modal field amplitudes defined by Eq.~\eqref{hfieldamplitude}. From this explicit form it is easy to check the following symmetry between the rates corresponding to the traced out NV ensemble lying above ($\eta=1$) or below the film ($\eta=-1$),
\begin{equation}
    G^2_{(n,\mathbf{k}_\parallel),(n',\mathbf{k}_\parallel')}\Big\vert_{\eta=-1} = (-1)^{n+n'}
    G^{2*}_{(n,-\mathbf{k}_\parallel),(n',-\mathbf{k}_\parallel')}\Big\vert_{\eta=1},
\end{equation}
from which similar relations for the frequency shift and decay rates $\delta_\beta$ and $\Gamma_\beta$ follow. Moreover, $G_{\beta\beta'}$ is maximized at $n=n'=0$. Similar properties also hold for $\Omega_{\beta\beta'}$. 

The two rates $G_{\beta\beta'}^2$ and $\Omega_{\beta\beta'}$ are proportional to a function of the following general form:
\begin{equation}\label{Fqlayerargument}
    F(\mathbf{q},q_0) \equiv \frac{1}{L^2}\sum_{j=1}^N e^{i\mathbf{q}\mathbf{r}_{\parallel j}} e^{-q_0l_j}.
\end{equation}
The above sum runs over the $N$ NV centres which, as discussed in the main text (see e.g.~\figref{Figure_band_modifications_1}[a]), are distributed in a volume delimited by $l_1$ and $l_2$ along the $x$ direction and infinitely extended on the directions parallel to the slab. For such configuration, we can compute the above function $F(\mathbf{q},q_0)$ following the strategy in Ref.~\cite{GonzalezTudelaPRL2013}, namely dividing the volumetric distribution of the NV centres into $N_L$ thin layers at positions $l_{0m}$ ($m=1,2,...N_L$), with $l_{01}=l_1$ and $l_{0N_L}=l_2$. In the limit of large number of NV centres, $N \to \infty$, we can choose a large number of layers $N_L \gg 1$, each containing a number $N_S \to \infty$ of NV centres. By grouping the NV centres in this way, and after multiplying and dividing by the separation between consecutive layers, namely $\Delta l \equiv (l_2-l_1)/N_L$, we can cast Eq.~\eqref{Fqlayerargument} in the form
\begin{equation}\label{Fqlayer2}
    F(\mathbf{q},q_0) = \varrho_{\rm ps} \sum_{m=1}^{N_L} (\Delta l) e^{-q_0l_{0m}} \frac{1}{N_S}\sum_{k\in\text{layer } m}^{N_S}e^{i\mathbf{q}\mathbf{r}_{\parallel k}},
\end{equation}
where we have defined the volumetric density of paramagnetic spins,
\begin{equation}
    \varrho_{\rm ps} \equiv \frac{N_S N_L}{L^2(l_2-l_1)}.
\end{equation}
In the above form, the function $F(\mathbf{q},q_0)$ can be exactly evaluated at $\mathbf{q}=0$, 
\begin{equation}\label{Fqlayer2b}
    F(0,q_0) = \varrho_{\rm ps} \sum_{l=l_1}^{l_2} (\Delta l) e^{-q_0l} \approx\varrho_{\rm ps} \int_{l_1}^{l_2} dl e^{-q_0l},
\end{equation}
where in the last step we have approximated the sum as an integral using the fact that $\Delta l \ll l_2-l_1$ and assuming that $q_0 \Delta l \ll 1$.
The above expression allows us to compute
 the coefficient $G_{\beta\beta}^2$, which determines the frequency shifts and decay rate modifications of the spin waves. For a slab of NV centres above the YIG film, this coefficient can be written using Eq.~\eqref{Gbetabetaprimeexpanded} as
\begin{multline}
    G_{\beta\beta}^2=\frac{4\pi^2v^2}{\omega_\beta/\omega_M} \left(1+\frac{k_y}{k_\parallel}\right)^2h_{\beta+0}^2 F(0,2 k_\parallel)
    \\
    = \frac{4\pi^2v^2\varrho_{\rm ps}}{\omega_\beta/\omega_M} \left(1+\frac{k_y}{k_\parallel}\right)^2h_{\beta+0}^2\int_{l1}^{l_2}dl e^{-2k_\parallel l}.
\end{multline}
 Combining the above expression, for $n=0$, with Eqs.~\eqref{deltabetadefAppendix} and \eqref{GammabetadefAppendix} we obtain the definitions Eq.~\eqref{deltabetaGammabetaexpressions} in the main text.

We now focus on the more general coefficients $G_{\beta\beta'}^2$, whose calculation requires evaluating the function $F(\mathbf{q},q_0)$ for $\mathbf{q}\ne 0$. This requires some further treatment as, by definition, the function as defined by Eq.~\eqref{Fqlayer2} depends on the values of the positions of all the NV centres. We thus assume that the NV centres are randomly distributed within each layer, and define the  normalized two-dimensional structure factor as
\begin{equation}
    s_{2D}(\mathbf{q}) \equiv \frac{1}{N_S}\sum_{m\text{ random}}^{N_S}e^{i\mathbf{q}\mathbf{r}_{\parallel m}}
\end{equation}
to write
\begin{equation}\label{Fqlayer3}
    F(\mathbf{q},q_0) = \varrho_{\rm ps} \sum_{l=l_1}^{l_2} (\Delta l) e^{-q_0l} s_{2D}^{(l)}(\mathbf{q}).
\end{equation}
Here, the quantity $s_{2D}^{(l)}(\mathbf{q})$ represents, for each $l$, a single value sampled from the probability distribution defined by the structure factor $s_{2D}(\mathbf{q})$. 
In the limit $N_S\to\infty$, the real and imaginary parts of such probability distribution tend to simple Gaussian distributions with mean value $0$
, as can be easily derived from the one-dimensional result~\cite{Barakat1988}. Thus, in this limit the average value of the function $F(\mathbf{q},q_0)$ vanishes exactly
\begin{multline}
   \lim_{N_S\to\infty} \langle F(\mathbf{q},q_0) \rangle= \\=  \lim_{N_S\to\infty} \varrho_{\rm ps}\sum_{l=l_1}^{l_2}
    (\Delta l) e^{-q_0l}
    \langle s_{2D}^{(l)}(\mathbf{q}) \rangle= 0.
\end{multline}
From the above expression we conclude that, in the thermodynamic limit, all the coupling rates $G_{\beta\beta'}$ become, in terms of statistical averaging over many random arrangements of NV centres, negligible for $\beta\ne\beta'$, and thus can be neglected. The same argument holds for the rate $\Omega_{\beta\beta'}$ which, since $\Omega_{\beta\beta}=0$, can be neglected for any $\beta$ and $\beta'$.

\subsection{Spin wave correlators and modification of the power spectral densities}\label{AppSWdynamics_PSDs}

Under the assumption $G_{\beta\beta'},\Omega_{\beta\beta'} \approx \delta_{\beta\beta'}$ justified above, the correlators of spin wave operators can be obtained from
Eqs.~\eqref{EOMsEFF}-\eqref{EOMsdagsEFF} using the quantum regression formula~\cite{CarmichaelBook}. For any operator $\hat{O}$ and for $s>0$ they read
\begin{multline}\label{SWcorrelatorsNVa}
    \langle \hat{s}_\beta(t+s)\hat{O}(t)\rangle_{\rm ss} = \langle \hat{s}_\beta\hat{O}\rangle_{\rm ss}
    \\
    \times\exp\left[(-i(\omega_\beta+\delta_\beta)-(\gamma_\beta + \Gamma_\beta)/2)s\right],
\end{multline}
\begin{multline}\label{SWcorrelatorsNVb}
    \langle \hat{s}_\beta^\dagger(t+s)\hat{O}(t)\rangle_{\rm ss} = \langle \hat{s}_\beta^\dagger\hat{O}\rangle_{\rm ss}
    \\
    \times\exp\left[(i(\omega_\beta+\delta_\beta)-(\gamma_\beta + \Gamma_\beta)/2)s\right].
\end{multline}
The steady-state occupations are also derived from Eqs.~\eqref{EOMsEFF}-\eqref{EOMsdagsEFF} and read
\begin{equation}
    \langle\hat{s}_\beta\rangle_{\rm ss} = \langle\hat{s}_\beta\hat{s}_{\beta'}\rangle_{\rm ss}=0,
\end{equation}
\begin{equation}
    \langle\hat{s}_\beta^\dagger\hat{s}_{\beta'}\rangle_{\rm ss} = \delta_{\beta\beta'}\frac{\gamma_\beta\bar{n}_\beta+\Gamma_{a\beta}}{\gamma_\beta+\Gamma_\beta}\equiv \delta_{\beta\beta'}N_\beta.
\end{equation}

\begin{figure*}[tbh!]
	\centering
	\includegraphics[width=\linewidth]{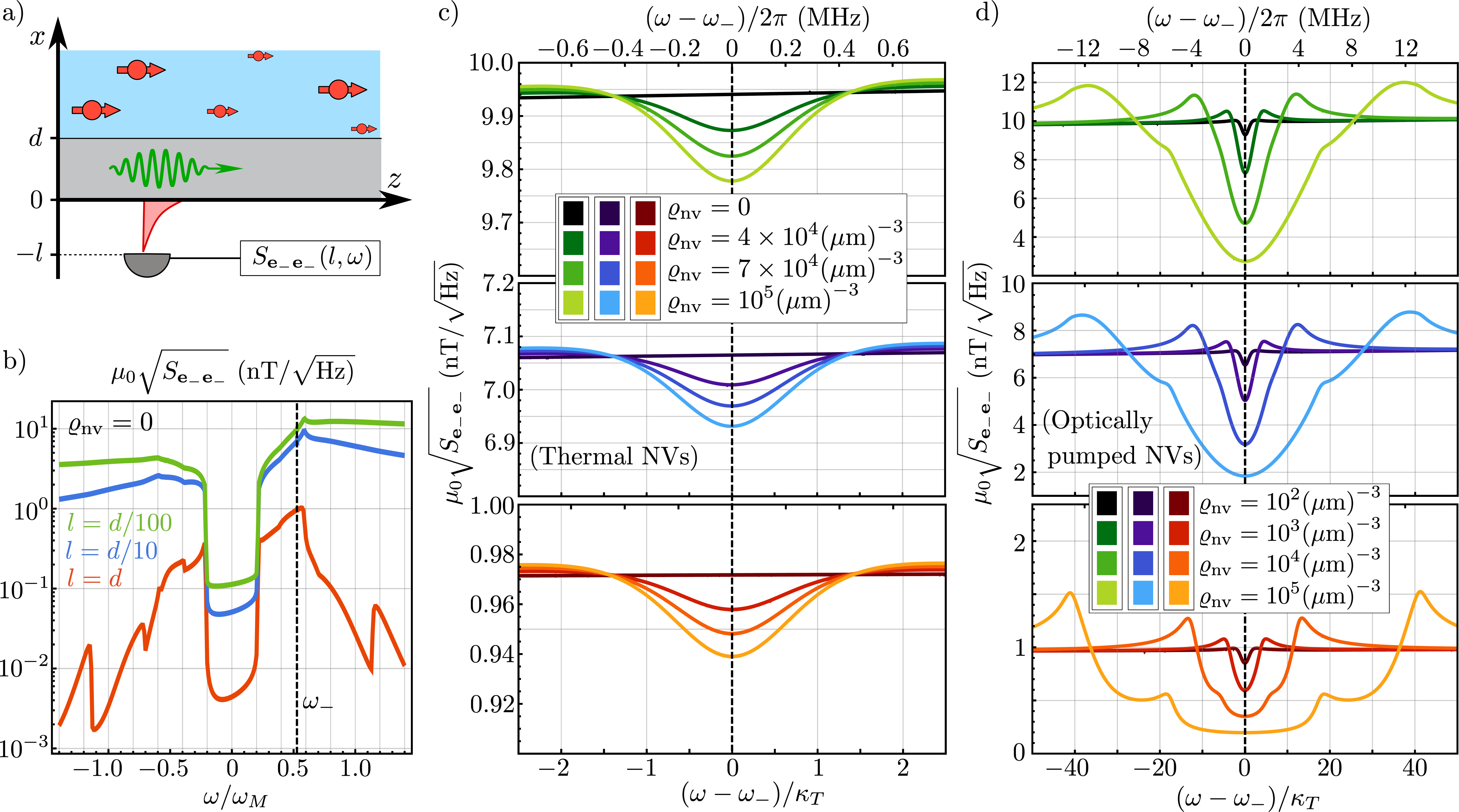}
	\caption{a) We consider the magnetic field power spectral density below the YIG film. b) Power spectral density in the absence of NV centres ($\varrho_{\rm nv}=0$) at different separations $l$ from the film surface. The dashed line indicates the NV transition frequency $\omega_-$. c) Power spectral density near $\omega=\omega_-$ for NV centres in a thermal state ($T=300$K). Panels with green, blue, and red-orange color schemes correspond to the same color scheme as in panel (a), i.e., to $l=10^{-2}d$, $l=10^{-1}d$, and $l=d$ respectively. The four lines on each set correspond to different densities of NV centres. d) Same as in panel (c), for optically pumped NV centres (optimal pumping conditions). In all the panels we take $\mu_0H_0=10$mT, $l_1=0$, $l_2=\infty$, $T=300$K, and the parameters of Table~\ref{tablePARAMS}.}\label{Fig_Modified_PSDs}
\end{figure*}

By using the above results we can compute the power spectral density of the magnetic field, Eq.~\eqref{twosidedPSDdef}, in an analogous way as for the uncoupled spin waves in Appendix~\ref{AppSW_PSD}. Specifically, one can show that for the above steady-state values, equation~\eqref{PSDintermsofcorrelators} holds also in the presence of NV centres. By introducing in this equation the corresponding correlators we obtain the following generalized expression for the power spectral density:
\begin{multline}
    S_{\mathbf{e}_j\mathbf{e}_k}(l,\omega) = \frac{1}{2\pi}\sum_\beta  \mathcal{M}_{0\beta}^2 h_{\beta\eta0}^2 e^{-2k_\parallel l} (\gamma_\beta+\Gamma_\beta)
    \\
    \times
    \Big[\bar{\Lambda}_{jk}(\eta,\mathbf{k}_\parallel)\frac{N_\beta+1}{(\omega-\omega_\beta-\delta_\beta)^2+((\gamma_\beta+\Gamma_\beta)/2)^2}
    \\
    +\bar{\Lambda}_{jk}(-\eta,\mathbf{k}_\parallel)\frac{N_\beta}{(\omega+\omega_\beta+\delta_\beta)^2+((\gamma_\beta+\Gamma_\beta)/2)^2}\Big],
\end{multline}
with the matrices $\bar{\Lambda}_{jk}(\eta,\mathbf{k}_\parallel)$ defined in Eq.~\eqref{BarLambdaMatrix}.
This expression is formally equivalent to the power spectral density in the absence of NV centres, Eq.~\eqref{PSDnoNVs}, under the substitutions $\omega_\beta \to \omega_\beta + \delta_\beta$, $\gamma_\beta \to \gamma_\beta+\Gamma_\beta$, and $\bar{n}_\beta \to N_\beta$. 
Hence, the above power spectral density satisfies the same symmetry properties of the power spectral density derived in the absence of spin waves: first,  the property Eq.~\eqref{PSDprops2} stems from the form of $\bar{\Lambda}_{jk}(\eta,\mathbf{k}_\parallel)$ which remains unaltered in the presence of NV centres; second, the properties Eqs.~\eqref{PSDprops1} and \eqref{PSDpropslast} hold due to the easily verifiable symmetries
\begin{equation}
    G_{\beta\beta}(-k_z) = G_{\beta\beta}(k_z),
\end{equation}
and
\begin{equation}
    G_{\beta\beta}(k_y)\vert_\eta =  G_{\beta\beta}(-k_y)\vert_{-\eta},
\end{equation}
respectively. Finally, the high-temperature parity properties Eqs.~\eqref{PSDprops3a}-\eqref{PSDprops3b} hold in the limit $N_\beta \gg 1$. Since for the parameters chosen in this work $N_\beta \approx \bar{n}_\beta$ we conclude (and have numerically checked) that these parity properties also hold, at room temperature, in the presence of NV centres.

We focus on the power spectral density $S_{\mathbf{e}_-\mathbf{e}_-}(l,\omega)\approx S_{\mathbf{e}_+\mathbf{e}_+}(l,-\omega)$, for which the NV-induced modification is the most pronounced. As schematically depicted in \figref{Fig_Modified_PSDs}(a), we consider this power spectral density in the region below the film ($\eta=-1$), where the diamond slab cannot hinder its detection. We first compute the power spectral density in the absence of NV centres in \figref{Fig_Modified_PSDs}(b), at room temperature and at three different separations $l$ from the YIG film. As discussed in Sec.~\ref{AppSW_PSD} the power spectral density is higher at shorter distances to the film $l$, where the thermal amplitude of the magnetic field is larger. It also becomes smoother at shorter distances as the contribution of many high-wavenumber spin waves becomes practically as relevant as that of low-wavenumber spin waves.
In the presence of NV centres, the only appreciable change in the power spectral density occurs in a narrow frequency range centered at the NV transition frequency, $\omega_-$, and with width of the order of $\kappa_T$. In \figref{Fig_Modified_PSDs}(c) we show the power spectral density within this relevant frequency range for un-pumped NV centres at room temperature thermal equilibrium. The spin waves that contribute most significantly to the power spectral density within this frequency range are those resonant to the NV transition, i.e., those fulfilling $\omega_\beta \approx\omega_-$. Since for these spin waves the linewidth is enhanced by a factor $\Gamma_\beta$ (see discussion in Sec.~\ref{SecSWmodification} in the main text), the steady-state spin wave correlators Eqs.~\eqref{SWcorrelatorsNVa}-\eqref{SWcorrelatorsNVb} decay faster (i.e., the fluctuations become smaller) than in the absence of the NV centres, resulting in a decrease of the power spectral density. Note that this decrease could be interpreted as a lower effective temperature for these spin waves, as the equilibrium power spectral density is proportional to the thermal occupation factor (see e.g. Eq.~\eqref{PSDintermsofcorrelators}). 
In this way the action of the NV centres is equivalent to an effective frequency-resolved cooling of the spin waves.
As shown by \figref{Fig_Modified_PSDs}(c), the dip in the power spectral density becomes more pronounced near the film, where the NV centres are more strongly coupled to the spin waves. Similar phenomenology is observed for optically pumped NV centres (\figref{Fig_Modified_PSDs}[d]). In this case, optical pumping enhances the impact of the NV centres on the power spectral density, which now shows much deeper minima (up to $\sim80\%$ decrease with respect to the case $\varrho_{\rm nv}=0$, as opposed to the $\sim 2\%$ in the un-pumped case) and modifications within a much wider frequency range (about $50\kappa_T$ as opposed to $\sim \kappa_T$). Although, in both the pumped and the un-pumped scenario, observing the dip in the power spectral density requires a lower magnetic noise floor and thus a higher sensitivity, this dip can still be experimentally measured.  Indeed, our results in \figref{Fig_Modified_PSDs}(c-d) for the most part remain within the sensitivity range ($\sim 10^{-4}-10^{-6}$nT/$\sqrt{\text{Hz}}$) of ultra-sensitive magnetometry techniques~\cite{WolfPRX2015,WasilewskiPRL2010,LedbetterPNAS2008,WrachtrupJMR2016}. As a final remark we note that the opposite effect, namely the electron spin-induced \emph{increase} of the magnetic fluctuations, could also be attained by driving the transition $\vert 0 \rangle\leftrightarrow \vert - \rangle$ into saturation or into population inversion. This procedure, which results in a reduction of the spin wave linewidth, is known as ``spectral hole burning'' and has been successfully used to increase the quality factor of mechanical resonators whose coherence is limited by two-level system impurities~\cite{AnderssonArxiv2020}. Exploring the potential of spectral hole burning for spintronics is an interesting outlook to this work, that can be studied with the theory we have developed.

 
\bibliographystyle{apsrev4-1}
\bibliography{bibliography}

\begin{thebibliography}{134}%
\makeatletter
\providecommand \@ifxundefined [1]{%
 \@ifx{#1\undefined}
}%
\providecommand \@ifnum [1]{%
 \ifnum #1\expandafter \@firstoftwo
 \else \expandafter \@secondoftwo
 \fi
}%
\providecommand \@ifx [1]{%
 \ifx #1\expandafter \@firstoftwo
 \else \expandafter \@secondoftwo
 \fi
}%
\providecommand \natexlab [1]{#1}%
\providecommand \enquote  [1]{``#1''}%
\providecommand \bibnamefont  [1]{#1}%
\providecommand \bibfnamefont [1]{#1}%
\providecommand \citenamefont [1]{#1}%
\providecommand \href@noop [0]{\@secondoftwo}%
\providecommand \href [0]{\begingroup \@sanitize@url \@href}%
\providecommand \@href[1]{\@@startlink{#1}\@@href}%
\providecommand \@@href[1]{\endgroup#1\@@endlink}%
\providecommand \@sanitize@url [0]{\catcode `\\12\catcode `\$12\catcode
  `\&12\catcode `\#12\catcode `\^12\catcode `\_12\catcode `\%12\relax}%
\providecommand \@@startlink[1]{}%
\providecommand \@@endlink[0]{}%
\providecommand \url  [0]{\begingroup\@sanitize@url \@url }%
\providecommand \@url [1]{\endgroup\@href {#1}{\urlprefix }}%
\providecommand \urlprefix  [0]{URL }%
\providecommand \Eprint [0]{\href }%
\providecommand \doibase [0]{http://dx.doi.org/}%
\providecommand \selectlanguage [0]{\@gobble}%
\providecommand \bibinfo  [0]{\@secondoftwo}%
\providecommand \bibfield  [0]{\@secondoftwo}%
\providecommand \translation [1]{[#1]}%
\providecommand \BibitemOpen [0]{}%
\providecommand \bibitemStop [0]{}%
\providecommand \bibitemNoStop [0]{.\EOS\space}%
\providecommand \EOS [0]{\spacefactor3000\relax}%
\providecommand \BibitemShut  [1]{\csname bibitem#1\endcsname}%
\let\auto@bib@innerbib\@empty
\bibitem [{\citenamefont {Tabuchi}\ \emph {et~al.}(2014)\citenamefont
  {Tabuchi}, \citenamefont {Ishino}, \citenamefont {Ishikawa}, \citenamefont
  {Yamazaki}, \citenamefont {Usami},\ and\ \citenamefont
  {Nakamura}}]{TabuchiPRL2014}%
  \BibitemOpen
  \bibfield  {author} {\bibinfo {author} {\bibfnamefont {Y.}~\bibnamefont
  {Tabuchi}}, \bibinfo {author} {\bibfnamefont {S.}~\bibnamefont {Ishino}},
  \bibinfo {author} {\bibfnamefont {T.}~\bibnamefont {Ishikawa}}, \bibinfo
  {author} {\bibfnamefont {R.}~\bibnamefont {Yamazaki}}, \bibinfo {author}
  {\bibfnamefont {K.}~\bibnamefont {Usami}}, \ and\ \bibinfo {author}
  {\bibfnamefont {Y.}~\bibnamefont {Nakamura}},\ }\href
  {https://link.aps.org/doi/10.1103/PhysRevLett.113.083603} {\bibfield
  {journal} {\bibinfo  {journal} {Phys. Rev. Lett.}\ }\textbf {\bibinfo
  {volume} {113}},\ \bibinfo {pages} {083603} (\bibinfo {year}
  {2014})}\BibitemShut {NoStop}%
\bibitem [{\citenamefont {Zhang}\ \emph {et~al.}(2014)\citenamefont {Zhang},
  \citenamefont {Zou}, \citenamefont {Jiang},\ and\ \citenamefont
  {Tang}}]{ZhangPRL2014}%
  \BibitemOpen
  \bibfield  {author} {\bibinfo {author} {\bibfnamefont {X.}~\bibnamefont
  {Zhang}}, \bibinfo {author} {\bibfnamefont {C.-L.}\ \bibnamefont {Zou}},
  \bibinfo {author} {\bibfnamefont {L.}~\bibnamefont {Jiang}}, \ and\ \bibinfo
  {author} {\bibfnamefont {H.~X.}\ \bibnamefont {Tang}},\ }\href
  {https://link.aps.org/doi/10.1103/PhysRevLett.113.156401} {\bibfield
  {journal} {\bibinfo  {journal} {Phys. Rev. Lett.}\ }\textbf {\bibinfo
  {volume} {113}},\ \bibinfo {pages} {156401} (\bibinfo {year}
  {2014})}\BibitemShut {NoStop}%
\bibitem [{\citenamefont {Haigh}\ \emph {et~al.}(2016)\citenamefont {Haigh},
  \citenamefont {Nunnenkamp}, \citenamefont {Ramsay},\ and\ \citenamefont
  {Ferguson}}]{HaighPRL2016}%
  \BibitemOpen
  \bibfield  {author} {\bibinfo {author} {\bibfnamefont {J.~A.}\ \bibnamefont
  {Haigh}}, \bibinfo {author} {\bibfnamefont {A.}~\bibnamefont {Nunnenkamp}},
  \bibinfo {author} {\bibfnamefont {A.~J.}\ \bibnamefont {Ramsay}}, \ and\
  \bibinfo {author} {\bibfnamefont {A.~J.}\ \bibnamefont {Ferguson}},\ }\href
  {https://link.aps.org/doi/10.1103/PhysRevLett.117.133602} {\bibfield
  {journal} {\bibinfo  {journal} {Phys. Rev. Lett.}\ }\textbf {\bibinfo
  {volume} {117}},\ \bibinfo {pages} {133602} (\bibinfo {year}
  {2016})}\BibitemShut {NoStop}%
\bibitem [{\citenamefont {Zhang}\ \emph {et~al.}(2016)\citenamefont {Zhang},
  \citenamefont {Zou}, \citenamefont {Jiang},\ and\ \citenamefont
  {Tang}}]{ZhangSciAdv2016}%
  \BibitemOpen
  \bibfield  {author} {\bibinfo {author} {\bibfnamefont {X.}~\bibnamefont
  {Zhang}}, \bibinfo {author} {\bibfnamefont {C.-L.}\ \bibnamefont {Zou}},
  \bibinfo {author} {\bibfnamefont {L.}~\bibnamefont {Jiang}}, \ and\ \bibinfo
  {author} {\bibfnamefont {H.~X.}\ \bibnamefont {Tang}},\ }\href
  {http://advances.sciencemag.org/content/2/3/e1501286} {\bibfield  {journal}
  {\bibinfo  {journal} {Sci. Adv.}\ }\textbf {\bibinfo {volume} {2}},\ \bibinfo
  {pages} {e1501286} (\bibinfo {year} {2016})}\BibitemShut {NoStop}%
\bibitem [{\citenamefont {Viola~Kusminskiy}\ \emph {et~al.}(2016)\citenamefont
  {Viola~Kusminskiy}, \citenamefont {Tang},\ and\ \citenamefont
  {Marquardt}}]{KusminskiyPRA2016}%
  \BibitemOpen
  \bibfield  {author} {\bibinfo {author} {\bibfnamefont {S.}~\bibnamefont
  {Viola~Kusminskiy}}, \bibinfo {author} {\bibfnamefont {H.~X.}\ \bibnamefont
  {Tang}}, \ and\ \bibinfo {author} {\bibfnamefont {F.}~\bibnamefont
  {Marquardt}},\ }\href {https://link.aps.org/doi/10.1103/PhysRevA.94.033821}
  {\bibfield  {journal} {\bibinfo  {journal} {Phys. Rev. A}\ }\textbf {\bibinfo
  {volume} {94}},\ \bibinfo {pages} {033821} (\bibinfo {year}
  {2016})}\BibitemShut {NoStop}%
\bibitem [{\citenamefont {Gonzalez-Ballestero}\ \emph
  {et~al.}(2020{\natexlab{a}})\citenamefont {Gonzalez-Ballestero},
  \citenamefont {Gieseler},\ and\ \citenamefont
  {Romero-Isart}}]{GonzalezBallesteroPRL2020}%
  \BibitemOpen
  \bibfield  {author} {\bibinfo {author} {\bibfnamefont {C.}~\bibnamefont
  {Gonzalez-Ballestero}}, \bibinfo {author} {\bibfnamefont {J.}~\bibnamefont
  {Gieseler}}, \ and\ \bibinfo {author} {\bibfnamefont {O.}~\bibnamefont
  {Romero-Isart}},\ }\href
  {https://link.aps.org/doi/10.1103/PhysRevLett.124.093602} {\bibfield
  {journal} {\bibinfo  {journal} {Phys. Rev. Lett.}\ }\textbf {\bibinfo
  {volume} {124}},\ \bibinfo {pages} {093602} (\bibinfo {year}
  {2020}{\natexlab{a}})}\BibitemShut {NoStop}%
\bibitem [{\citenamefont {Huebl}\ \emph {et~al.}(2013)\citenamefont {Huebl},
  \citenamefont {Zollitsch}, \citenamefont {Lotze}, \citenamefont {Hocke},
  \citenamefont {Greifenstein}, \citenamefont {Marx}, \citenamefont {Gross},\
  and\ \citenamefont {Goennenwein}}]{HueblPRL2013}%
  \BibitemOpen
  \bibfield  {author} {\bibinfo {author} {\bibfnamefont {H.}~\bibnamefont
  {Huebl}}, \bibinfo {author} {\bibfnamefont {C.~W.}\ \bibnamefont
  {Zollitsch}}, \bibinfo {author} {\bibfnamefont {J.}~\bibnamefont {Lotze}},
  \bibinfo {author} {\bibfnamefont {F.}~\bibnamefont {Hocke}}, \bibinfo
  {author} {\bibfnamefont {M.}~\bibnamefont {Greifenstein}}, \bibinfo {author}
  {\bibfnamefont {A.}~\bibnamefont {Marx}}, \bibinfo {author} {\bibfnamefont
  {R.}~\bibnamefont {Gross}}, \ and\ \bibinfo {author} {\bibfnamefont
  {S.~T.~B.}\ \bibnamefont {Goennenwein}},\ }\href
  {https://link.aps.org/doi/10.1103/PhysRevLett.111.127003} {\bibfield
  {journal} {\bibinfo  {journal} {Phys. Rev. Lett.}\ }\textbf {\bibinfo
  {volume} {111}},\ \bibinfo {pages} {127003} (\bibinfo {year}
  {2013})}\BibitemShut {NoStop}%
\bibitem [{\citenamefont {Li}\ \emph {et~al.}(2016)\citenamefont {Li},
  \citenamefont {Xu}, \citenamefont {Aldosary}, \citenamefont {Tang},
  \citenamefont {Lin}, \citenamefont {Zhang}, \citenamefont {Lake},\ and\
  \citenamefont {Shi}}]{LiNatComm2016}%
  \BibitemOpen
  \bibfield  {author} {\bibinfo {author} {\bibfnamefont {J.}~\bibnamefont
  {Li}}, \bibinfo {author} {\bibfnamefont {Y.}~\bibnamefont {Xu}}, \bibinfo
  {author} {\bibfnamefont {M.}~\bibnamefont {Aldosary}}, \bibinfo {author}
  {\bibfnamefont {C.}~\bibnamefont {Tang}}, \bibinfo {author} {\bibfnamefont
  {Z.}~\bibnamefont {Lin}}, \bibinfo {author} {\bibfnamefont {S.}~\bibnamefont
  {Zhang}}, \bibinfo {author} {\bibfnamefont {R.}~\bibnamefont {Lake}}, \ and\
  \bibinfo {author} {\bibfnamefont {J.}~\bibnamefont {Shi}},\ }\href
  {https://doi.org/10.1038/ncomms10858} {\bibfield  {journal} {\bibinfo
  {journal} {Nat Commun}\ }\textbf {\bibinfo {volume} {7}},\ \bibinfo {pages}
  {10858} (\bibinfo {year} {2016})}\BibitemShut {NoStop}%
\bibitem [{\citenamefont {Tabuchi}\ \emph {et~al.}(2015)\citenamefont
  {Tabuchi}, \citenamefont {Ishino}, \citenamefont {Noguchi}, \citenamefont
  {Ishikawa}, \citenamefont {Yamazaki}, \citenamefont {Usami},\ and\
  \citenamefont {Nakamura}}]{TabuchiScience2015}%
  \BibitemOpen
  \bibfield  {author} {\bibinfo {author} {\bibfnamefont {Y.}~\bibnamefont
  {Tabuchi}}, \bibinfo {author} {\bibfnamefont {S.}~\bibnamefont {Ishino}},
  \bibinfo {author} {\bibfnamefont {A.}~\bibnamefont {Noguchi}}, \bibinfo
  {author} {\bibfnamefont {T.}~\bibnamefont {Ishikawa}}, \bibinfo {author}
  {\bibfnamefont {R.}~\bibnamefont {Yamazaki}}, \bibinfo {author}
  {\bibfnamefont {K.}~\bibnamefont {Usami}}, \ and\ \bibinfo {author}
  {\bibfnamefont {Y.}~\bibnamefont {Nakamura}},\ }\href
  {https://science.sciencemag.org/content/349/6246/405} {\bibfield  {journal}
  {\bibinfo  {journal} {Science}\ }\textbf {\bibinfo {volume} {349}},\ \bibinfo
  {pages} {405} (\bibinfo {year} {2015})}\BibitemShut {NoStop}%
\bibitem [{\citenamefont {Marcos}\ \emph {et~al.}(2010)\citenamefont {Marcos},
  \citenamefont {Wubs}, \citenamefont {Taylor}, \citenamefont {Aguado},
  \citenamefont {Lukin},\ and\ \citenamefont {S\o{}rensen}}]{MarcosPRL2010}%
  \BibitemOpen
  \bibfield  {author} {\bibinfo {author} {\bibfnamefont {D.}~\bibnamefont
  {Marcos}}, \bibinfo {author} {\bibfnamefont {M.}~\bibnamefont {Wubs}},
  \bibinfo {author} {\bibfnamefont {J.~M.}\ \bibnamefont {Taylor}}, \bibinfo
  {author} {\bibfnamefont {R.}~\bibnamefont {Aguado}}, \bibinfo {author}
  {\bibfnamefont {M.~D.}\ \bibnamefont {Lukin}}, \ and\ \bibinfo {author}
  {\bibfnamefont {A.~S.}\ \bibnamefont {S\o{}rensen}},\ }\href
  {https://link.aps.org/doi/10.1103/PhysRevLett.105.210501} {\bibfield
  {journal} {\bibinfo  {journal} {Phys. Rev. Lett.}\ }\textbf {\bibinfo
  {volume} {105}},\ \bibinfo {pages} {210501} (\bibinfo {year}
  {2010})}\BibitemShut {NoStop}%
\bibitem [{\citenamefont {Zhu}\ \emph {et~al.}(2011)\citenamefont {Zhu},
  \citenamefont {Saito}, \citenamefont {Kemp}, \citenamefont {Kakuyanagi},
  \citenamefont {Karimoto}, \citenamefont {Nakano}, \citenamefont {Munro},
  \citenamefont {Tokura}, \citenamefont {Everitt}, \citenamefont {Nemoto},
  \citenamefont {Kasu}, \citenamefont {Mizuochi},\ and\ \citenamefont
  {Semba}}]{ZhuNature2011}%
  \BibitemOpen
  \bibfield  {author} {\bibinfo {author} {\bibfnamefont {X.}~\bibnamefont
  {Zhu}}, \bibinfo {author} {\bibfnamefont {S.}~\bibnamefont {Saito}}, \bibinfo
  {author} {\bibfnamefont {A.}~\bibnamefont {Kemp}}, \bibinfo {author}
  {\bibfnamefont {K.}~\bibnamefont {Kakuyanagi}}, \bibinfo {author}
  {\bibfnamefont {S.-i.}\ \bibnamefont {Karimoto}}, \bibinfo {author}
  {\bibfnamefont {H.}~\bibnamefont {Nakano}}, \bibinfo {author} {\bibfnamefont
  {W.~J.}\ \bibnamefont {Munro}}, \bibinfo {author} {\bibfnamefont
  {Y.}~\bibnamefont {Tokura}}, \bibinfo {author} {\bibfnamefont {M.~S.}\
  \bibnamefont {Everitt}}, \bibinfo {author} {\bibfnamefont {K.}~\bibnamefont
  {Nemoto}}, \bibinfo {author} {\bibfnamefont {M.}~\bibnamefont {Kasu}},
  \bibinfo {author} {\bibfnamefont {N.}~\bibnamefont {Mizuochi}}, \ and\
  \bibinfo {author} {\bibfnamefont {K.}~\bibnamefont {Semba}},\ }\href
  {https://doi.org/10.1038/nature10462} {\bibfield  {journal} {\bibinfo
  {journal} {Nature}\ }\textbf {\bibinfo {volume} {478}},\ \bibinfo {pages}
  {221} (\bibinfo {year} {2011})}\BibitemShut {NoStop}%
\bibitem [{\citenamefont {Lenk}\ \emph {et~al.}(2011)\citenamefont {Lenk},
  \citenamefont {Ulrichs}, \citenamefont {Garbs},\ and\ \citenamefont
  {M{\"u}nzenberg}}]{LenkPhysRep2011}%
  \BibitemOpen
  \bibfield  {author} {\bibinfo {author} {\bibfnamefont {B.}~\bibnamefont
  {Lenk}}, \bibinfo {author} {\bibfnamefont {H.}~\bibnamefont {Ulrichs}},
  \bibinfo {author} {\bibfnamefont {F.}~\bibnamefont {Garbs}}, \ and\ \bibinfo
  {author} {\bibfnamefont {M.}~\bibnamefont {M{\"u}nzenberg}},\ }\href
  {http://www.sciencedirect.com/science/article/pii/S0370157311001694}
  {\bibfield  {journal} {\bibinfo  {journal} {Phys Rep}\ }\textbf {\bibinfo
  {volume} {507}},\ \bibinfo {pages} {107 } (\bibinfo {year}
  {2011})}\BibitemShut {NoStop}%
\bibitem [{\citenamefont {Kruglyak}\ \emph {et~al.}(2010)\citenamefont
  {Kruglyak}, \citenamefont {Demokritov},\ and\ \citenamefont
  {Grundler}}]{KruglyakJPhysD2010}%
  \BibitemOpen
  \bibfield  {author} {\bibinfo {author} {\bibfnamefont {V.~V.}\ \bibnamefont
  {Kruglyak}}, \bibinfo {author} {\bibfnamefont {S.~O.}\ \bibnamefont
  {Demokritov}}, \ and\ \bibinfo {author} {\bibfnamefont {D.}~\bibnamefont
  {Grundler}},\ }\href
  {https://doi.org/10.1088\%2F0022-3727\%2F43\%2F26\%2F260301} {\bibfield
  {journal} {\bibinfo  {journal} {J Phys D Appl Phys}\ }\textbf {\bibinfo
  {volume} {43}},\ \bibinfo {pages} {260301} (\bibinfo {year}
  {2010})}\BibitemShut {NoStop}%
\bibitem [{\citenamefont {Chumak}\ \emph {et~al.}(2015)\citenamefont {Chumak},
  \citenamefont {Vasyuchka}, \citenamefont {Serga},\ and\ \citenamefont
  {Hillebrands}}]{ChumakNatPhys2015}%
  \BibitemOpen
  \bibfield  {author} {\bibinfo {author} {\bibfnamefont {A.~V.}\ \bibnamefont
  {Chumak}}, \bibinfo {author} {\bibfnamefont {V.~I.}\ \bibnamefont
  {Vasyuchka}}, \bibinfo {author} {\bibfnamefont {A.~A.}\ \bibnamefont
  {Serga}}, \ and\ \bibinfo {author} {\bibfnamefont {B.}~\bibnamefont
  {Hillebrands}},\ }\href {https://doi.org/10.1038/nphys3347} {\bibfield
  {journal} {\bibinfo  {journal} {Nat Physics}\ }\textbf {\bibinfo {volume}
  {11}},\ \bibinfo {pages} {453} (\bibinfo {year} {2015})}\BibitemShut
  {NoStop}%
\bibitem [{\citenamefont {Stamps}\ \emph {et~al.}(2014)\citenamefont {Stamps},
  \citenamefont {Breitkreutz}, \citenamefont {{\AA}kerman}, \citenamefont
  {Chumak}, \citenamefont {Otani}, \citenamefont {Bauer}, \citenamefont
  {Thiele}, \citenamefont {Bowen}, \citenamefont {Majetich}, \citenamefont
  {Kl{\"a}ui}, \citenamefont {Prejbeanu}, \citenamefont {Dieny}, \citenamefont
  {Dempsey},\ and\ \citenamefont {Hillebrands}}]{StampsJPhysD2014}%
  \BibitemOpen
  \bibfield  {author} {\bibinfo {author} {\bibfnamefont {R.~L.}\ \bibnamefont
  {Stamps}}, \bibinfo {author} {\bibfnamefont {S.}~\bibnamefont {Breitkreutz}},
  \bibinfo {author} {\bibfnamefont {J.}~\bibnamefont {{\AA}kerman}}, \bibinfo
  {author} {\bibfnamefont {A.~V.}\ \bibnamefont {Chumak}}, \bibinfo {author}
  {\bibfnamefont {Y.}~\bibnamefont {Otani}}, \bibinfo {author} {\bibfnamefont
  {G.~E.~W.}\ \bibnamefont {Bauer}}, \bibinfo {author} {\bibfnamefont {J.-U.}\
  \bibnamefont {Thiele}}, \bibinfo {author} {\bibfnamefont {M.}~\bibnamefont
  {Bowen}}, \bibinfo {author} {\bibfnamefont {S.~A.}\ \bibnamefont {Majetich}},
  \bibinfo {author} {\bibfnamefont {M.}~\bibnamefont {Kl{\"a}ui}}, \bibinfo
  {author} {\bibfnamefont {I.~L.}\ \bibnamefont {Prejbeanu}}, \bibinfo {author}
  {\bibfnamefont {B.}~\bibnamefont {Dieny}}, \bibinfo {author} {\bibfnamefont
  {N.~M.}\ \bibnamefont {Dempsey}}, \ and\ \bibinfo {author} {\bibfnamefont
  {B.}~\bibnamefont {Hillebrands}},\ }\href
  {https://doi.org/10.1088\%2F0022-3727\%2F47\%2F33\%2F333001} {\bibfield
  {journal} {\bibinfo  {journal} {J Phys D Appl Phys}\ }\textbf {\bibinfo
  {volume} {47}},\ \bibinfo {pages} {333001} (\bibinfo {year}
  {2014})}\BibitemShut {NoStop}%
\bibitem [{\citenamefont {Serga}\ \emph {et~al.}(2010)\citenamefont {Serga},
  \citenamefont {Chumak},\ and\ \citenamefont {Hillebrands}}]{SergaJPhysD2010}%
  \BibitemOpen
  \bibfield  {author} {\bibinfo {author} {\bibfnamefont {A.~A.}\ \bibnamefont
  {Serga}}, \bibinfo {author} {\bibfnamefont {A.~V.}\ \bibnamefont {Chumak}}, \
  and\ \bibinfo {author} {\bibfnamefont {B.}~\bibnamefont {Hillebrands}},\
  }\href {https://doi.org/10.1088\%2F0022-3727\%2F43\%2F26\%2F264002}
  {\bibfield  {journal} {\bibinfo  {journal} {J Phys D Appl Phys}\ }\textbf
  {\bibinfo {volume} {43}},\ \bibinfo {pages} {264002} (\bibinfo {year}
  {2010})}\BibitemShut {NoStop}%
\bibitem [{\citenamefont {Chumak}\ \emph {et~al.}(2014)\citenamefont {Chumak},
  \citenamefont {Serga},\ and\ \citenamefont
  {Hillebrands}}]{ChumakNatComm2014}%
  \BibitemOpen
  \bibfield  {author} {\bibinfo {author} {\bibfnamefont {A.~V.}\ \bibnamefont
  {Chumak}}, \bibinfo {author} {\bibfnamefont {A.~A.}\ \bibnamefont {Serga}}, \
  and\ \bibinfo {author} {\bibfnamefont {B.}~\bibnamefont {Hillebrands}},\
  }\href {https://doi.org/10.1038/ncomms5700} {\bibfield  {journal} {\bibinfo
  {journal} {Nat Commun}\ }\textbf {\bibinfo {volume} {5}},\ \bibinfo {pages}
  {4700} (\bibinfo {year} {2014})}\BibitemShut {NoStop}%
\bibitem [{\citenamefont {Gurevich}\ and\ \citenamefont
  {Melkov}(1996)}]{Gurevich1996magnetization}%
  \BibitemOpen
  \bibfield  {author} {\bibinfo {author} {\bibfnamefont {A.}~\bibnamefont
  {Gurevich}}\ and\ \bibinfo {author} {\bibfnamefont {G.}~\bibnamefont
  {Melkov}},\ }\href@noop {} {\emph {\bibinfo {title} {Magnetization
  Oscillations and Waves}}}\ (\bibinfo  {publisher} {Taylor \& Francis},\
  \bibinfo {year} {1996})\BibitemShut {NoStop}%
\bibitem [{\citenamefont {Stancil}\ and\ \citenamefont
  {Prabhakar}(2009)}]{StancilBook2009}%
  \BibitemOpen
  \bibfield  {author} {\bibinfo {author} {\bibfnamefont {D.}~\bibnamefont
  {Stancil}}\ and\ \bibinfo {author} {\bibfnamefont {A.}~\bibnamefont
  {Prabhakar}},\ }\href@noop {} {\emph {\bibinfo {title} {Spin Waves: Theory
  and Applications}}}\ (\bibinfo  {publisher} {Springer US},\ \bibinfo {year}
  {2009})\BibitemShut {NoStop}%
\bibitem [{\citenamefont {Nakata}\ \emph {et~al.}(2014)\citenamefont {Nakata},
  \citenamefont {van Hoogdalem}, \citenamefont {Simon},\ and\ \citenamefont
  {Loss}}]{NakataPRB2014}%
  \BibitemOpen
  \bibfield  {author} {\bibinfo {author} {\bibfnamefont {K.}~\bibnamefont
  {Nakata}}, \bibinfo {author} {\bibfnamefont {K.~A.}\ \bibnamefont {van
  Hoogdalem}}, \bibinfo {author} {\bibfnamefont {P.}~\bibnamefont {Simon}}, \
  and\ \bibinfo {author} {\bibfnamefont {D.}~\bibnamefont {Loss}},\ }\href
  {https://link.aps.org/doi/10.1103/PhysRevB.90.144419} {\bibfield  {journal}
  {\bibinfo  {journal} {Phys. Rev. B}\ }\textbf {\bibinfo {volume} {90}},\
  \bibinfo {pages} {144419} (\bibinfo {year} {2014})}\BibitemShut {NoStop}%
\bibitem [{\citenamefont {Takei}\ and\ \citenamefont
  {Tserkovnyak}(2014)}]{TakeiPRL2014}%
  \BibitemOpen
  \bibfield  {author} {\bibinfo {author} {\bibfnamefont {S.}~\bibnamefont
  {Takei}}\ and\ \bibinfo {author} {\bibfnamefont {Y.}~\bibnamefont
  {Tserkovnyak}},\ }\href
  {https://link.aps.org/doi/10.1103/PhysRevLett.112.227201} {\bibfield
  {journal} {\bibinfo  {journal} {Phys. Rev. Lett.}\ }\textbf {\bibinfo
  {volume} {112}},\ \bibinfo {pages} {227201} (\bibinfo {year}
  {2014})}\BibitemShut {NoStop}%
\bibitem [{\citenamefont {Lachance-Quirion}\ \emph {et~al.}(2019)\citenamefont
  {Lachance-Quirion}, \citenamefont {Tabuchi}, \citenamefont {Gloppe},
  \citenamefont {Usami},\ and\ \citenamefont
  {Nakamura}}]{LachanceQuirionAPE2019}%
  \BibitemOpen
  \bibfield  {author} {\bibinfo {author} {\bibfnamefont {D.}~\bibnamefont
  {Lachance-Quirion}}, \bibinfo {author} {\bibfnamefont {Y.}~\bibnamefont
  {Tabuchi}}, \bibinfo {author} {\bibfnamefont {A.}~\bibnamefont {Gloppe}},
  \bibinfo {author} {\bibfnamefont {K.}~\bibnamefont {Usami}}, \ and\ \bibinfo
  {author} {\bibfnamefont {Y.}~\bibnamefont {Nakamura}},\ }\href
  {https://doi.org/10.7567\%2F1882-0786\%2Fab248d} {\bibfield  {journal}
  {\bibinfo  {journal} {Appl Phys Express}\ }\textbf {\bibinfo {volume} {12}},\
  \bibinfo {pages} {070101} (\bibinfo {year} {2019})}\BibitemShut {NoStop}%
\bibitem [{\citenamefont {Gieseler}\ \emph {et~al.}(2020)\citenamefont
  {Gieseler}, \citenamefont {Kabcenell}, \citenamefont {Rosenfeld},
  \citenamefont {Schaefer}, \citenamefont {Safira}, \citenamefont {Schuetz},
  \citenamefont {Gonzalez-Ballestero}, \citenamefont {Rusconi}, \citenamefont
  {Romero-Isart},\ and\ \citenamefont {Lukin}}]{GieselerPRL2020}%
  \BibitemOpen
  \bibfield  {author} {\bibinfo {author} {\bibfnamefont {J.}~\bibnamefont
  {Gieseler}}, \bibinfo {author} {\bibfnamefont {A.}~\bibnamefont {Kabcenell}},
  \bibinfo {author} {\bibfnamefont {E.}~\bibnamefont {Rosenfeld}}, \bibinfo
  {author} {\bibfnamefont {J.~D.}\ \bibnamefont {Schaefer}}, \bibinfo {author}
  {\bibfnamefont {A.}~\bibnamefont {Safira}}, \bibinfo {author} {\bibfnamefont
  {M.~J.~A.}\ \bibnamefont {Schuetz}}, \bibinfo {author} {\bibfnamefont
  {C.}~\bibnamefont {Gonzalez-Ballestero}}, \bibinfo {author} {\bibfnamefont
  {C.~C.}\ \bibnamefont {Rusconi}}, \bibinfo {author} {\bibfnamefont
  {O.}~\bibnamefont {Romero-Isart}}, \ and\ \bibinfo {author} {\bibfnamefont
  {M.~D.}\ \bibnamefont {Lukin}},\ }\href
  {https://link.aps.org/doi/10.1103/PhysRevLett.124.163604} {\bibfield
  {journal} {\bibinfo  {journal} {Phys. Rev. Lett.}\ }\textbf {\bibinfo
  {volume} {124}},\ \bibinfo {pages} {163604} (\bibinfo {year}
  {2020})}\BibitemShut {NoStop}%
\bibitem [{\citenamefont {Bertelli}\ \emph {et~al.}(2020)\citenamefont
  {Bertelli}, \citenamefont {Carmiggelt}, \citenamefont {Yu}, \citenamefont
  {Simon}, \citenamefont {Pothoven}, \citenamefont {Bauer}, \citenamefont
  {Blanter}, \citenamefont {Aarts},\ and\ \citenamefont {van~der
  Sar}}]{BertelliSciAdv2020}%
  \BibitemOpen
  \bibfield  {author} {\bibinfo {author} {\bibfnamefont {I.}~\bibnamefont
  {Bertelli}}, \bibinfo {author} {\bibfnamefont {J.~J.}\ \bibnamefont
  {Carmiggelt}}, \bibinfo {author} {\bibfnamefont {T.}~\bibnamefont {Yu}},
  \bibinfo {author} {\bibfnamefont {B.~G.}\ \bibnamefont {Simon}}, \bibinfo
  {author} {\bibfnamefont {C.~C.}\ \bibnamefont {Pothoven}}, \bibinfo {author}
  {\bibfnamefont {G.~E.}\ \bibnamefont {Bauer}}, \bibinfo {author}
  {\bibfnamefont {Y.~M.}\ \bibnamefont {Blanter}}, \bibinfo {author}
  {\bibfnamefont {J.}~\bibnamefont {Aarts}}, \ and\ \bibinfo {author}
  {\bibfnamefont {T.}~\bibnamefont {van~der Sar}},\ }\href
  {https://advances.sciencemag.org/content/6/46/eabd3556} {\bibfield  {journal}
  {\bibinfo  {journal} {Sci. Adv.}\ }\textbf {\bibinfo {volume} {6}},\ \bibinfo
  {pages} {eabd3556} (\bibinfo {year} {2020})}\BibitemShut {NoStop}%
\bibitem [{\citenamefont {Lee-Wong}\ \emph {et~al.}(2020)\citenamefont
  {Lee-Wong}, \citenamefont {Xue}, \citenamefont {Ye}, \citenamefont {Kreisel},
  \citenamefont {van~der Sar}, \citenamefont {Yacoby},\ and\ \citenamefont
  {Du}}]{LeeWongNanoLetters2020}%
  \BibitemOpen
  \bibfield  {author} {\bibinfo {author} {\bibfnamefont {E.}~\bibnamefont
  {Lee-Wong}}, \bibinfo {author} {\bibfnamefont {R.}~\bibnamefont {Xue}},
  \bibinfo {author} {\bibfnamefont {F.}~\bibnamefont {Ye}}, \bibinfo {author}
  {\bibfnamefont {A.}~\bibnamefont {Kreisel}}, \bibinfo {author} {\bibfnamefont
  {T.}~\bibnamefont {van~der Sar}}, \bibinfo {author} {\bibfnamefont
  {A.}~\bibnamefont {Yacoby}}, \ and\ \bibinfo {author} {\bibfnamefont {C.~R.}\
  \bibnamefont {Du}},\ }\href {https://doi.org/10.1021/acs.nanolett.0c00085}
  {\bibfield  {journal} {\bibinfo  {journal} {Nano Lett}\ }\textbf {\bibinfo
  {volume} {20}},\ \bibinfo {pages} {3284} (\bibinfo {year}
  {2020})}\BibitemShut {NoStop}%
\bibitem [{\citenamefont {Wolfe}\ \emph {et~al.}(2014)\citenamefont {Wolfe},
  \citenamefont {Bhallamudi}, \citenamefont {Wang}, \citenamefont {Du},
  \citenamefont {Manuilov}, \citenamefont {Teeling-Smith}, \citenamefont
  {Berger}, \citenamefont {Adur}, \citenamefont {Yang},\ and\ \citenamefont
  {Hammel}}]{WolfePRB2014}%
  \BibitemOpen
  \bibfield  {author} {\bibinfo {author} {\bibfnamefont {C.~S.}\ \bibnamefont
  {Wolfe}}, \bibinfo {author} {\bibfnamefont {V.~P.}\ \bibnamefont
  {Bhallamudi}}, \bibinfo {author} {\bibfnamefont {H.~L.}\ \bibnamefont
  {Wang}}, \bibinfo {author} {\bibfnamefont {C.~H.}\ \bibnamefont {Du}},
  \bibinfo {author} {\bibfnamefont {S.}~\bibnamefont {Manuilov}}, \bibinfo
  {author} {\bibfnamefont {R.~M.}\ \bibnamefont {Teeling-Smith}}, \bibinfo
  {author} {\bibfnamefont {A.~J.}\ \bibnamefont {Berger}}, \bibinfo {author}
  {\bibfnamefont {R.}~\bibnamefont {Adur}}, \bibinfo {author} {\bibfnamefont
  {F.~Y.}\ \bibnamefont {Yang}}, \ and\ \bibinfo {author} {\bibfnamefont
  {P.~C.}\ \bibnamefont {Hammel}},\ }\href
  {http://link.aps.org/doi/10.1103/PhysRevB.89.180406} {\bibfield  {journal}
  {\bibinfo  {journal} {Phys. Rev. B}\ }\textbf {\bibinfo {volume} {89}},\
  \bibinfo {pages} {180406} (\bibinfo {year} {2014})}\BibitemShut {NoStop}%
\bibitem [{\citenamefont {Andrich}\ \emph {et~al.}(2017)\citenamefont
  {Andrich}, \citenamefont {de~las Casas}, \citenamefont {Liu}, \citenamefont
  {Bretscher}, \citenamefont {Berman}, \citenamefont {Heremans}, \citenamefont
  {Nealey},\ and\ \citenamefont {Awschalom}}]{AndrichNPJ2017}%
  \BibitemOpen
  \bibfield  {author} {\bibinfo {author} {\bibfnamefont {P.}~\bibnamefont
  {Andrich}}, \bibinfo {author} {\bibfnamefont {C.~F.}\ \bibnamefont {de~las
  Casas}}, \bibinfo {author} {\bibfnamefont {X.}~\bibnamefont {Liu}}, \bibinfo
  {author} {\bibfnamefont {H.~L.}\ \bibnamefont {Bretscher}}, \bibinfo {author}
  {\bibfnamefont {J.~R.}\ \bibnamefont {Berman}}, \bibinfo {author}
  {\bibfnamefont {F.~J.}\ \bibnamefont {Heremans}}, \bibinfo {author}
  {\bibfnamefont {P.~F.}\ \bibnamefont {Nealey}}, \ and\ \bibinfo {author}
  {\bibfnamefont {D.~D.}\ \bibnamefont {Awschalom}},\ }\href
  {https://doi.org/10.1038/s41534-017-0029-z} {\bibfield  {journal} {\bibinfo
  {journal} {npj Quantum Inf}\ }\textbf {\bibinfo {volume} {3}},\ \bibinfo
  {pages} {28} (\bibinfo {year} {2017})}\BibitemShut {NoStop}%
\bibitem [{\citenamefont {Page}\ \emph {et~al.}(2019)\citenamefont {Page},
  \citenamefont {McCullian}, \citenamefont {Purser}, \citenamefont {Schulze},
  \citenamefont {Nakatani}, \citenamefont {Wolfe}, \citenamefont {Childress},
  \citenamefont {McConney}, \citenamefont {Howe}, \citenamefont {Hammel},\ and\
  \citenamefont {Bhallamudi}}]{PageJAP2019}%
  \BibitemOpen
  \bibfield  {author} {\bibinfo {author} {\bibfnamefont {M.~R.}\ \bibnamefont
  {Page}}, \bibinfo {author} {\bibfnamefont {B.~A.}\ \bibnamefont {McCullian}},
  \bibinfo {author} {\bibfnamefont {C.~M.}\ \bibnamefont {Purser}}, \bibinfo
  {author} {\bibfnamefont {J.~G.}\ \bibnamefont {Schulze}}, \bibinfo {author}
  {\bibfnamefont {T.~M.}\ \bibnamefont {Nakatani}}, \bibinfo {author}
  {\bibfnamefont {C.~S.}\ \bibnamefont {Wolfe}}, \bibinfo {author}
  {\bibfnamefont {J.~R.}\ \bibnamefont {Childress}}, \bibinfo {author}
  {\bibfnamefont {M.~E.}\ \bibnamefont {McConney}}, \bibinfo {author}
  {\bibfnamefont {B.~M.}\ \bibnamefont {Howe}}, \bibinfo {author}
  {\bibfnamefont {P.~C.}\ \bibnamefont {Hammel}}, \ and\ \bibinfo {author}
  {\bibfnamefont {V.~P.}\ \bibnamefont {Bhallamudi}},\ }\href
  {https://doi.org/10.1063/1.5083991} {\bibfield  {journal} {\bibinfo
  {journal} {J appl Phys}\ }\textbf {\bibinfo {volume} {126}},\ \bibinfo
  {pages} {124902} (\bibinfo {year} {2019})}\BibitemShut {NoStop}%
\bibitem [{\citenamefont {Kikuchi}\ \emph {et~al.}(2017)\citenamefont
  {Kikuchi}, \citenamefont {Prananto}, \citenamefont {Hayashi}, \citenamefont
  {Laraoui}, \citenamefont {Mizuochi}, \citenamefont {Hatano}, \citenamefont
  {Saitoh}, \citenamefont {Kim}, \citenamefont {Meriles},\ and\ \citenamefont
  {An}}]{KikuchiAPExpress2017}%
  \BibitemOpen
  \bibfield  {author} {\bibinfo {author} {\bibfnamefont {D.}~\bibnamefont
  {Kikuchi}}, \bibinfo {author} {\bibfnamefont {D.}~\bibnamefont {Prananto}},
  \bibinfo {author} {\bibfnamefont {K.}~\bibnamefont {Hayashi}}, \bibinfo
  {author} {\bibfnamefont {A.}~\bibnamefont {Laraoui}}, \bibinfo {author}
  {\bibfnamefont {N.}~\bibnamefont {Mizuochi}}, \bibinfo {author}
  {\bibfnamefont {M.}~\bibnamefont {Hatano}}, \bibinfo {author} {\bibfnamefont
  {E.}~\bibnamefont {Saitoh}}, \bibinfo {author} {\bibfnamefont
  {Y.}~\bibnamefont {Kim}}, \bibinfo {author} {\bibfnamefont {C.~A.}\
  \bibnamefont {Meriles}}, \ and\ \bibinfo {author} {\bibfnamefont
  {T.}~\bibnamefont {An}},\ }\href {https://doi.org/10.7567\%2Fapex.10.103004}
  {\bibfield  {journal} {\bibinfo  {journal} {Appl Phys Express}\ }\textbf
  {\bibinfo {volume} {10}},\ \bibinfo {pages} {103004} (\bibinfo {year}
  {2017})}\BibitemShut {NoStop}%
\bibitem [{\citenamefont {Zhang}\ \emph {et~al.}(2020)\citenamefont {Zhang},
  \citenamefont {Ku}, \citenamefont {Casola}, \citenamefont {Du}, \citenamefont
  {van~der Sar}, \citenamefont {Onbasli}, \citenamefont {Ross}, \citenamefont
  {Tserkovnyak}, \citenamefont {Yacoby},\ and\ \citenamefont
  {Walsworth}}]{ZhangPRB2020}%
  \BibitemOpen
  \bibfield  {author} {\bibinfo {author} {\bibfnamefont {H.}~\bibnamefont
  {Zhang}}, \bibinfo {author} {\bibfnamefont {M.~J.~H.}\ \bibnamefont {Ku}},
  \bibinfo {author} {\bibfnamefont {F.}~\bibnamefont {Casola}}, \bibinfo
  {author} {\bibfnamefont {C.~H.~R.}\ \bibnamefont {Du}}, \bibinfo {author}
  {\bibfnamefont {T.}~\bibnamefont {van~der Sar}}, \bibinfo {author}
  {\bibfnamefont {M.~C.}\ \bibnamefont {Onbasli}}, \bibinfo {author}
  {\bibfnamefont {C.~A.}\ \bibnamefont {Ross}}, \bibinfo {author}
  {\bibfnamefont {Y.}~\bibnamefont {Tserkovnyak}}, \bibinfo {author}
  {\bibfnamefont {A.}~\bibnamefont {Yacoby}}, \ and\ \bibinfo {author}
  {\bibfnamefont {R.~L.}\ \bibnamefont {Walsworth}},\ }\href
  {https://link.aps.org/doi/10.1103/PhysRevB.102.024404} {\bibfield  {journal}
  {\bibinfo  {journal} {Phys. Rev. B}\ }\textbf {\bibinfo {volume} {102}},\
  \bibinfo {pages} {024404} (\bibinfo {year} {2020})}\BibitemShut {NoStop}%
\bibitem [{\citenamefont {Rodriguez}\ \emph {et~al.}(2013)\citenamefont
  {Rodriguez}, \citenamefont {Feist}, \citenamefont {Verschuuren},
  \citenamefont {Garcia~Vidal},\ and\ \citenamefont
  {G\'omez~Rivas}}]{RodriguezPRL2013}%
  \BibitemOpen
  \bibfield  {author} {\bibinfo {author} {\bibfnamefont {S.~R.~K.}\
  \bibnamefont {Rodriguez}}, \bibinfo {author} {\bibfnamefont {J.}~\bibnamefont
  {Feist}}, \bibinfo {author} {\bibfnamefont {M.~A.}\ \bibnamefont
  {Verschuuren}}, \bibinfo {author} {\bibfnamefont {F.~J.}\ \bibnamefont
  {Garcia~Vidal}}, \ and\ \bibinfo {author} {\bibfnamefont {J.}~\bibnamefont
  {G\'omez~Rivas}},\ }\href
  {https://link.aps.org/doi/10.1103/PhysRevLett.111.166802} {\bibfield
  {journal} {\bibinfo  {journal} {Phys. Rev. Lett.}\ }\textbf {\bibinfo
  {volume} {111}},\ \bibinfo {pages} {166802} (\bibinfo {year}
  {2013})}\BibitemShut {NoStop}%
\bibitem [{\citenamefont {Gehring}\ \emph {et~al.}(2006)\citenamefont
  {Gehring}, \citenamefont {Schweinsberg}, \citenamefont {Barsi}, \citenamefont
  {Kostinski},\ and\ \citenamefont {Boyd}}]{GehringScience2006}%
  \BibitemOpen
  \bibfield  {author} {\bibinfo {author} {\bibfnamefont {G.~M.}\ \bibnamefont
  {Gehring}}, \bibinfo {author} {\bibfnamefont {A.}~\bibnamefont
  {Schweinsberg}}, \bibinfo {author} {\bibfnamefont {C.}~\bibnamefont {Barsi}},
  \bibinfo {author} {\bibfnamefont {N.}~\bibnamefont {Kostinski}}, \ and\
  \bibinfo {author} {\bibfnamefont {R.~W.}\ \bibnamefont {Boyd}},\ }\href
  {\doibase 10.1126/science.1124524} {\bibfield  {journal} {\bibinfo  {journal}
  {Science}\ }\textbf {\bibinfo {volume} {312}},\ \bibinfo {pages} {895}
  (\bibinfo {year} {2006})}\BibitemShut {NoStop}%
\bibitem [{\citenamefont {Kang}\ \emph {et~al.}(2004)\citenamefont {Kang},
  \citenamefont {Hernandez},\ and\ \citenamefont {Zhu}}]{KangPRA2004}%
  \BibitemOpen
  \bibfield  {author} {\bibinfo {author} {\bibfnamefont {H.}~\bibnamefont
  {Kang}}, \bibinfo {author} {\bibfnamefont {G.}~\bibnamefont {Hernandez}}, \
  and\ \bibinfo {author} {\bibfnamefont {Y.}~\bibnamefont {Zhu}},\ }\href
  {https://link.aps.org/doi/10.1103/PhysRevA.70.011801} {\bibfield  {journal}
  {\bibinfo  {journal} {Phys. Rev. A}\ }\textbf {\bibinfo {volume} {70}},\
  \bibinfo {pages} {011801} (\bibinfo {year} {2004})}\BibitemShut {NoStop}%
\bibitem [{\citenamefont {Peyronel}\ \emph {et~al.}(2012)\citenamefont
  {Peyronel}, \citenamefont {Firstenberg}, \citenamefont {Liang}, \citenamefont
  {Hofferberth}, \citenamefont {Gorshkov}, \citenamefont {Pohl}, \citenamefont
  {Lukin},\ and\ \citenamefont {Vuleti{\'c}}}]{PayronelNature2012}%
  \BibitemOpen
  \bibfield  {author} {\bibinfo {author} {\bibfnamefont {T.}~\bibnamefont
  {Peyronel}}, \bibinfo {author} {\bibfnamefont {O.}~\bibnamefont
  {Firstenberg}}, \bibinfo {author} {\bibfnamefont {Q.-Y.}\ \bibnamefont
  {Liang}}, \bibinfo {author} {\bibfnamefont {S.}~\bibnamefont {Hofferberth}},
  \bibinfo {author} {\bibfnamefont {A.~V.}\ \bibnamefont {Gorshkov}}, \bibinfo
  {author} {\bibfnamefont {T.}~\bibnamefont {Pohl}}, \bibinfo {author}
  {\bibfnamefont {M.~D.}\ \bibnamefont {Lukin}}, \ and\ \bibinfo {author}
  {\bibfnamefont {V.}~\bibnamefont {Vuleti{\'c}}},\ }\href
  {https://doi.org/10.1038/nature11361} {\bibfield  {journal} {\bibinfo
  {journal} {Nature}\ }\textbf {\bibinfo {volume} {488}},\ \bibinfo {pages}
  {57} (\bibinfo {year} {2012})}\BibitemShut {NoStop}%
\bibitem [{\citenamefont {Murray}\ and\ \citenamefont
  {Pohl}(2016)}]{MurrayInColl2016}%
  \BibitemOpen
  \bibfield  {author} {\bibinfo {author} {\bibfnamefont {C.}~\bibnamefont
  {Murray}}\ and\ \bibinfo {author} {\bibfnamefont {T.}~\bibnamefont {Pohl}},\
  }in\ \href
  {http://www.sciencedirect.com/science/article/pii/S1049250X1630009X} {\emph
  {\bibinfo {booktitle} {Advances In Atomic, Molecular, and Optical
  Physics}}},\ Vol.~\bibinfo {volume} {65},\ \bibinfo {editor} {edited by\
  \bibinfo {editor} {\bibfnamefont {E.}~\bibnamefont {Arimondo}}, \bibinfo
  {editor} {\bibfnamefont {C.~C.}\ \bibnamefont {Lin}}, \ and\ \bibinfo
  {editor} {\bibfnamefont {S.~F.}\ \bibnamefont {Yelin}}}\ (\bibinfo
  {publisher} {Academic Press},\ \bibinfo {year} {2016})\ pp.\ \bibinfo {pages}
  {321 -- 372}\BibitemShut {NoStop}%
\bibitem [{\citenamefont {Cao}\ \emph {et~al.}(2018)\citenamefont {Cao},
  \citenamefont {Lin}, \citenamefont {Sun}, \citenamefont {Liang},\ and\
  \citenamefont {Song}}]{CaoNanophotonics2018}%
  \BibitemOpen
  \bibfield  {author} {\bibinfo {author} {\bibfnamefont {E.}~\bibnamefont
  {Cao}}, \bibinfo {author} {\bibfnamefont {W.}~\bibnamefont {Lin}}, \bibinfo
  {author} {\bibfnamefont {M.}~\bibnamefont {Sun}}, \bibinfo {author}
  {\bibfnamefont {W.}~\bibnamefont {Liang}}, \ and\ \bibinfo {author}
  {\bibfnamefont {Y.}~\bibnamefont {Song}},\ }\href
  {https://www.degruyter.com/view/journals/nanoph/7/1/article-p145.xml}
  {\bibfield  {journal} {\bibinfo  {journal} {Nanophotonics}\ }\textbf
  {\bibinfo {volume} {7}},\ \bibinfo {pages} {145 } (\bibinfo {year}
  {2018})}\BibitemShut {NoStop}%
\bibitem [{\citenamefont {Casola}\ \emph {et~al.}(2018)\citenamefont {Casola},
  \citenamefont {van~der Sar},\ and\ \citenamefont
  {Yacoby}}]{CasolaNatReviews2018}%
  \BibitemOpen
  \bibfield  {author} {\bibinfo {author} {\bibfnamefont {F.}~\bibnamefont
  {Casola}}, \bibinfo {author} {\bibfnamefont {T.}~\bibnamefont {van~der Sar}},
  \ and\ \bibinfo {author} {\bibfnamefont {A.}~\bibnamefont {Yacoby}},\ }\href
  {https://doi.org/10.1038/natrevmats.2017.88} {\bibfield  {journal} {\bibinfo
  {journal} {Nat Rev Materials}\ }\textbf {\bibinfo {volume} {3}},\ \bibinfo
  {pages} {17088} (\bibinfo {year} {2018})}\BibitemShut {NoStop}%
\bibitem [{\citenamefont {Au}\ \emph {et~al.}(2012)\citenamefont {Au},
  \citenamefont {Ahmad}, \citenamefont {Dmytriiev}, \citenamefont {Dvornik},
  \citenamefont {Davison},\ and\ \citenamefont {Kruglyak}}]{AuAPL2012}%
  \BibitemOpen
  \bibfield  {author} {\bibinfo {author} {\bibfnamefont {Y.}~\bibnamefont
  {Au}}, \bibinfo {author} {\bibfnamefont {E.}~\bibnamefont {Ahmad}}, \bibinfo
  {author} {\bibfnamefont {O.}~\bibnamefont {Dmytriiev}}, \bibinfo {author}
  {\bibfnamefont {M.}~\bibnamefont {Dvornik}}, \bibinfo {author} {\bibfnamefont
  {T.}~\bibnamefont {Davison}}, \ and\ \bibinfo {author} {\bibfnamefont
  {V.~V.}\ \bibnamefont {Kruglyak}},\ }\href
  {https://doi.org/10.1063/1.4711039} {\bibfield  {journal} {\bibinfo
  {journal} {Appl Phys Lett}\ }\textbf {\bibinfo {volume} {100}},\ \bibinfo
  {pages} {182404} (\bibinfo {year} {2012})}\BibitemShut {NoStop}%
\bibitem [{\citenamefont {Khitun}\ and\ \citenamefont
  {Wang}(2011)}]{KhitunJAP2011}%
  \BibitemOpen
  \bibfield  {author} {\bibinfo {author} {\bibfnamefont {A.}~\bibnamefont
  {Khitun}}\ and\ \bibinfo {author} {\bibfnamefont {K.~L.}\ \bibnamefont
  {Wang}},\ }\href {https://doi.org/10.1063/1.3609062} {\bibfield  {journal}
  {\bibinfo  {journal} {J appl Phys}\ }\textbf {\bibinfo {volume} {110}},\
  \bibinfo {pages} {034306} (\bibinfo {year} {2011})}\BibitemShut {NoStop}%
\bibitem [{\citenamefont {Vogt}\ \emph {et~al.}(2014)\citenamefont {Vogt},
  \citenamefont {Fradin}, \citenamefont {Pearson}, \citenamefont {Sebastian},
  \citenamefont {Bader}, \citenamefont {Hillebrands}, \citenamefont
  {Hoffmann},\ and\ \citenamefont {Schultheiss}}]{VogtNatComm2014}%
  \BibitemOpen
  \bibfield  {author} {\bibinfo {author} {\bibfnamefont {K.}~\bibnamefont
  {Vogt}}, \bibinfo {author} {\bibfnamefont {F.~Y.}\ \bibnamefont {Fradin}},
  \bibinfo {author} {\bibfnamefont {J.~E.}\ \bibnamefont {Pearson}}, \bibinfo
  {author} {\bibfnamefont {T.}~\bibnamefont {Sebastian}}, \bibinfo {author}
  {\bibfnamefont {S.~D.}\ \bibnamefont {Bader}}, \bibinfo {author}
  {\bibfnamefont {B.}~\bibnamefont {Hillebrands}}, \bibinfo {author}
  {\bibfnamefont {A.}~\bibnamefont {Hoffmann}}, \ and\ \bibinfo {author}
  {\bibfnamefont {H.}~\bibnamefont {Schultheiss}},\ }\href
  {https://doi.org/10.1038/ncomms4727} {\bibfield  {journal} {\bibinfo
  {journal} {Nat Commun}\ }\textbf {\bibinfo {volume} {5}},\ \bibinfo {pages}
  {3727} (\bibinfo {year} {2014})}\BibitemShut {NoStop}%
\bibitem [{\citenamefont {Doherty}\ \emph {et~al.}(2013)\citenamefont
  {Doherty}, \citenamefont {Manson}, \citenamefont {Delaney}, \citenamefont
  {Jelezko}, \citenamefont {Wrachtrup},\ and\ \citenamefont
  {Hollenberg}}]{DohertyPhysRep2013}%
  \BibitemOpen
  \bibfield  {author} {\bibinfo {author} {\bibfnamefont {M.~W.}\ \bibnamefont
  {Doherty}}, \bibinfo {author} {\bibfnamefont {N.~B.}\ \bibnamefont {Manson}},
  \bibinfo {author} {\bibfnamefont {P.}~\bibnamefont {Delaney}}, \bibinfo
  {author} {\bibfnamefont {F.}~\bibnamefont {Jelezko}}, \bibinfo {author}
  {\bibfnamefont {J.}~\bibnamefont {Wrachtrup}}, \ and\ \bibinfo {author}
  {\bibfnamefont {L.~C.}\ \bibnamefont {Hollenberg}},\ }\href
  {http://www.sciencedirect.com/science/article/pii/S0370157313000562}
  {\bibfield  {journal} {\bibinfo  {journal} {Phys Rep}\ }\textbf {\bibinfo
  {volume} {528}},\ \bibinfo {pages} {1 } (\bibinfo {year} {2013})}\BibitemShut
  {NoStop}%
\bibitem [{\citenamefont {Aharonovich}\ \emph {et~al.}(2011)\citenamefont
  {Aharonovich}, \citenamefont {Greentree},\ and\ \citenamefont
  {Prawer}}]{AharonovichNatPhot2011}%
  \BibitemOpen
  \bibfield  {author} {\bibinfo {author} {\bibfnamefont {I.}~\bibnamefont
  {Aharonovich}}, \bibinfo {author} {\bibfnamefont {A.~D.}\ \bibnamefont
  {Greentree}}, \ and\ \bibinfo {author} {\bibfnamefont {S.}~\bibnamefont
  {Prawer}},\ }\href {https://doi.org/10.1038/nphoton.2011.54} {\bibfield
  {journal} {\bibinfo  {journal} {Nat Photonics}\ }\textbf {\bibinfo {volume}
  {5}},\ \bibinfo {pages} {397} (\bibinfo {year} {2011})}\BibitemShut {NoStop}%
\bibitem [{\citenamefont {Atat{\"u}re}\ \emph {et~al.}(2018)\citenamefont
  {Atat{\"u}re}, \citenamefont {Englund}, \citenamefont {Vamivakas},
  \citenamefont {Lee},\ and\ \citenamefont {Wrachtrup}}]{AtatureNatRevMat2018}%
  \BibitemOpen
  \bibfield  {author} {\bibinfo {author} {\bibfnamefont {M.}~\bibnamefont
  {Atat{\"u}re}}, \bibinfo {author} {\bibfnamefont {D.}~\bibnamefont
  {Englund}}, \bibinfo {author} {\bibfnamefont {N.}~\bibnamefont {Vamivakas}},
  \bibinfo {author} {\bibfnamefont {S.-Y.}\ \bibnamefont {Lee}}, \ and\
  \bibinfo {author} {\bibfnamefont {J.}~\bibnamefont {Wrachtrup}},\ }\href
  {https://doi.org/10.1038/s41578-018-0008-9} {\bibfield  {journal} {\bibinfo
  {journal} {Nat Rev Materials}\ }\textbf {\bibinfo {volume} {3}},\ \bibinfo
  {pages} {38} (\bibinfo {year} {2018})}\BibitemShut {NoStop}%
\bibitem [{\citenamefont {Albrecht}\ \emph {et~al.}(2013)\citenamefont
  {Albrecht}, \citenamefont {Bommer}, \citenamefont {Deutsch}, \citenamefont
  {Reichel},\ and\ \citenamefont {Becher}}]{AlbrechtPRL2013}%
  \BibitemOpen
  \bibfield  {author} {\bibinfo {author} {\bibfnamefont {R.}~\bibnamefont
  {Albrecht}}, \bibinfo {author} {\bibfnamefont {A.}~\bibnamefont {Bommer}},
  \bibinfo {author} {\bibfnamefont {C.}~\bibnamefont {Deutsch}}, \bibinfo
  {author} {\bibfnamefont {J.}~\bibnamefont {Reichel}}, \ and\ \bibinfo
  {author} {\bibfnamefont {C.}~\bibnamefont {Becher}},\ }\href
  {https://link.aps.org/doi/10.1103/PhysRevLett.110.243602} {\bibfield
  {journal} {\bibinfo  {journal} {Phys. Rev. Lett.}\ }\textbf {\bibinfo
  {volume} {110}},\ \bibinfo {pages} {243602} (\bibinfo {year}
  {2013})}\BibitemShut {NoStop}%
\bibitem [{\citenamefont {Maze}\ \emph {et~al.}(2008)\citenamefont {Maze},
  \citenamefont {Stanwix}, \citenamefont {Hodges}, \citenamefont {Hong},
  \citenamefont {Taylor}, \citenamefont {Cappellaro}, \citenamefont {Jiang},
  \citenamefont {Dutt}, \citenamefont {Togan}, \citenamefont {Zibrov},
  \citenamefont {Yacoby}, \citenamefont {Walsworth},\ and\ \citenamefont
  {Lukin}}]{MazeNature2008}%
  \BibitemOpen
  \bibfield  {author} {\bibinfo {author} {\bibfnamefont {J.~R.}\ \bibnamefont
  {Maze}}, \bibinfo {author} {\bibfnamefont {P.~L.}\ \bibnamefont {Stanwix}},
  \bibinfo {author} {\bibfnamefont {J.~S.}\ \bibnamefont {Hodges}}, \bibinfo
  {author} {\bibfnamefont {S.}~\bibnamefont {Hong}}, \bibinfo {author}
  {\bibfnamefont {J.~M.}\ \bibnamefont {Taylor}}, \bibinfo {author}
  {\bibfnamefont {P.}~\bibnamefont {Cappellaro}}, \bibinfo {author}
  {\bibfnamefont {L.}~\bibnamefont {Jiang}}, \bibinfo {author} {\bibfnamefont
  {M.~V.~G.}\ \bibnamefont {Dutt}}, \bibinfo {author} {\bibfnamefont
  {E.}~\bibnamefont {Togan}}, \bibinfo {author} {\bibfnamefont {A.~S.}\
  \bibnamefont {Zibrov}}, \bibinfo {author} {\bibfnamefont {A.}~\bibnamefont
  {Yacoby}}, \bibinfo {author} {\bibfnamefont {R.~L.}\ \bibnamefont
  {Walsworth}}, \ and\ \bibinfo {author} {\bibfnamefont {M.~D.}\ \bibnamefont
  {Lukin}},\ }\href {https://doi.org/10.1038/nature07279} {\bibfield  {journal}
  {\bibinfo  {journal} {Nature}\ }\textbf {\bibinfo {volume} {455}},\ \bibinfo
  {pages} {644} (\bibinfo {year} {2008})}\BibitemShut {NoStop}%
\bibitem [{\citenamefont {Dolde}\ \emph {et~al.}(2011)\citenamefont {Dolde},
  \citenamefont {Fedder}, \citenamefont {Doherty}, \citenamefont {N{\"o}bauer},
  \citenamefont {Rempp}, \citenamefont {Balasubramanian}, \citenamefont {Wolf},
  \citenamefont {Reinhard}, \citenamefont {Hollenberg}, \citenamefont
  {Jelezko},\ and\ \citenamefont {Wrachtrup}}]{DoldeNatPhys2011}%
  \BibitemOpen
  \bibfield  {author} {\bibinfo {author} {\bibfnamefont {F.}~\bibnamefont
  {Dolde}}, \bibinfo {author} {\bibfnamefont {H.}~\bibnamefont {Fedder}},
  \bibinfo {author} {\bibfnamefont {M.~W.}\ \bibnamefont {Doherty}}, \bibinfo
  {author} {\bibfnamefont {T.}~\bibnamefont {N{\"o}bauer}}, \bibinfo {author}
  {\bibfnamefont {F.}~\bibnamefont {Rempp}}, \bibinfo {author} {\bibfnamefont
  {G.}~\bibnamefont {Balasubramanian}}, \bibinfo {author} {\bibfnamefont
  {T.}~\bibnamefont {Wolf}}, \bibinfo {author} {\bibfnamefont {F.}~\bibnamefont
  {Reinhard}}, \bibinfo {author} {\bibfnamefont {L.~C.~L.}\ \bibnamefont
  {Hollenberg}}, \bibinfo {author} {\bibfnamefont {F.}~\bibnamefont {Jelezko}},
  \ and\ \bibinfo {author} {\bibfnamefont {J.}~\bibnamefont {Wrachtrup}},\
  }\href {https://doi.org/10.1038/nphys1969} {\bibfield  {journal} {\bibinfo
  {journal} {Nat Physics}\ }\textbf {\bibinfo {volume} {7}},\ \bibinfo {pages}
  {459} (\bibinfo {year} {2011})}\BibitemShut {NoStop}%
\bibitem [{\citenamefont {Laraoui}\ \emph {et~al.}(2015)\citenamefont
  {Laraoui}, \citenamefont {Aycock-Rizzo}, \citenamefont {Gao}, \citenamefont
  {Lu}, \citenamefont {Riedo},\ and\ \citenamefont
  {Meriles}}]{LaraouiNatComm2015}%
  \BibitemOpen
  \bibfield  {author} {\bibinfo {author} {\bibfnamefont {A.}~\bibnamefont
  {Laraoui}}, \bibinfo {author} {\bibfnamefont {H.}~\bibnamefont
  {Aycock-Rizzo}}, \bibinfo {author} {\bibfnamefont {Y.}~\bibnamefont {Gao}},
  \bibinfo {author} {\bibfnamefont {X.}~\bibnamefont {Lu}}, \bibinfo {author}
  {\bibfnamefont {E.}~\bibnamefont {Riedo}}, \ and\ \bibinfo {author}
  {\bibfnamefont {C.~A.}\ \bibnamefont {Meriles}},\ }\href
  {https://doi.org/10.1038/ncomms9954} {\bibfield  {journal} {\bibinfo
  {journal} {Nat Commun}\ }\textbf {\bibinfo {volume} {6}},\ \bibinfo {pages}
  {8954} (\bibinfo {year} {2015})}\BibitemShut {NoStop}%
\bibitem [{\citenamefont {Grazioso}\ \emph {et~al.}(2013)\citenamefont
  {Grazioso}, \citenamefont {Patton}, \citenamefont {Delaney}, \citenamefont
  {Markham}, \citenamefont {Twitchen},\ and\ \citenamefont
  {Smith}}]{GraziosoAPL2013}%
  \BibitemOpen
  \bibfield  {author} {\bibinfo {author} {\bibfnamefont {F.}~\bibnamefont
  {Grazioso}}, \bibinfo {author} {\bibfnamefont {B.~R.}\ \bibnamefont
  {Patton}}, \bibinfo {author} {\bibfnamefont {P.}~\bibnamefont {Delaney}},
  \bibinfo {author} {\bibfnamefont {M.~L.}\ \bibnamefont {Markham}}, \bibinfo
  {author} {\bibfnamefont {D.~J.}\ \bibnamefont {Twitchen}}, \ and\ \bibinfo
  {author} {\bibfnamefont {J.~M.}\ \bibnamefont {Smith}},\ }\href
  {https://doi.org/10.1063/1.4819834} {\bibfield  {journal} {\bibinfo
  {journal} {Appl Phys Lett}\ }\textbf {\bibinfo {volume} {103}},\ \bibinfo
  {pages} {101905} (\bibinfo {year} {2013})}\BibitemShut {NoStop}%
\bibitem [{\citenamefont {Shao}\ \emph {et~al.}(2016)\citenamefont {Shao},
  \citenamefont {Zhang}, \citenamefont {Markham}, \citenamefont {Edmonds},\
  and\ \citenamefont {Lon\ifmmode~\check{c}\else
  \v{c}\fi{}ar}}]{ShaoPRApp2016}%
  \BibitemOpen
  \bibfield  {author} {\bibinfo {author} {\bibfnamefont {L.}~\bibnamefont
  {Shao}}, \bibinfo {author} {\bibfnamefont {M.}~\bibnamefont {Zhang}},
  \bibinfo {author} {\bibfnamefont {M.}~\bibnamefont {Markham}}, \bibinfo
  {author} {\bibfnamefont {A.~M.}\ \bibnamefont {Edmonds}}, \ and\ \bibinfo
  {author} {\bibfnamefont {M.}~\bibnamefont {Lon\ifmmode~\check{c}\else
  \v{c}\fi{}ar}},\ }\href
  {https://link.aps.org/doi/10.1103/PhysRevApplied.6.064008} {\bibfield
  {journal} {\bibinfo  {journal} {Phys. Rev. Appl.}\ }\textbf {\bibinfo
  {volume} {6}},\ \bibinfo {pages} {064008} (\bibinfo {year}
  {2016})}\BibitemShut {NoStop}%
\bibitem [{\citenamefont {Aharoni}(2000)}]{aharoni2000introduction}%
  \BibitemOpen
  \bibfield  {author} {\bibinfo {author} {\bibfnamefont {A.}~\bibnamefont
  {Aharoni}},\ }\href@noop {} {\emph {\bibinfo {title} {Introduction to the
  Theory of Ferromagnetism}}},\ International Series of Monogr\ (\bibinfo
  {publisher} {Clarendon Press},\ \bibinfo {year} {2000})\BibitemShut {NoStop}%
\bibitem [{\citenamefont {Kalinikos}(1981)}]{KalinikosSoviet1981}%
  \BibitemOpen
  \bibfield  {author} {\bibinfo {author} {\bibfnamefont {B.~A.}\ \bibnamefont
  {Kalinikos}},\ }\href {https://doi.org/10.1007/BF00941342} {\bibfield
  {journal} {\bibinfo  {journal} {Soviet Physics Journal}\ }\textbf {\bibinfo
  {volume} {24}},\ \bibinfo {pages} {718} (\bibinfo {year} {1981})}\BibitemShut
  {NoStop}%
\bibitem [{\citenamefont {Kalinikos}\ and\ \citenamefont
  {Slavin}(1986)}]{KalinikosJPHYSC1986}%
  \BibitemOpen
  \bibfield  {author} {\bibinfo {author} {\bibfnamefont {B.~A.}\ \bibnamefont
  {Kalinikos}}\ and\ \bibinfo {author} {\bibfnamefont {A.~N.}\ \bibnamefont
  {Slavin}},\ }\href {https://doi.org/10.1088\%2F0022-3719\%2F19\%2F35\%2F014}
  {\bibfield  {journal} {\bibinfo  {journal} {J Phys C Solid State}\ }\textbf
  {\bibinfo {volume} {19}},\ \bibinfo {pages} {7013} (\bibinfo {year}
  {1986})}\BibitemShut {NoStop}%
\bibitem [{\citenamefont {Kalinikos}(1994)}]{Kalinikos1994}%
  \BibitemOpen
  \bibfield  {author} {\bibinfo {author} {\bibfnamefont {B.~A.}\ \bibnamefont
  {Kalinikos}},\ }\enquote {\bibinfo {title} {Dipole-exchange spin-wave
  spectrum of magnetic films},}\ in\ \href
  {https://www.worldscientific.com/doi/abs/10.1142/9789814343121_0002} {\emph
  {\bibinfo {booktitle} {Linear and Nonlinear Spin Waves in Magnetic Films and
  Superlattices}}}\ (\bibinfo {year} {1994})\ pp.\ \bibinfo {pages}
  {89--156}\BibitemShut {NoStop}%
\bibitem [{\citenamefont {Patton}(1984)}]{PattonPHYSREP1984}%
  \BibitemOpen
  \bibfield  {author} {\bibinfo {author} {\bibfnamefont {C.~E.}\ \bibnamefont
  {Patton}},\ }\href
  {http://www.sciencedirect.com/science/article/pii/0370157384900231}
  {\bibfield  {journal} {\bibinfo  {journal} {Phys Rep}\ }\textbf {\bibinfo
  {volume} {103}},\ \bibinfo {pages} {251 } (\bibinfo {year}
  {1984})}\BibitemShut {NoStop}%
\bibitem [{\citenamefont {Kostylev}(2013)}]{KostylevJAP2013}%
  \BibitemOpen
  \bibfield  {author} {\bibinfo {author} {\bibfnamefont {M.}~\bibnamefont
  {Kostylev}},\ }\href {https://doi.org/10.1063/1.4789962} {\bibfield
  {journal} {\bibinfo  {journal} {J appl Phys}\ }\textbf {\bibinfo {volume}
  {113}},\ \bibinfo {pages} {053907} (\bibinfo {year} {2013})}\BibitemShut
  {NoStop}%
\bibitem [{\citenamefont {Mills}(2006)}]{Mills2006}%
  \BibitemOpen
  \bibfield  {author} {\bibinfo {author} {\bibfnamefont {D.}~\bibnamefont
  {Mills}},\ }\href
  {http://www.sciencedirect.com/science/article/pii/S0304885306003994}
  {\bibfield  {journal} {\bibinfo  {journal} {J. Magn. Magn. Mater.}\ }\textbf
  {\bibinfo {volume} {306}},\ \bibinfo {pages} {16 } (\bibinfo {year}
  {2006})}\BibitemShut {NoStop}%
\bibitem [{\citenamefont {Gonzalez-Ballestero}\ \emph
  {et~al.}(2020{\natexlab{b}})\citenamefont {Gonzalez-Ballestero},
  \citenamefont {H\"ummer}, \citenamefont {Gieseler},\ and\ \citenamefont
  {Romero-Isart}}]{GonzalezBallesteroPRB2020}%
  \BibitemOpen
  \bibfield  {author} {\bibinfo {author} {\bibfnamefont {C.}~\bibnamefont
  {Gonzalez-Ballestero}}, \bibinfo {author} {\bibfnamefont {D.}~\bibnamefont
  {H\"ummer}}, \bibinfo {author} {\bibfnamefont {J.}~\bibnamefont {Gieseler}},
  \ and\ \bibinfo {author} {\bibfnamefont {O.}~\bibnamefont {Romero-Isart}},\
  }\href {https://link.aps.org/doi/10.1103/PhysRevB.101.125404} {\bibfield
  {journal} {\bibinfo  {journal} {Phys. Rev. B}\ }\textbf {\bibinfo {volume}
  {101}},\ \bibinfo {pages} {125404} (\bibinfo {year}
  {2020}{\natexlab{b}})}\BibitemShut {NoStop}%
\bibitem [{\citenamefont {Carmichael}(2013)}]{CarmichaelBook}%
  \BibitemOpen
  \bibfield  {author} {\bibinfo {author} {\bibfnamefont {H.}~\bibnamefont
  {Carmichael}},\ }\href@noop {} {\emph {\bibinfo {title} {Statistical Methods
  in Quantum Optics 1: Master Equations and Fokker-Planck Equations}}},\
  Theoretical and Mathematical Physics\ (\bibinfo  {publisher} {Springer Berlin
  Heidelberg},\ \bibinfo {year} {2013})\BibitemShut {NoStop}%
\bibitem [{\citenamefont {Bar-Gill}\ \emph {et~al.}(2013)\citenamefont
  {Bar-Gill}, \citenamefont {Pham}, \citenamefont {Jarmola}, \citenamefont
  {Budker},\ and\ \citenamefont {Walsworth}}]{BarGillNatComm2013}%
  \BibitemOpen
  \bibfield  {author} {\bibinfo {author} {\bibfnamefont {N.}~\bibnamefont
  {Bar-Gill}}, \bibinfo {author} {\bibfnamefont {L.~M.}\ \bibnamefont {Pham}},
  \bibinfo {author} {\bibfnamefont {A.}~\bibnamefont {Jarmola}}, \bibinfo
  {author} {\bibfnamefont {D.}~\bibnamefont {Budker}}, \ and\ \bibinfo {author}
  {\bibfnamefont {R.~L.}\ \bibnamefont {Walsworth}},\ }\href
  {https://doi.org/10.1038/ncomms2771} {\bibfield  {journal} {\bibinfo
  {journal} {Nat Commun}\ }\textbf {\bibinfo {volume} {4}},\ \bibinfo {pages}
  {1743} (\bibinfo {year} {2013})}\BibitemShut {NoStop}%
\bibitem [{\citenamefont {Ajisaka}\ and\ \citenamefont
  {Band}(2016)}]{AjisakaPRB2016}%
  \BibitemOpen
  \bibfield  {author} {\bibinfo {author} {\bibfnamefont {S.}~\bibnamefont
  {Ajisaka}}\ and\ \bibinfo {author} {\bibfnamefont {Y.~B.}\ \bibnamefont
  {Band}},\ }\href {https://link.aps.org/doi/10.1103/PhysRevB.94.134107}
  {\bibfield  {journal} {\bibinfo  {journal} {Phys. Rev. B}\ }\textbf {\bibinfo
  {volume} {94}},\ \bibinfo {pages} {134107} (\bibinfo {year}
  {2016})}\BibitemShut {NoStop}%
\bibitem [{\citenamefont {Wang}\ \emph {et~al.}(2014)\citenamefont {Wang},
  \citenamefont {Cai}, \citenamefont {Retzker},\ and\ \citenamefont
  {Plenio}}]{WangNJP2014}%
  \BibitemOpen
  \bibfield  {author} {\bibinfo {author} {\bibfnamefont {Z.-Y.}\ \bibnamefont
  {Wang}}, \bibinfo {author} {\bibfnamefont {J.-M.}\ \bibnamefont {Cai}},
  \bibinfo {author} {\bibfnamefont {A.}~\bibnamefont {Retzker}}, \ and\
  \bibinfo {author} {\bibfnamefont {M.~B.}\ \bibnamefont {Plenio}},\ }\href
  {https://doi.org/10.1088\%2F1367-2630\%2F16\%2F8\%2F083033} {\bibfield
  {journal} {\bibinfo  {journal} {New J Phys}\ }\textbf {\bibinfo {volume}
  {16}},\ \bibinfo {pages} {083033} (\bibinfo {year} {2014})}\BibitemShut
  {NoStop}%
\bibitem [{\citenamefont {H{\"u}mmer}\ \emph {et~al.}(2020)\citenamefont
  {H{\"u}mmer}, \citenamefont {Romero-Isart}, \citenamefont {Rauschenbeutel},\
  and\ \citenamefont {Schneeweiss}}]{HummerArxiv2020}%
  \BibitemOpen
  \bibfield  {author} {\bibinfo {author} {\bibfnamefont {D.}~\bibnamefont
  {H{\"u}mmer}}, \bibinfo {author} {\bibfnamefont {O.}~\bibnamefont
  {Romero-Isart}}, \bibinfo {author} {\bibfnamefont {A.}~\bibnamefont
  {Rauschenbeutel}}, \ and\ \bibinfo {author} {\bibfnamefont {P.}~\bibnamefont
  {Schneeweiss}},\ }\href {https://arxiv.org/abs/2006.12855} {\bibfield
  {journal} {\bibinfo  {journal} {arXiv preprint arXiv:2006.12855}\ } (\bibinfo
  {year} {2020})}\BibitemShut {NoStop}%
\bibitem [{\citenamefont {Kalarickal}\ \emph {et~al.}(2006)\citenamefont
  {Kalarickal}, \citenamefont {Krivosik}, \citenamefont {Wu}, \citenamefont
  {Patton}, \citenamefont {Schneider}, \citenamefont {Kabos}, \citenamefont
  {Silva},\ and\ \citenamefont {Nibarger}}]{KalarickalJAP2006}%
  \BibitemOpen
  \bibfield  {author} {\bibinfo {author} {\bibfnamefont {S.~S.}\ \bibnamefont
  {Kalarickal}}, \bibinfo {author} {\bibfnamefont {P.}~\bibnamefont
  {Krivosik}}, \bibinfo {author} {\bibfnamefont {M.}~\bibnamefont {Wu}},
  \bibinfo {author} {\bibfnamefont {C.~E.}\ \bibnamefont {Patton}}, \bibinfo
  {author} {\bibfnamefont {M.~L.}\ \bibnamefont {Schneider}}, \bibinfo {author}
  {\bibfnamefont {P.}~\bibnamefont {Kabos}}, \bibinfo {author} {\bibfnamefont
  {T.~J.}\ \bibnamefont {Silva}}, \ and\ \bibinfo {author} {\bibfnamefont
  {J.~P.}\ \bibnamefont {Nibarger}},\ }\href
  {https://doi.org/10.1063/1.2197087} {\bibfield  {journal} {\bibinfo
  {journal} {J appl Phys}\ }\textbf {\bibinfo {volume} {99}},\ \bibinfo {pages}
  {093909} (\bibinfo {year} {2006})}\BibitemShut {NoStop}%
\bibitem [{\citenamefont {Breuer}\ \emph {et~al.}(2002)\citenamefont {Breuer},
  \citenamefont {Petruccione} \emph {et~al.}}]{breuer2002theory}%
  \BibitemOpen
  \bibfield  {author} {\bibinfo {author} {\bibfnamefont {H.-P.}\ \bibnamefont
  {Breuer}}, \bibinfo {author} {\bibfnamefont {F.}~\bibnamefont {Petruccione}},
   \emph {et~al.},\ }\href@noop {} {\emph {\bibinfo {title} {The theory of open
  quantum systems}}}\ (\bibinfo  {publisher} {Oxford University Press on
  Demand},\ \bibinfo {year} {2002})\BibitemShut {NoStop}%
\bibitem [{\citenamefont {Chumak}(2019)}]{ChumakInBook}%
  \BibitemOpen
  \bibfield  {author} {\bibinfo {author} {\bibfnamefont {A.~V.}\ \bibnamefont
  {Chumak}},\ }\enquote {\bibinfo {title} {Magnon spintronics},}\ in\ \href
  {https://www.routledgehandbooks.com/doi/10.1201/9780429423079-6} {\emph
  {\bibinfo {booktitle} {Spintronics Handbook: Spin Transport and Magnetism,
  Second Edition}}}\ (\bibinfo  {publisher} {CRC Press},\ \bibinfo {year}
  {2019})\ Chap.~\bibinfo {chapter} {6}\BibitemShut {NoStop}%
\bibitem [{\citenamefont {Stancil}(1986)}]{StancilJAP1986}%
  \BibitemOpen
  \bibfield  {author} {\bibinfo {author} {\bibfnamefont {D.~D.}\ \bibnamefont
  {Stancil}},\ }\href {https://doi.org/10.1063/1.336867} {\bibfield  {journal}
  {\bibinfo  {journal} {J appl Phys}\ }\textbf {\bibinfo {volume} {59}},\
  \bibinfo {pages} {218} (\bibinfo {year} {1986})}\BibitemShut {NoStop}%
\bibitem [{\citenamefont {Tetienne}\ \emph {et~al.}(2013)\citenamefont
  {Tetienne}, \citenamefont {Hingant}, \citenamefont {Rondin}, \citenamefont
  {Cavaill\`es}, \citenamefont {Mayer}, \citenamefont {Dantelle}, \citenamefont
  {Gacoin}, \citenamefont {Wrachtrup}, \citenamefont {Roch},\ and\
  \citenamefont {Jacques}}]{TetiennePRB2013}%
  \BibitemOpen
  \bibfield  {author} {\bibinfo {author} {\bibfnamefont {J.-P.}\ \bibnamefont
  {Tetienne}}, \bibinfo {author} {\bibfnamefont {T.}~\bibnamefont {Hingant}},
  \bibinfo {author} {\bibfnamefont {L.}~\bibnamefont {Rondin}}, \bibinfo
  {author} {\bibfnamefont {A.}~\bibnamefont {Cavaill\`es}}, \bibinfo {author}
  {\bibfnamefont {L.}~\bibnamefont {Mayer}}, \bibinfo {author} {\bibfnamefont
  {G.}~\bibnamefont {Dantelle}}, \bibinfo {author} {\bibfnamefont
  {T.}~\bibnamefont {Gacoin}}, \bibinfo {author} {\bibfnamefont
  {J.}~\bibnamefont {Wrachtrup}}, \bibinfo {author} {\bibfnamefont {J.-F.}\
  \bibnamefont {Roch}}, \ and\ \bibinfo {author} {\bibfnamefont
  {V.}~\bibnamefont {Jacques}},\ }\href
  {https://link.aps.org/doi/10.1103/PhysRevB.87.235436} {\bibfield  {journal}
  {\bibinfo  {journal} {Phys. Rev. B}\ }\textbf {\bibinfo {volume} {87}},\
  \bibinfo {pages} {235436} (\bibinfo {year} {2013})}\BibitemShut {NoStop}%
\bibitem [{\citenamefont {Myers}\ \emph {et~al.}(2017)\citenamefont {Myers},
  \citenamefont {Ariyaratne},\ and\ \citenamefont {Jayich}}]{MyersPRL2017}%
  \BibitemOpen
  \bibfield  {author} {\bibinfo {author} {\bibfnamefont {B.~A.}\ \bibnamefont
  {Myers}}, \bibinfo {author} {\bibfnamefont {A.}~\bibnamefont {Ariyaratne}}, \
  and\ \bibinfo {author} {\bibfnamefont {A.~C.~B.}\ \bibnamefont {Jayich}},\
  }\href {https://link.aps.org/doi/10.1103/PhysRevLett.118.197201} {\bibfield
  {journal} {\bibinfo  {journal} {Phys. Rev. Lett.}\ }\textbf {\bibinfo
  {volume} {118}},\ \bibinfo {pages} {197201} (\bibinfo {year}
  {2017})}\BibitemShut {NoStop}%
\bibitem [{\citenamefont {Herbschleb}\ \emph {et~al.}(2019)\citenamefont
  {Herbschleb}, \citenamefont {Kato}, \citenamefont {Maruyama}, \citenamefont
  {Danjo}, \citenamefont {Makino}, \citenamefont {Yamasaki}, \citenamefont
  {Ohki}, \citenamefont {Hayashi}, \citenamefont {Morishita}, \citenamefont
  {Fujiwara},\ and\ \citenamefont {Mizuochi}}]{HerbschlebNatComm2019}%
  \BibitemOpen
  \bibfield  {author} {\bibinfo {author} {\bibfnamefont {E.~D.}\ \bibnamefont
  {Herbschleb}}, \bibinfo {author} {\bibfnamefont {H.}~\bibnamefont {Kato}},
  \bibinfo {author} {\bibfnamefont {Y.}~\bibnamefont {Maruyama}}, \bibinfo
  {author} {\bibfnamefont {T.}~\bibnamefont {Danjo}}, \bibinfo {author}
  {\bibfnamefont {T.}~\bibnamefont {Makino}}, \bibinfo {author} {\bibfnamefont
  {S.}~\bibnamefont {Yamasaki}}, \bibinfo {author} {\bibfnamefont
  {I.}~\bibnamefont {Ohki}}, \bibinfo {author} {\bibfnamefont {K.}~\bibnamefont
  {Hayashi}}, \bibinfo {author} {\bibfnamefont {H.}~\bibnamefont {Morishita}},
  \bibinfo {author} {\bibfnamefont {M.}~\bibnamefont {Fujiwara}}, \ and\
  \bibinfo {author} {\bibfnamefont {N.}~\bibnamefont {Mizuochi}},\ }\href
  {https://doi.org/10.1038/s41467-019-11776-8} {\bibfield  {journal} {\bibinfo
  {journal} {Nat Commun}\ }\textbf {\bibinfo {volume} {10}},\ \bibinfo {pages}
  {3766} (\bibinfo {year} {2019})}\BibitemShut {NoStop}%
\bibitem [{\citenamefont {Hanson}\ \emph {et~al.}(2008)\citenamefont {Hanson},
  \citenamefont {Dobrovitski}, \citenamefont {Feiguin}, \citenamefont {Gywat},\
  and\ \citenamefont {Awschalom}}]{HansonScience2008}%
  \BibitemOpen
  \bibfield  {author} {\bibinfo {author} {\bibfnamefont {R.}~\bibnamefont
  {Hanson}}, \bibinfo {author} {\bibfnamefont {V.~V.}\ \bibnamefont
  {Dobrovitski}}, \bibinfo {author} {\bibfnamefont {A.~E.}\ \bibnamefont
  {Feiguin}}, \bibinfo {author} {\bibfnamefont {O.}~\bibnamefont {Gywat}}, \
  and\ \bibinfo {author} {\bibfnamefont {D.~D.}\ \bibnamefont {Awschalom}},\
  }\href {https://science.sciencemag.org/content/320/5874/352} {\bibfield
  {journal} {\bibinfo  {journal} {Science}\ }\textbf {\bibinfo {volume}
  {320}},\ \bibinfo {pages} {352} (\bibinfo {year} {2008})}\BibitemShut
  {NoStop}%
\bibitem [{\citenamefont {de~Lange}\ \emph {et~al.}(2010)\citenamefont
  {de~Lange}, \citenamefont {Wang}, \citenamefont {Rist{\`e}}, \citenamefont
  {Dobrovitski},\ and\ \citenamefont {Hanson}}]{deLangeScience2010}%
  \BibitemOpen
  \bibfield  {author} {\bibinfo {author} {\bibfnamefont {G.}~\bibnamefont
  {de~Lange}}, \bibinfo {author} {\bibfnamefont {Z.~H.}\ \bibnamefont {Wang}},
  \bibinfo {author} {\bibfnamefont {D.}~\bibnamefont {Rist{\`e}}}, \bibinfo
  {author} {\bibfnamefont {V.~V.}\ \bibnamefont {Dobrovitski}}, \ and\ \bibinfo
  {author} {\bibfnamefont {R.}~\bibnamefont {Hanson}},\ }\href
  {https://science.sciencemag.org/content/330/6000/60} {\bibfield  {journal}
  {\bibinfo  {journal} {Science}\ }\textbf {\bibinfo {volume} {330}},\ \bibinfo
  {pages} {60} (\bibinfo {year} {2010})}\BibitemShut {NoStop}%
\bibitem [{\citenamefont {Kleinsasser}\ \emph {et~al.}(2016)\citenamefont
  {Kleinsasser}, \citenamefont {Stanfield}, \citenamefont {Banks},
  \citenamefont {Zhu}, \citenamefont {Li}, \citenamefont {Acosta},
  \citenamefont {Watanabe}, \citenamefont {Itoh},\ and\ \citenamefont
  {Fu}}]{KleinsasserAPL2016}%
  \BibitemOpen
  \bibfield  {author} {\bibinfo {author} {\bibfnamefont {E.~E.}\ \bibnamefont
  {Kleinsasser}}, \bibinfo {author} {\bibfnamefont {M.~M.}\ \bibnamefont
  {Stanfield}}, \bibinfo {author} {\bibfnamefont {J.~K.~Q.}\ \bibnamefont
  {Banks}}, \bibinfo {author} {\bibfnamefont {Z.}~\bibnamefont {Zhu}}, \bibinfo
  {author} {\bibfnamefont {W.-D.}\ \bibnamefont {Li}}, \bibinfo {author}
  {\bibfnamefont {V.~M.}\ \bibnamefont {Acosta}}, \bibinfo {author}
  {\bibfnamefont {H.}~\bibnamefont {Watanabe}}, \bibinfo {author}
  {\bibfnamefont {K.~M.}\ \bibnamefont {Itoh}}, \ and\ \bibinfo {author}
  {\bibfnamefont {K.-M.~C.}\ \bibnamefont {Fu}},\ }\href
  {https://doi.org/10.1063/1.4949357} {\bibfield  {journal} {\bibinfo
  {journal} {Appl Phys Lett}\ }\textbf {\bibinfo {volume} {108}},\ \bibinfo
  {pages} {202401} (\bibinfo {year} {2016})}\BibitemShut {NoStop}%
\bibitem [{\citenamefont {Bauch}\ \emph {et~al.}(2018)\citenamefont {Bauch},
  \citenamefont {Hart}, \citenamefont {Schloss}, \citenamefont {Turner},
  \citenamefont {Barry}, \citenamefont {Kehayias}, \citenamefont {Singh},\ and\
  \citenamefont {Walsworth}}]{BauchPRX2018}%
  \BibitemOpen
  \bibfield  {author} {\bibinfo {author} {\bibfnamefont {E.}~\bibnamefont
  {Bauch}}, \bibinfo {author} {\bibfnamefont {C.~A.}\ \bibnamefont {Hart}},
  \bibinfo {author} {\bibfnamefont {J.~M.}\ \bibnamefont {Schloss}}, \bibinfo
  {author} {\bibfnamefont {M.~J.}\ \bibnamefont {Turner}}, \bibinfo {author}
  {\bibfnamefont {J.~F.}\ \bibnamefont {Barry}}, \bibinfo {author}
  {\bibfnamefont {P.}~\bibnamefont {Kehayias}}, \bibinfo {author}
  {\bibfnamefont {S.}~\bibnamefont {Singh}}, \ and\ \bibinfo {author}
  {\bibfnamefont {R.~L.}\ \bibnamefont {Walsworth}},\ }\href
  {https://link.aps.org/doi/10.1103/PhysRevX.8.031025} {\bibfield  {journal}
  {\bibinfo  {journal} {Phys. Rev. X}\ }\textbf {\bibinfo {volume} {8}},\
  \bibinfo {pages} {031025} (\bibinfo {year} {2018})}\BibitemShut {NoStop}%
\bibitem [{\citenamefont {Robledo}\ \emph {et~al.}(2011)\citenamefont
  {Robledo}, \citenamefont {Bernien}, \citenamefont {van~der Sar},\ and\
  \citenamefont {Hanson}}]{RobledoNJP2011}%
  \BibitemOpen
  \bibfield  {author} {\bibinfo {author} {\bibfnamefont {L.}~\bibnamefont
  {Robledo}}, \bibinfo {author} {\bibfnamefont {H.}~\bibnamefont {Bernien}},
  \bibinfo {author} {\bibfnamefont {T.}~\bibnamefont {van~der Sar}}, \ and\
  \bibinfo {author} {\bibfnamefont {R.}~\bibnamefont {Hanson}},\ }\href
  {\doibase 10.1088/1367-2630/13/2/025013} {\bibfield  {journal} {\bibinfo
  {journal} {New J Phys}\ }\textbf {\bibinfo {volume} {13}},\ \bibinfo {pages}
  {025013} (\bibinfo {year} {2011})}\BibitemShut {NoStop}%
\bibitem [{\citenamefont {Wang}\ and\ \citenamefont
  {Yang}(2015)}]{WangNJP2015}%
  \BibitemOpen
  \bibfield  {author} {\bibinfo {author} {\bibfnamefont {P.}~\bibnamefont
  {Wang}}\ and\ \bibinfo {author} {\bibfnamefont {W.}~\bibnamefont {Yang}},\
  }\href {\doibase 10.1088/1367-2630/17/11/113041} {\bibfield  {journal}
  {\bibinfo  {journal} {New J Phys}\ }\textbf {\bibinfo {volume} {17}},\
  \bibinfo {pages} {113041} (\bibinfo {year} {2015})}\BibitemShut {NoStop}%
\bibitem [{\citenamefont {Meirzada}\ \emph {et~al.}(2018)\citenamefont
  {Meirzada}, \citenamefont {Hovav}, \citenamefont {Wolf},\ and\ \citenamefont
  {Bar-Gill}}]{MeirzadaPRB2018}%
  \BibitemOpen
  \bibfield  {author} {\bibinfo {author} {\bibfnamefont {I.}~\bibnamefont
  {Meirzada}}, \bibinfo {author} {\bibfnamefont {Y.}~\bibnamefont {Hovav}},
  \bibinfo {author} {\bibfnamefont {S.~A.}\ \bibnamefont {Wolf}}, \ and\
  \bibinfo {author} {\bibfnamefont {N.}~\bibnamefont {Bar-Gill}},\ }\href
  {\doibase 10.1103/PhysRevB.98.245411} {\bibfield  {journal} {\bibinfo
  {journal} {Phys. Rev. B}\ }\textbf {\bibinfo {volume} {98}},\ \bibinfo
  {pages} {245411} (\bibinfo {year} {2018})}\BibitemShut {NoStop}%
\bibitem [{\citenamefont {Mildren}\ and\ \citenamefont
  {Rabeau}(2013)}]{MildrenBook}%
  \BibitemOpen
  \bibfield  {author} {\bibinfo {author} {\bibfnamefont {R.}~\bibnamefont
  {Mildren}}\ and\ \bibinfo {author} {\bibfnamefont {J.}~\bibnamefont
  {Rabeau}},\ }\href@noop {} {\emph {\bibinfo {title} {Optical Engineering of
  Diamond}}}\ (\bibinfo  {publisher} {Wiley},\ \bibinfo {year}
  {2013})\BibitemShut {NoStop}%
\bibitem [{\citenamefont {Roberts}\ \emph {et~al.}(2019)\citenamefont
  {Roberts}, \citenamefont {Juan},\ and\ \citenamefont
  {Molina-Terriza}}]{RobertsPRB2019}%
  \BibitemOpen
  \bibfield  {author} {\bibinfo {author} {\bibfnamefont {R.~P.}\ \bibnamefont
  {Roberts}}, \bibinfo {author} {\bibfnamefont {M.~L.}\ \bibnamefont {Juan}}, \
  and\ \bibinfo {author} {\bibfnamefont {G.}~\bibnamefont {Molina-Terriza}},\
  }\href {\doibase 10.1103/PhysRevB.99.174307} {\bibfield  {journal} {\bibinfo
  {journal} {Phys. Rev. B}\ }\textbf {\bibinfo {volume} {99}},\ \bibinfo
  {pages} {174307} (\bibinfo {year} {2019})}\BibitemShut {NoStop}%
\bibitem [{\citenamefont {Tetienne}\ \emph {et~al.}(2012)\citenamefont
  {Tetienne}, \citenamefont {Rondin}, \citenamefont {Spinicelli}, \citenamefont
  {Chipaux}, \citenamefont {Debuisschert}, \citenamefont {Roch},\ and\
  \citenamefont {Jacques}}]{TetienneNJP2012}%
  \BibitemOpen
  \bibfield  {author} {\bibinfo {author} {\bibfnamefont {J.-P.}\ \bibnamefont
  {Tetienne}}, \bibinfo {author} {\bibfnamefont {L.}~\bibnamefont {Rondin}},
  \bibinfo {author} {\bibfnamefont {P.}~\bibnamefont {Spinicelli}}, \bibinfo
  {author} {\bibfnamefont {M.}~\bibnamefont {Chipaux}}, \bibinfo {author}
  {\bibfnamefont {T.}~\bibnamefont {Debuisschert}}, \bibinfo {author}
  {\bibfnamefont {J.-F.}\ \bibnamefont {Roch}}, \ and\ \bibinfo {author}
  {\bibfnamefont {V.}~\bibnamefont {Jacques}},\ }\href
  {https://doi.org/10.1088\%2F1367-2630\%2F14\%2F10\%2F103033} {\bibfield
  {journal} {\bibinfo  {journal} {New J Phys}\ }\textbf {\bibinfo {volume}
  {14}},\ \bibinfo {pages} {103033} (\bibinfo {year} {2012})}\BibitemShut
  {NoStop}%
\bibitem [{\citenamefont {Hatano}\ \emph {et~al.}(2018)\citenamefont {Hatano},
  \citenamefont {Sekiguchi}, \citenamefont {Iwasaki}, \citenamefont {Hatano},\
  and\ \citenamefont {Harada}}]{HatanoPhysStatusSolidi2018}%
  \BibitemOpen
  \bibfield  {author} {\bibinfo {author} {\bibfnamefont {Y.}~\bibnamefont
  {Hatano}}, \bibinfo {author} {\bibfnamefont {T.}~\bibnamefont {Sekiguchi}},
  \bibinfo {author} {\bibfnamefont {T.}~\bibnamefont {Iwasaki}}, \bibinfo
  {author} {\bibfnamefont {M.}~\bibnamefont {Hatano}}, \ and\ \bibinfo {author}
  {\bibfnamefont {Y.}~\bibnamefont {Harada}},\ }\href
  {https://onlinelibrary.wiley.com/doi/abs/10.1002/pssa.201800254} {\bibfield
  {journal} {\bibinfo  {journal} {Phys Status solidi A}\ }\textbf {\bibinfo
  {volume} {215}},\ \bibinfo {pages} {1800254} (\bibinfo {year}
  {2018})}\BibitemShut {NoStop}%
\bibitem [{\citenamefont {Ozawa}\ \emph {et~al.}(2019)\citenamefont {Ozawa},
  \citenamefont {Hatano}, \citenamefont {Iwasaki}, \citenamefont {Harada},\
  and\ \citenamefont {Hatano}}]{OzawaJapJAPhys2019}%
  \BibitemOpen
  \bibfield  {author} {\bibinfo {author} {\bibfnamefont {H.}~\bibnamefont
  {Ozawa}}, \bibinfo {author} {\bibfnamefont {Y.}~\bibnamefont {Hatano}},
  \bibinfo {author} {\bibfnamefont {T.}~\bibnamefont {Iwasaki}}, \bibinfo
  {author} {\bibfnamefont {Y.}~\bibnamefont {Harada}}, \ and\ \bibinfo {author}
  {\bibfnamefont {M.}~\bibnamefont {Hatano}},\ }\href
  {https://doi.org/10.7567\%2F1347-4065\%2Fab203c} {\bibfield  {journal}
  {\bibinfo  {journal} {Jpn J appl Phys}\ }\textbf {\bibinfo {volume} {58}},\
  \bibinfo {pages} {SIIB26} (\bibinfo {year} {2019})}\BibitemShut {NoStop}%
\bibitem [{\citenamefont {Acosta}\ \emph {et~al.}(2009)\citenamefont {Acosta},
  \citenamefont {Bauch}, \citenamefont {Ledbetter}, \citenamefont {Santori},
  \citenamefont {Fu}, \citenamefont {Barclay}, \citenamefont {Beausoleil},
  \citenamefont {Linget}, \citenamefont {Roch}, \citenamefont {Treussart},
  \citenamefont {Chemerisov}, \citenamefont {Gawlik},\ and\ \citenamefont
  {Budker}}]{AcostaPRB2009}%
  \BibitemOpen
  \bibfield  {author} {\bibinfo {author} {\bibfnamefont {V.~M.}\ \bibnamefont
  {Acosta}}, \bibinfo {author} {\bibfnamefont {E.}~\bibnamefont {Bauch}},
  \bibinfo {author} {\bibfnamefont {M.~P.}\ \bibnamefont {Ledbetter}}, \bibinfo
  {author} {\bibfnamefont {C.}~\bibnamefont {Santori}}, \bibinfo {author}
  {\bibfnamefont {K.-M.~C.}\ \bibnamefont {Fu}}, \bibinfo {author}
  {\bibfnamefont {P.~E.}\ \bibnamefont {Barclay}}, \bibinfo {author}
  {\bibfnamefont {R.~G.}\ \bibnamefont {Beausoleil}}, \bibinfo {author}
  {\bibfnamefont {H.}~\bibnamefont {Linget}}, \bibinfo {author} {\bibfnamefont
  {J.~F.}\ \bibnamefont {Roch}}, \bibinfo {author} {\bibfnamefont
  {F.}~\bibnamefont {Treussart}}, \bibinfo {author} {\bibfnamefont
  {S.}~\bibnamefont {Chemerisov}}, \bibinfo {author} {\bibfnamefont
  {W.}~\bibnamefont {Gawlik}}, \ and\ \bibinfo {author} {\bibfnamefont
  {D.}~\bibnamefont {Budker}},\ }\href
  {https://link.aps.org/doi/10.1103/PhysRevB.80.115202} {\bibfield  {journal}
  {\bibinfo  {journal} {Phys. Rev. B}\ }\textbf {\bibinfo {volume} {80}},\
  \bibinfo {pages} {115202} (\bibinfo {year} {2009})}\BibitemShut {NoStop}%
\bibitem [{\citenamefont {Kucsko}\ \emph {et~al.}(2018)\citenamefont {Kucsko},
  \citenamefont {Choi}, \citenamefont {Choi}, \citenamefont {Maurer},
  \citenamefont {Zhou}, \citenamefont {Landig}, \citenamefont {Sumiya},
  \citenamefont {Onoda}, \citenamefont {Isoya}, \citenamefont {Jelezko},
  \citenamefont {Demler}, \citenamefont {Yao},\ and\ \citenamefont
  {Lukin}}]{KucskoPRL2018}%
  \BibitemOpen
  \bibfield  {author} {\bibinfo {author} {\bibfnamefont {G.}~\bibnamefont
  {Kucsko}}, \bibinfo {author} {\bibfnamefont {S.}~\bibnamefont {Choi}},
  \bibinfo {author} {\bibfnamefont {J.}~\bibnamefont {Choi}}, \bibinfo {author}
  {\bibfnamefont {P.~C.}\ \bibnamefont {Maurer}}, \bibinfo {author}
  {\bibfnamefont {H.}~\bibnamefont {Zhou}}, \bibinfo {author} {\bibfnamefont
  {R.}~\bibnamefont {Landig}}, \bibinfo {author} {\bibfnamefont
  {H.}~\bibnamefont {Sumiya}}, \bibinfo {author} {\bibfnamefont
  {S.}~\bibnamefont {Onoda}}, \bibinfo {author} {\bibfnamefont
  {J.}~\bibnamefont {Isoya}}, \bibinfo {author} {\bibfnamefont
  {F.}~\bibnamefont {Jelezko}}, \bibinfo {author} {\bibfnamefont
  {E.}~\bibnamefont {Demler}}, \bibinfo {author} {\bibfnamefont {N.~Y.}\
  \bibnamefont {Yao}}, \ and\ \bibinfo {author} {\bibfnamefont {M.~D.}\
  \bibnamefont {Lukin}},\ }\href
  {https://link.aps.org/doi/10.1103/PhysRevLett.121.023601} {\bibfield
  {journal} {\bibinfo  {journal} {Phys. Rev. Lett.}\ }\textbf {\bibinfo
  {volume} {121}},\ \bibinfo {pages} {023601} (\bibinfo {year}
  {2018})}\BibitemShut {NoStop}%
\bibitem [{\citenamefont {{Mojahedi}}\ \emph {et~al.}(2003)\citenamefont
  {{Mojahedi}}, \citenamefont {{Malloy}}, \citenamefont {{Eleftheriades}},
  \citenamefont {{Woodley}},\ and\ \citenamefont {{Chiao}}}]{MojahediIEEE2003}%
  \BibitemOpen
  \bibfield  {author} {\bibinfo {author} {\bibfnamefont {M.}~\bibnamefont
  {{Mojahedi}}}, \bibinfo {author} {\bibfnamefont {K.~J.}\ \bibnamefont
  {{Malloy}}}, \bibinfo {author} {\bibfnamefont {G.~V.}\ \bibnamefont
  {{Eleftheriades}}}, \bibinfo {author} {\bibfnamefont {J.}~\bibnamefont
  {{Woodley}}}, \ and\ \bibinfo {author} {\bibfnamefont {R.~Y.}\ \bibnamefont
  {{Chiao}}},\ }\href {https://ieeexplore.ieee.org/document/1193071} {\bibfield
   {journal} {\bibinfo  {journal} {IEEE J Sel Top Quant}\ }\textbf {\bibinfo
  {volume} {9}},\ \bibinfo {pages} {30} (\bibinfo {year} {2003})}\BibitemShut
  {NoStop}%
\bibitem [{\citenamefont {Casimir}\ and\ \citenamefont
  {Polder}(1948)}]{CasimirPhysRev1948}%
  \BibitemOpen
  \bibfield  {author} {\bibinfo {author} {\bibfnamefont {H.~B.~G.}\
  \bibnamefont {Casimir}}\ and\ \bibinfo {author} {\bibfnamefont
  {D.}~\bibnamefont {Polder}},\ }\href
  {https://link.aps.org/doi/10.1103/PhysRev.73.360} {\bibfield  {journal}
  {\bibinfo  {journal} {Phys. Rev.}\ }\textbf {\bibinfo {volume} {73}},\
  \bibinfo {pages} {360} (\bibinfo {year} {1948})}\BibitemShut {NoStop}%
\bibitem [{\citenamefont {Buhmann}\ and\ \citenamefont
  {Welsch}(2007)}]{BuhmannPQE2007}%
  \BibitemOpen
  \bibfield  {author} {\bibinfo {author} {\bibfnamefont {S.~Y.}\ \bibnamefont
  {Buhmann}}\ and\ \bibinfo {author} {\bibfnamefont {D.-G.}\ \bibnamefont
  {Welsch}},\ }\href
  {http://www.sciencedirect.com/science/article/pii/S0079672707000249}
  {\bibfield  {journal} {\bibinfo  {journal} {Prog Quant Electron}\ }\textbf
  {\bibinfo {volume} {31}},\ \bibinfo {pages} {51 } (\bibinfo {year}
  {2007})}\BibitemShut {NoStop}%
\bibitem [{\citenamefont {Buhmann}\ and\ \citenamefont
  {Scheel}(2008)}]{BuhmannPRL2008}%
  \BibitemOpen
  \bibfield  {author} {\bibinfo {author} {\bibfnamefont {S.~Y.}\ \bibnamefont
  {Buhmann}}\ and\ \bibinfo {author} {\bibfnamefont {S.}~\bibnamefont
  {Scheel}},\ }\href {https://link.aps.org/doi/10.1103/PhysRevLett.100.253201}
  {\bibfield  {journal} {\bibinfo  {journal} {Phys. Rev. Lett.}\ }\textbf
  {\bibinfo {volume} {100}},\ \bibinfo {pages} {253201} (\bibinfo {year}
  {2008})}\BibitemShut {NoStop}%
\bibitem [{\citenamefont {Skagerstam}\ \emph {et~al.}(2009)\citenamefont
  {Skagerstam}, \citenamefont {Rekdal},\ and\ \citenamefont
  {Vaskinn}}]{SkagerstamPRA2009}%
  \BibitemOpen
  \bibfield  {author} {\bibinfo {author} {\bibfnamefont {B.-S.}\ \bibnamefont
  {Skagerstam}}, \bibinfo {author} {\bibfnamefont {P.~K.}\ \bibnamefont
  {Rekdal}}, \ and\ \bibinfo {author} {\bibfnamefont {A.~H.}\ \bibnamefont
  {Vaskinn}},\ }\href {https://link.aps.org/doi/10.1103/PhysRevA.80.022902}
  {\bibfield  {journal} {\bibinfo  {journal} {Phys. Rev. A}\ }\textbf {\bibinfo
  {volume} {80}},\ \bibinfo {pages} {022902} (\bibinfo {year}
  {2009})}\BibitemShut {NoStop}%
\bibitem [{\citenamefont {Haakh}\ \emph {et~al.}(2009)\citenamefont {Haakh},
  \citenamefont {Intravaia}, \citenamefont {Henkel}, \citenamefont {Spagnolo},
  \citenamefont {Passante}, \citenamefont {Power},\ and\ \citenamefont
  {Sols}}]{HaakhPRA2009}%
  \BibitemOpen
  \bibfield  {author} {\bibinfo {author} {\bibfnamefont {H.}~\bibnamefont
  {Haakh}}, \bibinfo {author} {\bibfnamefont {F.}~\bibnamefont {Intravaia}},
  \bibinfo {author} {\bibfnamefont {C.}~\bibnamefont {Henkel}}, \bibinfo
  {author} {\bibfnamefont {S.}~\bibnamefont {Spagnolo}}, \bibinfo {author}
  {\bibfnamefont {R.}~\bibnamefont {Passante}}, \bibinfo {author}
  {\bibfnamefont {B.}~\bibnamefont {Power}}, \ and\ \bibinfo {author}
  {\bibfnamefont {F.}~\bibnamefont {Sols}},\ }\href
  {https://link.aps.org/doi/10.1103/PhysRevA.80.062905} {\bibfield  {journal}
  {\bibinfo  {journal} {Phys. Rev. A}\ }\textbf {\bibinfo {volume} {80}},\
  \bibinfo {pages} {062905} (\bibinfo {year} {2009})}\BibitemShut {NoStop}%
\bibitem [{\citenamefont {Vinante}\ \emph {et~al.}(2011)\citenamefont
  {Vinante}, \citenamefont {Wijts}, \citenamefont {Usenko}, \citenamefont
  {Schinkelshoek},\ and\ \citenamefont {Oosterkamp}}]{VinanteNatComm2011}%
  \BibitemOpen
  \bibfield  {author} {\bibinfo {author} {\bibfnamefont {A.}~\bibnamefont
  {Vinante}}, \bibinfo {author} {\bibfnamefont {G.}~\bibnamefont {Wijts}},
  \bibinfo {author} {\bibfnamefont {O.}~\bibnamefont {Usenko}}, \bibinfo
  {author} {\bibfnamefont {L.}~\bibnamefont {Schinkelshoek}}, \ and\ \bibinfo
  {author} {\bibfnamefont {T.~H.}\ \bibnamefont {Oosterkamp}},\ }\href
  {https://doi.org/10.1038/ncomms1581} {\bibfield  {journal} {\bibinfo
  {journal} {Nat Commun}\ }\textbf {\bibinfo {volume} {2}},\ \bibinfo {pages}
  {572} (\bibinfo {year} {2011})}\BibitemShut {NoStop}%
\bibitem [{\citenamefont {Bunch}\ \emph {et~al.}(2007)\citenamefont {Bunch},
  \citenamefont {van~der Zande}, \citenamefont {Verbridge}, \citenamefont
  {Frank}, \citenamefont {Tanenbaum}, \citenamefont {Parpia}, \citenamefont
  {Craighead},\ and\ \citenamefont {McEuen}}]{BunchScience2007}%
  \BibitemOpen
  \bibfield  {author} {\bibinfo {author} {\bibfnamefont {J.~S.}\ \bibnamefont
  {Bunch}}, \bibinfo {author} {\bibfnamefont {A.~M.}\ \bibnamefont {van~der
  Zande}}, \bibinfo {author} {\bibfnamefont {S.~S.}\ \bibnamefont {Verbridge}},
  \bibinfo {author} {\bibfnamefont {I.~W.}\ \bibnamefont {Frank}}, \bibinfo
  {author} {\bibfnamefont {D.~M.}\ \bibnamefont {Tanenbaum}}, \bibinfo {author}
  {\bibfnamefont {J.~M.}\ \bibnamefont {Parpia}}, \bibinfo {author}
  {\bibfnamefont {H.~G.}\ \bibnamefont {Craighead}}, \ and\ \bibinfo {author}
  {\bibfnamefont {P.~L.}\ \bibnamefont {McEuen}},\ }\href
  {https://science.sciencemag.org/content/315/5811/490} {\bibfield  {journal}
  {\bibinfo  {journal} {Science}\ }\textbf {\bibinfo {volume} {315}},\ \bibinfo
  {pages} {490} (\bibinfo {year} {2007})}\BibitemShut {NoStop}%
\bibitem [{\citenamefont {Weber}\ \emph {et~al.}(2016)\citenamefont {Weber},
  \citenamefont {G{\"u}ttinger}, \citenamefont {Noury}, \citenamefont
  {Vergara-Cruz},\ and\ \citenamefont {Bachtold}}]{WeberNatComm2016}%
  \BibitemOpen
  \bibfield  {author} {\bibinfo {author} {\bibfnamefont {P.}~\bibnamefont
  {Weber}}, \bibinfo {author} {\bibfnamefont {J.}~\bibnamefont
  {G{\"u}ttinger}}, \bibinfo {author} {\bibfnamefont {A.}~\bibnamefont
  {Noury}}, \bibinfo {author} {\bibfnamefont {J.}~\bibnamefont {Vergara-Cruz}},
  \ and\ \bibinfo {author} {\bibfnamefont {A.}~\bibnamefont {Bachtold}},\
  }\href {https://doi.org/10.1038/ncomms12496} {\bibfield  {journal} {\bibinfo
  {journal} {Nat Commun}\ }\textbf {\bibinfo {volume} {7}},\ \bibinfo {pages}
  {12496} (\bibinfo {year} {2016})}\BibitemShut {NoStop}%
\bibitem [{\citenamefont {Fischer}\ \emph {et~al.}(2019)\citenamefont
  {Fischer}, \citenamefont {McNally}, \citenamefont {Reetz}, \citenamefont
  {Assump{\c{c}}{\~{a}}o}, \citenamefont {Knief}, \citenamefont {Lin},\ and\
  \citenamefont {Regal}}]{FischerNJP2019}%
  \BibitemOpen
  \bibfield  {author} {\bibinfo {author} {\bibfnamefont {R.}~\bibnamefont
  {Fischer}}, \bibinfo {author} {\bibfnamefont {D.~P.}\ \bibnamefont
  {McNally}}, \bibinfo {author} {\bibfnamefont {C.}~\bibnamefont {Reetz}},
  \bibinfo {author} {\bibfnamefont {G.~G.~T.}\ \bibnamefont
  {Assump{\c{c}}{\~{a}}o}}, \bibinfo {author} {\bibfnamefont {T.}~\bibnamefont
  {Knief}}, \bibinfo {author} {\bibfnamefont {Y.}~\bibnamefont {Lin}}, \ and\
  \bibinfo {author} {\bibfnamefont {C.~A.}\ \bibnamefont {Regal}},\ }\href
  {https://doi.org/10.1088\%2F1367-2630\%2Fab117a} {\bibfield  {journal}
  {\bibinfo  {journal} {New J Phys}\ }\textbf {\bibinfo {volume} {21}},\
  \bibinfo {pages} {043049} (\bibinfo {year} {2019})}\BibitemShut {NoStop}%
\bibitem [{\citenamefont {Li}\ \emph {et~al.}(2007)\citenamefont {Li},
  \citenamefont {Tang},\ and\ \citenamefont {Roukes}}]{LiNatNano2007}%
  \BibitemOpen
  \bibfield  {author} {\bibinfo {author} {\bibfnamefont {M.}~\bibnamefont
  {Li}}, \bibinfo {author} {\bibfnamefont {H.~X.}\ \bibnamefont {Tang}}, \ and\
  \bibinfo {author} {\bibfnamefont {M.~L.}\ \bibnamefont {Roukes}},\ }\href
  {https://doi.org/10.1038/nnano.2006.208} {\bibfield  {journal} {\bibinfo
  {journal} {Nat Nanotechnol}\ }\textbf {\bibinfo {volume} {2}},\ \bibinfo
  {pages} {114} (\bibinfo {year} {2007})}\BibitemShut {NoStop}%
\bibitem [{\citenamefont {Sazonova}\ \emph {et~al.}(2004)\citenamefont
  {Sazonova}, \citenamefont {Yaish}, \citenamefont {{\"U}st{\"u}nel},
  \citenamefont {Roundy}, \citenamefont {Arias},\ and\ \citenamefont
  {McEuen}}]{SazonovaNature2004}%
  \BibitemOpen
  \bibfield  {author} {\bibinfo {author} {\bibfnamefont {V.}~\bibnamefont
  {Sazonova}}, \bibinfo {author} {\bibfnamefont {Y.}~\bibnamefont {Yaish}},
  \bibinfo {author} {\bibfnamefont {H.}~\bibnamefont {{\"U}st{\"u}nel}},
  \bibinfo {author} {\bibfnamefont {D.}~\bibnamefont {Roundy}}, \bibinfo
  {author} {\bibfnamefont {T.~A.}\ \bibnamefont {Arias}}, \ and\ \bibinfo
  {author} {\bibfnamefont {P.~L.}\ \bibnamefont {McEuen}},\ }\href
  {https://doi.org/10.1038/nature02905} {\bibfield  {journal} {\bibinfo
  {journal} {Nature}\ }\textbf {\bibinfo {volume} {431}},\ \bibinfo {pages}
  {284} (\bibinfo {year} {2004})}\BibitemShut {NoStop}%
\bibitem [{\citenamefont {Moser}\ \emph {et~al.}(2013)\citenamefont {Moser},
  \citenamefont {G{\"u}ttinger}, \citenamefont {Eichler}, \citenamefont
  {Esplandiu}, \citenamefont {Liu}, \citenamefont {Dykman},\ and\ \citenamefont
  {Bachtold}}]{MoserNatNano2013}%
  \BibitemOpen
  \bibfield  {author} {\bibinfo {author} {\bibfnamefont {J.}~\bibnamefont
  {Moser}}, \bibinfo {author} {\bibfnamefont {J.}~\bibnamefont
  {G{\"u}ttinger}}, \bibinfo {author} {\bibfnamefont {A.}~\bibnamefont
  {Eichler}}, \bibinfo {author} {\bibfnamefont {M.~J.}\ \bibnamefont
  {Esplandiu}}, \bibinfo {author} {\bibfnamefont {D.~E.}\ \bibnamefont {Liu}},
  \bibinfo {author} {\bibfnamefont {M.~I.}\ \bibnamefont {Dykman}}, \ and\
  \bibinfo {author} {\bibfnamefont {A.}~\bibnamefont {Bachtold}},\ }\href
  {https://doi.org/10.1038/nnano.2013.97} {\bibfield  {journal} {\bibinfo
  {journal} {Nat Nanotechnol}\ }\textbf {\bibinfo {volume} {8}},\ \bibinfo
  {pages} {493} (\bibinfo {year} {2013})}\BibitemShut {NoStop}%
\bibitem [{\citenamefont {Nichol}\ \emph {et~al.}(2012)\citenamefont {Nichol},
  \citenamefont {Hemesath}, \citenamefont {Lauhon},\ and\ \citenamefont
  {Budakian}}]{NicholPRB2012}%
  \BibitemOpen
  \bibfield  {author} {\bibinfo {author} {\bibfnamefont {J.~M.}\ \bibnamefont
  {Nichol}}, \bibinfo {author} {\bibfnamefont {E.~R.}\ \bibnamefont
  {Hemesath}}, \bibinfo {author} {\bibfnamefont {L.~J.}\ \bibnamefont
  {Lauhon}}, \ and\ \bibinfo {author} {\bibfnamefont {R.}~\bibnamefont
  {Budakian}},\ }\href {https://link.aps.org/doi/10.1103/PhysRevB.85.054414}
  {\bibfield  {journal} {\bibinfo  {journal} {Phys. Rev. B}\ }\textbf {\bibinfo
  {volume} {85}},\ \bibinfo {pages} {054414} (\bibinfo {year}
  {2012})}\BibitemShut {NoStop}%
\bibitem [{\citenamefont {Gloppe}\ \emph {et~al.}(2014)\citenamefont {Gloppe},
  \citenamefont {Verlot}, \citenamefont {Dupont-Ferrier}, \citenamefont
  {Siria}, \citenamefont {Poncharal}, \citenamefont {Bachelier}, \citenamefont
  {Vincent},\ and\ \citenamefont {Arcizet}}]{GloppeNatNano2014}%
  \BibitemOpen
  \bibfield  {author} {\bibinfo {author} {\bibfnamefont {A.}~\bibnamefont
  {Gloppe}}, \bibinfo {author} {\bibfnamefont {P.}~\bibnamefont {Verlot}},
  \bibinfo {author} {\bibfnamefont {E.}~\bibnamefont {Dupont-Ferrier}},
  \bibinfo {author} {\bibfnamefont {A.}~\bibnamefont {Siria}}, \bibinfo
  {author} {\bibfnamefont {P.}~\bibnamefont {Poncharal}}, \bibinfo {author}
  {\bibfnamefont {G.}~\bibnamefont {Bachelier}}, \bibinfo {author}
  {\bibfnamefont {P.}~\bibnamefont {Vincent}}, \ and\ \bibinfo {author}
  {\bibfnamefont {O.}~\bibnamefont {Arcizet}},\ }\href
  {https://doi.org/10.1038/nnano.2014.189} {\bibfield  {journal} {\bibinfo
  {journal} {Nat Nanotechnol}\ }\textbf {\bibinfo {volume} {9}},\ \bibinfo
  {pages} {920} (\bibinfo {year} {2014})}\BibitemShut {NoStop}%
\bibitem [{\citenamefont {Rossi}\ \emph {et~al.}(2017)\citenamefont {Rossi},
  \citenamefont {Braakman}, \citenamefont {Cadeddu}, \citenamefont {Vasyukov},
  \citenamefont {T{\"u}t{\"u}nc{\"u}oglu}, \citenamefont {Fontcuberta~i
  Morral},\ and\ \citenamefont {Poggio}}]{RossiNatNano2017}%
  \BibitemOpen
  \bibfield  {author} {\bibinfo {author} {\bibfnamefont {N.}~\bibnamefont
  {Rossi}}, \bibinfo {author} {\bibfnamefont {F.~R.}\ \bibnamefont {Braakman}},
  \bibinfo {author} {\bibfnamefont {D.}~\bibnamefont {Cadeddu}}, \bibinfo
  {author} {\bibfnamefont {D.}~\bibnamefont {Vasyukov}}, \bibinfo {author}
  {\bibfnamefont {G.}~\bibnamefont {T{\"u}t{\"u}nc{\"u}oglu}}, \bibinfo
  {author} {\bibfnamefont {A.}~\bibnamefont {Fontcuberta~i Morral}}, \ and\
  \bibinfo {author} {\bibfnamefont {M.}~\bibnamefont {Poggio}},\ }\href
  {https://doi.org/10.1038/nnano.2016.189} {\bibfield  {journal} {\bibinfo
  {journal} {Nat Nanotechnol}\ }\textbf {\bibinfo {volume} {12}},\ \bibinfo
  {pages} {150} (\bibinfo {year} {2017})}\BibitemShut {NoStop}%
\bibitem [{\citenamefont {Braakman}\ and\ \citenamefont
  {Poggio}(2019)}]{BraakmanNanotechnology2019}%
  \BibitemOpen
  \bibfield  {author} {\bibinfo {author} {\bibfnamefont {F.~R.}\ \bibnamefont
  {Braakman}}\ and\ \bibinfo {author} {\bibfnamefont {M.}~\bibnamefont
  {Poggio}},\ }\href {https://doi.org/10.1088\%2F1361-6528\%2Fab19cf}
  {\bibfield  {journal} {\bibinfo  {journal} {Nanotechnology}\ }\textbf
  {\bibinfo {volume} {30}},\ \bibinfo {pages} {332001} (\bibinfo {year}
  {2019})}\BibitemShut {NoStop}%
\bibitem [{\citenamefont {H{\'e}ritier}\ \emph {et~al.}(2018)\citenamefont
  {H{\'e}ritier}, \citenamefont {Eichler}, \citenamefont {Pan}, \citenamefont
  {Grob}, \citenamefont {Shorubalko}, \citenamefont {Krass}, \citenamefont
  {Tao},\ and\ \citenamefont {Degen}}]{HeritierNanoLett2018}%
  \BibitemOpen
  \bibfield  {author} {\bibinfo {author} {\bibfnamefont {M.}~\bibnamefont
  {H{\'e}ritier}}, \bibinfo {author} {\bibfnamefont {A.}~\bibnamefont
  {Eichler}}, \bibinfo {author} {\bibfnamefont {Y.}~\bibnamefont {Pan}},
  \bibinfo {author} {\bibfnamefont {U.}~\bibnamefont {Grob}}, \bibinfo {author}
  {\bibfnamefont {I.}~\bibnamefont {Shorubalko}}, \bibinfo {author}
  {\bibfnamefont {M.~D.}\ \bibnamefont {Krass}}, \bibinfo {author}
  {\bibfnamefont {Y.}~\bibnamefont {Tao}}, \ and\ \bibinfo {author}
  {\bibfnamefont {C.~L.}\ \bibnamefont {Degen}},\ }\href
  {https://doi.org/10.1021/acs.nanolett.7b05035} {\bibfield  {journal}
  {\bibinfo  {journal} {Nano Lett}\ }\textbf {\bibinfo {volume} {18}},\
  \bibinfo {pages} {1814} (\bibinfo {year} {2018})}\BibitemShut {NoStop}%
\bibitem [{\citenamefont {Jaskula}\ \emph {et~al.}(2017)\citenamefont
  {Jaskula}, \citenamefont {Bauch}, \citenamefont {Arroyo-Camejo},
  \citenamefont {Lukin}, \citenamefont {Hell}, \citenamefont {Trifonov},\ and\
  \citenamefont {Walsworth}}]{JaskulaOptExpress2017}%
  \BibitemOpen
  \bibfield  {author} {\bibinfo {author} {\bibfnamefont {J.-C.}\ \bibnamefont
  {Jaskula}}, \bibinfo {author} {\bibfnamefont {E.}~\bibnamefont {Bauch}},
  \bibinfo {author} {\bibfnamefont {S.}~\bibnamefont {Arroyo-Camejo}}, \bibinfo
  {author} {\bibfnamefont {M.~D.}\ \bibnamefont {Lukin}}, \bibinfo {author}
  {\bibfnamefont {S.~W.}\ \bibnamefont {Hell}}, \bibinfo {author}
  {\bibfnamefont {A.~S.}\ \bibnamefont {Trifonov}}, \ and\ \bibinfo {author}
  {\bibfnamefont {R.~L.}\ \bibnamefont {Walsworth}},\ }\href
  {http://www.opticsexpress.org/abstract.cfm?URI=oe-25-10-11048} {\bibfield
  {journal} {\bibinfo  {journal} {Opt. Express}\ }\textbf {\bibinfo {volume}
  {25}},\ \bibinfo {pages} {11048} (\bibinfo {year} {2017})}\BibitemShut
  {NoStop}%
\bibitem [{\citenamefont {Zoepfl}\ \emph {et~al.}(2020)\citenamefont {Zoepfl},
  \citenamefont {Juan}, \citenamefont {Schneider},\ and\ \citenamefont
  {Kirchmair}}]{ZoepflPRL2020}%
  \BibitemOpen
  \bibfield  {author} {\bibinfo {author} {\bibfnamefont {D.}~\bibnamefont
  {Zoepfl}}, \bibinfo {author} {\bibfnamefont {M.~L.}\ \bibnamefont {Juan}},
  \bibinfo {author} {\bibfnamefont {C.~M.~F.}\ \bibnamefont {Schneider}}, \
  and\ \bibinfo {author} {\bibfnamefont {G.}~\bibnamefont {Kirchmair}},\ }\href
  {https://link.aps.org/doi/10.1103/PhysRevLett.125.023601} {\bibfield
  {journal} {\bibinfo  {journal} {Phys. Rev. Lett.}\ }\textbf {\bibinfo
  {volume} {125}},\ \bibinfo {pages} {023601} (\bibinfo {year}
  {2020})}\BibitemShut {NoStop}%
\bibitem [{\citenamefont {Rabl}\ \emph {et~al.}(2009)\citenamefont {Rabl},
  \citenamefont {Cappellaro}, \citenamefont {Dutt}, \citenamefont {Jiang},
  \citenamefont {Maze},\ and\ \citenamefont {Lukin}}]{RablPRB2009}%
  \BibitemOpen
  \bibfield  {author} {\bibinfo {author} {\bibfnamefont {P.}~\bibnamefont
  {Rabl}}, \bibinfo {author} {\bibfnamefont {P.}~\bibnamefont {Cappellaro}},
  \bibinfo {author} {\bibfnamefont {M.~V.~G.}\ \bibnamefont {Dutt}}, \bibinfo
  {author} {\bibfnamefont {L.}~\bibnamefont {Jiang}}, \bibinfo {author}
  {\bibfnamefont {J.~R.}\ \bibnamefont {Maze}}, \ and\ \bibinfo {author}
  {\bibfnamefont {M.~D.}\ \bibnamefont {Lukin}},\ }\href
  {https://link.aps.org/doi/10.1103/PhysRevB.79.041302} {\bibfield  {journal}
  {\bibinfo  {journal} {Phys. Rev. B}\ }\textbf {\bibinfo {volume} {79}},\
  \bibinfo {pages} {041302} (\bibinfo {year} {2009})}\BibitemShut {NoStop}%
\bibitem [{\citenamefont {Via}\ \emph {et~al.}(2015)\citenamefont {Via},
  \citenamefont {Kirchmair},\ and\ \citenamefont {Romero-Isart}}]{ViaPRL2015}%
  \BibitemOpen
  \bibfield  {author} {\bibinfo {author} {\bibfnamefont {G.}~\bibnamefont
  {Via}}, \bibinfo {author} {\bibfnamefont {G.}~\bibnamefont {Kirchmair}}, \
  and\ \bibinfo {author} {\bibfnamefont {O.}~\bibnamefont {Romero-Isart}},\
  }\href {https://link.aps.org/doi/10.1103/PhysRevLett.114.143602} {\bibfield
  {journal} {\bibinfo  {journal} {Phys. Rev. Lett.}\ }\textbf {\bibinfo
  {volume} {114}},\ \bibinfo {pages} {143602} (\bibinfo {year}
  {2015})}\BibitemShut {NoStop}%
\bibitem [{\citenamefont {Jackson}(1975)}]{Jackson1975classical}%
  \BibitemOpen
  \bibfield  {author} {\bibinfo {author} {\bibfnamefont {J.}~\bibnamefont
  {Jackson}},\ }\href@noop {} {\emph {\bibinfo {title} {Classical
  electrodynamics}}}\ (\bibinfo  {publisher} {Wiley},\ \bibinfo {year}
  {1975})\BibitemShut {NoStop}%
\bibitem [{\citenamefont {Wohlfarth}(1986)}]{wohlfarth1986handbook}%
  \BibitemOpen
  \bibfield  {author} {\bibinfo {author} {\bibfnamefont {E.}~\bibnamefont
  {Wohlfarth}},\ }\href@noop {} {\emph {\bibinfo {title} {Handbook of Magnetic
  Materials}}},\ \bibinfo {series} {Ferromagnetic materials : a handbook on the
  properties of magnetically ordered substances}\ No.\ \bibinfo {number} {v.
  2}\ (\bibinfo  {publisher} {Elsevier Science},\ \bibinfo {year}
  {1986})\BibitemShut {NoStop}%
\bibitem [{\citenamefont {Krysztofik}\ \emph {et~al.}(2017)\citenamefont
  {Krysztofik}, \citenamefont {G{\l}owi{\'{n}}ski}, \citenamefont
  {Ku{\'{s}}wik}, \citenamefont {Zi{\k{e}}tek}, \citenamefont {Coy},
  \citenamefont {Rych{\l}y}, \citenamefont {Jurga}, \citenamefont {Stobiecki},\
  and\ \citenamefont {Dubowik}}]{KrysztofikJPhysD2017}%
  \BibitemOpen
  \bibfield  {author} {\bibinfo {author} {\bibfnamefont {A.}~\bibnamefont
  {Krysztofik}}, \bibinfo {author} {\bibfnamefont {H.}~\bibnamefont
  {G{\l}owi{\'{n}}ski}}, \bibinfo {author} {\bibfnamefont {P.}~\bibnamefont
  {Ku{\'{s}}wik}}, \bibinfo {author} {\bibfnamefont {S.}~\bibnamefont
  {Zi{\k{e}}tek}}, \bibinfo {author} {\bibfnamefont {L.~E.}\ \bibnamefont
  {Coy}}, \bibinfo {author} {\bibfnamefont {J.~N.}\ \bibnamefont {Rych{\l}y}},
  \bibinfo {author} {\bibfnamefont {S.}~\bibnamefont {Jurga}}, \bibinfo
  {author} {\bibfnamefont {T.~W.}\ \bibnamefont {Stobiecki}}, \ and\ \bibinfo
  {author} {\bibfnamefont {J.}~\bibnamefont {Dubowik}},\ }\href
  {https://doi.org/10.1088\%2F1361-6463\%2Faa6df0} {\bibfield  {journal}
  {\bibinfo  {journal} {J Phys D Appl Phys}\ }\textbf {\bibinfo {volume}
  {50}},\ \bibinfo {pages} {235004} (\bibinfo {year} {2017})}\BibitemShut
  {NoStop}%
\bibitem [{\citenamefont {Kalinikos}\ \emph {et~al.}(1990)\citenamefont
  {Kalinikos}, \citenamefont {Kostylev}, \citenamefont {Kozhus},\ and\
  \citenamefont {Slavin}}]{KalinikosJPhysCondMat1990}%
  \BibitemOpen
  \bibfield  {author} {\bibinfo {author} {\bibfnamefont {B.~A.}\ \bibnamefont
  {Kalinikos}}, \bibinfo {author} {\bibfnamefont {M.~P.}\ \bibnamefont
  {Kostylev}}, \bibinfo {author} {\bibfnamefont {N.~V.}\ \bibnamefont
  {Kozhus}}, \ and\ \bibinfo {author} {\bibfnamefont {A.~N.}\ \bibnamefont
  {Slavin}},\ }\href {https://doi.org/10.1088\%2F0953-8984\%2F2\%2F49\%2F012}
  {\bibfield  {journal} {\bibinfo  {journal} {J Phys-Condens Mat}\ }\textbf
  {\bibinfo {volume} {2}},\ \bibinfo {pages} {9861} (\bibinfo {year}
  {1990})}\BibitemShut {NoStop}%
\bibitem [{\citenamefont {Rado}\ and\ \citenamefont
  {Weertman}(1959)}]{RadoJPCS1959}%
  \BibitemOpen
  \bibfield  {author} {\bibinfo {author} {\bibfnamefont {G.}~\bibnamefont
  {Rado}}\ and\ \bibinfo {author} {\bibfnamefont {J.}~\bibnamefont
  {Weertman}},\ }\href
  {https://www.sciencedirect.com/science/article/pii/0022369759902331}
  {\bibfield  {journal} {\bibinfo  {journal} {J Phys Chem Solids}\ }\textbf
  {\bibinfo {volume} {11}},\ \bibinfo {pages} {315 } (\bibinfo {year}
  {1959})}\BibitemShut {NoStop}%
\bibitem [{\citenamefont {Wang}\ \emph {et~al.}(2019)\citenamefont {Wang},
  \citenamefont {Heinz}, \citenamefont {Verba}, \citenamefont {Kewenig},
  \citenamefont {Pirro}, \citenamefont {Schneider}, \citenamefont {Meyer},
  \citenamefont {L\"agel}, \citenamefont {Dubs}, \citenamefont {Br\"acher},\
  and\ \citenamefont {Chumak}}]{WangPRL2019}%
  \BibitemOpen
  \bibfield  {author} {\bibinfo {author} {\bibfnamefont {Q.}~\bibnamefont
  {Wang}}, \bibinfo {author} {\bibfnamefont {B.}~\bibnamefont {Heinz}},
  \bibinfo {author} {\bibfnamefont {R.}~\bibnamefont {Verba}}, \bibinfo
  {author} {\bibfnamefont {M.}~\bibnamefont {Kewenig}}, \bibinfo {author}
  {\bibfnamefont {P.}~\bibnamefont {Pirro}}, \bibinfo {author} {\bibfnamefont
  {M.}~\bibnamefont {Schneider}}, \bibinfo {author} {\bibfnamefont
  {T.}~\bibnamefont {Meyer}}, \bibinfo {author} {\bibfnamefont
  {B.}~\bibnamefont {L\"agel}}, \bibinfo {author} {\bibfnamefont
  {C.}~\bibnamefont {Dubs}}, \bibinfo {author} {\bibfnamefont {T.}~\bibnamefont
  {Br\"acher}}, \ and\ \bibinfo {author} {\bibfnamefont {A.~V.}\ \bibnamefont
  {Chumak}},\ }\href {https://link.aps.org/doi/10.1103/PhysRevLett.122.247202}
  {\bibfield  {journal} {\bibinfo  {journal} {Phys. Rev. Lett.}\ }\textbf
  {\bibinfo {volume} {122}},\ \bibinfo {pages} {247202} (\bibinfo {year}
  {2019})}\BibitemShut {NoStop}%
\bibitem [{\citenamefont {De~Wames}\ and\ \citenamefont
  {Wolfram}(1969)}]{deWamesAPS1969}%
  \BibitemOpen
  \bibfield  {author} {\bibinfo {author} {\bibfnamefont {R.~E.}\ \bibnamefont
  {De~Wames}}\ and\ \bibinfo {author} {\bibfnamefont {T.}~\bibnamefont
  {Wolfram}},\ }\href {https://doi.org/10.1063/1.1653006} {\bibfield  {journal}
  {\bibinfo  {journal} {Appl Phys Lett}\ }\textbf {\bibinfo {volume} {15}},\
  \bibinfo {pages} {297} (\bibinfo {year} {1969})}\BibitemShut {NoStop}%
\bibitem [{\citenamefont {De~Wames}\ and\ \citenamefont
  {Wolfram}(1970)}]{DeWamesJAP1970}%
  \BibitemOpen
  \bibfield  {author} {\bibinfo {author} {\bibfnamefont {R.~E.}\ \bibnamefont
  {De~Wames}}\ and\ \bibinfo {author} {\bibfnamefont {T.}~\bibnamefont
  {Wolfram}},\ }\href {https://doi.org/10.1063/1.1659049} {\bibfield  {journal}
  {\bibinfo  {journal} {J appl Phys}\ }\textbf {\bibinfo {volume} {41}},\
  \bibinfo {pages} {987} (\bibinfo {year} {1970})}\BibitemShut {NoStop}%
\bibitem [{\citenamefont {De~Wames}\ and\ \citenamefont
  {Wolfram}(1971)}]{deWamesPRL1971}%
  \BibitemOpen
  \bibfield  {author} {\bibinfo {author} {\bibfnamefont {R.~E.}\ \bibnamefont
  {De~Wames}}\ and\ \bibinfo {author} {\bibfnamefont {T.}~\bibnamefont
  {Wolfram}},\ }\href {https://link.aps.org/doi/10.1103/PhysRevLett.26.1445}
  {\bibfield  {journal} {\bibinfo  {journal} {Phys. Rev. Lett.}\ }\textbf
  {\bibinfo {volume} {26}},\ \bibinfo {pages} {1445} (\bibinfo {year}
  {1971})}\BibitemShut {NoStop}%
\bibitem [{\citenamefont {Wolfram}\ and\ \citenamefont
  {De~Wames}(1970)}]{WolframPRL1970}%
  \BibitemOpen
  \bibfield  {author} {\bibinfo {author} {\bibfnamefont {T.}~\bibnamefont
  {Wolfram}}\ and\ \bibinfo {author} {\bibfnamefont {R.~E.}\ \bibnamefont
  {De~Wames}},\ }\href {https://link.aps.org/doi/10.1103/PhysRevLett.24.1489}
  {\bibfield  {journal} {\bibinfo  {journal} {Phys. Rev. Lett.}\ }\textbf
  {\bibinfo {volume} {24}},\ \bibinfo {pages} {1489} (\bibinfo {year}
  {1970})}\BibitemShut {NoStop}%
\bibitem [{\citenamefont {Yu}\ \emph {et~al.}(2019)\citenamefont {Yu},
  \citenamefont {Liu}, \citenamefont {Yu}, \citenamefont {Blanter},\ and\
  \citenamefont {Bauer}}]{YuPRB2019}%
  \BibitemOpen
  \bibfield  {author} {\bibinfo {author} {\bibfnamefont {T.}~\bibnamefont
  {Yu}}, \bibinfo {author} {\bibfnamefont {C.}~\bibnamefont {Liu}}, \bibinfo
  {author} {\bibfnamefont {H.}~\bibnamefont {Yu}}, \bibinfo {author}
  {\bibfnamefont {Y.~M.}\ \bibnamefont {Blanter}}, \ and\ \bibinfo {author}
  {\bibfnamefont {G.~E.~W.}\ \bibnamefont {Bauer}},\ }\href
  {https://link.aps.org/doi/10.1103/PhysRevB.99.134424} {\bibfield  {journal}
  {\bibinfo  {journal} {Phys. Rev. B}\ }\textbf {\bibinfo {volume} {99}},\
  \bibinfo {pages} {134424} (\bibinfo {year} {2019})}\BibitemShut {NoStop}%
\bibitem [{\citenamefont {Kustura}\ \emph {et~al.}(pear)\citenamefont
  {Kustura}, \citenamefont {Romero-Isart},\ and\ \citenamefont
  {Gonzalez-Ballestero}}]{KusturaToAppear2020}%
  \BibitemOpen
  \bibfield  {author} {\bibinfo {author} {\bibfnamefont {K.}~\bibnamefont
  {Kustura}}, \bibinfo {author} {\bibfnamefont {O.}~\bibnamefont
  {Romero-Isart}}, \ and\ \bibinfo {author} {\bibfnamefont {C.}~\bibnamefont
  {Gonzalez-Ballestero}},\ }\href@noop {} {\enquote {\bibinfo {title} {Master
  equation description of driven spin baths},}\ } (\bibinfo {year} {2020 (To
  Appear)})\BibitemShut {NoStop}%
\bibitem [{\citenamefont {Liu}\ \emph {et~al.}(2007)\citenamefont {Liu},
  \citenamefont {Giesen}, \citenamefont {Zhu}, \citenamefont {Sydora},\ and\
  \citenamefont {Freeman}}]{LiuPRL2007}%
  \BibitemOpen
  \bibfield  {author} {\bibinfo {author} {\bibfnamefont {Z.}~\bibnamefont
  {Liu}}, \bibinfo {author} {\bibfnamefont {F.}~\bibnamefont {Giesen}},
  \bibinfo {author} {\bibfnamefont {X.}~\bibnamefont {Zhu}}, \bibinfo {author}
  {\bibfnamefont {R.~D.}\ \bibnamefont {Sydora}}, \ and\ \bibinfo {author}
  {\bibfnamefont {M.~R.}\ \bibnamefont {Freeman}},\ }\href
  {https://link.aps.org/doi/10.1103/PhysRevLett.98.087201} {\bibfield
  {journal} {\bibinfo  {journal} {Phys. Rev. Lett.}\ }\textbf {\bibinfo
  {volume} {98}},\ \bibinfo {pages} {087201} (\bibinfo {year}
  {2007})}\BibitemShut {NoStop}%
\bibitem [{\citenamefont {Sebastian}\ \emph {et~al.}(2015)\citenamefont
  {Sebastian}, \citenamefont {Schultheiss}, \citenamefont {Obry}, \citenamefont
  {Hillebrands},\ and\ \citenamefont {Schultheiss}}]{SebastianFrontiers2015}%
  \BibitemOpen
  \bibfield  {author} {\bibinfo {author} {\bibfnamefont {T.}~\bibnamefont
  {Sebastian}}, \bibinfo {author} {\bibfnamefont {K.}~\bibnamefont
  {Schultheiss}}, \bibinfo {author} {\bibfnamefont {B.}~\bibnamefont {Obry}},
  \bibinfo {author} {\bibfnamefont {B.}~\bibnamefont {Hillebrands}}, \ and\
  \bibinfo {author} {\bibfnamefont {H.}~\bibnamefont {Schultheiss}},\ }\href
  {https://www.frontiersin.org/article/10.3389/fphy.2015.00035} {\bibfield
  {journal} {\bibinfo  {journal} {Front Phys}\ }\textbf {\bibinfo {volume}
  {3}},\ \bibinfo {pages} {35} (\bibinfo {year} {2015})}\BibitemShut {NoStop}%
\bibitem [{\citenamefont {Acremann}\ \emph {et~al.}(2000)\citenamefont
  {Acremann}, \citenamefont {Back}, \citenamefont {Buess}, \citenamefont
  {Portmann}, \citenamefont {Vaterlaus}, \citenamefont {Pescia},\ and\
  \citenamefont {Melchior}}]{AcremannScience2000}%
  \BibitemOpen
  \bibfield  {author} {\bibinfo {author} {\bibfnamefont {Y.}~\bibnamefont
  {Acremann}}, \bibinfo {author} {\bibfnamefont {C.~H.}\ \bibnamefont {Back}},
  \bibinfo {author} {\bibfnamefont {M.}~\bibnamefont {Buess}}, \bibinfo
  {author} {\bibfnamefont {O.}~\bibnamefont {Portmann}}, \bibinfo {author}
  {\bibfnamefont {A.}~\bibnamefont {Vaterlaus}}, \bibinfo {author}
  {\bibfnamefont {D.}~\bibnamefont {Pescia}}, \ and\ \bibinfo {author}
  {\bibfnamefont {H.}~\bibnamefont {Melchior}},\ }\href
  {https://science.sciencemag.org/content/290/5491/492} {\bibfield  {journal}
  {\bibinfo  {journal} {Science}\ }\textbf {\bibinfo {volume} {290}},\ \bibinfo
  {pages} {492} (\bibinfo {year} {2000})}\BibitemShut {NoStop}%
\bibitem [{\citenamefont {Wolf}\ \emph {et~al.}(2015)\citenamefont {Wolf},
  \citenamefont {Neumann}, \citenamefont {Nakamura}, \citenamefont {Sumiya},
  \citenamefont {Ohshima}, \citenamefont {Isoya},\ and\ \citenamefont
  {Wrachtrup}}]{WolfPRX2015}%
  \BibitemOpen
  \bibfield  {author} {\bibinfo {author} {\bibfnamefont {T.}~\bibnamefont
  {Wolf}}, \bibinfo {author} {\bibfnamefont {P.}~\bibnamefont {Neumann}},
  \bibinfo {author} {\bibfnamefont {K.}~\bibnamefont {Nakamura}}, \bibinfo
  {author} {\bibfnamefont {H.}~\bibnamefont {Sumiya}}, \bibinfo {author}
  {\bibfnamefont {T.}~\bibnamefont {Ohshima}}, \bibinfo {author} {\bibfnamefont
  {J.}~\bibnamefont {Isoya}}, \ and\ \bibinfo {author} {\bibfnamefont
  {J.}~\bibnamefont {Wrachtrup}},\ }\href
  {https://link.aps.org/doi/10.1103/PhysRevX.5.041001} {\bibfield  {journal}
  {\bibinfo  {journal} {Phys. Rev. X}\ }\textbf {\bibinfo {volume} {5}},\
  \bibinfo {pages} {041001} (\bibinfo {year} {2015})}\BibitemShut {NoStop}%
\bibitem [{\citenamefont {Wrachtrup}\ and\ \citenamefont
  {Finkler}(2016)}]{WrachtrupJMR2016}%
  \BibitemOpen
  \bibfield  {author} {\bibinfo {author} {\bibfnamefont {J.}~\bibnamefont
  {Wrachtrup}}\ and\ \bibinfo {author} {\bibfnamefont {A.}~\bibnamefont
  {Finkler}},\ }\href
  {http://www.sciencedirect.com/science/article/pii/S1090780716300933}
  {\bibfield  {journal} {\bibinfo  {journal} {J Magn Reson}\ }\textbf {\bibinfo
  {volume} {269}},\ \bibinfo {pages} {225 } (\bibinfo {year}
  {2016})}\BibitemShut {NoStop}%
\bibitem [{\citenamefont {Wasilewski}\ \emph {et~al.}(2010)\citenamefont
  {Wasilewski}, \citenamefont {Jensen}, \citenamefont {Krauter}, \citenamefont
  {Renema}, \citenamefont {Balabas},\ and\ \citenamefont
  {Polzik}}]{WasilewskiPRL2010}%
  \BibitemOpen
  \bibfield  {author} {\bibinfo {author} {\bibfnamefont {W.}~\bibnamefont
  {Wasilewski}}, \bibinfo {author} {\bibfnamefont {K.}~\bibnamefont {Jensen}},
  \bibinfo {author} {\bibfnamefont {H.}~\bibnamefont {Krauter}}, \bibinfo
  {author} {\bibfnamefont {J.~J.}\ \bibnamefont {Renema}}, \bibinfo {author}
  {\bibfnamefont {M.~V.}\ \bibnamefont {Balabas}}, \ and\ \bibinfo {author}
  {\bibfnamefont {E.~S.}\ \bibnamefont {Polzik}},\ }\href
  {https://link.aps.org/doi/10.1103/PhysRevLett.104.133601} {\bibfield
  {journal} {\bibinfo  {journal} {Phys. Rev. Lett.}\ }\textbf {\bibinfo
  {volume} {104}},\ \bibinfo {pages} {133601} (\bibinfo {year}
  {2010})}\BibitemShut {NoStop}%
\bibitem [{\citenamefont {Ledbetter}\ \emph {et~al.}(2008)\citenamefont
  {Ledbetter}, \citenamefont {Savukov}, \citenamefont {Budker}, \citenamefont
  {Shah}, \citenamefont {Knappe}, \citenamefont {Kitching}, \citenamefont
  {Michalak}, \citenamefont {Xu},\ and\ \citenamefont
  {Pines}}]{LedbetterPNAS2008}%
  \BibitemOpen
  \bibfield  {author} {\bibinfo {author} {\bibfnamefont {M.~P.}\ \bibnamefont
  {Ledbetter}}, \bibinfo {author} {\bibfnamefont {I.~M.}\ \bibnamefont
  {Savukov}}, \bibinfo {author} {\bibfnamefont {D.}~\bibnamefont {Budker}},
  \bibinfo {author} {\bibfnamefont {V.}~\bibnamefont {Shah}}, \bibinfo {author}
  {\bibfnamefont {S.}~\bibnamefont {Knappe}}, \bibinfo {author} {\bibfnamefont
  {J.}~\bibnamefont {Kitching}}, \bibinfo {author} {\bibfnamefont {D.~J.}\
  \bibnamefont {Michalak}}, \bibinfo {author} {\bibfnamefont {S.}~\bibnamefont
  {Xu}}, \ and\ \bibinfo {author} {\bibfnamefont {A.}~\bibnamefont {Pines}},\
  }\href {https://www.pnas.org/content/105/7/2286} {\bibfield  {journal}
  {\bibinfo  {journal} {Proc Natl Acad Sci U.S.A.}\ }\textbf {\bibinfo {volume}
  {105}},\ \bibinfo {pages} {2286} (\bibinfo {year} {2008})}\BibitemShut
  {NoStop}%
\bibitem [{\citenamefont {Montoya-Castillo}\ \emph {et~al.}(2015)\citenamefont
  {Montoya-Castillo}, \citenamefont {Berkelbach},\ and\ \citenamefont
  {Reichman}}]{MontoyaCastilloJchemPhys2015}%
  \BibitemOpen
  \bibfield  {author} {\bibinfo {author} {\bibfnamefont {A.}~\bibnamefont
  {Montoya-Castillo}}, \bibinfo {author} {\bibfnamefont {T.~C.}\ \bibnamefont
  {Berkelbach}}, \ and\ \bibinfo {author} {\bibfnamefont {D.~R.}\ \bibnamefont
  {Reichman}},\ }\href {\doibase 10.1063/1.4935443} {\bibfield  {journal}
  {\bibinfo  {journal} {J Chem Phys}\ }\textbf {\bibinfo {volume} {143}},\
  \bibinfo {pages} {194108} (\bibinfo {year} {2015})}\BibitemShut {NoStop}%
\bibitem [{\citenamefont {Campisi}\ \emph {et~al.}(2012)\citenamefont
  {Campisi}, \citenamefont {Denisov},\ and\ \citenamefont
  {H\"anggi}}]{CampisiPRA2012}%
  \BibitemOpen
  \bibfield  {author} {\bibinfo {author} {\bibfnamefont {M.}~\bibnamefont
  {Campisi}}, \bibinfo {author} {\bibfnamefont {S.}~\bibnamefont {Denisov}}, \
  and\ \bibinfo {author} {\bibfnamefont {P.}~\bibnamefont {H\"anggi}},\ }\href
  {https://link.aps.org/doi/10.1103/PhysRevA.86.032114} {\bibfield  {journal}
  {\bibinfo  {journal} {Phys. Rev. A}\ }\textbf {\bibinfo {volume} {86}},\
  \bibinfo {pages} {032114} (\bibinfo {year} {2012})}\BibitemShut {NoStop}%
\bibitem [{\citenamefont {Liu}\ and\ \citenamefont
  {Houck}(2017)}]{LiuNatPhys2017}%
  \BibitemOpen
  \bibfield  {author} {\bibinfo {author} {\bibfnamefont {Y.}~\bibnamefont
  {Liu}}\ and\ \bibinfo {author} {\bibfnamefont {A.~A.}\ \bibnamefont
  {Houck}},\ }\href {https://doi.org/10.1038/nphys3834} {\bibfield  {journal}
  {\bibinfo  {journal} {Nat Physics}\ }\textbf {\bibinfo {volume} {13}},\
  \bibinfo {pages} {48} (\bibinfo {year} {2017})}\BibitemShut {NoStop}%
\bibitem [{\citenamefont {Novotny}\ and\ \citenamefont
  {Hecht}(2006)}]{novotny2006principles}%
  \BibitemOpen
  \bibfield  {author} {\bibinfo {author} {\bibfnamefont {L.}~\bibnamefont
  {Novotny}}\ and\ \bibinfo {author} {\bibfnamefont {B.}~\bibnamefont
  {Hecht}},\ }\href@noop {} {\emph {\bibinfo {title} {Principles of
  Nano-Optics}}}\ (\bibinfo  {publisher} {Cambridge University Press},\
  \bibinfo {year} {2006})\BibitemShut {NoStop}%
\bibitem [{\citenamefont {Marocico}\ and\ \citenamefont
  {Knoester}(2011)}]{MarocicoPRA2011}%
  \BibitemOpen
  \bibfield  {author} {\bibinfo {author} {\bibfnamefont {C.~A.}\ \bibnamefont
  {Marocico}}\ and\ \bibinfo {author} {\bibfnamefont {J.}~\bibnamefont
  {Knoester}},\ }\href {https://link.aps.org/doi/10.1103/PhysRevA.84.053824}
  {\bibfield  {journal} {\bibinfo  {journal} {Phys. Rev. A}\ }\textbf {\bibinfo
  {volume} {84}},\ \bibinfo {pages} {053824} (\bibinfo {year}
  {2011})}\BibitemShut {NoStop}%
\bibitem [{\citenamefont {Huidobro}\ \emph {et~al.}(2012)\citenamefont
  {Huidobro}, \citenamefont {Nikitin}, \citenamefont {Gonz\'alez-Ballestero},
  \citenamefont {Mart\'{\i}n-Moreno},\ and\ \citenamefont
  {Garc\'{\i}a-Vidal}}]{HuidobroPRB2012}%
  \BibitemOpen
  \bibfield  {author} {\bibinfo {author} {\bibfnamefont {P.~A.}\ \bibnamefont
  {Huidobro}}, \bibinfo {author} {\bibfnamefont {A.~Y.}\ \bibnamefont
  {Nikitin}}, \bibinfo {author} {\bibfnamefont {C.}~\bibnamefont
  {Gonz\'alez-Ballestero}}, \bibinfo {author} {\bibfnamefont {L.}~\bibnamefont
  {Mart\'{\i}n-Moreno}}, \ and\ \bibinfo {author} {\bibfnamefont {F.~J.}\
  \bibnamefont {Garc\'{\i}a-Vidal}},\ }\href
  {https://link.aps.org/doi/10.1103/PhysRevB.85.155438} {\bibfield  {journal}
  {\bibinfo  {journal} {Phys. Rev. B}\ }\textbf {\bibinfo {volume} {85}},\
  \bibinfo {pages} {155438} (\bibinfo {year} {2012})}\BibitemShut {NoStop}%
\bibitem [{\citenamefont {Lodahl}\ \emph {et~al.}(2017)\citenamefont {Lodahl},
  \citenamefont {Mahmoodian}, \citenamefont {Stobbe}, \citenamefont
  {Rauschenbeutel}, \citenamefont {Schneeweiss}, \citenamefont {Volz},
  \citenamefont {Pichler},\ and\ \citenamefont {Zoller}}]{LodahlNature2017}%
  \BibitemOpen
  \bibfield  {author} {\bibinfo {author} {\bibfnamefont {P.}~\bibnamefont
  {Lodahl}}, \bibinfo {author} {\bibfnamefont {S.}~\bibnamefont {Mahmoodian}},
  \bibinfo {author} {\bibfnamefont {S.}~\bibnamefont {Stobbe}}, \bibinfo
  {author} {\bibfnamefont {A.}~\bibnamefont {Rauschenbeutel}}, \bibinfo
  {author} {\bibfnamefont {P.}~\bibnamefont {Schneeweiss}}, \bibinfo {author}
  {\bibfnamefont {J.}~\bibnamefont {Volz}}, \bibinfo {author} {\bibfnamefont
  {H.}~\bibnamefont {Pichler}}, \ and\ \bibinfo {author} {\bibfnamefont
  {P.}~\bibnamefont {Zoller}},\ }\href {https://doi.org/10.1038/nature21037}
  {\bibfield  {journal} {\bibinfo  {journal} {Nature}\ }\textbf {\bibinfo
  {volume} {541}},\ \bibinfo {pages} {473} (\bibinfo {year}
  {2017})}\BibitemShut {NoStop}%
\bibitem [{\citenamefont {Gonz\'alez-Tudela}\ \emph {et~al.}(2013)\citenamefont
  {Gonz\'alez-Tudela}, \citenamefont {Huidobro}, \citenamefont
  {Mart\'{\i}n-Moreno}, \citenamefont {Tejedor},\ and\ \citenamefont
  {Garc\'{\i}a-Vidal}}]{GonzalezTudelaPRL2013}%
  \BibitemOpen
  \bibfield  {author} {\bibinfo {author} {\bibfnamefont {A.}~\bibnamefont
  {Gonz\'alez-Tudela}}, \bibinfo {author} {\bibfnamefont {P.~A.}\ \bibnamefont
  {Huidobro}}, \bibinfo {author} {\bibfnamefont {L.}~\bibnamefont
  {Mart\'{\i}n-Moreno}}, \bibinfo {author} {\bibfnamefont {C.}~\bibnamefont
  {Tejedor}}, \ and\ \bibinfo {author} {\bibfnamefont {F.~J.}\ \bibnamefont
  {Garc\'{\i}a-Vidal}},\ }\href {\doibase 10.1103/PhysRevLett.110.126801}
  {\bibfield  {journal} {\bibinfo  {journal} {Phys. Rev. Lett.}\ }\textbf
  {\bibinfo {volume} {110}},\ \bibinfo {pages} {126801} (\bibinfo {year}
  {2013})}\BibitemShut {NoStop}%
\bibitem [{\citenamefont {Barakat}(1988)}]{Barakat1988}%
  \BibitemOpen
  \bibfield  {author} {\bibinfo {author} {\bibfnamefont {R.}~\bibnamefont
  {Barakat}},\ }\href {\doibase 10.1121/1.396046} {\bibfield  {journal}
  {\bibinfo  {journal} {J Acoust Soc Am}\ }\textbf {\bibinfo {volume} {83}},\
  \bibinfo {pages} {1014} (\bibinfo {year} {1988})}\BibitemShut {NoStop}%
\bibitem [{\citenamefont {Andersson}\ \emph {et~al.}(2020)\citenamefont
  {Andersson}, \citenamefont {Bilobran}, \citenamefont {Scigliuzzo},
  \citenamefont {de~Lima}, \citenamefont {Cole},\ and\ \citenamefont
  {Delsing}}]{AnderssonArxiv2020}%
  \BibitemOpen
  \bibfield  {author} {\bibinfo {author} {\bibfnamefont {G.}~\bibnamefont
  {Andersson}}, \bibinfo {author} {\bibfnamefont {A.~L.~O.}\ \bibnamefont
  {Bilobran}}, \bibinfo {author} {\bibfnamefont {M.}~\bibnamefont
  {Scigliuzzo}}, \bibinfo {author} {\bibfnamefont {M.~M.}\ \bibnamefont
  {de~Lima}}, \bibinfo {author} {\bibfnamefont {J.~H.}\ \bibnamefont {Cole}}, \
  and\ \bibinfo {author} {\bibfnamefont {P.}~\bibnamefont {Delsing}},\ }\href
  {https://arxiv.org/abs/2002.09389} {\bibfield  {journal} {\bibinfo  {journal}
  {arXiv preprint arXiv:2002.09389}\ } (\bibinfo {year} {2020})}\BibitemShut
  {NoStop}%
\end{thebibliography}%

\end{document}